\documentclass{aa}

\usepackage[varg]{txfonts} 
\usepackage{color}

\usepackage[breaklinks,bookmarks,bookmarksnumbered, colorlinks=true, linkcolor=blue, citecolor=blue]{hyperref}
\usepackage{mathabx}
\usepackage{amsmath}
\usepackage{natbib}

\title{Physical properties of dusty protoplanetary disks in Lupus: evidence for viscous evolution?} 

\author{
M.~Tazzari\inst{\ref{IoA}, \ref{ESO}, \ref{ExCluster}}
L.~Testi\inst{\ref{ESO}, \ref{ExCluster}, \ref{Arcetri}},
A. Natta\inst{ \ref{Arcetri}, \ref{DIAS}},
M.~Ansdell\inst{\ref{IfAHawaii}},
J.~Carpenter\inst{\ref{JointALMA}}, 
G.~Guidi\inst{\ref{Arcetri}, \ref{IfAHawaii}},
M.~Hogerheijde\inst{\ref{Leiden}}, 
C.~F.~Manara\inst{\ref{ESTEC}},
A.~Miotello\inst{\ref{Leiden}},  
N.~van~der~Marel\inst{\ref{IfAHawaii}}, 
E.~F.~van~Dishoeck\inst{\ref{Leiden}, \ref{MPE}},
J.~P.~Williams\inst{\ref{IfAHawaii}}
}

\institute{
Institute of Astronomy, University of Cambridge, Madingley Road, CB3 0HA,  Cambridge, UK\\ \email{mtazzari@ast.cam.ac.uk} \label{IoA}
\and
European Southern Observatory, Karl-Schwarzschild-Str. 2, D-85748 Garching, Germany; \email{mtazzari@eso.org} \label{ESO}
\and
Excellence Cluster Universe, Boltzmannstr. 2, D-85748 Garching, Germany \label{ExCluster}
\and
INAF-Osservatorio Astrofisico di Arcetri, Largo E. Fermi 5, I-50125 Firenze, Italy \label{Arcetri}
\and
School of Cosmic Physics, Dublin Institute for Advanced Studies, 31 Fitzwilliams Place, 2 Dublin, Ireland \label{DIAS}
\and
Institute for Astronomy, University of Hawaii at Manoa, 2680 Woodlawn dr., Honolulu, HI, 96822, USA \label{IfAHawaii}
\and
Joint ALMA Observatory, Av. Alonso de Córdova 3107, Vitacura, Santiago, Chile \label{JointALMA}
\and
Leiden Observatory, Leiden University, Niels Bohrweg 2, NL-2333 CA Leiden, The Netherlands \label{Leiden}
\and
Scientific Support Office, Directorate of Science, European Space Research and Technology Centre (ESA/ESTEC), Keplerlaan 1, 2201AZ Noordwijk, The Netherlands \label{ESTEC}
\and
Max-Plank-Institut für Extraterrestrische Physik, Giessenbachstraße 1, D-85748 Garching, Germany \label{MPE}
}

\authorrunning{Tazzari, M. et al.}

\definecolor{linkcolor}{rgb}{0,0,1.}
\newcommand{\chapter}{Article}

\renewcommand{\deg}{$^\circ$}

\newcommand{\Msun}{\,M_{\odot}}

\renewcommand{\u}[1]{\,\textrm{#1}}
\newcommand{\tbref}[1]{Table~\ref{#1}}
\newcommand{\figref}[1]{Figure~\ref{#1}}

\newcommand{\Lstar}{L_{\star}}
\newcommand{\Mearth}{M_\oplus}
\newcommand{\Lsun}{L_\odot}
\newcommand{\simless}{\mathbin{\lower 3pt\hbox
      {$\rlap{\raise 5pt\hbox{$\char'074$}}\mathchar"7218$}}}
\newcommand{\simgreat}{\mathbin{\lower 3pt\hbox
     {$\rlap{\raise 5pt\hbox{$\char'076$}}\mathchar"7218$}}}

\newcommand{\intflux}{F_{\mathrm{cont}}}

\graphicspath{{./}}

\abstract
{The formation of planets strongly depends on the total amount as well as on the spatial distribution of solids in protoplanetary disks. Thanks to the improvements in resolution and sensitivity provided by ALMA, measurements of the surface density of mm-sized grains are now possible on large samples of disks. Such measurements provide statistical constraints that can be used to inform our understanding of the initial conditions of planet formation.}
{We analyze spatially resolved observations of 36 protoplanetary disks in the Lupus star forming complex from our ALMA survey at 890$\mu$m, aiming to determine physical properties such as the dust surface density, the disk mass and size and to provide a constraint on the temperature profile.}
{We fit the observations directly in the uv-plane using a two-layer disk model that computes the $890\u{$\mu$m}$ emission by solving the energy balance at each disk radius.}
{For 22 out of 36 protoplanetary disks we derive robust estimates of their physical properties. The sample covers stellar masses between $\sim$0.1 and $\sim 2 \Msun$, and we find no trend between the average disk temperatures and the stellar parameters. We find, instead, a correlation between the integrated sub-mm flux (a proxy for the disk mass) and the exponential cut-off radii (a proxy of the disk size) of the Lupus disks. Comparing these results with observations at similar angular resolution of Taurus-Auriga/Ophiuchus disks found in literature and scaling them to the same distance, we observe that the Lupus disks are generally fainter and larger at a high level of statistical significance. Considering the 1-2\,Myr age difference between these regions, it is possible to tentatively explain the offset in the disk mass/disk size relation with viscous spreading, however with the current measurements other mechanisms cannot be ruled out.}
{}

\begin{document}

\maketitle

\section{Introduction}
Planets form in the circumstellar disks orbiting young pre-main-sequence stars. In the last decade the number of known exoplanetary systems has increased exponentially, uncovering a large diversity in their architectures as well as in the physical properties - mass, size and average density - of the single planets \citep{Lissauer:2014rr,Winn:2015aa}.

The number and location of planets that can form around a star strongly depend not only on the total mass of the parent protoplanetary disk but also on how it is spatially distributed \citep{Mordasini:2012aa,Alibert:2011aa}. 

In the core accretion theory \citep{safronov:1972}, planet formation starts with the growth of the sub-micron interstellar dust grains that initially populate the disk to mm/cm-sized pebbles via pair-wise collisions and subsequently to km-sized planetesimals via gravitational interaction \citep[][as reviews]{Testi:2014kx,Birnstiel:2016ve}. The population of planetesimals is then  assembled into a rocky core that rapidly accrete gaseous material. The efficiency of the formation of planetesimals is affected by the local conditions within the disk and is tightly linked to the available mass in solids \citep[][and references therein]{Chiang:2010fk}. Characterizing how solids are spatially distributed in protoplanetary disks thus provides an excellent probe of the initial conditions of the planet formation process.

Optical and near- to mid-infrared observations have been used to study the dust emission in protoplanetary disks \citep{Bouwman:2001qv,van-Boekel:2003nr, Juhasz:2010fj, Miotello:2012fv} but they effectively trace only the emission of the micron-sized grains in the upper layers of the disk structure. In these regions the dust emission is largely optically thick and therefore provides us with a measurement of the disk surface temperature rather than of the total dust mass. Measurement of the disk mass in solids can be obtained from millimeter and sub-millimeter continuum observations, which are sensitive to the thermal emission of mm-sized grains located in the disk midplane. At these wavelengths the continuum emission
is optically thin and it can thus be used - given an assumption on the dust opacity and temperature - to infer the dust mass \citep{Beckwith:1990qf,Beckwith:1991pd}. 

In the last two decades, the development of sub-mm/mm interferometers provided us with the first spatially resolved images of protoplanetary disks which have typical angular sizes of 1- 2'' at the distance of the nearby star forming regions (150-200\u{pc}).  Using sub-mm/mm resolved observations and a proper modeling of temperature and opacity it is possible to infer the radial profile of the dust surface density \cite[][and references therein]{Williams:2011jk}. The radial profile generally used is a power law  $\Sigma(R)=\Sigma_0(R/R_0)^{-\gamma}$, which can be either truncated at a radius $R_\mathrm{out}$ or - following the arguments of accretion theory \citep{1974MNRAS.168..603L,Hartmann:1998qy} - exponentially tapered with an e-folding radius $R_c$. The temperature profile is usually parametrized as a power-law or in some cases derived self-consistently with a simplified physical model for irradiated disks \citep{Isella:2009qy}. Simultaneous fit of photometric infrared fluxes is also used to provide additional constraint on the vertical structure of the disk \citep{Andrews:2009zr}.

The first sub-arcsecond constraints on the dust surface density profiles obtained for disks in the Taurus-Auriga \citep{Isella:2009qy,2011A&amp;A...529A.105G} and Ophiuchus \citep{Andrews:2009zr,Andrews:2010fk} clouds relied on interferometric observations at 870\,$\mu$m and 1.3\,mm and indicated that disks are described well from an exponentially tapered profile and generally appear to have flat interiors ($\gamma\sim 1$) and sharp outer edges. 
These studies provided the first constraints on the dust distribution on scales of 30-40\u{au}, but are limited in terms of sample size (10-15 objects) and result difficult to compare due to the different modeling techniques. The determination of a profile able to describe the overall dust surface density in protoplanetary disks
still needs an homogeneous analysis of larger samples of disks and is now actively investigated. 

In the last years, the advent of the Atacama large millimeter and sub-millimeter array (ALMA) marked the beginning of a new era for protoplanetary disk studies. On the one hand, thanks to its unprecedented angular resolution, it is revealing structures in disks at $\sim$\,au scales that provide excellent testbeds for our understanding of the physical processes involved in planet formation \citep{2041-8205-808-1-L3,Perez:2016aa,Andrews:2016lr, Isella:2016ww}. On the other hand, its increased collecting area allows us to target large samples of disks at high resolution and sensitivity in a modest amount of time. The recent surveys at 890\,$\mu$m in the Lupus \citep{Ansdell:2016qf}, Upper Scorpius \citep{Barenfeld:2016lr} and Chamaeleon I \citep{Pascucci:2016fk} star forming regions targeted hundreds of disks with comparable sensitivity and resolution (0.3-0.5'', 15-25\u{au} in radius). These observational data sets constitute an exceptionally large and homogeneous sample that can be used to infer fundamental physical properties of protoplanetary disks. 

In this work we focus on the determination of the physical properties of the disks in the Lupus star forming region, analyzing the ALMA observations presented in \citet{Ansdell:2016qf}, which is a near complete (96\% completeness) survey of Class~II disks in the Lupus I-IV clouds. By modeling the continuum emission with a physical disk model for passively irradiated disks \citep{Chiang:2010fk,2001ApJ...560..957D} we determine the surface density profile of more than 20 disks and we put constraints on their midplane temperature. We perform the fit of the interferometric visibilities directly in the uv-plane and we notice that the surface brightness profiles of most disks can be described remarkably well with a smooth exponentially tapered surface density profile, with the exception of two disks (IM~Lup and Sz~98) that exhibit significant ring-like excesses (van Terwisga et al., in prep.).

By comparing the inferred properties of disks in star forming regions of different mean age, it is possible to trace their time evolution. In particular, we are interested at the time evolution of the distribution of solids, which in first instance can be characterized in simple terms of radial extent and total mass. In this work we use the inferred surface density cut-off radius $R_c$ and the 890\,$\mu$m integrated flux as proxies for these two quantities. By comparing the mass-size correlation that we obtain for the Lupus disks with that obtained from literature observations of disks in the Taurus-Auriga/Ophiuchus clouds, we find a significant difference that might be tentatively explained with viscous disk spreading. The present measurements do not allow us to rule out other mechanisms (e.g., radial drift) that could possibly explain such a discrepancy, and a complete survey of the Taurus-Auriga/Ophiuchus disks sample (at present probably observed in its high-mass end) is needed. 

The paper is organized as follows. In Section~\ref{sec:sample} we present the sample selection, in Section~\ref{sec:modeling} the modeling details and in Section~\ref{sec:results} the results, with a comparison with previous results by \citet{Ansdell:2016qf}. The results of the analysis are discussed in Section~\ref{sec:discussion} and in Section~\ref{sec:conclusions} we draw the conclusions. In Appendix~\ref{app:fits} we report all the fit results for all the disks and in Appendix~\ref{app:kelly} the detailed results of Bayesian linear regressions.

\section{Observations and sample selection}
\label{sec:sample}

In this paper we analyze a sub-sample of the ALMA Cycle 2 continuum observations at 890\u{$\mu$m} of the Lupus disks presented by \citet{Ansdell:2016qf} (id: ADS/JAO.ALMA\#2013.1.00220.S). 

The Lupus disks in the \citet{Ansdell:2016qf} sample have been observed with the an array configuration covering baselines between 21.4 and 783.5m, corresponding to an average beam size (for the continuum) of 0.34''$\times$0.28'', i.e. $\sim 50\times$ 40\u{au} at 150\u{pc}. The bandwidth-weighted average frequency of the continuum observations is 335.8\u{GHz} (890\u{$\mu$m}) and the average rms between 0.25 and 0.41 mJy\,beam$^{-1}$. The flux calibration error of these observations is estimated to be 10\%. For additional details on the observational setup and the data reduction we refer to \citet{Ansdell:2016qf}.

The observations in \citet{Ansdell:2016qf} targeted a nearly complete (96\% completeness) sample of the sources with Class II or Flat IR spectra in the Lupus star forming complex (I to IV clouds), for a total of 89 sources out of 93, detecting 61 of them in the continuum with $>3\sigma$ significance. 
The sub-mm observations are complemented by the VLT/X-Shooter spectroscopic survey by \citet{Alcala:2014uq, Alcala:2016ys} which derive fundamental stellar parameters for all the Class~II sources of the region. 

The sub-sample considered in this study has been selected from the total sample of 61 detected sources \citep{Ansdell:2016qf} by excluding: the edge-on disks (J16070854-3914075, Sz~118), the disks with clearly resolved gaps or holes (J16083070-3828268, RY~Lup, Sz~111), the resolved and unresolved binaries (V856~Sco, Sz~74, Sz~123A), the sources for which we have no information on the stellar parameters (J15450634-3417378, J16011549-4152351) and two sources that exhibit an elongated irregular continuum emission (J16090141-3925119, J16070384-3911113, which could potentially be  partially resolved binaries, although there is no spectroscopic confirmation, see Fig.~2 in \citealt{Ansdell:2016qf}). We also exclude 14 sources whose Band~7 observations exhibit a very low signal-to-noise ratio at large uv-distances which makes the deprojected visibility profile compatible with being unresolved: these sources result to have an integrated flux $\intflux<4\u{mJy}$ (Sz~106, J16002612-4153553, J16000060-4221567, J16085529-3848481, J16084940-3905393, V1192~Sco, Sz~104,  Sz 112, J16073773-3921388, J16080017-3902595, J16085373-3914367, J16075475-3915446, J16092697-3836269, J16134410-3736462). This results in a sub-sample of 35 sources. In addition to these sources, we analyze ALMA 890\,$\mu$m observations of Sz~82 (IM~Lup) that have been taken by another observing program (PI: Cleeves, I.; id: ADS/JAO.ALMA\#2013.1.00226.S). with comparable resolution and rms noise. After calibrating the IM~Lup data set with the script provided by the ALMA Observatory, we performed self calibration with two rounds of phase calibration and one round of amplitude and phase.

\begin{figure}
\centering
\resizebox{0.9\hsize}{!}{\includegraphics{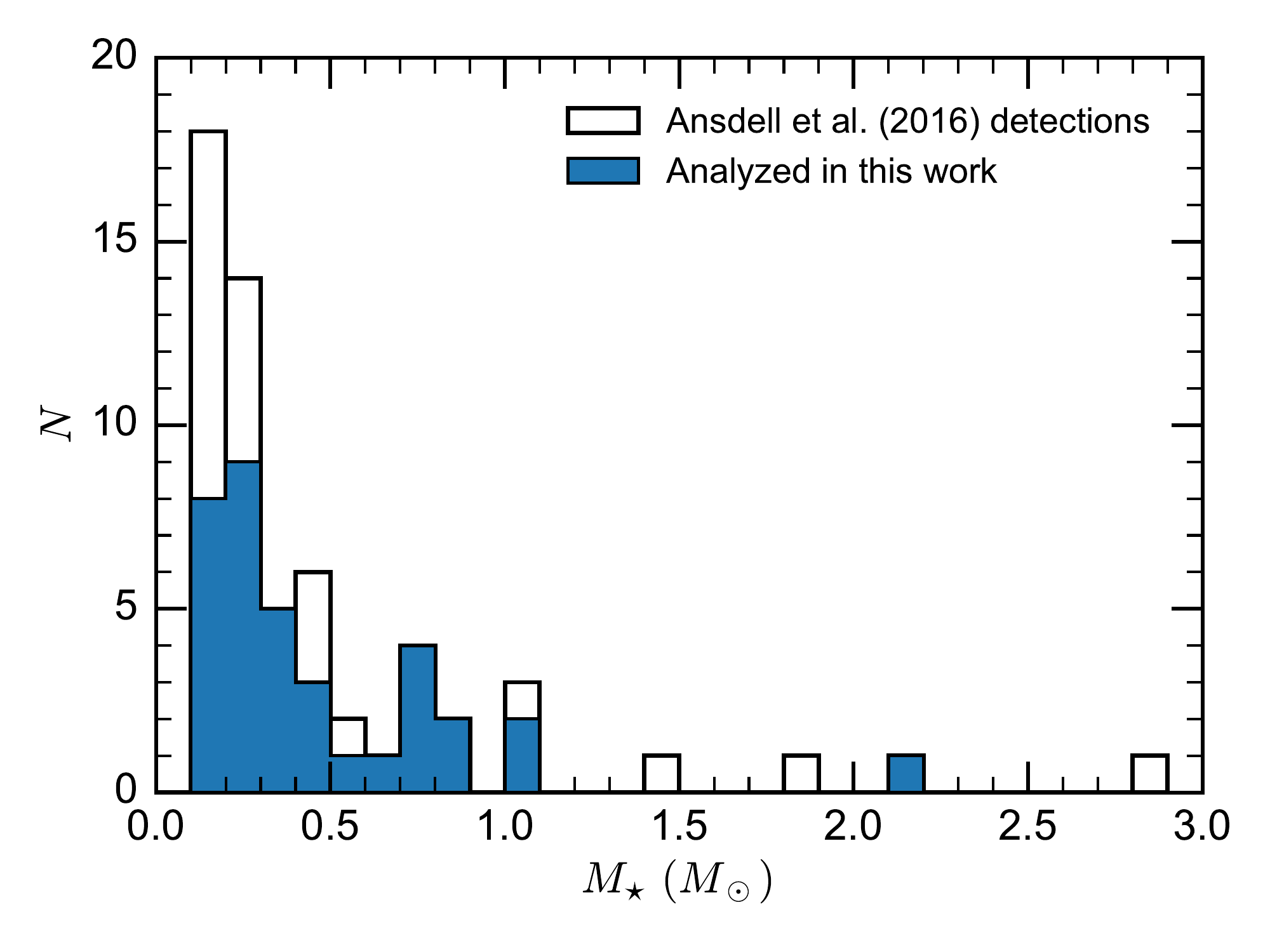}}\\
\resizebox{0.9\hsize}{!}{\includegraphics{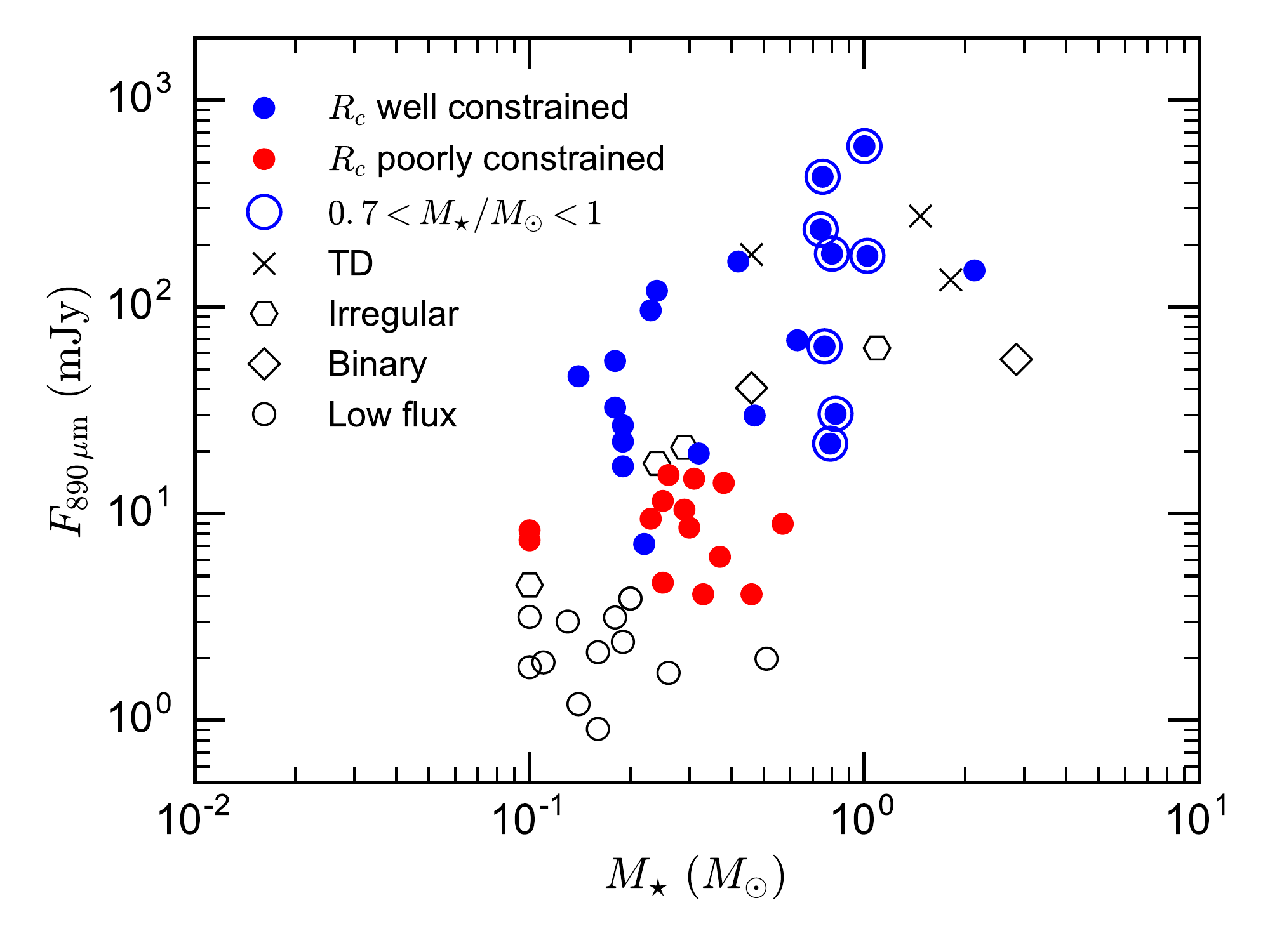}}
\caption{\textit{(top)}: distribution of stellar masses for the full Lupus sample from \citet{Ansdell:2016qf} (white bars) and for the sub-sample analyzed in this work (blue bars). \textit{(bottom)}:  observed integrated continuum flux at 890\u{$\mu$m} \citep{Ansdell:2016qf} as a function of stellar mass. Blue dots represent sources whose analysis is presented in this work. Red dots represent sources that appear unresolved and have been discarded in the rest of the analysis due low signal-to-noise ratio. In this and subsequent plots, the circled dots highlight sources in the $0.7-1\,\Msun$ mass bin. We also report transition disks (``x'' symbols), sources with irregular shape (hexagons),  binaries  (diamonds), and the sources with integrated flux $\intflux<4\u{mJy}$ that we excluded from our analysis (empty circles). 
}
\label{fig:sample.properties}
\end{figure}

\tbref{tb:sources} summarizes the properties of the 36 sources analyzed in this work, including their stellar properties and integrated 890$\mu$m flux and rms.
In \figref{fig:sample.properties} we highlight the properties of the sub-sample in comparison to the complete sample from \citet{Ansdell:2016qf}. In the top panel, we show the distribution of stellar masses, ranging between $0.1\Msun$ and $3\Msun$: it is noteworthy that the sub-sample selected for this analysis includes all the stars in the $0.7-1\,\Msun$ mass bin except J16090141-3925119 that has an irregular shape: in this and in future plots the sources in this mass bin are identified with circled dots. In the bottom panel, we present the integrated continuum flux at $890\u{$\mu$m}$ as a function of stellar mass, differentiating the sources whose analysis is presented in this paper (blue symbols) from the sources that we excluded (black symbols) according to the criteria listed above. The 14 red dots represent disks that were initially part of the 36 analyzed objects, but resulted in having a signal-to-noise ratio at long baselines too small to allow for a robust estimate of their disk structure, compatible with being unresolved. An exception among the red dots is J16102955-3922144 (marked in blue) on whose structure we have been able to obtain a marginal constraint.

\begin{table*}
\caption[Source properties]{Source properties.}
\centering
\small
\begin{tabular}{lrrrrrrrrr}
\hline
\hline
Name &	RA (J2000)  & Dec (J2000)  & $F_{890\mu\mathrm{m}}$&	rms &	$d$                 &	$M_{\star}$             &	$L_{\star}$             &	SpT                 &	$T_{\mathrm{eff}}$  \\ 
& & & (mJy) & (mJy/beam) & (pc) & ($\Msun$) & ($\Lsun$) &  & (K)\\
  \hline
Sz 65&	15:39:27.75&	-34:46:17.56&	     64.49&	      0.32&	       150&	      0.76&	     0.832&	K7        &	      4060\\ 
J15450887-3417333&	15:45:08.85&	-34:17:33.81&	     46.27&	       0.5&	       150&	      0.14&	     0.058&	M5.5      &	      3060\\ 
Sz 68&	15:45:12.84&	-34:17:30.98&	    150.37&	      0.46&	       150&	      2.13&	     5.129&	K2        &	      4900\\ 
Sz 69&	15:45:17.39&	-34:18:28.66&	     16.96&	      0.28&	       150&	      0.19&	     0.088&	M4.5      &	      3197\\ 
Sz 71&	15:46:44.71&	-34:30:36.05&	    166.04&	      0.63&	       150&	      0.42&	     0.309&	M1.5      &	      3632\\ 
Sz 73&	15:47:56.92&	-35:14:35.15&	     30.43&	      0.55&	       150&	      0.82&	     0.419&	K7        &	      4060\\ 
IM Lup&	15:56:09.18&	-37:56:06.12&	     600.0&	      0.13&	       150&	       1.0&	      1.65&	M0        &	      3850\\ 
Sz 83&	15:56:42.29&	-37:49:15.82&	     426.9&	      0.72&	       150&	      0.75&	     1.313&	K7        &	      4060\\ 
Sz 84&	15:58:02.50&	-37:36:03.08&	     32.64&	       0.4&	       150&	      0.18&	     0.122&	M5.0      &	      3125\\ 
Sz 129&	15:59:16.45&	-41:57:10.66&	    181.12&	      0.52&	       150&	       0.8&	     0.372&	K7        &	      4060\\ 
J16000236-4222145&	16:00:02.34&	-42:22:14.99&	    119.85&	      0.63&	       150&	      0.24&	     0.148&	M4        &	      3270\\ 
MY Lup&	16:00:44.50&	-41:55:31.27&	    176.81&	      0.76&	       150&	      1.02&	     0.776&	K0        &	      5100\\ 
Sz 133&	16:03:29.37&	-41:40:02.14&	     69.05&	      0.77&	       150&	      0.63&	      0.07&	K2        &	      4900\\ 
Sz 90&	16:07:10.05&	-39:11:03.64&	     21.83&	      0.46&	       200&	      0.79&	     0.661&	K7        &	      4060\\ 
Sz 98&	16:08:22.48&	-39:04:46.81&	    237.29&	      1.42&	       200&	      0.74&	     2.512&	K7        &	      4060\\ 
Sz 100&	16:08:25.74&	-39:06:1.63&	     54.85&	      0.58&	       200&	      0.18&	     0.169&	M5.5      &	      3057\\ 
Sz 108B&	16:08:42.86&	-39:06:15.04&	     26.77&	      0.34&	       200&	      0.19&	     0.151&	M5        &	      3125\\ 
J16085324-3914401&	16:08:53.22&	-39:14:40.53&	     19.57&	      0.28&	       200&	      0.32&	     0.302&	M3        &	      3415\\ 
Sz 113&	16:08:57.78&	-39:02:23.21&	     22.35&	      0.27&	       200&	      0.19&	     0.064&	M4.5      &	      3197\\ 
Sz 114&	16:09:01.83&	-39:05:12.79&	     96.41&	      0.41&	       200&	      0.23&	     0.312&	M4.8      &	      3175\\ 
J16102955-3922144&	16:10:29.53&	-39:22:14.83&	      7.14&	      0.35&	       200&	      0.22&	     0.158&	M4.5      &	      3200\\ 
J16124373-3815031&	16:12:43.73&	-38:15:03.40&	     29.88&	      0.49&	       200&	      0.47&	     0.617&	M1        &	      3705\\ 
\hline
\end{tabular}
\begin{flushleft}
\textbf{Notes.} $F_{890\mu\mathrm{m}}$ is the integrated flux and rms is the root mean square noise from \citet{Ansdell:2016qf}. $d$ is the distance to the source: for sources in the Lupus I, II and IV clouds $d=150\u{pc}$, while for stars in the Lupus III cloud $d=200\u{pc}$ \citep[see][]{Comeron:2008qy}. Note that recent measurements from the \emph{Gaia} space telescope might suggest a distance of 150\u{pc} for the Lupus III cloud. $M_\star$ is the stellar mass as in \citet{Ansdell:2016qf}, obtained using evolutionary tracks by \citet{Siess:2000qy}. Stellar bolometric luminosity $\Lstar$, effective temperature $T_{\mathrm{eff}}$ and spectral type SpT are derived from X-Shooter measurements by \citep{Alcala:2014uq, Alcala:2016ys}.
\end{flushleft}
\label{tb:sources}
\end{table*}

\section{Modeling}
\label{sec:modeling}
To study the structure of the disks we fit their continuum emission with a disk model that is based on the two-layer approximation \citep{1997ApJ...490..368C,2001ApJ...560..957D}. In the following we introduce the fundamental quantities that are needed to compute the disk emission and for more details we refer to \citet{Tazzari:2016qy}. 

\subsection{Disk model}
\label{sec:disk.model}
Under the basic assumptions that, at each radius, the disk is vertically isothermal and in  hydrostatic equilibrium, the two-layer approximation allows us to compute the disk continuum emission given the properties of the central star, a surface density profile and a dust grain size distribution. At mm wavelengths, most of the disk emission originates from the dust grains residing in the dense and geometrically thin disk midplane. Based on the conservation of the energy delivered by the star onto the disk surface and on its propagation to the disk midplane, the two-layer approximation is thus  appropriate for reproducing the disk emission at mm wavelengths \citep{Isella:2009qy,Ricci:2010eu,2013A&amp;A...558A..64T, Testi:2016jk}.

Following previous studies \citep{Andrews:2009zr, 2013A&amp;A...558A..64T,Tazzari:2016qy}, we parametrize the gas surface density with a self-similar solution for an accretion disk \citep{1974MNRAS.168..603L,Hartmann:1998qy}, using the parametrization:
\begin{equation}
\label{eq:surface.density}
\Sigma_{\mathrm{g}}(R) = \Sigma_{0}\left(\frac{R}{R_{0}} \right)^{-\gamma} \exp\left[-\left(\frac{R}{R_{c}} \right)^{2-\gamma} \right]\,,
\end{equation}
where $\Sigma_{0}$ is a normalization factor, $R_{0}$ is a characteristic radius that we  keep fixed to 10\u{au}, $\gamma$ is the power-law slope and $R_{c}$ is the exponential cut-off radius. The dust surface density is given by:
\begin{equation}
\Sigma_{\mathrm{d}}(R)=\zeta\, \Sigma_{\mathrm{g}}(R)\,,
\end{equation}
where $\zeta$ is the dust-to-gas mass ratio, assumed to be constant and equal to the typical ISM value $\zeta=0.01$. The choice of the profile in Eq.~\eqref{eq:surface.density} and of a constant dust-to-gas ratio are a clear simplification of reality, in which we expect $\zeta$ to change across the disk from both observational \citep{de-Gregorio-Monsalvo:2013qf} and theoretical \citep{Birnstiel:2014zr} arguments. However, since in this study we are only analyzing the dust continuum emission, we cannot pose any constraint on the actual gas-to-dust variations in the disks. These choices are useful as they provide us with a simple parametrization of the dust distribution that we can directly compare to other studies. As a result, there are three free parameters describing the surface density: $\Sigma_{0}$, $\gamma$ and $R_{c}$. 

To compute the continuum emission we compute the dust opacity of the grain population using the Mie Theory, which allows us to compute the emissivity of a single spherical grain, and the Bruggeman mixing theory \citep{Bruggeman:1935ys}, which allows us to compute the effective dielectric constants for composite grains. For both the disk surface and midplane we use a MRN-like grain size distribution \citep{Mathis:1977yq}, i.e. a number density $n(a)\propto a^{-q}$ for $a_{\mathrm{min}}\leq a \leq a_{\mathrm{max}}$ and $n(a) = 0$ otherwise, where $a$ is the grain radius. In the surface layer we use $a_{\mathrm{min}}=10\u{nm}$ and $a_{\mathrm{max}}=1\u{$\mu$m}$ with $q=3.5$, typical of a population of small grains. This ensures that the surface layer is optically thick to the stellar radiation. In the disk midplane, where dust coagulation and settling are expected to occur \citep{Dullemond:2004ly}, we use $a_{\mathrm{min}}=10\u{nm}$ and $a_{\mathrm{max}}=1.023\u{cm}$ with $q=3.0$, which corresponds to a population of larger grains. The particular value of $a_{\mathrm{max}}$ has been chosen for continuity with previous analysis and reproduces the same opacity (per gram of dust) used by \citet{Ansdell:2016qf} $\kappa_{890\u{$\mu$m}}=3.37\u{cm}^{2}\u{g}^{-1}$. Similarly to \citet{2013A&amp;A...558A..64T} and \citet{Tazzari:2016qy}, we assume spherical grains and we adopt the simplified volume fractional abundances found by \citet{Pollack:1994vn}: 20.6\% carbonaceous materials, 5.4\% astronomical silicates, 44\% water ice and 30\% vacuum, for an  average grain density of $0.9\u{g}\u{cm}^{-3}$. 

Finally, the disk appearance on sky is set by the disk inclination along the line of sight, defined as $i=0$\deg\ for a face-on disk and $i=90$\deg\ for an edge-on disk, and by the disk Position Angle, defined East-of-North from $P.A.=0$\deg\ to $P.A.=180$\deg. 

\subsection{Disk flaring}
Computing a realistic dust temperature profile is key for a reliable estimate of their sub-mm continuum emission and therefore of their mass. The self-consistent fully-flared models based on the two-layer approximation \citep{1997ApJ...490..368C,2001ApJ...560..957D} are typically too vertically thick and do not properly reproduce the spectral energy distribution in the far-IR. This is confirmed by theoretical and observational studies that require a reduced disk flaring (i.e., some degree of dust settling) in order to reconcile the far-IR and the sub-mm fluxes \citep[e.g.][]{Rodgers-Lee:2014aa,Daemgen:2016fk}. These authors use the ratio between the fluxes in the far-IR and the J-band (a good proxy for the stellar photospheric emission) to estimate the disk flaring. Indeed, spectroscopic studies of disks around very low mass T Tauri stars and brown dwarfs have shown that the dust temperature strongly depends on the disk flaring (in addition to stellar luminosity), and to a lesser degree on the disk mass and other disk properties. 
In this work we use the spectral slope between the far-IR and the J-band to obtain a rough estimate of the disk flaring that characterizes the disks in our sample. Since the spectroscopic measurements are not available for all the sources in our sample, we derive an average disk flaring reduction factor that we use for all the fitted sources.

\begin{figure*}
\resizebox{\hsize}{!}{\includegraphics{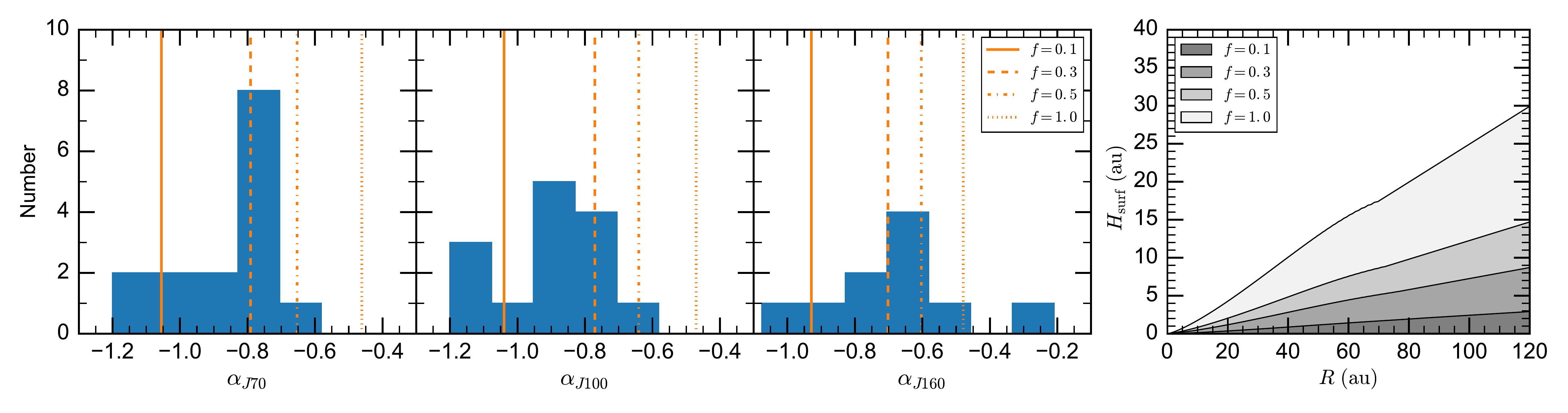}}
\caption[Spectral indices between J-band and FIR with relative disk scale height]{\textit{Left:} spectral indices for the sources in our sample measured between J-band (2MASS, 1.235\u{$\mu$m}) and, respectively, 70\u{$\mu$m}, 100\u{$\mu$m}, 160\u{$\mu$m} (Herschel/PACS) using the definition in Eq. \eqref{eq:spectral.slope.def}. The vertical lines show the spectral indices obtained with our disk model for different values of the flaring  reduction parameter $f$, from a fully-flared model ($f=1$, dotted line) to less flared models ($f=0.5, 0.3, 01.$, respectively dash-dotted, dashed and solid line). 
\textit{Right:} scale-height of the disk surface layer computed by the model. }
\label{fig:sources.spectral.indices}
\end{figure*}

To this purpose, we use \emph{Herschel}/PACS measurements at 70, 100 and 160\u{$\mu$m} obtained from the \emph{Herschel} Science Archive and performing the photometry as in \citet{van-der-Marel:2016aa}. We compute the spectral slope between the Herschel/PACS bands and the 2MASS J-band (1.235\u{$\mu$m}) as \citep{Adams:1987jk,Daemgen:2016fk}:
\begin{equation}
\label{eq:spectral.slope.def}
\alpha_{\lambda_{1}\lambda_{2}} = 
\frac
{\log(\lambda_{1}F_{\lambda_{1}}/\lambda_{2}F_{\lambda_{2}})}
{\log(\lambda_{1}/\lambda_{2})}\,,
\end{equation}
which provides a model-independent estimate of the far-IR dust emission. 
In \figref{fig:sources.spectral.indices} we present the histograms of the computed spectral slopes. 
On the same histograms the orange vertical lines represent the spectral slopes computed from our disk model for different values of the flaring reduction parameter $f$.  By varying $f$, we can manually reduce the disk $H_\mathrm{surf}/R$ from a fully-flared profile ($f=1$), which produces flatter spectral slopes ($\alpha \geq -0.5$), to progressively less flared models ($f<1$), which produce progressively steeper spectral slopes ($\alpha \leq -0.5$). 
Given that Herschel measurements are available only for a subset of the sources in our sample and the approximate approach for deriving the flaring reduction factor, we decided to apply the same average value of $f$ to all disks in the sample.
By comparing  synthetic and  observed spectral slopes (Fig.~\ref{fig:sources.spectral.indices}) we find that a disk flaring reduced by a factor  $\approx3$ ($f=0.3$) gives a good representation of the slopes at all the three bands simultaneously. We adopt this value for all the fits that we perform in this study. To assess the impact of this choice on the constrained disk structure we have repeated the fits for different $f$ values of 0.1, 0.3 and 1.0. We do not present these detailed checks here, but we observe that the average temperature (and therefore surface density normalization) change by a factor 1.5 at most with $f$ varying between 0.1 and 1.0, while the surface density profile remains mostly unaltered with the cut-off radius $R_c$ varying by 10\% at most.

\subsection{Modeling methodology}
\label{sec:mcmc}
We perform the fits with a Bayesian approach, which produces probability distribution functions (PDFs) for the free parameters of the model by means of a Markov Chain Monte Carlo (MCMC) algorithm.   The free parameters are seven: $\Sigma_{0}$, $\gamma$ and $R_{c}$ which define the surface density, $i$ and $P.A.$ which define the disk appearance, $\Delta\alpha$ and $\Delta\delta$ which define the (R.A., dec) offset of the disk center with respect to the phase center of the observations. We consider these last two parameters as \textit{nuisance parameters}: for each disk we fit the disk center to achieve a better matching between model and observations, however the information encoded in such offset is not relevant for the aims of this study\footnote{In principle it is possible to relate the fitted offsets to the proper motion of each single object as shown in \citet{Tazzari:2016qy}, but this is beyond the scope of this paper.}. 

Following the implementation developed by \citet{Tazzari:2016qy}, for a given set of values of the free parameters the model produces a synthetic image of the disk that is then Fourier-transformed and sampled in the same $(u,v)-$locations of the observed visibilities. We finally compute the $\chi^{2}$ as:
\begin{equation}
\chi^{2} = \sum_{i=0}^{N}w_{i}|V^{\mathrm{obs}}(u_{i}, v_{i})-V^{\mathrm{mod}}(u_{i}, v_{i})|^{2}\,,
\end{equation}
where $V^{\mathrm{obs}}$ and $V^{\mathrm{mod}}$ are the observed and the synthetic visibilities, respectively, $w_{i}$ is the weight associated with the observed visibilities at the $(u_{i},v_{i})$ location and $N$ is the total number of $(u,v)-$locations. As in \citet{Tazzari:2016qy}, the theoretical weights are calculated according to \citet{Wrobel:1999gf} and then re-scaled to ensure that $\sum_{i}w_{i}=1$. The posterior PDF is computed as $\exp(-\chi^{2}/2)$ within a rectangular domain in the parameter space (the region of interest), and zero outside such domain. The ranges defining the domain of exploration of the parameter space are detailed in \tbref{table:space.parameters}. 
\begin{table}[h!]
\caption[Parameter space explored by the MCMC]{Parameter space explored by the Markov chains.}
\begin{center}
\begin{tabular}{ccc}
\hline
\hline
Parameter		&	Min	&	Max\\
\hline
${\gamma}$		&	-2	&	2 \\
${\Sigma_{0}}$	&	0.05\u{g}\u{cm}${^{-2}}$& 400\u{g}\u{cm}${^{-2}}$\\
${R_{\mathrm{c}}}$ & 2\u{au}	&	400\u{au}	\\
$i$ & 0\deg	&	90\deg		\\
$P.A.$ & 0\deg	&	180\deg	\\
$\Delta\alpha$& -2"	&	+2"	\\
$\Delta\delta$& -2"	&	+2"	 \\
\hline
\end{tabular}
\end{center}
\label{table:space.parameters}
\end{table}%

The region of interest in the 7-dimensional parameter space is explored using an ensemble of Markov chains that evolve simultaneously according to the affine-invariant MCMC algorithm by \citep{Goodman:2010}. Two of the main advantages of using this algorithm are: first, several chains are initialized at random locations across the domain of interest, thus ensuring that the PDFs that we derive after they have converged do not depend on their initialization. Second, the algorithm enables a massive parallelization of the computation
: in a reasonable amount of time it allows us to achieve a rather solid sampling of the posterior PDF out of which we can derive reliable PDFs for all the parameters.
 
In this study, we perform the fits using one hundred chains, which is a reasonable number for a seven-dimensional parameter space (approx. 10-20 chains
per parameter). We initialize the chains in random positions in the domain of interest, making sure that they are not initialized too close to the borders in order to avoid computational issues. After initialization, we let the chains evolve for a burn-in phase of 1000-1500 steps (the actual number varies from source to source) and then we take 4000-5000 steps to achieve a good sampling of the posterior PDF. This results in approximately $\sim 5-7\times 10^5$ evaluations of the posterior for the fit of each disk. 

The product of a MCMC fit is the chain that results from collecting the locations of all the walkers throughout their evolution. Each element of the chain represents a sample of the posterior PDF. Therefore, to give an adequate representation of the results of the fit we always provide a plot of the chain, projected in the various dimensions. By marginalizing\footnote{Integrating over all but the one parameter of interest.} the chain over one parameter we obtain an estimate of its PDF, out of which we derive its value as the median and uncertainty as the central interval (between 16\% and 84\% percentiles). By marginalizing the chain over two parameters of interest we obtain the 2D distribution of the samples from which we can study their correlation. 
To perform the fit we use the implementation of the MCMC algorithm provided by the Python package \texttt{emcee} \citep{2013PASP..125..306F}, which allows us to exploit the massively parallel nature of the algorithm by running the fits on many cores simultaneously.

\section{Derived disk properties}
\label{sec:results}
In this section we present the results of the fits. In order to present the analysis performed for each disk, here we illustrate the fit results for Sz~71, and in Appendix~\ref{app:fits} we report the detailed plots for all the other sources. 

\begin{figure}
\centering
\resizebox{\hsize}{!}{\includegraphics{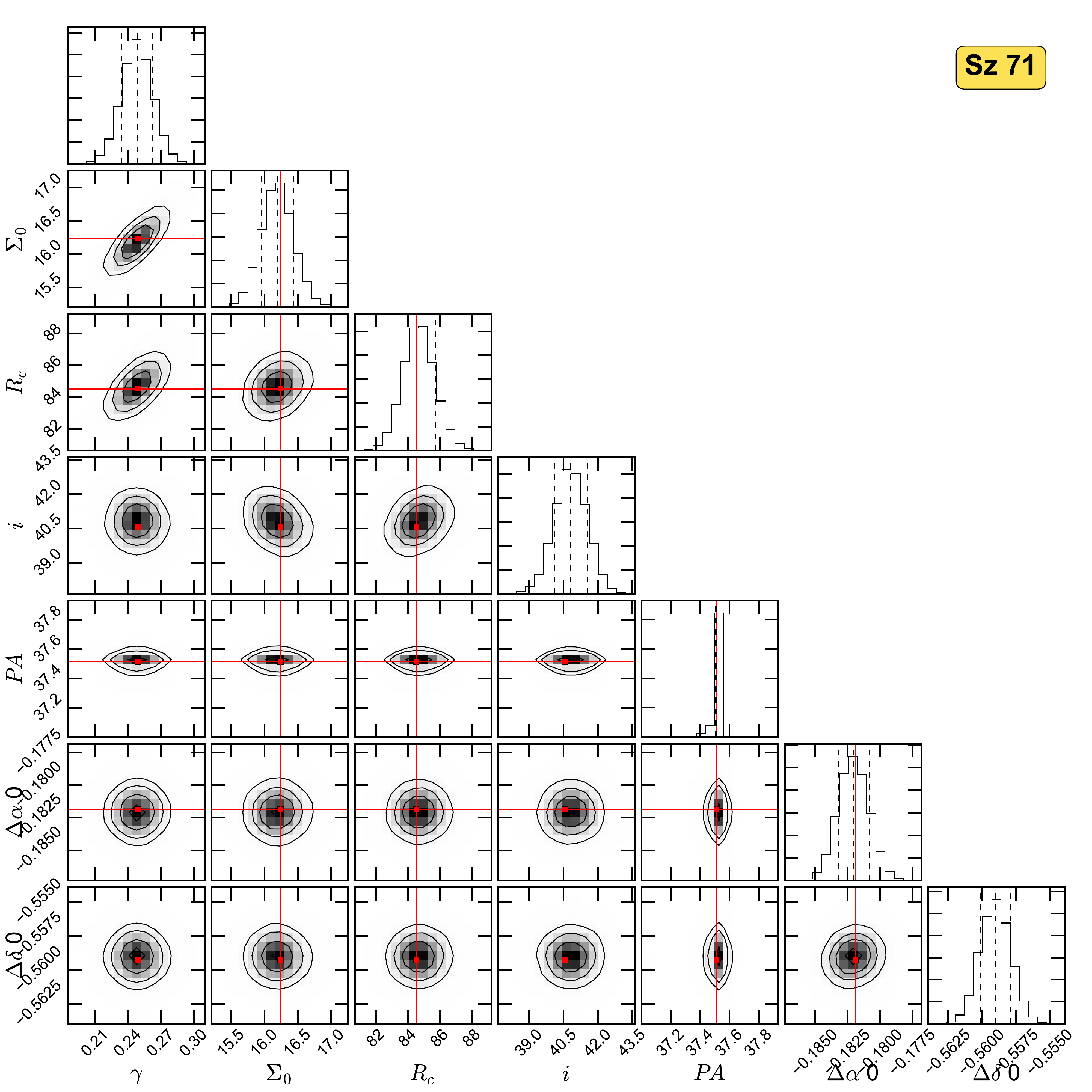}}
\caption[Staircase plot of the chain resulting from the MCMC fit of Sz~71]{Staircase plot of the chain resulting from the MCMC fit of Sz~71. The histograms on the main diagonal are the marginalized distributions of each parameter: from left to right, $\gamma$, $\Sigma_{0}$, $R_{c}$, $i$ and $P.A.$, $\Delta\alpha$, $\Delta\delta$. The vertical dashed lines indicate the 16\%, 50\% and 84\% percentiles. The off-diagonal 2D plots show the correlation between each couple of parameter, with contour lines showing 0.5$\sigma$ increments. The solid red lines highlight the coordinates of the best-fit model.}
\label{fig:samplemodel.triangle}
\end{figure}

The ``staircase'' plot in \figref{fig:samplemodel.triangle} shows the Markov chain resulting from the fit of Sz~71. On the main diagonal we show the histograms of the marginalized distribution, with vertical dashed lines indicating the 16th, 50th and 84th percentiles. The distributions are single-peaked, with profiles very close to Gaussians. For Sz~71 we find the following best fit parameters, corresponding to the model with maximum likelihood: $\gamma=0.27\pm0.01$, $\Sigma_{0}=16.2\pm0.25\u{g cm$^{-2}$}$, $R_{c}=85\pm1\u{au}$, $i=40.8\pm 0.7^{\circ}$, $P.A.=37.5\pm 0.1^{\circ}$., $\Delta \alpha=-0.182''\pm0.002''$ and $\Delta \delta-0.559'' \pm 0.002''$. The red lines in \figref{fig:samplemodel.triangle} highlight the location of the best-fit model. The off-diagonal 2D plots represent the bivariate distributions between each pair of parameters which provide an immediate estimate of their correlation. The staircase plots for the other disks are  presented in Appendix~\ref{app:fits}.

\begin{figure*}
\centering
\resizebox{\hsize}{!}{\includegraphics[width=0.3\textwidth]{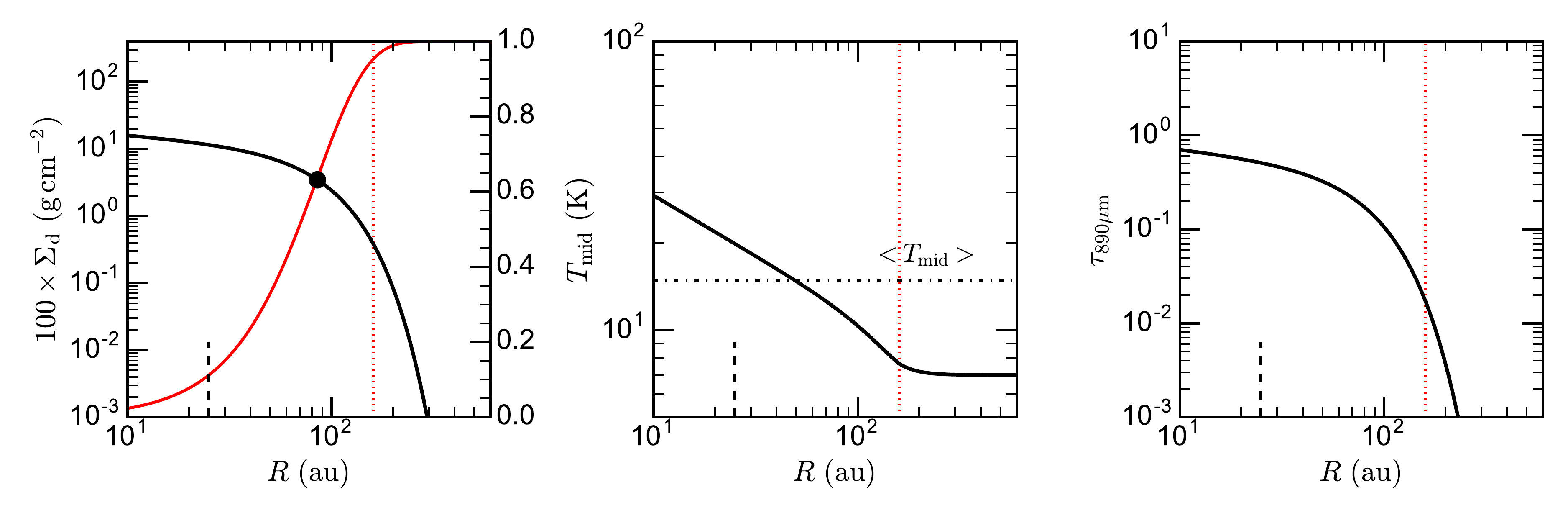}}
\caption[Physical structure of the best-fit model describing Sz~71]{Physical structure of the best-fit model describing Sz~71. \textit{(left)}: gas surface density profile computed as $100\times \Sigma_{\mathrm{d}}(R)$, where 100 is the gas-to-dust ratio. The solid red line is the cumulative mass (right y axis) and the black filled circle highlights the cut-off radius $R_{c}\simeq85\u{au}$. In all three plots, the dotted red line indicates the radius containing 95\% of mass and the vertical dashed black line half the synthesized beam size ($\simeq 25\,$)au. \textit{(middle)}: dust temperature profile $T_{\mathrm{mid}}(R)$ derived self-consistently at each radius from the disk model. The dot-dashed horizontal line represents the mass-averaged dust temperature $<T_{\mathrm{mid}}>$, which for Sz~71 is $\simeq 14\u{K}$. \textit{(right)}: optical depth of the disk midplane at the observing wavelength: the sub-mm continuum emission is expected to be optically thick only within $R<2\u{au}$.}
\label{fig:samplemodel.structure}
\end{figure*}

A description of the physical structure of the best fit model for Sz~71 can be found in  \figref{fig:samplemodel.structure}, where we plot the gas surface density and cumulative mass (left panel), the midplane temperature (middle panel) and the optical depth of the disk midplane at the observing wavelength (right panel) as a function of the distance from the star. In all plots, we highlight the radius containing the 95\% of the mass as a vertical dotted red line. The disk model for Sz~71 has a very flat ($\gamma\sim 0$) surface density profile in the inner disk and a sharp exponential cut-off at $R_{c}\sim85\u{au}$, with 95\% of the mass contained within 150\u{au}. The disk is optically thin at 890\u{$\mu$m} almost everywhere, except in the inner $R<2.6\u{au}$ region which, anyway, gives a negligible contribution to the total mass. The midplane temperature decreases monotonically from 325\u{K} in the innermost disk region ($R\sim 0.1\u{au}$) to $\sim10\u{K}$ at 100\u{au} and then levels to 7\u{K}, which is the minimum temperature allowed in the model, chosen to give a simple realization of the typical interstellar radiation field. 

\begin{figure}
\centering
\resizebox{0.8\hsize}{!}{\includegraphics[width=0.3\textwidth]{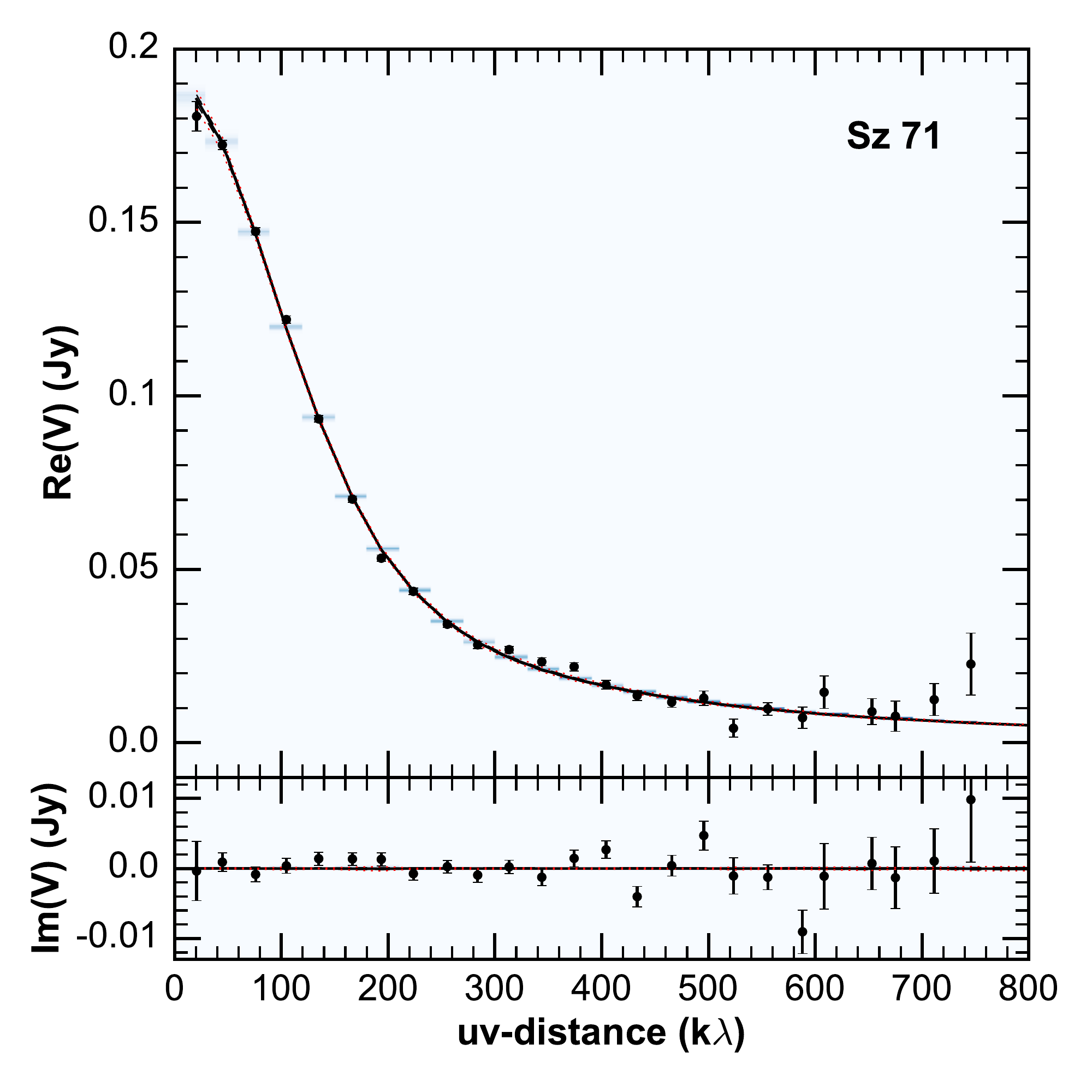}}
\caption[Comparison between model and observed visibilities of Sz~71]{Comparison between model and observed visibilities of Sz~71 as a function of deprojected baseline length (uv-distance). The panel above (below) shows the Real (Imaginary) part of the visibilities.
Black dots show the observed visibilities (binned every 30k$\lambda$), the blue density indicators represent the PDF of the model visibilities in each uv-bin. The black solid line corresponds to the median of the model visibilities PDF, the black dashed lines the 16th and 84th percentiles, the red dotted lines the 2.7th and 97.7th percentiles.}
\label{fig:samplemodel.uvplot}
\end{figure}

\figref{fig:samplemodel.uvplot} is a visual representation of how the distribution of models (i.e. the posterior samples shown in \figref{fig:samplemodel.triangle}) compare with the observations. The top (bottom) panel shows the Real (Imaginary) part of the deprojected observation and model visibilities as a function of baseline length. The observation and model visibilities have been centered on the disk centroid (according to the fitted offsets $\Delta\alpha$ and $\Delta\delta$) and then de-projected assuming the fitted values of $i$ and $P.A.$. Of the model visibilities we show the posterior PDF (as the blue density indicator for each uv-bin), the median (black solid line), the 1$\sigma$ central interval (the black dashed lines) and the 2$\sigma$ central interval (the red dotted lines). In the case of Sz~71 the posterior PDF of the model visibilities has a very narrow peak, therefore these lines are very close to the median (cfr. with the broader model visibilities PDF derived for some disks in Appendix~\ref{app:fits}). The Real part of the observation and model visibilities match almost perfectly up to 300\u{k$\lambda$} and are compatible within $2\sigma$ at higher uv-distances. The Imaginary part of the observed visibilities is on average 0 (as it should be for a centered azimuthally symmetric surface brightness distribution) with a residual oscillating behaviour at very low signal to noise level, probably due to some sort of asymmetry in the disk that cannot be described by our axisymmetric disk model.

\begin{figure}
\resizebox{\hsize}{!}{\includegraphics{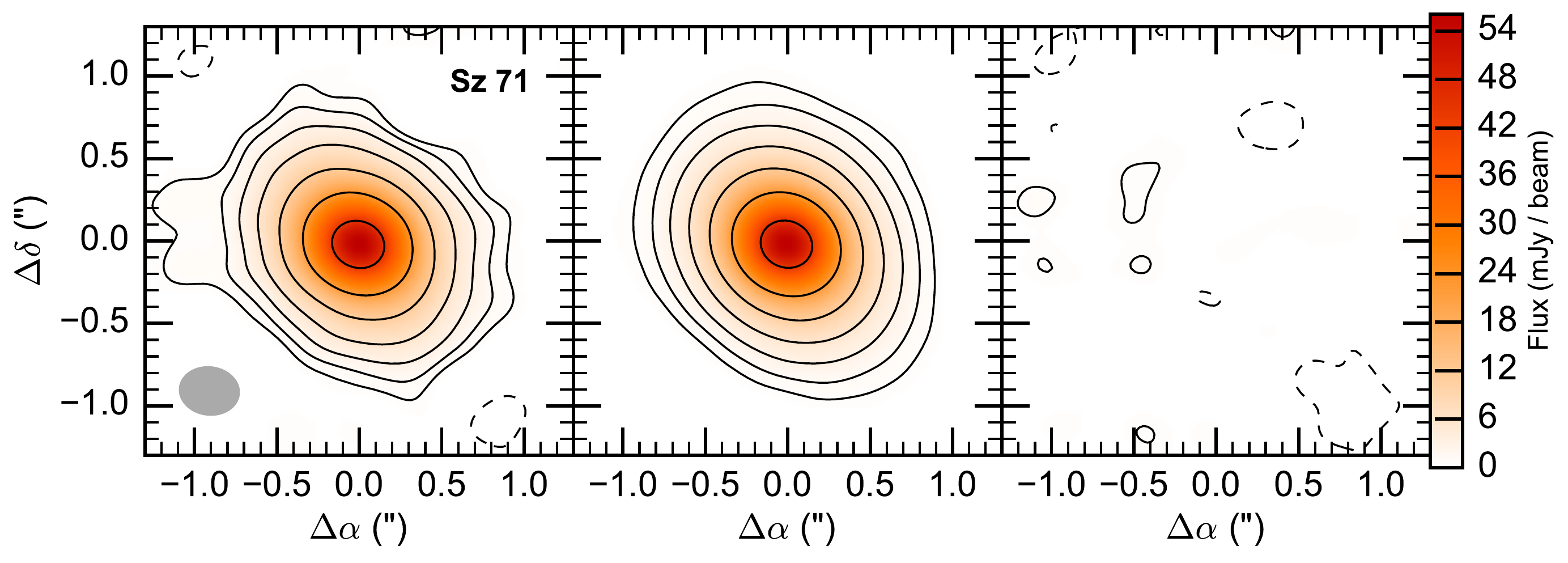}}
\caption[Comparison between observed and best-fit model continuum maps of Sz~71]{Comparison between the observed \textit{(left)} and the best-fit model \textit{(middle)} images for Sz~71. Residuals are shown in the right panel. The best-fit model has the following parameters $\gamma=0.25\pm0.01,\ \Sigma_0=(16.2\pm 0.2)\mathrm{g}/\mathrm{cm}^2,\ R_c=(85\pm 1)\mathrm{au}$. The three images have been produced by applying the CLEAN deconvolution algorithm with natural weighting to the observed, best-fit model and residual visibilities, respectively. Contour levels refer to -3 (dashed), 3, 6, 12, 24, 48, 50, 100, 150, etc. multiple of the rms, which here is $\sigma=0.3$\u{mJy/beam}. The FWHM of the synthesized beam is shown as a gray ellipse in the left panel.}
\label{fig:samplemodel.maps}
\end{figure}

In \figref{fig:samplemodel.maps} we compare the observation to the best-fit model images. The three panels illustrate the images of the observations, of the model and of the residuals, derived from the respective visibilities with the CLEAN algorithm \citep{Clark:1980lr} and a natural weighting scheme using the software CASA 4.5.0. The best-fit model (whose location in the parameter space is highlighted in \figref{fig:samplemodel.triangle}) reproduces the observations extremely well (the residuals being lower than the 3$\sigma$ level).

We note that our smooth model describes well all the disks (see Appendix \ref{app:fits}), with the exceptions of IM~Lup and Sz~98 where we find systematic residuals. These residuals are, respectively, 14\% and 4\% of the total intensity and are probably related to the presence of radial inhomogeneities like rings or spirals \citep[see e.g.][]{2041-8205-808-1-L3, Guidi:2016sf, Isella:2016ww, Perez:2016aa}. These are minor discrepancies related to a very small fraction of disks in our sample and, while interesting to follow up in the future (van Terwisga et al. in prep.), do not affect the results we discuss in this paper.

In \tbref{tb:fits.results} we report the values of the parameters derived for Sz~71 and for all the other disks. For each disk, we provide estimates of the free parameters $\gamma$, $\Sigma_{0}$, $R_{c}$, $i$ and $P.A.$, derived from the Markov chains as described in Section~\ref{sec:mcmc}. For the 22 disks in the sample the Markov chains converged to single-peaked distributions with moderate to absent degeneracy. In all these cases, the best-fit model has usually a normalized chi-square of $1.0\pm0.2$ and the residuals are at very low signal-to-noise levels. From the chains, we compute some derived quantities such as the disk outer radius $R_{\mathrm{out}}$, defined as the radius containing 95\% of the model flux, and the total dust mass $M_{\mathrm{dust}}$, computed by integrating the dust surface density:
\begin{equation}
\label{eq:dust.mass.integral}
M_{\mathrm{dust}} = \int_{R_{\mathrm{in}}}^{R_{\mathrm{out}}} \Sigma_{d} (R)\, 2\pi R\, \mathrm{d}R\,,
\end{equation}
where $R_{\mathrm{in}}=0.1\u{au}$ is the inner edge of the radial grid of the model (fixed for all the disks). The derived quantities are estimated as the median of their derived PDF, assigning an uncertainty that corresponds to the central interval between the 16th and 84th percentile.

\begin{sidewaystable*}
\caption{Fit results: free parameters and derived quantities.}
\begin{center}
\resizebox{\hsize}{!}{
\begin{tabular}{lrrrrr|rr}
\hline 
\hline
Name                    & $\gamma$	              	&$\Sigma_0$	             	& $R_c$	                 	& $i$	                   	& $P.A.$				&$R_{\mathrm{out}}$			& $M_{\mathrm{dust}}$	   \\[1mm]
	                    & 		 	             	&$\u{g}\u{cm}^{-2}$	        & $\u{au}$	                & \deg	                   	& \deg	    			&$\u{au}$   				& $\Mearth$         \\[1mm]
\hline 
Sz 65                   & $0.12\pm{0.24}$	& $20.71\pm{2.95}$	& $28.03\pm{0.56}$	& $61.46\pm{0.88}$	& $108.63\pm{0.37}$	& $42.39\pm{2.96}$	& $16.07\pm{0.61}$	 \\[1mm] 
J15450887-3417333       & $-0.72\pm{0.70}$	& $104.05\pm{42.70}$	& $16.13\pm{1.08}$	& $36.30\pm{5.56}$	& $2.41\pm{2.53}$	& $23.43\pm{1.23}$	& $32.19\pm{7.86}$	 \\[1mm] 
Sz 68                   & $-0.39\pm{0.27}$	& $188.99\pm{80.18}$	& $14.04\pm{0.96}$	& $32.89\pm{3.32}$	& $175.78\pm{3.13}$	& $23.03\pm{0.40}$	& $40.05\pm{12.32}$	 \\[1mm] 
Sz 69                   & $0.72\pm{0.21}$	& $50.24\pm{18.02}$	& $9.11\pm{2.00}$	& $43.53\pm{8.65}$	& $124.28\pm{17.10}$	& $18.04\pm{1.60}$	& $7.59\pm{1.84}$	 \\[1mm] 
Sz 71                   & $0.25\pm{0.01}$	& $16.19\pm{0.24}$	& $84.68\pm{1.00}$	& $40.82\pm{0.71}$	& $37.51\pm{0.01}$	& $127.14\pm{1.12}$	& $80.21\pm{1.01}$	 \\[1mm] 
Sz 73                   & $1.00\pm{0.17}$	& $10.94\pm{1.10}$	& $43.47\pm{7.43}$	& $49.76\pm{3.95}$	& $94.71\pm{5.17}$	& $79.41\pm{19.36}$	& $9.32\pm{1.20}$	 \\[1mm] 
IM Lup                  & $0.65\pm{0.01}$	& $22.25\pm{0.10}$	& $431.74\pm{6.81}$	& $48.40\pm{0.11}$	& $144.37\pm{0.10}$	& $504.01\pm{0.10}$	& $439.46\pm{3.76}$	 \\[1mm] 
Sz 83                   & $0.40\pm{0.09}$	& $151.83\pm{28.03}$	& $36.18\pm{1.32}$	& $3.31\pm{2.90}$	& $163.76\pm{5.94}$	& $64.42\pm{1.12}$	& $160.76\pm{11.49}$	 \\[1mm] 
Sz 84                   & $-0.98\pm{0.20}$	& $3.04\pm{0.46}$	& $40.57\pm{1.10}$	& $73.99\pm{1.56}$	& $167.31\pm{0.77}$	& $52.26\pm{2.26}$	& $13.83\pm{0.54}$	 \\[1mm] 
Sz 129                  & $-0.33\pm{0.02}$	& $18.43\pm{0.38}$	& $50.25\pm{0.39}$	& $31.74\pm{0.75}$	& $154.94\pm{0.43}$	& $70.29\pm{1.23}$	& $70.94\pm{0.56}$	 \\[1mm] 
J16000236-4222145       & $-0.20\pm{0.02}$	& $6.13\pm{0.16}$	& $89.60\pm{0.91}$	& $65.71\pm{0.36}$	& $160.45\pm{0.02}$	& $129.38\pm{2.26}$	& $72.56\pm{0.75}$	 \\[1mm] 
MY Lup                  & $-0.59\pm{0.06}$	& $9.85\pm{0.66}$	& $60.67\pm{0.62}$	& $72.98\pm{0.35}$	& $58.94\pm{0.12}$	& $85.14\pm{0.74}$	& $86.05\pm{2.06}$	 \\[1mm] 
Sz 133                  & $-0.17\pm{0.16}$	& $14.25\pm{6.01}$	& $66.77\pm{2.39}$	& $78.53\pm{0.65}$	& $126.29\pm{0.09}$	& $108.68\pm{4.78}$	& $91.73\pm{11.82}$	 \\[1mm] 
Sz 90                   & $-0.68\pm{0.80}$	& $6.36\pm{2.73}$	& $29.07\pm{2.21}$	& $61.31\pm{5.34}$	& $123.00\pm{4.86}$	& $38.18\pm{5.23}$	& $8.31\pm{0.46}$	 \\[1mm] 
Sz 98                   & $0.11\pm{0.01}$	& $5.06\pm{0.10}$	& $198.93\pm{2.80}$	& $47.10\pm{0.70}$	& $111.58\pm{0.06}$	& $278.62\pm{4.86}$	& $153.91\pm{2.75}$	 \\[1mm] 
Sz 100                  & $-1.52\pm{0.01}$	& $1.31\pm{0.03}$	& $59.99\pm{0.65}$	& $45.11\pm{0.97}$	& $60.20\pm{0.06}$	& $74.06\pm{1.29}$	& $42.53\pm{0.61}$	 \\[1mm] 
Sz 108B                 & $-0.06\pm{0.52}$	& $9.08\pm{2.57}$	& $40.12\pm{4.93}$	& $49.09\pm{5.34}$	& $151.76\pm{6.05}$	& $55.55\pm{10.59}$	& $15.64\pm{0.88}$	 \\[1mm] 
J16085324-3914401       & $0.20\pm{0.86}$	& $46.56\pm{49.69}$	& $15.76\pm{3.76}$	& $60.72\pm{4.00}$	& $100.31\pm{5.47}$	& $25.57\pm{3.18}$	& $14.60\pm{11.15}$	 \\[1mm] 
Sz 113                  & $-0.16\pm{0.59}$	& $81.09\pm{38.54}$	& $14.96\pm{1.85}$	& $10.78\pm{9.19}$	& $147.36\pm{14.20}$	& $22.24\pm{1.37}$	& $19.95\pm{4.61}$	 \\[1mm] 
Sz 114                  & $0.23\pm{0.03}$	& $21.01\pm{0.54}$	& $60.71\pm{0.87}$	& $15.84\pm{3.39}$	& $148.73\pm{6.87}$	& $89.71\pm{1.56}$	& $58.30\pm{0.73}$	 \\[1mm] 
J16102955-3922144       & $-0.52\pm{0.64}$	& $0.72\pm{0.46}$	& $62.54\pm{16.18}$	& $66.54\pm{9.21}$	& $118.86\pm{9.49}$	& $80.80\pm{27.55}$	& $5.40\pm{1.07}$	 \\[1mm] 
J16124373-3815031       & $0.65\pm{0.83}$	& $24.85\pm{15.94}$	& $21.85\pm{4.32}$	& $43.69\pm{7.39}$	& $22.99\pm{8.85}$	& $37.52\pm{7.64}$	& $11.44\pm{3.67}$	 \\[1mm] 
\hline 
\end{tabular}
}
\end{center}
\begin{flushleft}
\textbf{Notes.}
Columns 1-5: free parameters, estimated from the corresponding marginalized PDF: the parameter value is estimated as the median, the uncertainties represent the extent of the central interval between 16th and 84th percentiles. Columns $R_{\mathrm{out}}$ and $M_{\mathrm{dust}}$ report two derived quantities: the disk outer radius (defined as the radius containing 95\% of the model flux) and the total dust mass (computed as in Eq.~\ref{eq:dust.mass.integral}), respectively.\\
\end{flushleft}
\label{tb:fits.results}
\end{sidewaystable*}

\subsection{Comparison with \citet{Ansdell:2016qf}}
\label{sec:results.comparison.with.ansdell}
We now compare our results with those by \citet{Ansdell:2016qf}, who derived dust masses and disk inclinations with a simplified method. 

\begin{figure*}
\resizebox{0.49\hsize}{!}{\includegraphics{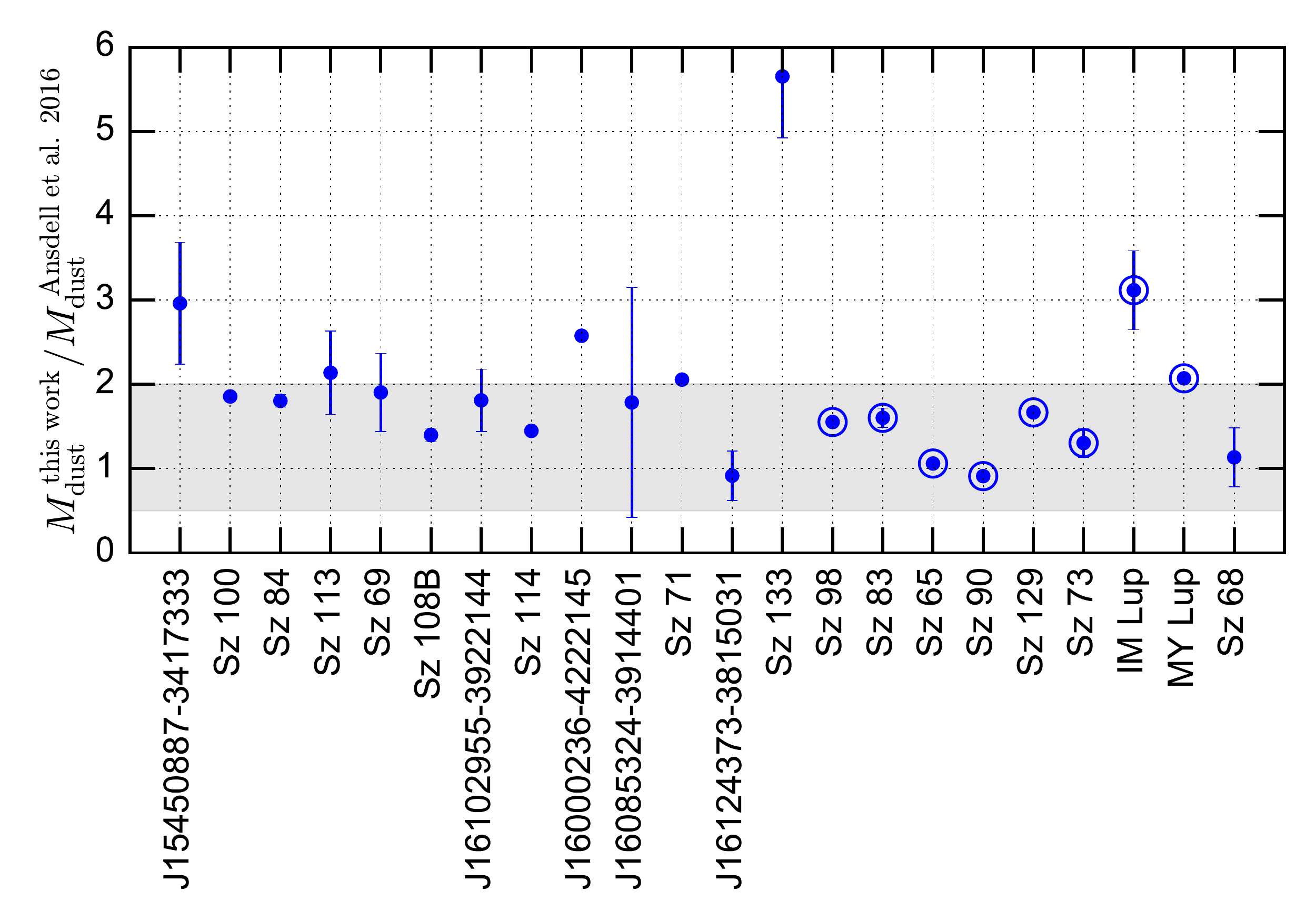}}
\resizebox{0.49\hsize}{!}{\includegraphics{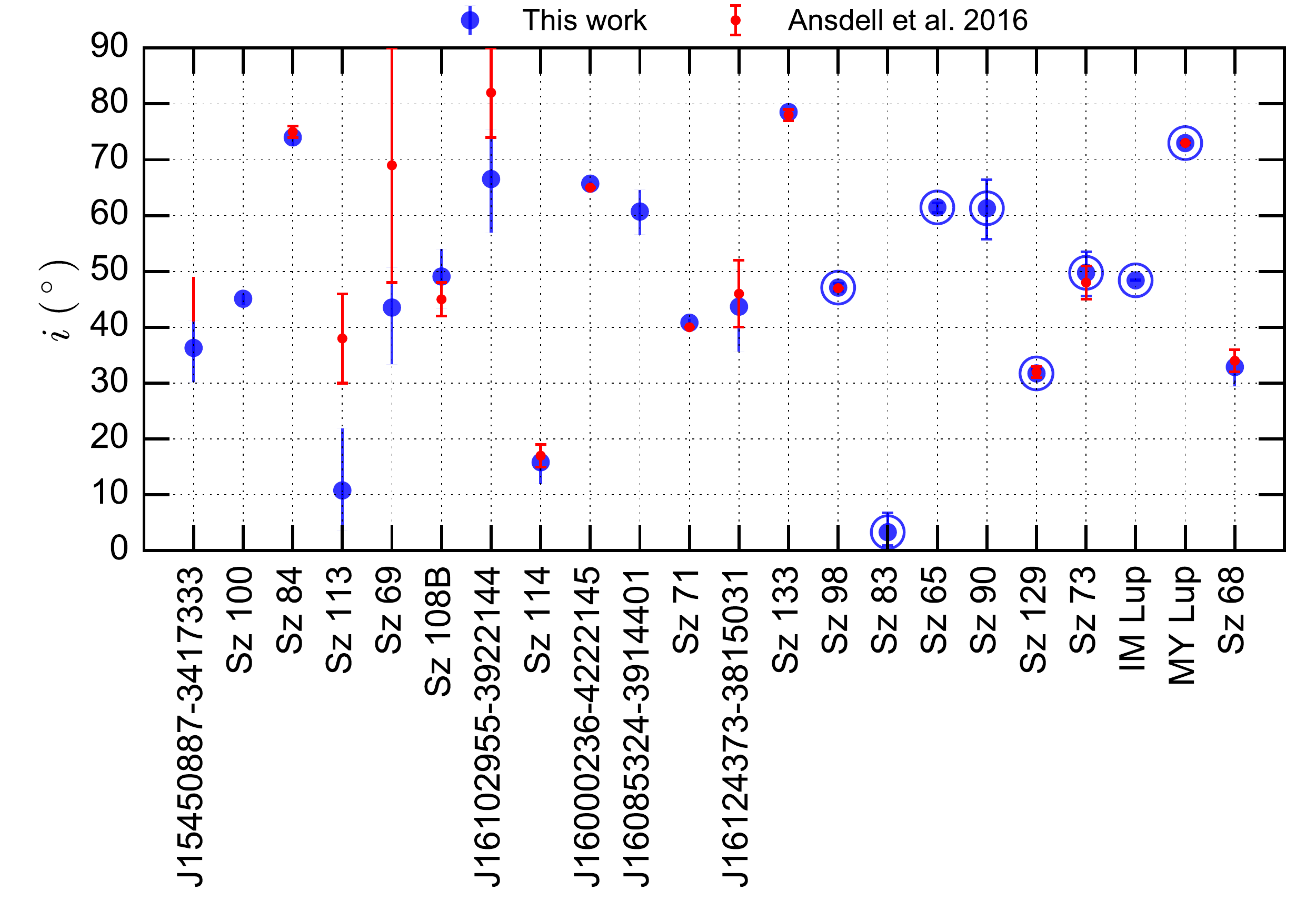}}
\caption[]{
\textit{(left)}: Ratio between the dust masses derived in this work and those derived by \citet{Ansdell:2016qf}. Although derived with very different methods (for a detailed description, see the text), the masses are within a factor of 2 (gray area) for the vast majority of disks. 
\textit{(right)}: 
Comparison between the disk inclinations \textit{i} (along the line of sight) obtained in this work and those derived by \citet{Ansdell:2016qf}. In both plots, the disks are sorted from left to right by increasing stellar mass and the circled dots highlight sources with $0.7\leq M_{\star}/M_\odot\leq 1$.
}
\label{fig:comparison.inc.mdust}
\end{figure*}
First, in \figref{fig:comparison.inc.mdust} we compare the dust masses: 
the plot shows the ratio between the masses derived from our fits and those derived by \citet{Ansdell:2016qf} by converting the spatially-integrated sub-mm continuum flux $F_{\nu}$ into dust mass:
\begin{equation}
\label{eq:conversion.flux.mass}
M_{\mathrm{dust}}' = \frac{d^{2}\,F_{\nu}}{\kappa_{890\u{$\mu$m}} B_{\nu}(T_{\mathrm{d}})}\,,
\end{equation}
where $d$ is the distance, $\kappa_{890\u{$\mu$m}}=3.37\u{cm$^{2}$}\u{g}^{-1}$ (per gram of dust) is the dust opacity, $B_{\nu}(T)$ is the black-body brightness at the temperature $T$, and $T_{\mathrm{d}}=20\u{K}$ is the dust temperature. 

The dust masses derived by \citet{Ansdell:2016qf} with the simple conversion formula of Eq~\eqref{eq:conversion.flux.mass} and a constant temperature of $T_\mathrm{d}=20$~K are accurate within a factor of 2 at a 1$\sigma$ level for the majority of disks (15 out of 20).

\begin{figure}[htpb]
\centering
\resizebox{\hsize}{!}{\includegraphics{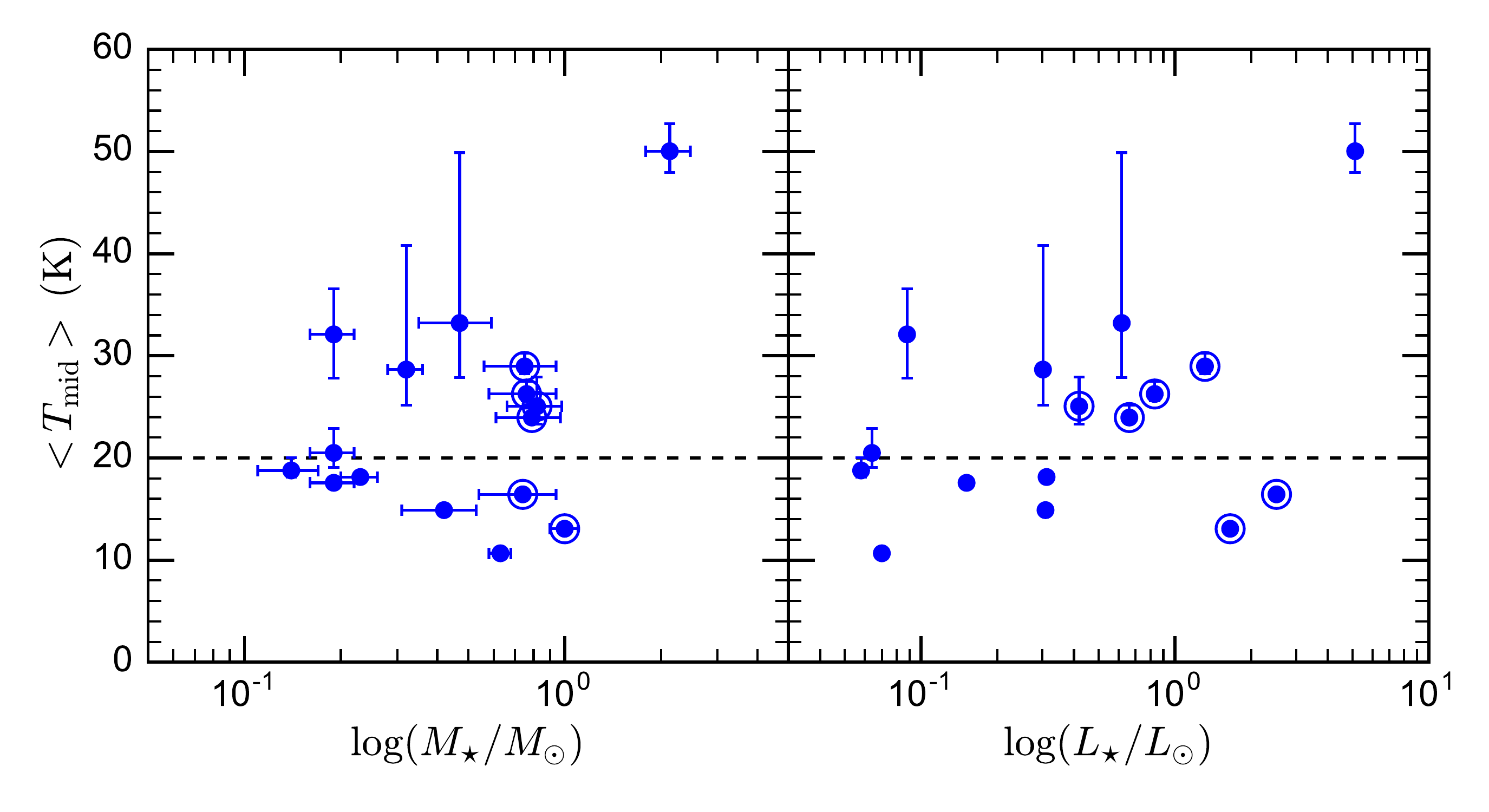}}
\caption[]{Mass-averaged midplane temperature as a function of stellar mass \textit{(left)} and luminosity \textit{(right)} for the disks for which we derive a reliable disk structure. Error bars on the y-axis reflect the distribution of models obtained from the fits. In more than half disks we find a disk temperature smaller than $T_{\mathrm{d}}=20\u{K}$ used by \citet{Ansdell:2016qf}.}
\label{fig:temp.vs.mstar}
\end{figure}

From \figref{fig:comparison.inc.mdust} it is also clear that our mass estimates are systematically larger than those by \citet{Ansdell:2016qf}. This discrepancy is not caused by a different dust opacity, as
in our fits we assumed exactly the same opacity used by \citet{Ansdell:2016qf}. Rather, we interpret the discrepancy to originate from a different assumption on the dust temperature: while \citet{Ansdell:2016qf} assumed a disk average temperature of 20\u{K} for all the disks in the sample (regardless the spectral type, the mass and the luminosity of the central star), in our fits we use a physical model based on the two layer approximation (see Section~\ref{sec:disk.model}) that takes into account the stellar properties and derives the radial profile of the midplane temperature $T_{\mathrm{mid}}(R)$ by solving the energy balance at each radius under the assumption of vertical hydrostatic equilibrium and some degree of dust settling set by the flaring parameter $f$. By checking the temperature profiles resulting from our fits, we find that in many cases (see \figref{fig:temp.vs.mstar}) the disk-averaged temperature derived by our physical model ($<T_\mathrm{mid}>$) is smaller than $20\u{K}$ (but never smaller than 7\u{K} by definition), thus explaining the tendency towards larger masses that characterizes our estimates. 
Moreover, since we do not observe a systematic trend of $T_{\mathrm{mid}}$ as a function of stellar mass and luminosity, we conclude that the assumption of a constant $T_{\mathrm{d}}$ \citep{Ansdell:2016qf} has not introduced a bias in the determination of $M_{\mathrm{d}}$. Conversely the assumptions of \citet{Andrews:2013qy} and \citet{van-der-Plas:2016lr} on the dependence of the average $T_{d}$ on the stellar parameters may introduce a spurious dependence on the derived M$_{d}$~vs. M$_{\star}$ relationship.

In the right panel of \figref{fig:comparison.inc.mdust} we compare the inclinations derived for all  disks. In all cases for which \citet{Ansdell:2016qf} provide a measurement of inclination, their estimate is in very good agreement with ours. Moreover, in many cases we are able to put a more stringent constraint on the disk inclination, as shown by the smaller error bars. The improvement in the estimate of the disk inclination is likely due to the fit procedure, which in the case of \citet{Ansdell:2016qf} is based on the CASA procedure \textit{uvmodelfit} with the assumption of a gaussian bightness profile, while in our case benefits of a more extended exploration of the parameter space.

\subsection{Distribution of surface density profiles and outer disk radii}
\begin{figure*}
\centering
\resizebox{0.32\hsize}{!}{\includegraphics{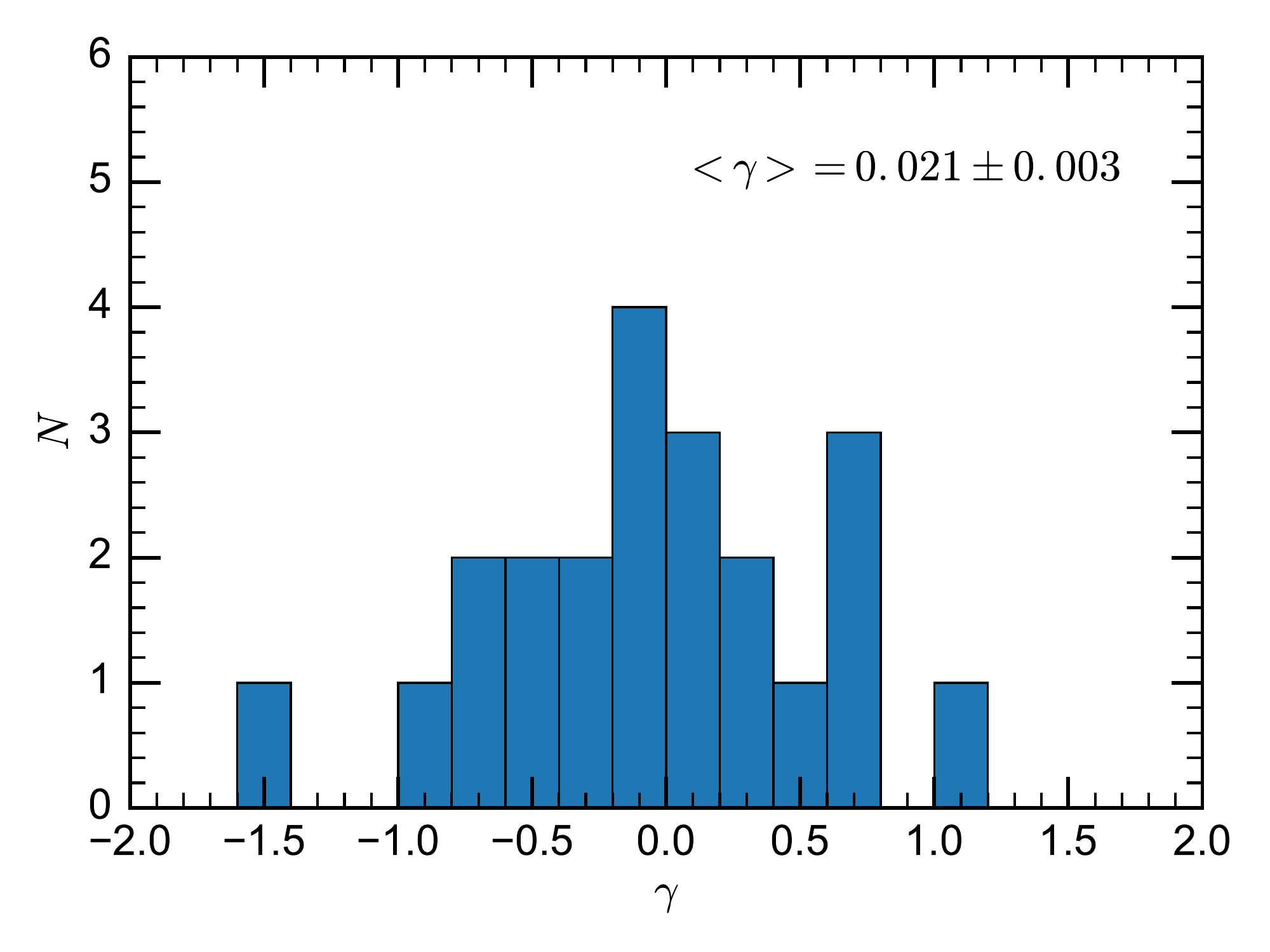}}
\resizebox{0.32\hsize}{!}{\includegraphics{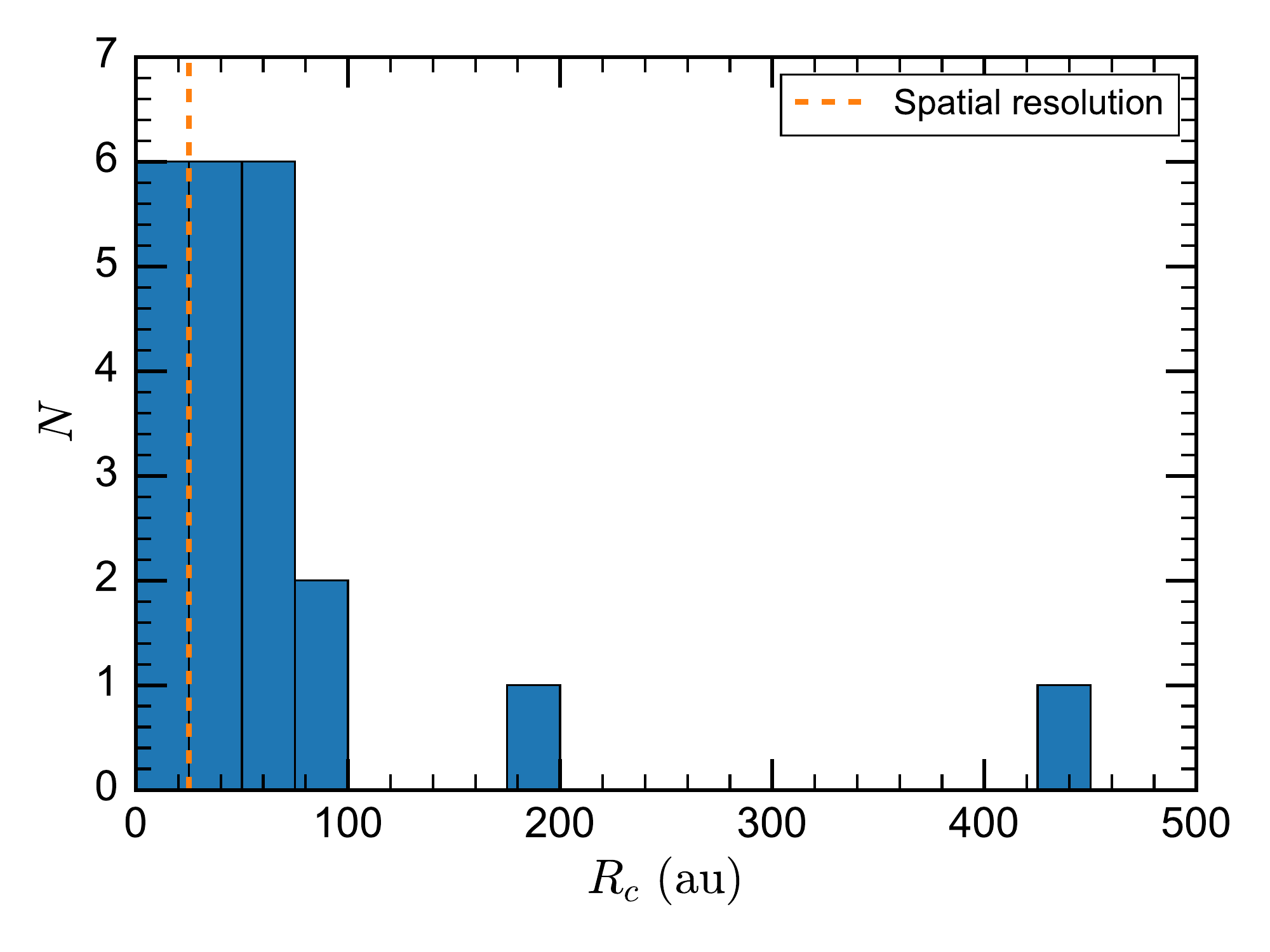}}
\resizebox{0.32\hsize}{!}{\includegraphics{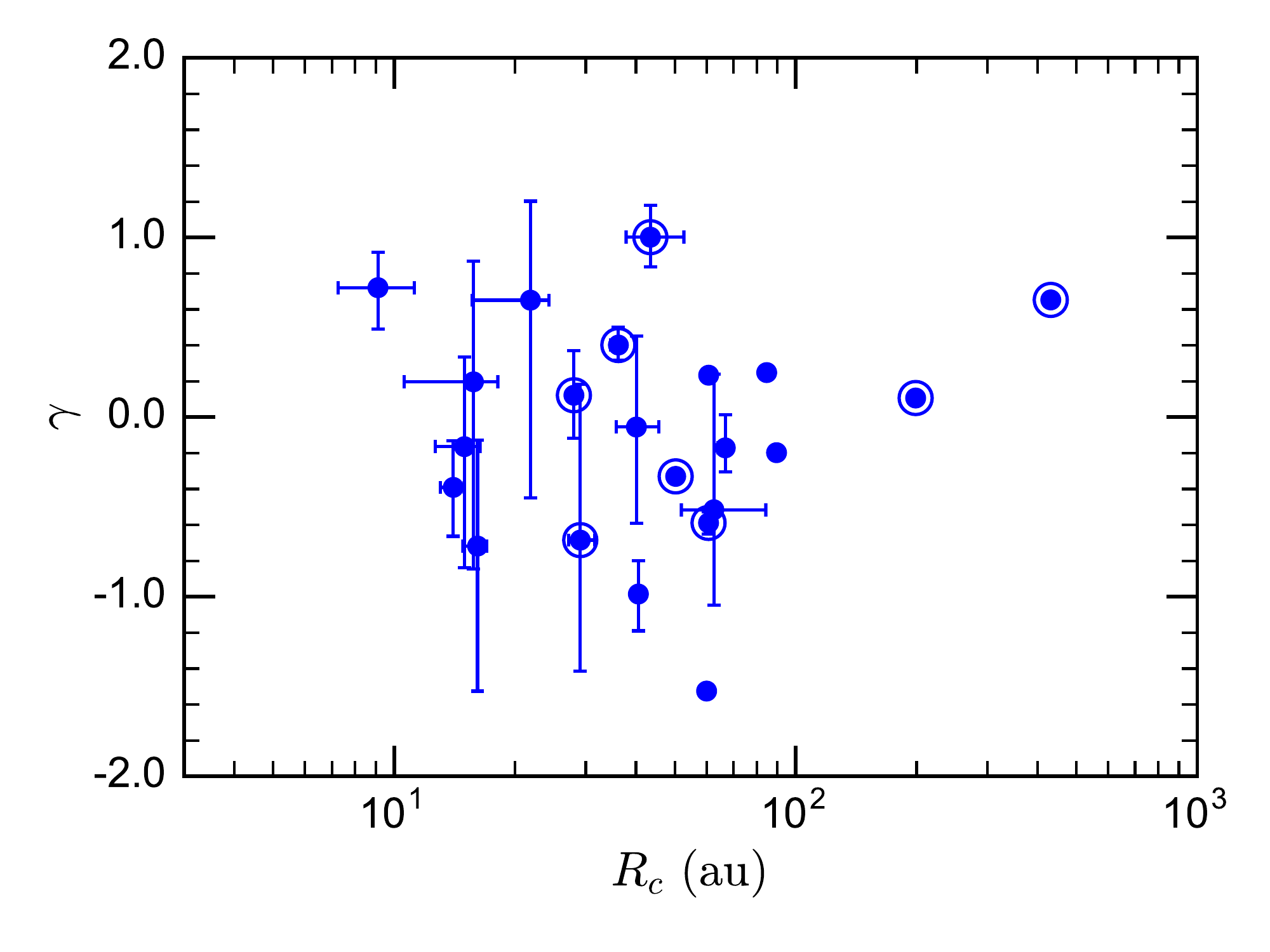}}
\caption[Distribution of parameter values obtained from the fits]{
Distribution of surface density slopes $\gamma$ \textit{(left)}, exponential cut-off radii $R_{c}$ \textit{(center)} and the correlation between them \textit{(right)} derived for the disks in the sample. The vertical dashed line in the top and middle panels gives a visual representation of the spatial resolution of the observations, estimated as half of the synthesized beam size (in our case $\sim 25\u{au}$ at a distance of 150\u{pc}). There is no apparent correlation between $\gamma$ and  $R_{c}$.
}
\label{fig:histograms.gamma.rc}
\end{figure*}

In \figref{fig:histograms.gamma.rc} we show the distribution of slopes $\gamma$ (left panel), cut-off radii $R_{c}$ (middle) and their correlation (right) obtained from our fits. 
The distribution of cut-off radii $R_{c}$ has 6 disks with $R_{c}< 25\u{au}$, 14 disks with ${25\u{au}\leq R_{c} \leq 100\u{au} }$ and two much larger disks with $R_{c}\sim 200\u{au}$ (Sz~98) and $R_{c}\sim 430\u{au}$ (IM~Lup). The distribution of disk sizes ($R_{\mathrm{out}}$, plot not shown) is similar to that of $R_{c}$: 4 disks are compatible with being unresolved (the inferred size is smaller than the spatial resolution), 14 disks have a size between 25\u{au} and 100\u{au}, and two largest disks Sz~98 and IM~Lup have a size of $\sim280\u{au}$ and $\sim500\u{au}$, respectively. Our inferred range of values of $R_{c}$, in line with the findings of \citet{Andrews:2009zr,Andrews:2010fk} in Ophiuchus ($14\u{au}\leq R_{c}\leq 200\u{au}$) and \citet{Isella:2009qy} in Taurus-Auriga ($30\u{au}\leq Rc\leq 230\u{au}$) who fitted sub-mm observations of several bright disks using an exponentially-tapered power law surface density profile as we do here. 

The distribution of $\gamma$ (\figref{fig:histograms.gamma.rc}, left panel) is centrally peaked around $\gamma=0$ and has a standard deviation of $0.6$. 
In the right panel of \figref{fig:histograms.gamma.rc} we show the distribution of $\gamma$ as a function of $R_{c}$ obtained for each disk, showing that there is not any particular trend between these two quantities. 
We notice that 10 disks are compatible with $\gamma>0$ and other 10 disks with $\gamma<0$. Interestingly, among the disks with negative $\gamma$, several of them are characterized by relatively small cut-off radii ($40-70\u{au}$), which could be a signature of a partially resolved cavity. 

These findings compare well with the trend that has been emerging in last years thanks to the significant improvement in the observational capabilities in the (sub-)mm window \citep{Williams:2011jk}: while low angular resolution observations were usually compatible with large and positive $\gamma$ values (e.g., $\langle \gamma \rangle \sim 0.9$, \citealt{Andrews:2009zr,Andrews:2010fk}; $\gamma\sim~0.9$, \citealt{Hughes:2008lr}) mostly set by the fall-off of the outer disk (the only part clearly spatially resolved by such observations), higher angular resolution observations (capable of resolving the inner 100~\u{au} region) reveal that smaller $\gamma$ values ($\langle \gamma \rangle \sim 0.1$, \citealt{Isella:2009qy};  $\gamma\sim0.1$, \citealt{de-Gregorio-Monsalvo:2013qf}) are required in order to reproduce the radial profile of the continuum brightness distribution consistently in the inner and outer disk. 

It is worth noting that the surface density profiles have been determined assuming a dust opacity $\kappa_{\lambda}$ that is constant with radius, choice that is justified by the current lack of information about the radial changes of the dust grain properties. Since a combination of grain growth and radial drift are expected to produce a size-sorting effect in the grain radial distribution \citep{Birnstiel:2010jk} with a consequent gradient in the dust opacity, we caution that the slope of $\Sigma(R)$ derived here might be shallower than the real one \citep[see the discussion in][]{Banzatti:2011ff,2013A&amp;A...558A..64T}. As shown by \citet{Tazzari:2016qy}, such degeneracy can be broken by forward-modeling spatially resolved multi-wavelength observations.

\begin{figure}[tbp]
\centering
\resizebox{\hsize}{!}{\includegraphics{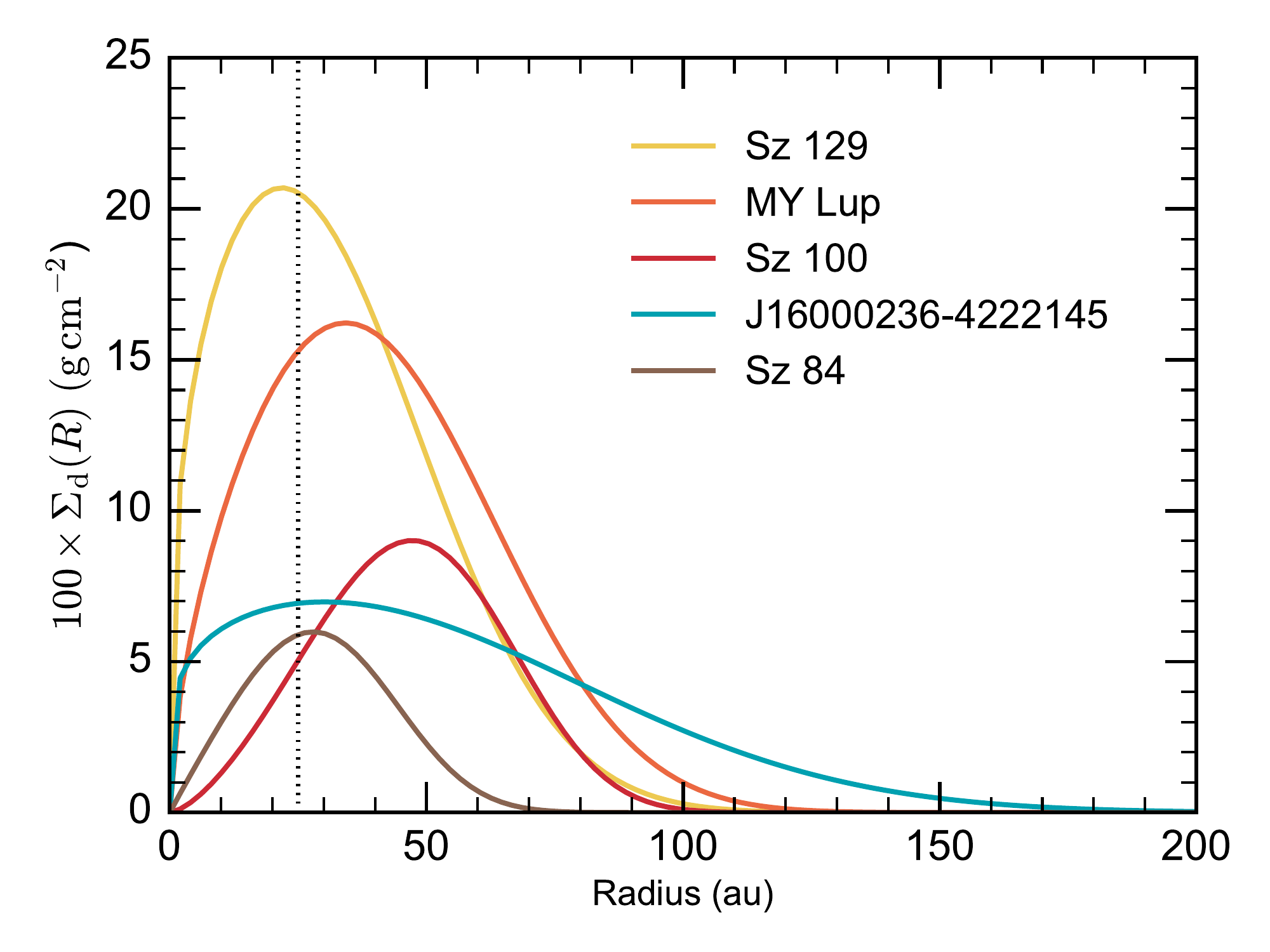}}
\caption[Surface density profiles of the five disks having $\gamma< 0$]{Surface density profiles of the five disks having $\gamma< 0$ at a significance level larger than $3\sigma$. The vertical dashed line gives a visual representation of half the beam size (25\u{au} at 150\u{pc}). }
\label{fig:surf.dens.negative.gammas}
\end{figure}

\subsection{Transition disks}
\label{sec:transition.disks}
Transition disks are protoplanetary disks that exhibit inner cavities or gaps in the distribution of their dust emission and, in some cases, also in that of the gas. In the observations by \citet{Ansdell:2016qf} three disks already classified as TD (see references below Table~\ref{tb:transition.disks}) were detected with clearly resolved gaps in the continuum emission at $\sim 50$\,au resolution (J16083070-3828268, RY~Lup, Sz~111), three other sources (Sz 123A, Sz 100 and J16070854-3914075) showed marginal evidence for cavities with diameter of 0.4'' and six sources (Sz~84, MY~Lup, Sz~112, J16011549-4152351, J16102955-3922144, and J16081497-3857145) previously classified as TD did not exhibit any cavity or hole in the synthesized maps at ~60\u{au} resolution.  

In \tbref{tb:transition.disks} we report the findings of the fits of these disks. As explained in Section~\ref{sec:sample}, we do not fit the three disks with clearly resolved gaps since the huge density depletion and the considerable size of such gaps is likely to have changed the heating of the dust in the remainder of the disk, which cannot be explained with a simple model based on the two-layer approximation. These disks will be analyzed in detail in van der Marel et al., in prep. Conversely, we apply our modeling to all the disks in the sample for which a gap is not observed in the synthesized maps: among these disks, five of them are found to have surface densities inwardly decreasing that are compatible with partially depleted gaps. 
\begin{table}
\centering
\small
\caption{Summary of fit results for disks with cavities.}
\resizebox{0.5\textwidth}{!}{
\begin{tabular}{lllll}
\hline
\hline
\multicolumn{5}{c}{Previously known TDs}\\
Name	&	$\gamma$	&	$R_{\mathrm{hole}}$ 	&	reason & ref.\\
		&				&	(au) 						&	&  \\
\hline
MY~Lup	&	$-0.59\pm0.01$		&	34						& 	& 1 \\
Sz~100	&	$-1.5\pm0.1$		&	46						& 	& 2 \\
Sz~84	&	$-1.0\pm0.2$		&	41						& 	& 3 \\
J16102955-3922144	&	$-0.5\pm0.64$		&	31			& 	& 4 \\[3mm]
\hline
\multicolumn{5}{c}{Disks with tentatively new evidence of a cavity}\\
Sz~129	&	$-0.33\pm0.02$		&	22						& 	& 5	\\
J16000236-4222145	&	$-0.20\pm0.02$		&	30			& 	& 6	\\[3mm]
\hline
\multicolumn{5}{c}{Not fitted}\\
J16083070-3828268 	&			&				& large cavity		&   2 		\\
RY~Lup	&			&							& large cavity		&	7		\\
Sz~111	&			&							& large cavity		&	2		\\
Sz~123A	&			&							& binary			&	2, 4		\\
J16070854-3914075	&			&				& edge-on			&	2, 4		\\
Sz~112	&			&							& $F_{890\mu\mathrm{m}}<4\,$mJy	 & 5 \\
J16011549-4152351	&			&				& no stellar parameters	&	5	\\
J16081497-3857145	&			&				& unresolved	&	(4)	\\
\hline
\end{tabular}
}
\begin{flushleft}
{\textbf{Notes\ } For four previously known TDs \textit{(top)} and for two disks for which we found new evidence of a cavity \textit{(middle)} we report the radial slope $\gamma$, the hole radius $R_\mathrm{hole}$ (estimated as the the $\Sigma_\mathrm{d}(R)$ peak radius). We also list the disks that we have not fitted \textit{(bottom)} with the reason for which we have been unable to fit them.}\\
{\textbf{References\ } (1) \cite{Romero:2012qf}, (2) \cite{Merin:2008xy}, (3) \cite{Merin:2010ul}, (4) \cite{Bustamante:2015ve} (5), \cite{van-der-Marel:2016aa}, (7) see discussion in \cite{Ansdell:2016qf}.}
\end{flushleft}
\label{tb:transition.disks}
\end{table}

For Sz 100, one of the three sources with possible cavities, we obtain a robust fit with $\gamma=-1.5$ that confirms the presence of an inner hole with radius $R_{\mathrm{hole}}\approx 46\u{au}$, where we defined $R_{\mathrm{hole}}$ as the radius where $\Sigma_\mathrm{d}(R)$ peaks. Unlike Sz 100, the other two sources with possible cavities Sz~123A and J16070854-3914075 were excluded from our sample, the former because it is a binary, the latter because it is edge-on. Finally, four out of the six disks classified as TD but with no evidence of cavities in the continuum maps were included in our sample: for Sz~84 and MY~Lup we derive a surface density profile with a clear hole ($\gamma=-1.0$ and $\gamma=-0.8$) located respectively at $R_{\mathrm{hole}}\approx 41\u{au}$ and $R_{\mathrm{hole}}\approx 34\u{au}$, for J16102955-3922144 the fit is more uncertain (uncertainty on $\gamma$ is large), with a marginal evidence of a cavity at $R_{\mathrm{hole}}\approx 31\u{au}$). The findings of our fits are confirmed from the fact that all these disks exhibit visibility profiles (cfr. deprojected visibility plots in in Appendix~\ref{app:fits}) whose real part tends toward (and in some cases reaches) negative values, compatible with the surface brightness of a disk with a central cavity.
The other two disks classified as TD (Sz~112 and J16011549-4152351) were excluded from our sample due to a low integrated flux (below $4\u{mJy}$) and due to the lack of stellar parameters, respectively. 

In addition to these disks, we also find evidence for the presence of holes in other two disks not classified as TD (J16000236-4222145 and Sz 129) for which we find robust estimates of negative $\gamma$ values, respectively $\gamma=-0.20\pm 0.02$ and $\gamma=-0.33\pm 0.02$, and hole sizes of $R_{\mathrm{hole}}=30\u{au}$ and $R_{\mathrm{hole}}=22\u{au}$ (comparable with the spatial resolution of the observations). The surface density profiles corresponding to such $\gamma$ values imply the presence of inner holes but the depletion factor inside $R_{\mathrm{hole}}$ is expected to be not as high as for $\gamma\leq -1$. We thus conclude that for these two disks the evidence for an inner hole is tentative and to be confirmed with higher angular resolution observations. For a visual representation of the surface density profiles of these disks, see \figref{fig:surf.dens.negative.gammas}. 

\section{Discussion}
\label{sec:discussion}

\subsection{Disk mass-size relation}

During the evolution of a protoplanetary disk, the spatial distribution of its mass and angular momentum  change dramatically. While the material gets accreted onto the star, the disk mass is a decreasing function of time, while the disk size may increase as well as remain constant, depending on the mechanism driving the angular momentum redistribution. While a viscosity-driven disk would increase in size as a result of the diffusive evolution \citep{1974MNRAS.168..603L,Pringle:1981uq}, if the disk angular momentum is lost through MHD winds, then partial suppression of the disk spreading can take place \citep{Armitage:2013vn, Bai:2013yq, Bai:2016kx, Suzuki:2014rt}. In the previous Section we have constrained the mass and the size of 22 Lupus disks and here we use such measurements to gain an insight on their evolutionary stage. 

Recent (sub-)mm observations of protoplanetary disks at an angular resolution  high enough to resolve their spatial structure ($\simless0.75^{\prime\prime}$ for nearby SFRs) seem to suggest that fainter disks are also more compact \citep{Andrews:2010fk,Pietu:2014bh,Andrews:2015pd}. Since at sub-mm wavelengths the disk emission is substantially optically thin, such trend can be interpreted in terms of a disk size - disk mass correlation. 
\begin{figure}
\centering
\resizebox{\hsize}{!}{\includegraphics{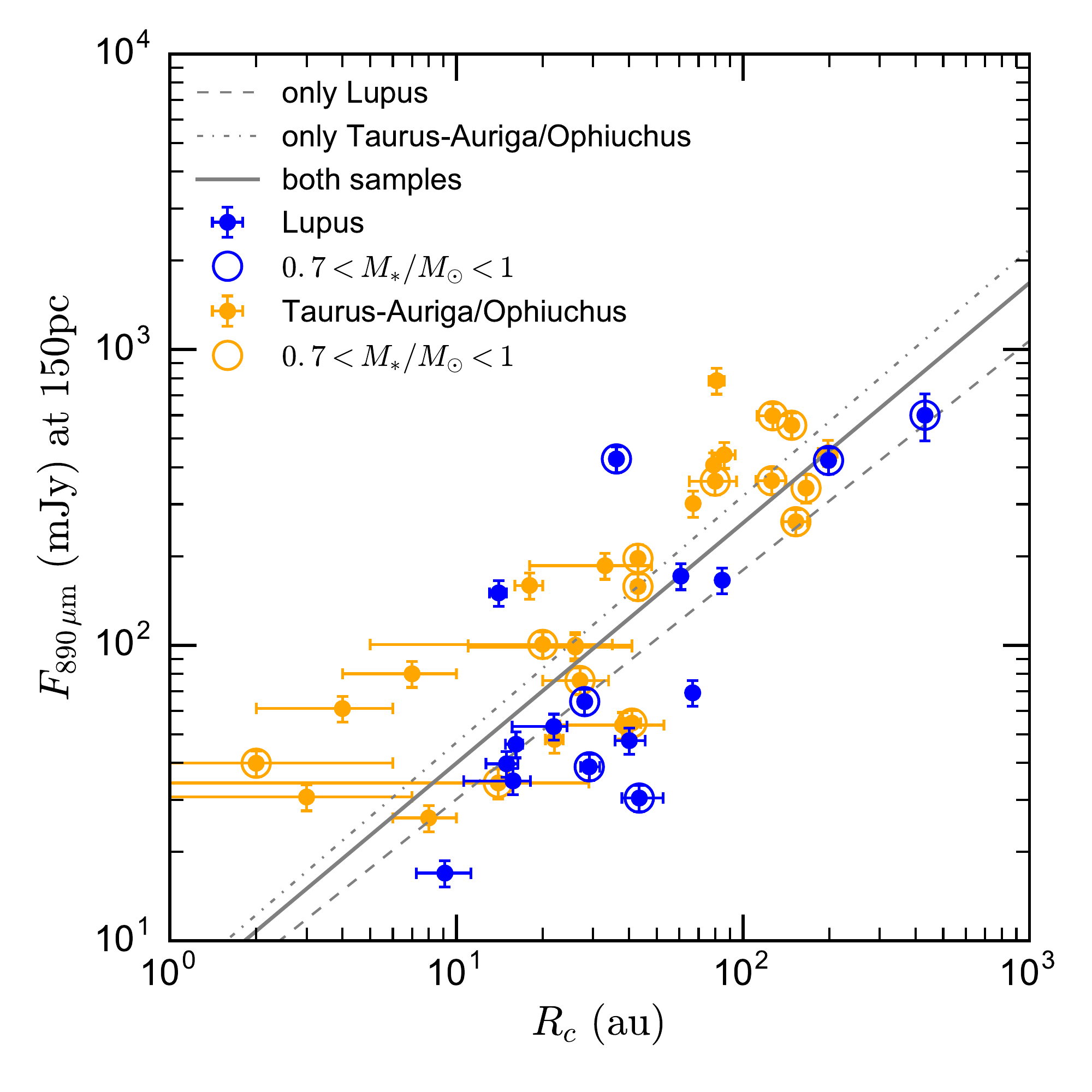}}
\caption{890$\mu$m integrated flux (normalized to a common distance of 150\u{pc}) as a function of exponential cut-off radius, constrained by fitting spatially resolved observations with angular resolution better than 0.75''. For some disks, the 890$\mu$m flux was extrapolated from the measured ${1.3\u{mm}}$ flux assuming an average spectral index $\alpha=3.0$. Yellow elements are Taurus-Auriga/Ophiuchus disks from the \citet{Andrews:2015pd} compilation; blue elements are Lupus disks analyzed in this paper. Gray lines represent the correlations found by the Bayesian linear regression for the Taurus-Auriga/Ophiuchus disks (dash-dotted), for the Lupus disks (dashed), and for both the samples taken together (solid).
}
\label{fig:flux.vs.rc}
\end{figure}
In \figref{fig:flux.vs.rc} we show the 890\u{$\mu$m} integrated flux (normalized to a common distance of 150\u{pc}) as a function of cut-off radius for the sample of Taurus-Auriga/Ophiuchus disks (yellow points). The data have been collected from previous observations with angular resolution better than 0.75'' from \citet{Andrews:2009zr,Andrews:2010fk,Isella:2010dq, 2011A&amp;A...529A.105G, Pietu:2014bh} and considers only \textit{full} disks (binaries and transition disks are excluded). 
We note that the size measurements of this sample have been determined homogeneously by fitting the spatially resolved observations with an exponentially tapered power-law profile like the one in Eq.~\ref{eq:surface.density} and correspond to the exponential cut-off radius $R_c$.
For some disks in Taurus-Auriga we extrapolated the flux at 890\u{$\mu$m} from the observed 1.3~mm flux using an average spectral index $\alpha=3.0$ that corresponds to moderate grain growth. We note that this choice might have a minor effect in the inferred correlation, and future spatially resolved observations at several sub-mm/mm wavelengths will help removing this assumption.

To test the presence of a size-mass correlation in the Taurus-Auriga/Ophiuchus sample, we perform a linear regression using the Bayesian algorithm by \cite{Kelly:2007lq}\footnote{We use the \texttt{linmix} Python package which implements \citet{Kelly:2007lq} and is available here: \texttt{https://github.com/jmeyers314/linmix}.} which allows the uncertainties on both axes to be included in the computation. For the present case we assume uniform priors on the correlation coefficients and we take the medians of the marginalized posterior as their best-fit values (see full posterior in Appendix~\ref{app:fits}). We obtain the following correlation:
\begin{equation}
\label{eq:corr.taurus}
\log F_{890\,\mu\mathrm{m}} = (0.8\pm0.1)\log R_{c} + (0.84\pm0.2)\,,
\end{equation}
with a correlation coefficient of $0.86\pm 0.06$ and a standard deviation of $0.06\pm 0.02$.
Considering this sample alone, it is difficult to understand whether the observed correlation represents an evolutionary sequence or rather reflects the time evolution of disks with intrinsically different viscous timescales and/or initial disk conditions. Indeed, the size-mass measurements of the Taurus-Auriga/Ophiuchus disks in \figref{fig:flux.vs.rc} give us a snapshot of their structure at a given moment of their evolution, and the large uncertainties in the relative ages of the stars in the sample do not allow us to distinguish between these three possible causes. 

In order to shed light on the observed size-mass correlation, in \figref{fig:flux.vs.rc} we add the results of this study, reporting the size measurements of 16 (out of 22 analyzed) Lupus disks. Since the Taurus-Auriga/Ophiuchus samples consider only full disks and exclude binaries and transition disks, we removed from the plotted Lupus sample the six disks with cavities described in the previous Section. In terms of stellar spectral types the Taurus-Auriga/Ophiuchus and the Lupus samples are not identical, but comparable: the former sample is mostly made of stars between K5 to M1 spectral types, while the latter  extends to slightly later types, mostly between K7 to M5 types. 

The resulting luminosity-size plot in \figref{fig:flux.vs.rc} shows that, for a given integrated (sub-)mm flux, the Lupus disks tend to be slightly larger than the disks in Taurus-Auriga/Ophiuchus,  with the exception of two disks, Sz~68 and Sz~83, which appear smaller than the average Lupus disks with a comparable flux. This evidence seems consistent throughout the range of cut-off radii between 10 and 200\u{au}. We note that the Lupus disks in the $0.7-1.0\Msun$ mass range (circled blue dots) in which our sample is complete appear to be distributed randomly, with no signs of a particular correlation. 
If we apply to the Lupus sample the same linear regression used above, we find that also the Lupus disks exhibit a similar size-mass correlation:
\begin{equation}
\log F_{890\,\mu\mathrm{m}} = (0.8\pm 0.2)\log R_{c} + (0.7\pm 0.3)\,,
\end{equation}
with correlation coefficient of $0.73\pm0.15$ and a standard deviation of $0.12\pm0.05$. The slopes of the two linear correlations are similar within the uncertainties, both suggesting that larger disks appear brighter and smaller disks appear fainter. The Lupus correlation results  systematically \textit{below} the Taurus-Auriga/Ophiuchus one,  confirming that the Lupus disks tend to be  larger and fainter than the Taurus-Auriga/Ophiuchus ones. In this study we assumed a distance of 200\,pc for the disks in the Lupus~III cloud, however recent measurements of the \emph{Gaia} space telescope suggest they could be closer, at a distance of $\sim 150\,$pc, which would make them even fainter.

\begin{figure}
\centering
\resizebox{\hsize}{!}{\includegraphics{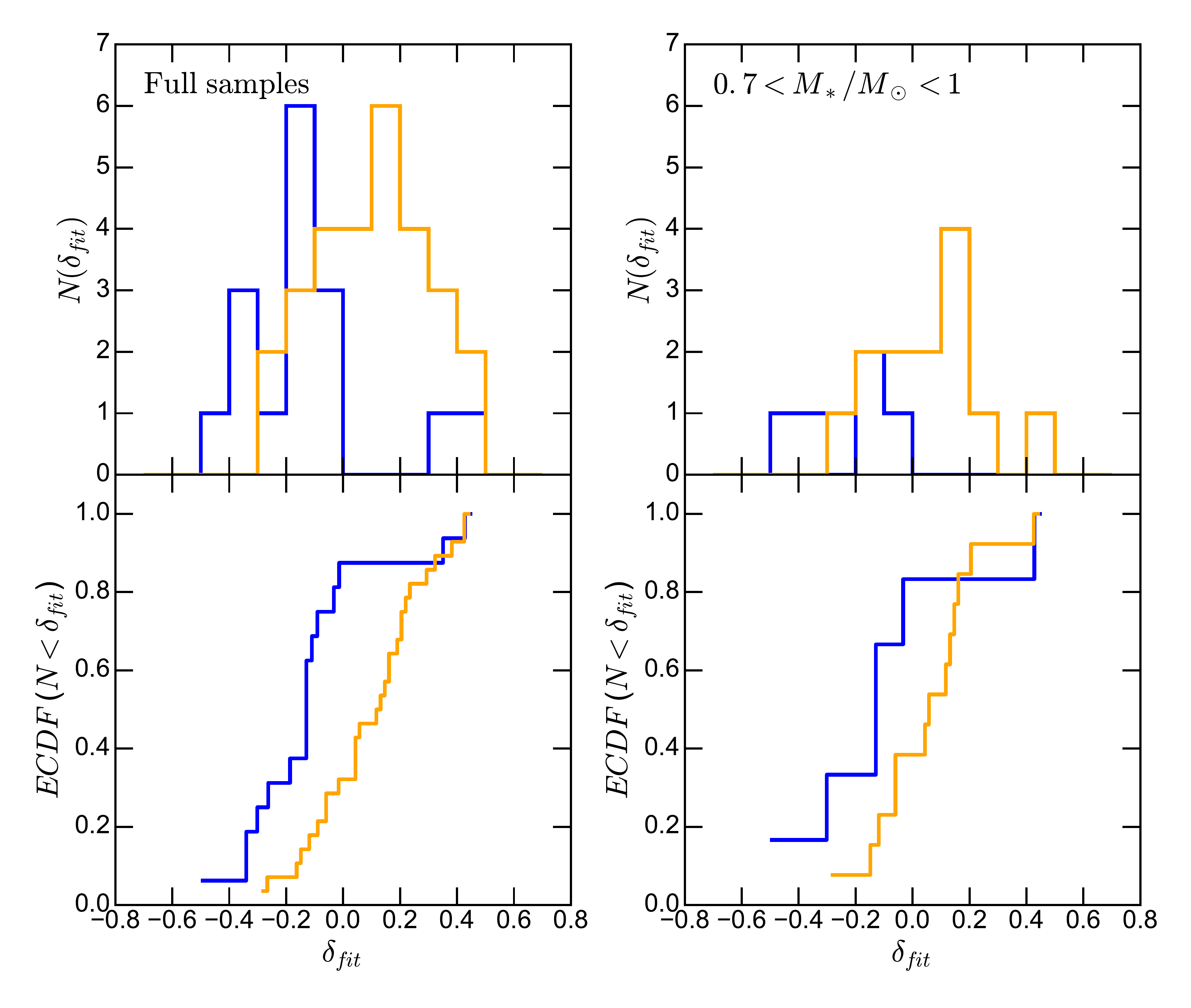}}
\caption{Top panels: distribution of the parameter $\delta_{fit}$ (see text) for the Lupus (blue) and Taurus-Auriga/Ophiuchus (yellow) samples. In the right panel we show the full samples, while in the right one we restrict to the range $0.7 \le {\rm M}_\star/{\rm M}_\odot\le 1$. Bottom panels: empirical cumulative distribution functions for the values of $\delta_{fit}$ for the same samples as in the upper panels.}
\label{fig:flux.vs.rc.adtest}
\end{figure}

In order to test whether the Lupus disks are effectively less massive and more extended than the Taurus-Auriga/Ophiuchus ones, we need to check whether the two samples are distinguishable along a direction perpendicular to the inferred average correlation. To do that, we proceed as follows. First, we consider the two samples as a single sample and we perform a linear regression (same Bayesian algorithm used above), finding a correlation:
\begin{equation}
\label{eq:corr.both.samples}
\log F_{890\,\mu\mathrm{m}} = (0.8\pm 0.2)\log R_{c} + (0.8\pm 0.2)\,,
\end{equation}
with correlation coefficient of $0.80\pm 0.06$ and a standard deviation of $0.08\pm 0.02$. As expected, the correlation results parallel to those derived for the two samples considered separately and with a vertical offset intermediate between them. 
Then, for each disk, we computed the distance from the fit $\delta_{fit}$, which is defined to be positive for disks more massive and smaller than the linear relationship in Eq.~\ref{eq:corr.both.samples}, and negative for disks that are less massive and larger.
In Fig.~\ref{fig:flux.vs.rc.adtest} (upper panels) , we show the distributions of ($\delta_{fit}$) for the two samples  of Lupus (blue histogram) and Taurus-Auriga/Ophiuchus (yellow histogram); the left plot is for the full sample, while the right one is for the subsample of stars between 0.7 and 1~M$_\odot$.
In the bottom panels of Fig.~\ref{fig:flux.vs.rc.adtest} we show the empirical cumulative distribution functions (ECDF) of the two samples.
We performed an Anderson-Darling (AD) two-sample test\footnote{We used the implementation of the test as provided in {\tt scipy.stats.anderson\_ksamp}, which is based on \citet{CIS-74828}} to check whether the two populations in Lupus and in Taurus-Auriga/Ophiuchus are consistent with being drawn from the same distribution of $\delta_{fit}$. The AD test on the full sample gives a very low probability that the two samples are drawn by the same parent distribution ($\le 0.2$\%). When we restrict to the sample in the $0.7 \le {\rm M}_\star/{\rm M}_\odot\le 1$ range, then the null hypothesis cannot be excluded (p$\sim$8\%).

\subsection{Evidence for viscous evolution?}
Considering that the typical disk dispersal time scale is 5-10 Myr  and that the Lupus SFR is slightly older (by $~1-2$ Myr) than the Taurus-Auriga and Ophiuchus ones \citep{Hernandez:2007fk, Fedele:2010fj}, viscous evolution could be a candidate mechanism to explain the systematic difference found in the two disk populations. 
Indeed, in the context of viscously evolving disks, while the inner disk material accretes onto the star, the outer disk radius spreads outwards. For the typical self-similar solutions adopted in this study (Eq.~\ref{eq:surface.density}), the disk mass and the exponential cut-off radius evolve such that, at any given time, $M_{\mathrm{disk}}\propto R_{c}^{-1/2}$, where $M_{\mathrm{disk}}$ is the total disk mass. As a result, the total disk mass decreases with time, while the disk size increases, compatible with the difference that we observe between the Lupus and the Taurus-Auriga/Ophiuchus disks in \figref{fig:flux.vs.rc}. 

We note that while a viscously evolving disk model would explain the direction of the mass-size offset between the Lupus and the Taurus-Auriga/Ophiuchus populations in Figure~\ref{fig:flux.vs.rc} (with older disks being larger and fainter), estimating whether the extent of such offset is compatible with the age difference remains problematic as it would require making assumptions on the viscosity and on the evolutionary stage (early or late phase w.r.t the viscous time scale) at least of one of the two populations. 

\subsubsection{Disk mm fluxes: caveats}
Viscous disk evolution may not be the only mechanism that could produce a time evolution in the luminosity-size plot \figref{fig:flux.vs.rc}. Other processes such as disk photoevaporation, grain growth, radial drift and planet formation itself might induce time changes in the observed integrated sub-mm flux and dust outer edge. As an example, the generation of pressure maxima in the gas disk (e.g., induced by tidal interaction with forming planets) would substantially change the growth efficiency (and therefore the opacity) and migration rate of disk solids w.r.t. to a simpler scenario of a full disk with continuous mass distribution and a radially decreasing pressure profile. However, a detailed estimate of the signatures of these effects on the luminosity-size plot would require accurate numerical simulations of global disk evolution, which is beyond the scope of this paper. The current observations at 890$\mu$m do not allow us to rule out these alternative processes. However, in the future, the application of a homogeneous analysis to observations of disks in other star forming regions with different mean ages will help in gaining a global view of disk evolution. In addition, the collection of observations at multiple sub-mm/mm wavelengths would reveal the grain growth level in these disks \citep{Tazzari:2016qy}, providing important constraints on the dust emissivity that could be used to break the degeneracies between the different scenarios.

Theoretically the self-similar solutions (and their diffusive behaviour) characterize the evolution of the gaseous component, while in this work we have constrained disk masses and sizes from the dust continuum emission. It is thus important to understand what are the potential biases in our estimates when comparing to the theoretical expectations.
Regarding the disk masses, recent work by \citet{Manara:2016lr} on the same Lupus data set found a  correlation between the disk masses (inferred from the dust continuum emission) and the mass accretion rates onto the star (derived from UV excess measurements), in agreement with the theoretical expectations of viscous theory \citep{Hartmann:1998qy} for which $M_{\mathrm{disk}}/\dot M_{\mathrm{acc}}\approx t_{\mathrm{age}}$. This broad correlation suggests that, despite the decoupling of the dust and gas in disks, the mm-continuum observations that we have discussed here provide a useful proxy of the total disk mass. 

\subsubsection{Outer radius: caveats}
In this work we derived disk sizes assuming that the 890$\mu$m dust continuum emission is tracing most of the disk material, namely that the dust is co-located to the gas, while in general this might not be the case. At this sub-mm wavelength the observations are mostly sensitive to the thermal emission of the large mm-sized grains and therefore might not be recovering the full spatial extent of the smaller dust particles. 
Indeed, while the small dust particles are tightly coupled to the gas and therefore are expected to be brought to large radii from the gas viscous spreading (in this respect, they would be good tracers of the gas distribution), as soon as they grow they become less coupled to the gas and start being subject to the inward pointing radial drift \citep{Weidenschilling:1977lr,Brauer:2008kq}. 
The cut-off radius of the 890$\mu$m dust continuum distribution is therefore a result of these two effects and it is not trivial to assess which one is dominating as they both depend on the grain size as well as on local gas properties (density, pressure, turbulence) that change with time. 

\cite{Birnstiel:2014zr} demonstrated that the combination of radial drift and viscous gas drag generates a sharp cut-off at the outer edge of the dust distribution already in the very early phases of disk evolution (before grain growth has had time to take place) and that such a feature is preserved throughout the disk viscous evolution. They also showed that in the late phases of disk evolution the dust outer edge follows the gas edge, which is roughly a factor 1.5x larger. This globally supports our finding of $\gamma\sim 0$ in many cases (i.e. disks with flat  interior and a sharp cut-off) but is still unclear how the \cite{Birnstiel:2014zr} results translate for a population of mm-sized grains. However, a detailed study of the time evolution of the outer edge of different grain size populations in viscously evolving disks is still missing.

\subsubsection{Comparison with the minimum mass Solar nebula}
\begin{figure*}
\centering
\resizebox{0.49\hsize}{!}{\includegraphics{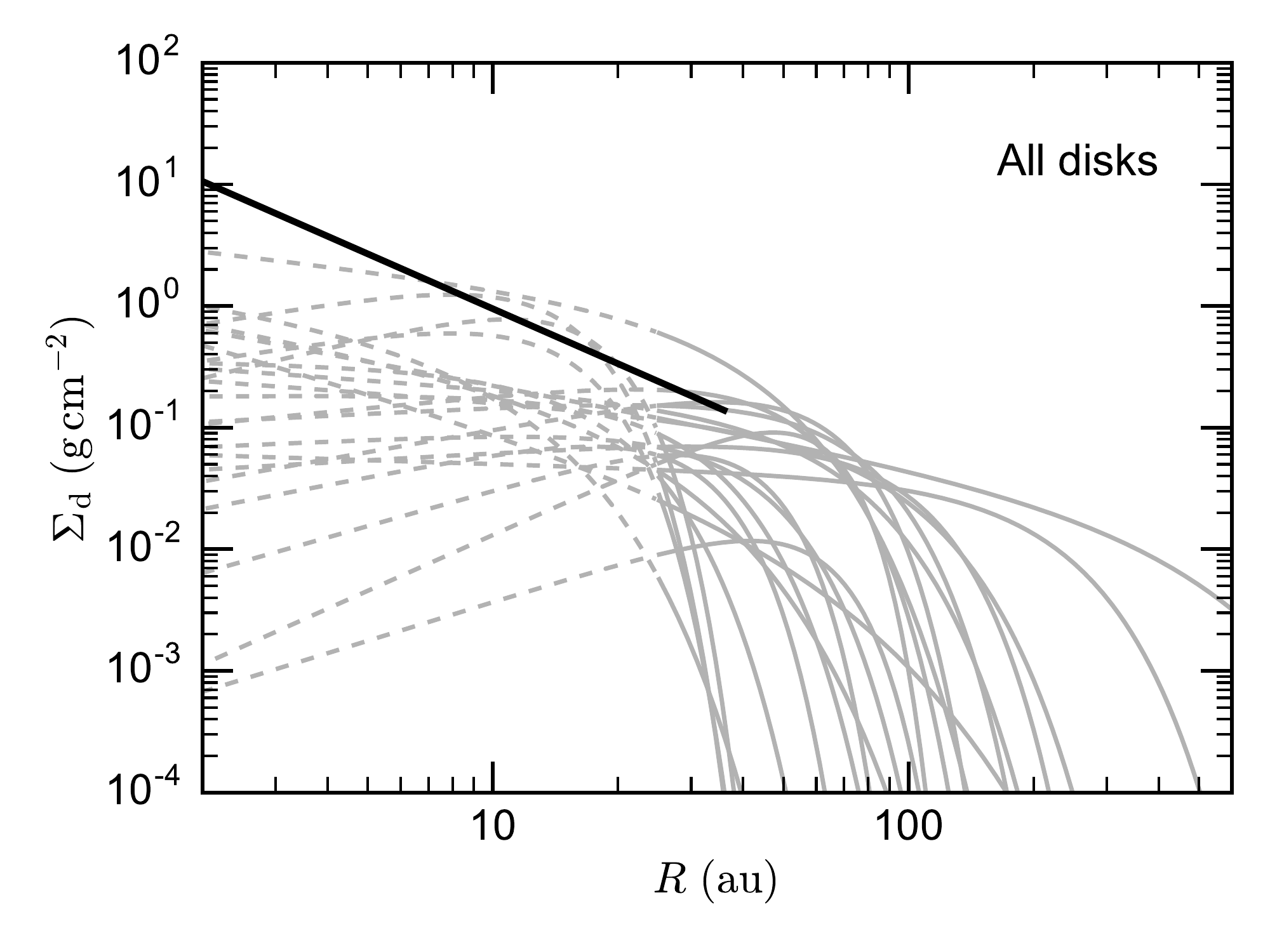}}
\resizebox{0.49\hsize}{!}{\includegraphics{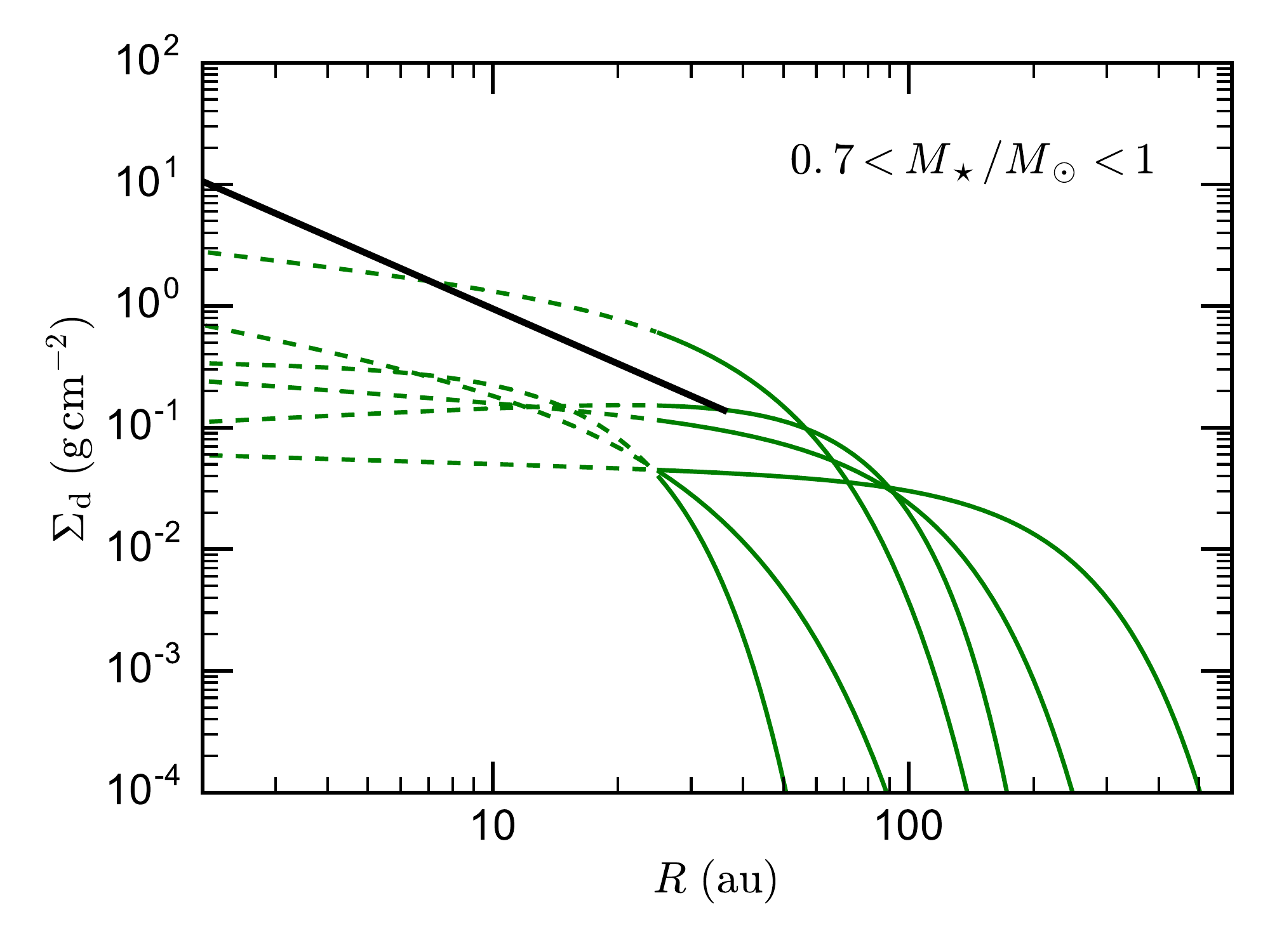}}\\
\caption{\textit{(left)}: Dust surface density profiles inferred for all the disks in our Lupus sample (gray curves) compared to the \citet{Hayashi:1981qy} dust surface density model for the Solar System (black thick line). The inferred surface density profiles are dashed where they are not spatially resolved, i.e. for $R<25\u{au}$. \textit{(right)}: Dust surface density profiles for the sources in the 0.7-1$\Msun$ stellar mass bin. 
}
\label{fig:surf.dens.hayashi}
\end{figure*}

In \figref{fig:surf.dens.hayashi} we collect the dust surface density profiles obtained for the fitted disks. In the left plot we report all the inferred profiles, in the right plot only those of the disks orbiting stars with mass in the range $0.7<M_\star/M_\odot<1$. We compare these profiles with the surface density in solids of the minimum mass solar nebula (MMSN, \citealt{Weidenschilling:1977lr}) which estimates the dust surface density profile of the Solar System primordial disk. In particular, we use the MMSN normalization by \citet{Hayashi:1981qy}.

The surface density profiles of the Lupus disks appear generally less massive than the MMSN, only a few of them having a comparable or larger mass. We also note that the MMSN profile is smaller in size and has a steeper profile than the Lupus disks: considering that the Lupus disks are still young by planet formation time scales, this might suggest that the Solar System disk was probably much more compact, or that the planets (or even the planetesimals) migrated very far inwards.
Following the fact that the Lupus disks orbit stars with $M_\star<M_\odot$, it is also possible that the difference with the MMSN  reflects an intrinsically lower initial disk mass distribution or a different - mass-dependent - time evolution of the disk surface density.
For comparison, the surface density profiles of the disks in Taurus-Auriga/Ophiuchus \citep{Andrews:2015pd} result more massive than the MMSN, in many cases by a factor between 4 and 10. This might in part be due to the bias towards higher masses of the Taurus-Auriga/Ophiuchus sample, or might be another signature that the Lupus disks are more evolved - and therefore dust depleted - than those in Taurus-Auriga/Ophiuchus.   

\section{Conclusions}
\label{sec:conclusions}
In this paper we have analyzed the 890\u{$\mu$m} continuum emission of 22 disks in the Lupus SFR which have been observed with ALMA at $\sim$0.3'' ($\sim$50\u{au}) resolution \citep{Ansdell:2016qf}. 
\begin{enumerate}
\item 
We fit the spatially resolved disk continuum emission with a self-consistent disk model based on the two layer approximation and a realistic dust opacity computation. The fits are performed directly in the uv-plane.
\item
For each disk, we derive the dust surface density profile $\Sigma_{\mathrm{d}}(R)$, the midplane temperature profile $T_{\mathrm{mid}}(R)$, and we constrain the disk inclination $i$ and position angle $PA$. The disk (gas + dust) surface density profile is inferred with a constant gas-to-dust ratio of 100.
\item
The disk masses are computed by integrating $100\times \Sigma_{\mathrm{d}}(R)$ between an inner radius of 0.1\u{au} and an outer radius $R_\mathrm{out}$ that contains 95\% of the continuum flux. The masses are compatible within a factor of 2 with those derived by \citet{Ansdell:2016qf} assuming a constant $T_\mathrm{d}=20\u{K}$. Our masses tend to be slightly larger as our disk-averaged temperatures $<T_\mathrm{mid}>$ are generally smaller than 20\u{K}. We observed no trend between $T_\mathrm{mid}$ and the stellar mass and luminosity.
\item
The surface density profiles assumed self-similar solution of viscously evolving disks. The average radial slope for the sample of 22 disks is $<\gamma> = 0\pm 0.6$, calling for a flat disk interior and sharp outer disk edges. 18 out of 22 disks have cut-off radii $R_c<75\u{au}$. We observe no correlation between $\gamma$ and $R_c$.
\item
The spatially resolved continuum emission of almost all the disks is reproduced very well (residuals less than $3\sigma$) by a smooth, monotonically decreasing, self-similar $\Sigma_\mathrm{d}(R)$, except for Sz~98 and IM~Lup, for which we find a ring-like residual emission with flux $\lesssim 15\%$ of the integrated flux.
\item
We find a correlation between the sub-mm integrated flux and the cut-off radius inferred for the disks, which can be interpreted in terms of a disk mass-size correlation. A similar correlation was claimed for disks in the slightly younger regions of Taurus-Auriga/Ophiuchus. We observe that the Lupus disks appear generally larger and fainter (less massive) than those in Taurus-Auriga/Ophiuchus. The spreading of the disk material (gas and subsequently dust) induced by the viscous evolution can be a possible explanation of such difference but we could not rule out other processes that might be at play.
\item
The $\Sigma_\mathrm{d}$ profiles inferred for the Lupus disks are generally less massive than the Solar System protoplanetary disk. Also, the Lupus disks around stars with mass $\sim 1\,M_\odot$ have a shallower  $\Sigma_\mathrm{d}$ and a larger outer radius than the Solar System protoplanetary disk, possibly suggesting that the planets/planetesimals in the Solar nebula migrated very far inward during or after their formation. 
\end{enumerate}

\begin{acknowledgements}
MT and LT acknowledge support by the DFG cluster of excellence
Origin and Structure of the Universe (\href{http://www.universe-cluster.de}{www.universe-cluster.de}). This work has been supported by the DISCSIM project, grant agreement 341137 funded by the European Research Council under ERC-2013-ADG. MT and LT thank Eva Wirstr\"om, the Gothenburg Centre for Advanced Studies in Science and Technology and the participants of the workshop \textit{Origins of Habitable Planets} hosted at Chalmers (Gothenburg) in June 2016 where considerable part of this work was fruitfully discussed. MT is grateful to Cathie Clarke for insightful discussions.
The fits have been carried out on the computing facilities of the Computational Center for Particle and Astrophysics (C2PAP) as part of the approved project ``Dust evolution in protoplanetary disks''. MT acknowledges usage of the Max Planck Gesellschaft Hydra computing cluster for code testing, for which is grateful to Paola Caselli. Figures have been generated using the Python-based \texttt{matplotlib} package \citep{Hunter:2007fk}. Staircase plots of PDFs have been generated with a user-modified version of the Python-based \texttt{triangle} package \citep{dan_foreman_mackey_2014_11020}. This research has made use of the SIMBAD database, operated at CDS, Strasbourg, France.
This work was partly supported by the Italian Ministero dell\'\,Istruzione, Universit\`a e Ricerca through the grant Progetti Premiali 2012 -- iALMA (CUP C52I13000140001). AN acknowledges funding from Science Foundation Ireland (Grant 13/ERC/I2907). CFM acknowledges an ESA Research Fellowship. This paper makes use of the following ALMA data: ADS/JAO.ALMA\#2013.1.00220.S, ADS/JAO.ALMA\#2013.1.00226.S. ALMA is a partnership of ESO (representing its member states), NSF (USA) and NINS (Japan), together with NRC (Canada), NSC and ASIAA (Taiwan), and KASI (Republic of Korea), in cooperation with the Republic of Chile. The Joint ALMA Observatory is operated by ESO, AUI/NRAO and NAOJ

\end{acknowledgements}

\bibliographystyle{aa}
\bibliography{./mt_disks}

\begin{thebibliography}{80}
\expandafter\ifx\csname natexlab\endcsname\relax\def\natexlab#1{#1}\fi

\bibitem[{{Adams} {et~al.}(1987){Adams}, {Lada}, \& {Shu}}]{Adams:1987jk}
{Adams}, F.~C., {Lada}, C.~J., \& {Shu}, F.~H. 1987, \apj, 312, 788

\bibitem[{{Alcala'} {et~al.}(2016){Alcala'}, {Manara}, {Natta}, {Frasca},
  {Testi}, {Nisini}, {Stelzer}, {Williams}, {Antoniucci}, {BIazzo}, {Covino},
  {Esposito}, {Getman}, \& {Rigliaco}}]{Alcala:2016ys}
{Alcala'}, J.~M., {Manara}, C.~F., {Natta}, A., {et~al.} 2016, ArXiv e-prints
  [\eprint[arXiv]{1612.07054}]

\bibitem[{{Alcal{\'a}} {et~al.}(2014){Alcal{\'a}}, {Natta}, {Manara}, {Spezzi},
  {Stelzer}, {Frasca}, {Biazzo}, {Covino}, {Randich}, {Rigliaco}, {Testi},
  {Comer{\'o}n}, {Cupani}, \& {D'Elia}}]{Alcala:2014uq}
{Alcal{\'a}}, J.~M., {Natta}, A., {Manara}, C.~F., {et~al.} 2014, \aap, 561, A2

\bibitem[{{Alibert} {et~al.}(2011){Alibert}, {Mordasini}, \&
  {Benz}}]{Alibert:2011aa}
{Alibert}, Y., {Mordasini}, C., \& {Benz}, W. 2011, \aap, 526, A63

\bibitem[{{ALMA~Partnership} {et~al.}(2015){ALMA~Partnership}, Brogan,
  P{\'e}rez, Hunter, Dent, Hales, Hills, Corder, Fomalont, Vlahakis, Asaki,
  Barkats, Hirota, Hodge, Impellizzeri, Kneissl, Liuzzo, Lucas, Marcelino,
  Matsushita, Nakanishi, Phillips, Richards, Toledo, Aladro, Broguiere, Cortes,
  Cortes, Espada, Galarza, Garcia-Appadoo, Guzman-Ramirez, Humphreys, Jung,
  Kameno, Laing, Leon, Marconi, Mignano, Nikolic, Nyman, Radiszcz, Remijan,
  Rod{\'o}n, Sawada, Takahashi, Tilanus, Vilaro, Watson, Wiklind, Akiyama,
  Chapillon, de~Gregorio-Monsalvo, Francesco, Gueth, Kawamura, Lee, Luong,
  Mangum, Pietu, Sanhueza, Saigo, Takakuwa, Ubach, van Kempen, Wootten,
  Castro-Carrizo, Francke, Gallardo, Garcia, Gonzalez, Hill, Kaminski, Kurono,
  Liu, Lopez, Morales, Plarre, Schieven, Testi, Videla, Villard, Andreani,
  Hibbard, \& Tatematsu}]{2041-8205-808-1-L3}
{ALMA~Partnership}, A., Brogan, C.~L., P{\'e}rez, L.~M., {et~al.} 2015, The
  Astrophysical Journal Letters, 808, L3

\bibitem[{{Andrews}(2015)}]{Andrews:2015pd}
{Andrews}, S.~M. 2015, \pasp, 127, 961

\bibitem[{{Andrews} {et~al.}(2013){Andrews}, {Rosenfeld}, {Kraus}, \&
  {Wilner}}]{Andrews:2013qy}
{Andrews}, S.~M., {Rosenfeld}, K.~A., {Kraus}, A.~L., \& {Wilner}, D.~J. 2013,
  \apj, 771, 129

\bibitem[{{Andrews} {et~al.}(2009){Andrews}, {Wilner}, {Hughes}, {Qi}, \&
  {Dullemond}}]{Andrews:2009zr}
{Andrews}, S.~M., {Wilner}, D.~J., {Hughes}, A.~M., {Qi}, C., \& {Dullemond},
  C.~P. 2009, \apj, 700, 1502

\bibitem[{{Andrews} {et~al.}(2010){Andrews}, {Wilner}, {Hughes}, {Qi}, \&
  {Dullemond}}]{Andrews:2010fk}
{Andrews}, S.~M., {Wilner}, D.~J., {Hughes}, A.~M., {Qi}, C., \& {Dullemond},
  C.~P. 2010, \apj, 723, 1241

\bibitem[{{Andrews} {et~al.}(2016){Andrews}, {Wilner}, {Zhu}, {Birnstiel},
  {Carpenter}, {P{\'e}rez}, {Bai}, {{\"O}berg}, {Hughes}, {Isella}, \&
  {Ricci}}]{Andrews:2016lr}
{Andrews}, S.~M., {Wilner}, D.~J., {Zhu}, Z., {et~al.} 2016, \apjl, 820, L40

\bibitem[{{Ansdell} {et~al.}(2016){Ansdell}, {Williams}, {van der Marel},
  {Carpenter}, {Guidi}, {Hogerheijde}, {Mathews}, {Manara}, {Miotello},
  {Natta}, {Oliveira}, {Tazzari}, {Testi}, {van Dishoeck}, \& {van
  Terwisga}}]{Ansdell:2016qf}
{Ansdell}, M., {Williams}, J.~P., {van der Marel}, N., {et~al.} 2016, \apj,
  828, 46

\bibitem[{{Armitage} {et~al.}(2013){Armitage}, {Simon}, \&
  {Martin}}]{Armitage:2013vn}
{Armitage}, P.~J., {Simon}, J.~B., \& {Martin}, R.~G. 2013, \apjl, 778, L14

\bibitem[{{Bai}(2016)}]{Bai:2016kx}
{Bai}, X.-N. 2016, \apj, 821, 80

\bibitem[{{Bai} \& {Stone}(2013)}]{Bai:2013yq}
{Bai}, X.-N. \& {Stone}, J.~M. 2013, \apj, 769, 76

\bibitem[{{Banzatti} {et~al.}(2011){Banzatti}, {Testi}, {Isella}, {Natta},
  {Neri}, \& {Wilner}}]{Banzatti:2011ff}
{Banzatti}, A., {Testi}, L., {Isella}, A., {et~al.} 2011, \aap, 525, A12

\bibitem[{{Barenfeld} {et~al.}(2016){Barenfeld}, {Carpenter}, {Ricci}, \&
  {Isella}}]{Barenfeld:2016lr}
{Barenfeld}, S.~A., {Carpenter}, J.~M., {Ricci}, L., \& {Isella}, A. 2016,
  \apj, 827, 142

\bibitem[{{Beckwith} \& {Sargent}(1991)}]{Beckwith:1991pd}
{Beckwith}, S.~V.~W. \& {Sargent}, A.~I. 1991, \apj, 381, 250

\bibitem[{{Beckwith} {et~al.}(1990){Beckwith}, {Sargent}, {Chini}, \&
  {Guesten}}]{Beckwith:1990qf}
{Beckwith}, S.~V.~W., {Sargent}, A.~I., {Chini}, R.~S., \& {Guesten}, R. 1990,
  \aj, 99, 924

\bibitem[{{Birnstiel} \& {Andrews}(2014)}]{Birnstiel:2014zr}
{Birnstiel}, T. \& {Andrews}, S.~M. 2014, \apj, 780, 153

\bibitem[{{Birnstiel} {et~al.}(2010){Birnstiel}, {Dullemond}, \&
  {Brauer}}]{Birnstiel:2010jk}
{Birnstiel}, T., {Dullemond}, C.~P., \& {Brauer}, F. 2010, \aap, 513, A79

\bibitem[{{Birnstiel} {et~al.}(2016){Birnstiel}, {Fang}, \&
  {Johansen}}]{Birnstiel:2016ve}
{Birnstiel}, T., {Fang}, M., \& {Johansen}, A. 2016, \ssr
  [\eprint[arXiv]{1604.02952}]

\bibitem[{{Bouwman} {et~al.}(2001){Bouwman}, {Meeus}, {de Koter}, {Hony},
  {Dominik}, \& {Waters}}]{Bouwman:2001qv}
{Bouwman}, J., {Meeus}, G., {de Koter}, A., {et~al.} 2001, \aap, 375, 950

\bibitem[{{Brauer} {et~al.}(2008){Brauer}, {Dullemond}, \&
  {Henning}}]{Brauer:2008kq}
{Brauer}, F., {Dullemond}, C.~P., \& {Henning}, T. 2008, \aap, 480, 859

\bibitem[{{Bruggeman}(1935)}]{Bruggeman:1935ys}
{Bruggeman}, D.~A.~G. 1935, Annalen der Physik, 416, 636

\bibitem[{{Bustamante} {et~al.}(2015){Bustamante}, {Mer{\'{\i}}n}, {Ribas},
  {Bouy}, {Prusti}, {Pilbratt}, \& {Andr{\'e}}}]{Bustamante:2015ve}
{Bustamante}, I., {Mer{\'{\i}}n}, B., {Ribas}, {\'A}., {et~al.} 2015, \aap,
  578, A23

\bibitem[{{Chiang} \& {Youdin}(2010)}]{Chiang:2010fk}
{Chiang}, E. \& {Youdin}, A.~N. 2010, Annual Review of Earth and Planetary
  Sciences, 38, 493

\bibitem[{{Chiang} \& {Goldreich}(1997)}]{1997ApJ...490..368C}
{Chiang}, E.~I. \& {Goldreich}, P. 1997, \apj, 490, 368

\bibitem[{{Clark}(1980)}]{Clark:1980lr}
{Clark}, B.~G. 1980, \aap, 89, 377

\bibitem[{{Comer{\'o}n}(2008)}]{Comeron:2008qy}
{Comer{\'o}n}, F. 2008, {The Lupus Clouds}, ed. B.~{Reipurth}, 295

\bibitem[{{Daemgen} {et~al.}(2016){Daemgen}, {Natta}, {Scholz}, {Testi},
  {Jayawardhana}, {Greaves}, \& {Eastwood}}]{Daemgen:2016fk}
{Daemgen}, S., {Natta}, A., {Scholz}, A., {et~al.} 2016, ArXiv e-prints
  [\eprint[arXiv]{1607.07458}]

\bibitem[{{de Gregorio-Monsalvo} {et~al.}(2013){de Gregorio-Monsalvo},
  {M{\'e}nard}, {Dent}, {Pinte}, {L{\'o}pez}, {Klaassen}, {Hales},
  {Cort{\'e}s}, {Rawlings}, {Tachihara}, {Testi}, {Takahashi}, {Chapillon},
  {Mathews}, {Juhasz}, {Akiyama}, {Higuchi}, {Saito}, {Nyman}, {Phillips},
  {Rod{\'o}n}, {Corder}, \& {Van Kempen}}]{de-Gregorio-Monsalvo:2013qf}
{de Gregorio-Monsalvo}, I., {M{\'e}nard}, F., {Dent}, W., {et~al.} 2013, \aap,
  557, A133

\bibitem[{{Dullemond} \& {Dominik}(2004)}]{Dullemond:2004ly}
{Dullemond}, C.~P. \& {Dominik}, C. 2004, \aap, 421, 1075

\bibitem[{{Dullemond} {et~al.}(2001){Dullemond}, {Dominik}, \&
  {Natta}}]{2001ApJ...560..957D}
{Dullemond}, C.~P., {Dominik}, C., \& {Natta}, A. 2001, \apj, 560, 957

\bibitem[{{Fedele} {et~al.}(2010){Fedele}, {van den Ancker}, {Henning},
  {Jayawardhana}, \& {Oliveira}}]{Fedele:2010fj}
{Fedele}, D., {van den Ancker}, M.~E., {Henning}, T., {Jayawardhana}, R., \&
  {Oliveira}, J.~M. 2010, \aap, 510, A72

\bibitem[{{Foreman-Mackey} {et~al.}(2013){Foreman-Mackey}, {Hogg}, {Lang}, \&
  {Goodman}}]{2013PASP..125..306F}
{Foreman-Mackey}, D., {Hogg}, D.~W., {Lang}, D., \& {Goodman}, J. 2013, \pasp,
  125, 306

\bibitem[{Foreman-Mackey {et~al.}(2014)Foreman-Mackey, Price-Whelan, Ryan,
  Emily, Smith, Barbary, Hogg, \& Brewer}]{dan_foreman_mackey_2014_11020}
Foreman-Mackey, D., Price-Whelan, A., Ryan, G., {et~al.} 2014, triangle.py
  v0.1.1

\bibitem[{{Goodman} \& {Weare}(2010)}]{Goodman:2010}
{Goodman}, J. \& {Weare}, J. 2010, Comm. App. Math. Comp. Sci., 5, 65

\bibitem[{{Guidi} {et~al.}(2016){Guidi}, {Tazzari}, {Testi}, {de
  Gregorio-Monsalvo}, {Chandler}, {P{\'e}rez}, {Isella}, {Natta}, {Ortolani},
  {Henning}, {Corder}, {Linz}, {Andrews}, {Wilner}, {Ricci}, {Carpenter},
  {Sargent}, {Mundy}, {Storm}, {Calvet}, {Dullemond}, {Greaves}, {Lazio},
  {Deller}, \& {Kwon}}]{Guidi:2016sf}
{Guidi}, G., {Tazzari}, M., {Testi}, L., {et~al.} 2016, \aap, 588, A112

\bibitem[{{Guilloteau} {et~al.}(2011){Guilloteau}, {Dutrey}, {Pi{\'e}tu}, \&
  {Boehler}}]{2011A&amp;A...529A.105G}
{Guilloteau}, S., {Dutrey}, A., {Pi{\'e}tu}, V., \& {Boehler}, Y. 2011, \aap,
  529, A105

\bibitem[{{Hartmann} {et~al.}(1998){Hartmann}, {Calvet}, {Gullbring}, \&
  {D'Alessio}}]{Hartmann:1998qy}
{Hartmann}, L., {Calvet}, N., {Gullbring}, E., \& {D'Alessio}, P. 1998, \apj,
  495, 385

\bibitem[{{Hayashi}(1981)}]{Hayashi:1981qy}
{Hayashi}, C. 1981, Progress of Theoretical Physics Supplement, 70, 35

\bibitem[{{Hern{\'a}ndez} {et~al.}(2007){Hern{\'a}ndez}, {Hartmann}, {Megeath},
  {Gutermuth}, {Muzerolle}, {Calvet}, {Vivas}, {Brice{\~n}o}, {Allen},
  {Stauffer}, {Young}, \& {Fazio}}]{Hernandez:2007fk}
{Hern{\'a}ndez}, J., {Hartmann}, L., {Megeath}, T., {et~al.} 2007, \apj, 662,
  1067

\bibitem[{{Hughes} {et~al.}(2008){Hughes}, {Wilner}, {Qi}, \&
  {Hogerheijde}}]{Hughes:2008lr}
{Hughes}, A.~M., {Wilner}, D.~J., {Qi}, C., \& {Hogerheijde}, M.~R. 2008, \apj,
  678, 1119

\bibitem[{{Hunter}(2007)}]{Hunter:2007fk}
{Hunter}, J.~D. 2007, Computing in Science and Engineering, 9, 90

\bibitem[{{Isella} {et~al.}(2009){Isella}, {Carpenter}, \&
  {Sargent}}]{Isella:2009qy}
{Isella}, A., {Carpenter}, J.~M., \& {Sargent}, A.~I. 2009, \apj, 701, 260

\bibitem[{{Isella} {et~al.}(2010){Isella}, {Carpenter}, \&
  {Sargent}}]{Isella:2010dq}
{Isella}, A., {Carpenter}, J.~M., \& {Sargent}, A.~I. 2010, \apj, 714, 1746

\bibitem[{Isella {et~al.}(2016)Isella, Guidi, Testi, Liu, Li, Li, Weaver,
  Boehler, Carperter, De~Gregorio-Monsalvo, Manara, Natta, P\'erez, Ricci,
  Sargent, Tazzari, \& Turner}]{Isella:2016ww}
Isella, A., Guidi, G., Testi, L., {et~al.} 2016, Phys. Rev. Lett., 117, 251101

\bibitem[{{Juh{\'a}sz} {et~al.}(2010){Juh{\'a}sz}, {Bouwman}, {Henning},
  {Acke}, {van den Ancker}, {Meeus}, {Dominik}, {Min}, {Tielens}, \&
  {Waters}}]{Juhasz:2010fj}
{Juh{\'a}sz}, A., {Bouwman}, J., {Henning}, T., {et~al.} 2010, \apj, 721, 431

\bibitem[{{Kelly}(2007)}]{Kelly:2007lq}
{Kelly}, B.~C. 2007, \apj, 665, 1489

\bibitem[{{Lissauer} {et~al.}(2014){Lissauer}, {Dawson}, \&
  {Tremaine}}]{Lissauer:2014rr}
{Lissauer}, J.~J., {Dawson}, R.~I., \& {Tremaine}, S. 2014, \nat, 513, 336

\bibitem[{{Lynden-Bell} \& {Pringle}(1974)}]{1974MNRAS.168..603L}
{Lynden-Bell}, D. \& {Pringle}, J.~E. 1974, \mnras, 168, 603

\bibitem[{{Manara} {et~al.}(2016){Manara}, {Rosotti}, {Testi}, {Natta},
  {Alcal{\'a}}, {Williams}, {Ansdell}, {Miotello}, {van der Marel}, {Tazzari},
  {Carpenter}, {Guidi}, {Mathews}, {Oliveira}, {Prusti}, \& {van
  Dishoeck}}]{Manara:2016lr}
{Manara}, C.~F., {Rosotti}, G., {Testi}, L., {et~al.} 2016, \aap, 591, L3

\bibitem[{{Mathis} {et~al.}(1977){Mathis}, {Rumpl}, \&
  {Nordsieck}}]{Mathis:1977yq}
{Mathis}, J.~S., {Rumpl}, W., \& {Nordsieck}, K.~H. 1977, \apj, 217, 425

\bibitem[{{Mer{\'{\i}}n} {et~al.}(2010){Mer{\'{\i}}n}, {Brown}, {Oliveira},
  {Herczeg}, {van Dishoeck}, {Bottinelli}, {Evans}, {Cieza}, {Spezzi},
  {Alcal{\'a}}, {Harvey}, {Blake}, {Bayo}, {Geers}, {Lahuis}, {Prusti},
  {Augereau}, {Olofsson}, {Walter}, \& {Chiu}}]{Merin:2010ul}
{Mer{\'{\i}}n}, B., {Brown}, J.~M., {Oliveira}, I., {et~al.} 2010, \apj, 718,
  1200

\bibitem[{{Mer{\'{\i}}n} {et~al.}(2008){Mer{\'{\i}}n}, {J{\o}rgensen},
  {Spezzi}, {Alcal{\'a}}, {Evans}, {Harvey}, {Prusti}, {Chapman}, {Huard}, {van
  Dishoeck}, \& {Comer{\'o}n}}]{Merin:2008xy}
{Mer{\'{\i}}n}, B., {J{\o}rgensen}, J., {Spezzi}, L., {et~al.} 2008, \apjs,
  177, 551

\bibitem[{{Miotello} {et~al.}(2012){Miotello}, {Robberto}, {Potenza}, \&
  {Ricci}}]{Miotello:2012fv}
{Miotello}, A., {Robberto}, M., {Potenza}, M.~A.~C., \& {Ricci}, L. 2012, \apj,
  757, 78

\bibitem[{{Mordasini} {et~al.}(2012){Mordasini}, {Alibert}, {Benz}, {Klahr}, \&
  {Henning}}]{Mordasini:2012aa}
{Mordasini}, C., {Alibert}, Y., {Benz}, W., {Klahr}, H., \& {Henning}, T. 2012,
  \aap, 541, A97

\bibitem[{{Pascucci} {et~al.}(2016){Pascucci}, {Testi}, {Herczeg}, {Long},
  {Manara}, {Hendler}, {Mulders}, {Krijt}, {Ciesla}, {Henning}, {Mohanty},
  {Drabek-Maunder}, {Apai}, {Szucs}, {Sacco}, \& {Olofsson}}]{Pascucci:2016fk}
{Pascucci}, I., {Testi}, L., {Herczeg}, G.~J., {et~al.} 2016, ArXiv e-prints
  [\eprint[arXiv]{1608.03621}]

\bibitem[{{P{\'e}rez} {et~al.}(2016){P{\'e}rez}, {Carpenter}, {Andrews},
  {Ricci}, {Isella}, {Linz}, {Sargent}, {Wilner}, {Henning}, {Deller},
  {Chandler}, {Dullemond}, {Lazio}, {Menten}, {Corder}, {Storm}, {Testi},
  {Tazzari}, {Kwon}, {Calvet}, {Greaves}, {Harris}, \& {Mundy}}]{Perez:2016aa}
{P{\'e}rez}, L.~M., {Carpenter}, J.~M., {Andrews}, S.~M., {et~al.} 2016,
  Science, 353, 1519

\bibitem[{{Pi{\'e}tu} {et~al.}(2014){Pi{\'e}tu}, {Guilloteau}, {Di Folco},
  {Dutrey}, \& {Boehler}}]{Pietu:2014bh}
{Pi{\'e}tu}, V., {Guilloteau}, S., {Di Folco}, E., {Dutrey}, A., \& {Boehler},
  Y. 2014, \aap, 564, A95

\bibitem[{{Pollack} {et~al.}(1994){Pollack}, {Hollenbach}, {Beckwith},
  {Simonelli}, {Roush}, \& {Fong}}]{Pollack:1994vn}
{Pollack}, J.~B., {Hollenbach}, D., {Beckwith}, S., {et~al.} 1994, \apj, 421,
  615

\bibitem[{{Pringle}(1981)}]{Pringle:1981uq}
{Pringle}, J.~E. 1981, \araa, 19, 137

\bibitem[{{Ricci} {et~al.}(2010){Ricci}, {Testi}, {Natta}, {Neri}, {Cabrit}, \&
  {Herczeg}}]{Ricci:2010eu}
{Ricci}, L., {Testi}, L., {Natta}, A., {et~al.} 2010, \aap, 512, A15

\bibitem[{{Rodgers-Lee} {et~al.}(2014){Rodgers-Lee}, {Scholz}, {Natta}, \&
  {Ray}}]{Rodgers-Lee:2014aa}
{Rodgers-Lee}, D., {Scholz}, A., {Natta}, A., \& {Ray}, T. 2014, \mnras, 443,
  1587

\bibitem[{{Romero} {et~al.}(2012){Romero}, {Schreiber}, {Cieza},
  {Rebassa-Mansergas}, {Mer{\'{\i}}n}, {Smith Castelli}, {Allen}, \&
  {Morrell}}]{Romero:2012qf}
{Romero}, G.~A., {Schreiber}, M.~R., {Cieza}, L.~A., {et~al.} 2012, \apj, 749,
  79

\bibitem[{{Safronov}(1972)}]{safronov:1972}
{Safronov}, V.~S. 1972, NASA TTF, 677

\bibitem[{Scholz \& Stephens(1987)}]{CIS-74828}
Scholz, F.~W. \& Stephens, M.~A. 1987, Journal of the American Statistical
  Association, 82, 918

\bibitem[{{Siess} {et~al.}(2000){Siess}, {Dufour}, \&
  {Forestini}}]{Siess:2000qy}
{Siess}, L., {Dufour}, E., \& {Forestini}, M. 2000, \aap, 358, 593

\bibitem[{{Suzuki} \& {Inutsuka}(2014)}]{Suzuki:2014rt}
{Suzuki}, T.~K. \& {Inutsuka}, S.-i. 2014, \apj, 784, 121

\bibitem[{{Tazzari} {et~al.}(2016){Tazzari}, {Testi}, {Ercolano}, {Natta},
  {Isella}, {Chandler}, {P{\'e}rez}, {Andrews}, {Wilner}, {Ricci}, {Henning},
  {Linz}, {Kwon}, {Corder}, {Dullemond}, {Carpenter}, {Sargent}, {Mundy},
  {Storm}, {Calvet}, {Greaves}, {Lazio}, \& {Deller}}]{Tazzari:2016qy}
{Tazzari}, M., {Testi}, L., {Ercolano}, B., {et~al.} 2016, \aap, 588, A53

\bibitem[{{Testi} {et~al.}(2014){Testi}, {Birnstiel}, {Ricci}, {Andrews},
  {Blum}, {Carpenter}, {Dominik}, {Isella}, {Natta}, {Williams}, \&
  {Wilner}}]{Testi:2014kx}
{Testi}, L., {Birnstiel}, T., {Ricci}, L., {et~al.} 2014, Protostars and
  Planets VI, 339

\bibitem[{{Testi} {et~al.}(2016){Testi}, {Natta}, {Scholz}, {Tazzari}, {Ricci},
  \& {de Gregorio Monsalvo}}]{Testi:2016jk}
{Testi}, L., {Natta}, A., {Scholz}, A., {et~al.} 2016, ArXiv e-prints
  [\eprint[arXiv]{1606.06448}]

\bibitem[{{Trotta} {et~al.}(2013){Trotta}, {Testi}, {Natta}, {Isella}, \&
  {Ricci}}]{2013A&amp;A...558A..64T}
{Trotta}, F., {Testi}, L., {Natta}, A., {Isella}, A., \& {Ricci}, L. 2013,
  \aap, 558, A64

\bibitem[{{van Boekel} {et~al.}(2003){van Boekel}, {Waters}, {Dominik},
  {Bouwman}, {de Koter}, {Dullemond}, \& {Paresce}}]{van-Boekel:2003nr}
{van Boekel}, R., {Waters}, L.~B.~F.~M., {Dominik}, C., {et~al.} 2003, \aap,
  400, L21

\bibitem[{{van der Marel} {et~al.}(2016){van der Marel}, {Verhaar}, {van
  Terwisga}, {Mer{\'{\i}}n}, {Herczeg}, {Ligterink}, \& {van
  Dishoeck}}]{van-der-Marel:2016aa}
{van der Marel}, N., {Verhaar}, B.~W., {van Terwisga}, S., {et~al.} 2016, \aap,
  592, A126

\bibitem[{{van der Plas} {et~al.}(2016){van der Plas}, {M{\'e}nard},
  {Ward-Duong}, {Bulger}, {Harvey}, {Pinte}, {Patience}, {Hales}, \&
  {Casassus}}]{van-der-Plas:2016lr}
{van der Plas}, G., {M{\'e}nard}, F., {Ward-Duong}, K., {et~al.} 2016, \apj,
  819, 102

\bibitem[{{Weidenschilling}(1977)}]{Weidenschilling:1977lr}
{Weidenschilling}, S.~J. 1977, \mnras, 180, 57

\bibitem[{{Williams} \& {Cieza}(2011)}]{Williams:2011jk}
{Williams}, J.~P. \& {Cieza}, L.~A. 2011, \araa, 49, 67

\bibitem[{{Winn} \& {Fabrycky}(2015)}]{Winn:2015aa}
{Winn}, J.~N. \& {Fabrycky}, D.~C. 2015, \araa, 53, 409

\bibitem[{{Wrobel} \& {Walker}(1999)}]{Wrobel:1999gf}
{Wrobel}, J.~M. \& {Walker}, R.~C. 1999, in Astronomical Society of the Pacific
  Conference Series, Vol. 180, Synthesis Imaging in Radio Astronomy II, ed.
  G.~B. {Taylor}, C.~L. {Carilli}, \& R.~A. {Perley}, 171

\end{thebibliography}

\begin{appendix}

\section{Fits of the individual sources}
\label{app:fits}
Here we report the results of the fits for the individual sources, following the order in \tbref{tb:fits.results}. For each disk we report:
\begin{enumerate}
\item 
a triangle plot showing the MCMC results, providing estimates of the parameters as median of the marginalized posterior;
\item
a deprojected uv-plot showing a comparison between observed and best-fit model visibilities;
\item
radial profiles of the total (dust + gas) surface density $100\times\Sigma_\mathrm{d}(R)$, midplane temperature $T_\mathrm{mid}(R)$, optical depth $\tau_{890\mu\mathrm{m}}.$;
\item
synthesized maps of the observations, of the best-fit model and of the residuals.
\end{enumerate} 

All the models fit well the observations with negligible residuals, except for Sz~98 and IM~Lup where a smoothly decreasing surface density profile is not sufficient to explain the brightness distribution. In these cases ring-like residuals are left, respectively 14\% and 4\% of the total intensity, and are likely to be due to the presence of inhomogeneities like rings or spirals \citep{2041-8205-808-1-L3, Guidi:2016sf,Isella:2016ww}.

\pagebreak
\begin{figure*}
\centering
\Large\textbf{Sz 65\vspace{1cm}}
\resizebox{\hsize}{!}{\includegraphics[scale=0.5]{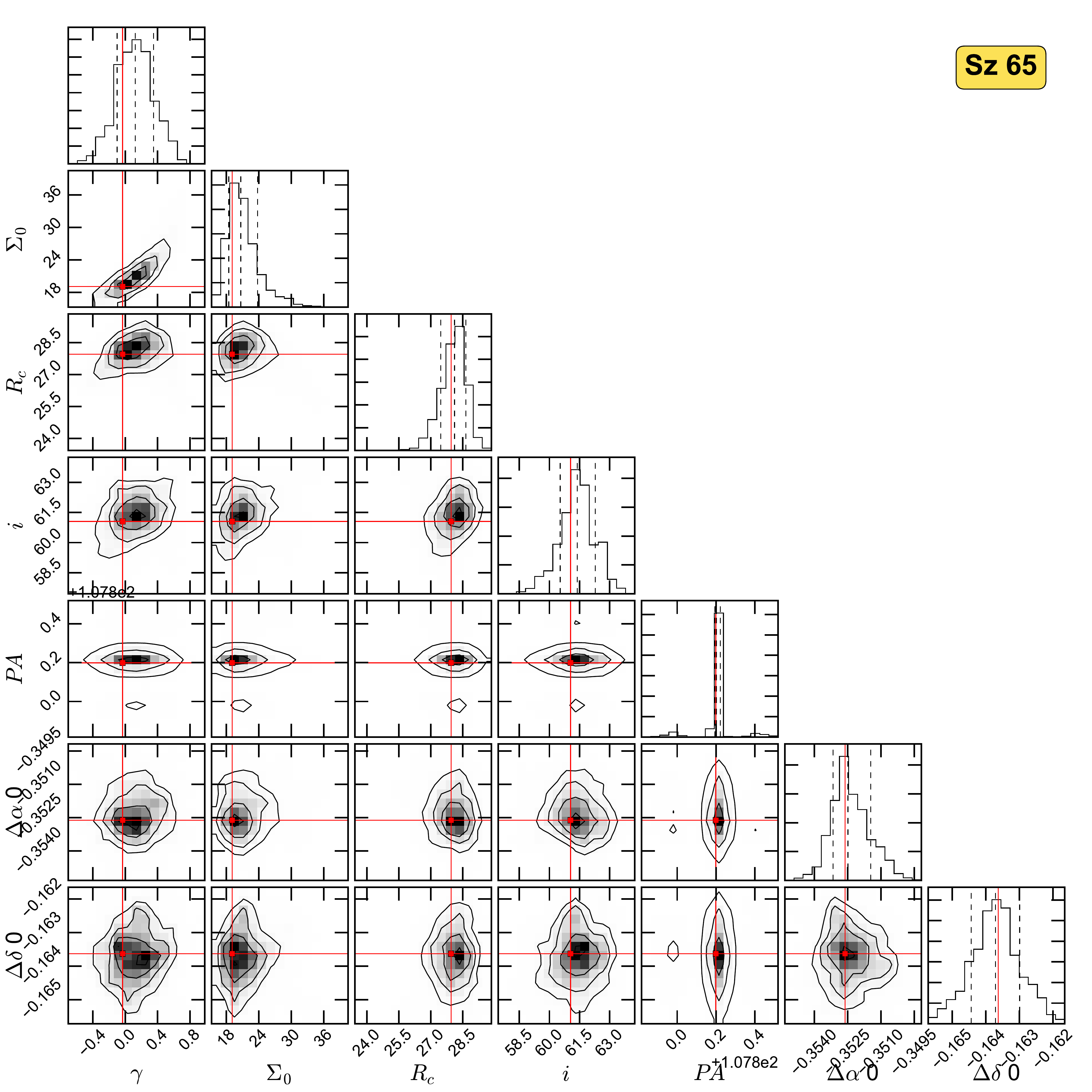}\includegraphics{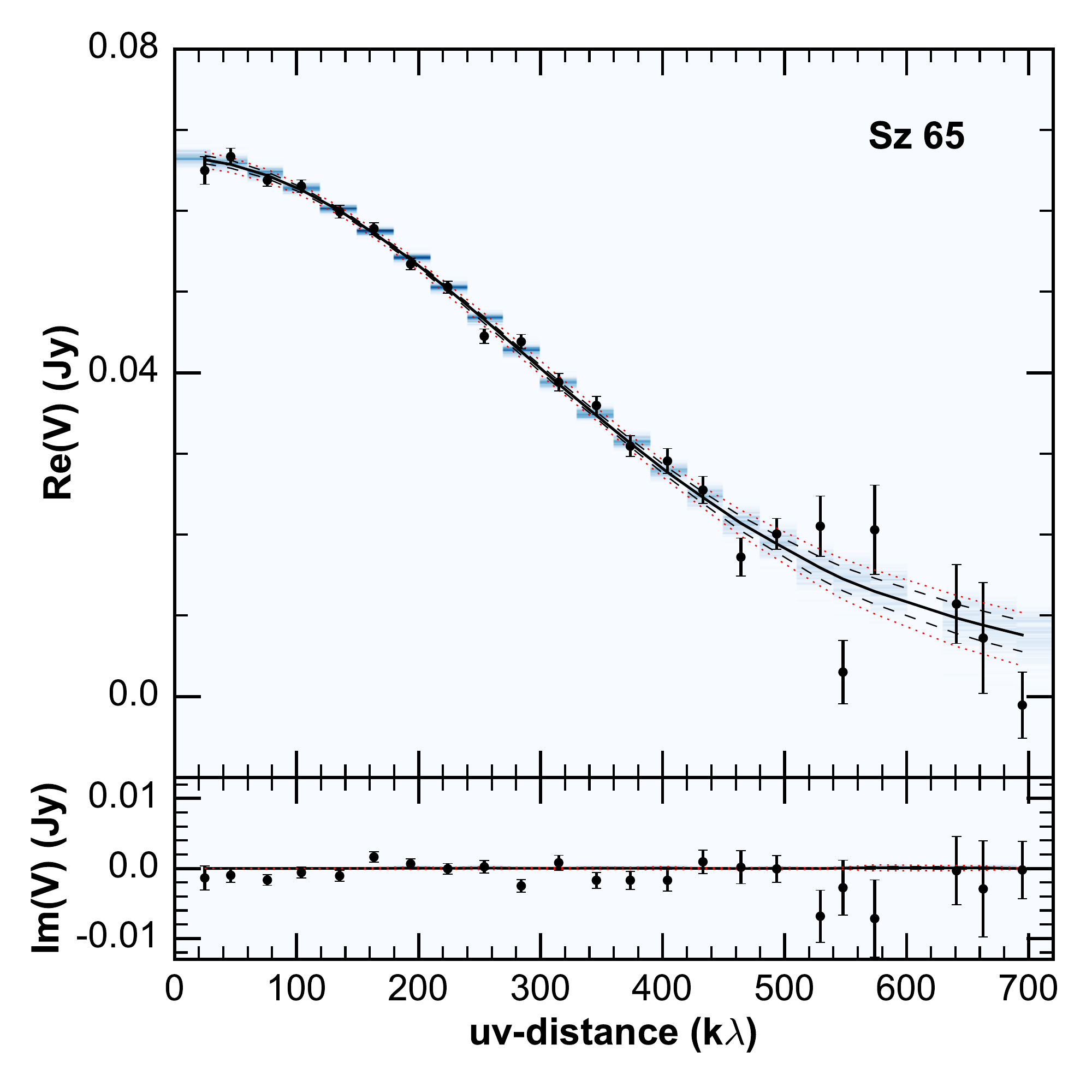}}\\[0.4cm]
\resizebox{\hsize}{!}{\includegraphics{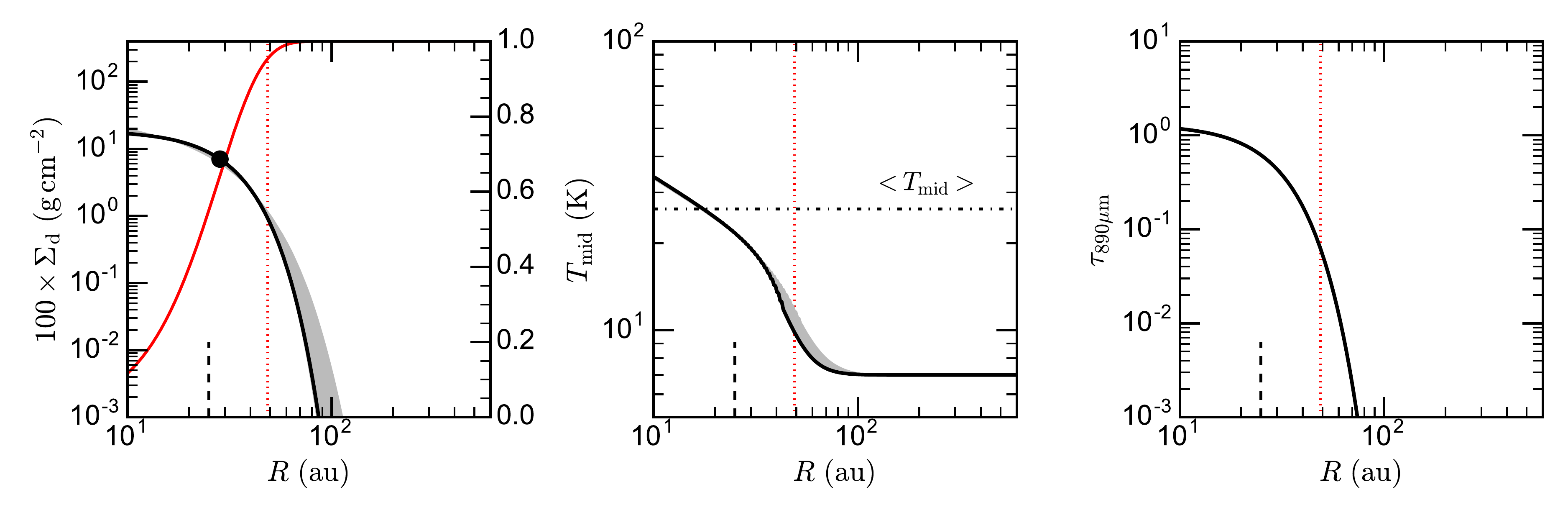}}\\[0.4cm]
\resizebox{0.8\hsize}{!}{\includegraphics{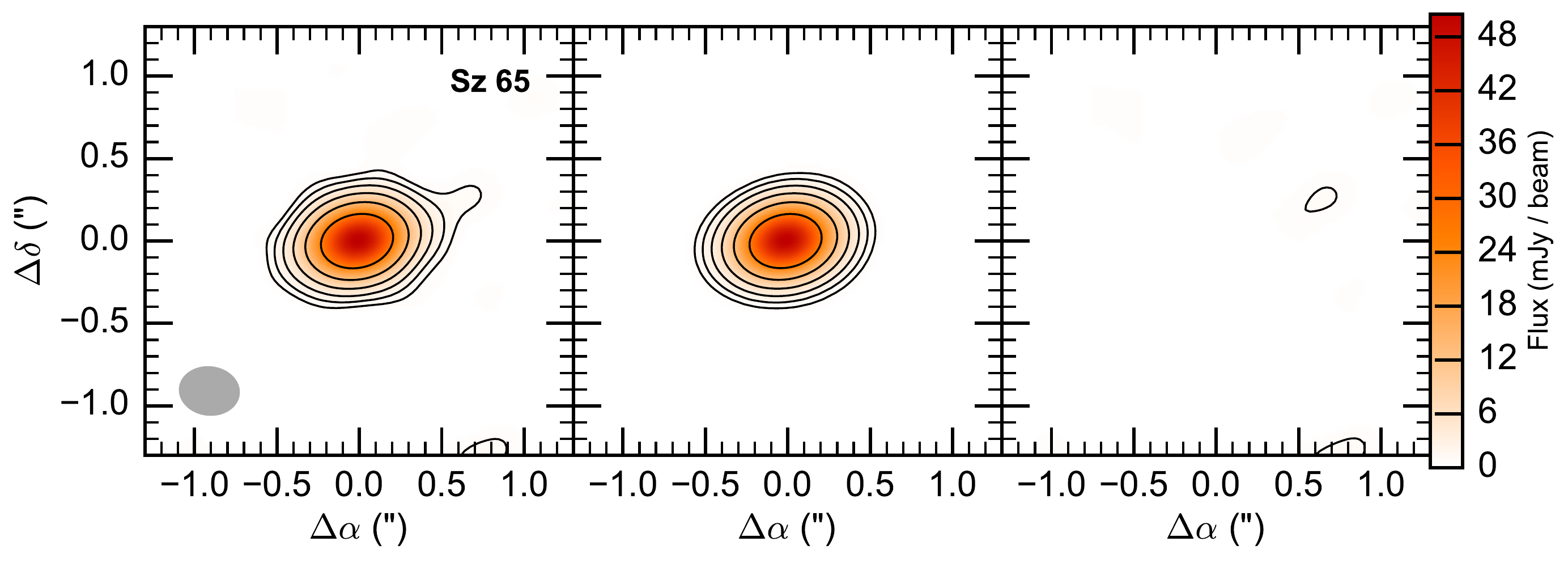}}\\[0.4cm]
\caption{Fit results for Sz 65. \textit{Top row:} on the left, the staircase plot showing the MCMC chains as in \figref{fig:samplemodel.triangle}; on the right, the comparison of the observations and model deprojected visibilities as a function of $uv-$distance as in \figref{fig:samplemodel.uvplot}. \textit{Middle row:} plots showing the physical structure of the disk as in \figref{fig:samplemodel.structure}. \textit{Bottom row:} synthesized images of observations, model and residual visibilities as in \figref{fig:samplemodel.maps}. In the images $\sigma=0.27\u{mJy/beam}$.}
\label{fig:reference.fit.results}
\end{figure*}

\pagebreak
\begin{figure*}
\centering
\Large\textbf{J15450887-3417333\vspace{1cm}}
\resizebox{\hsize}{!}{\includegraphics[scale=0.5]{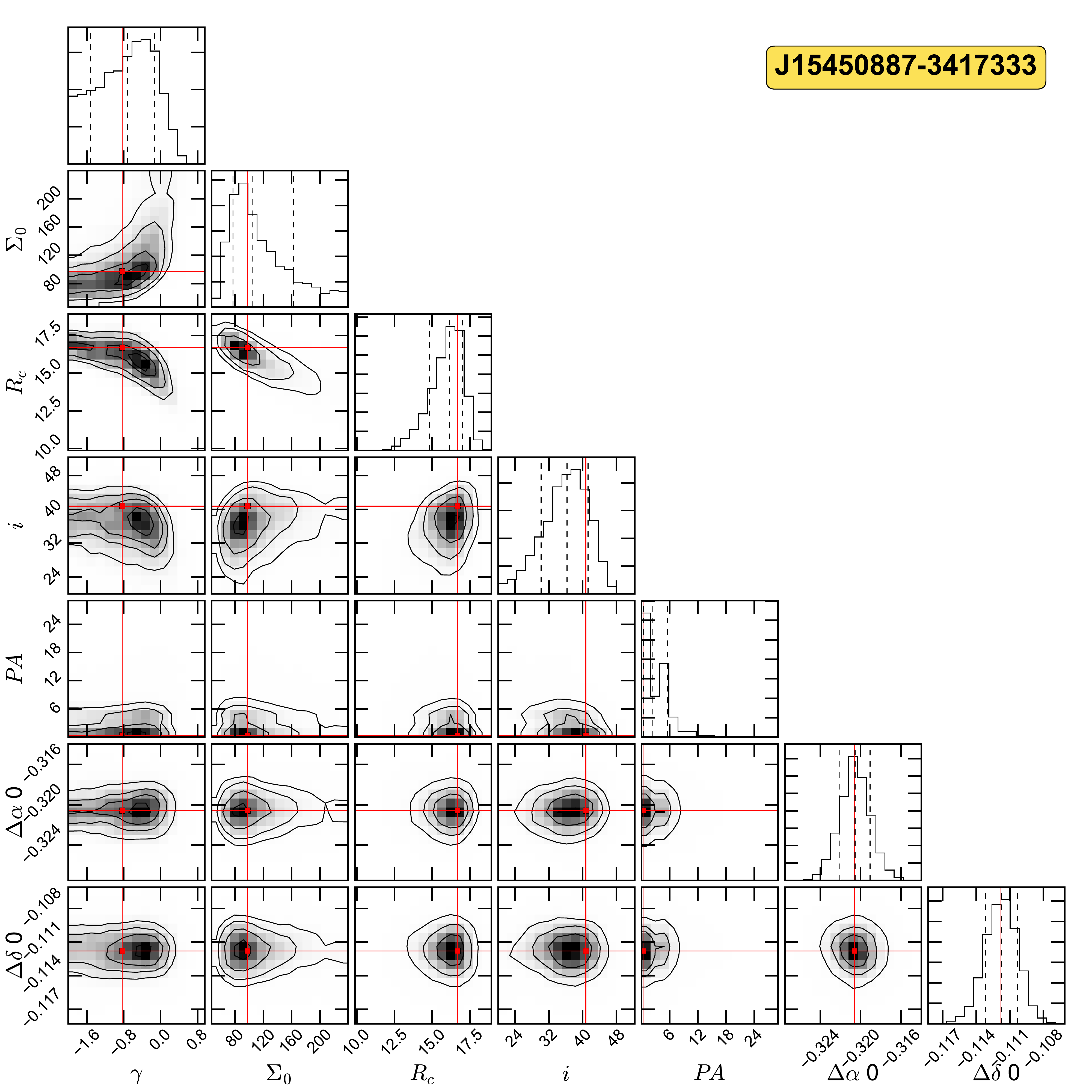}\includegraphics{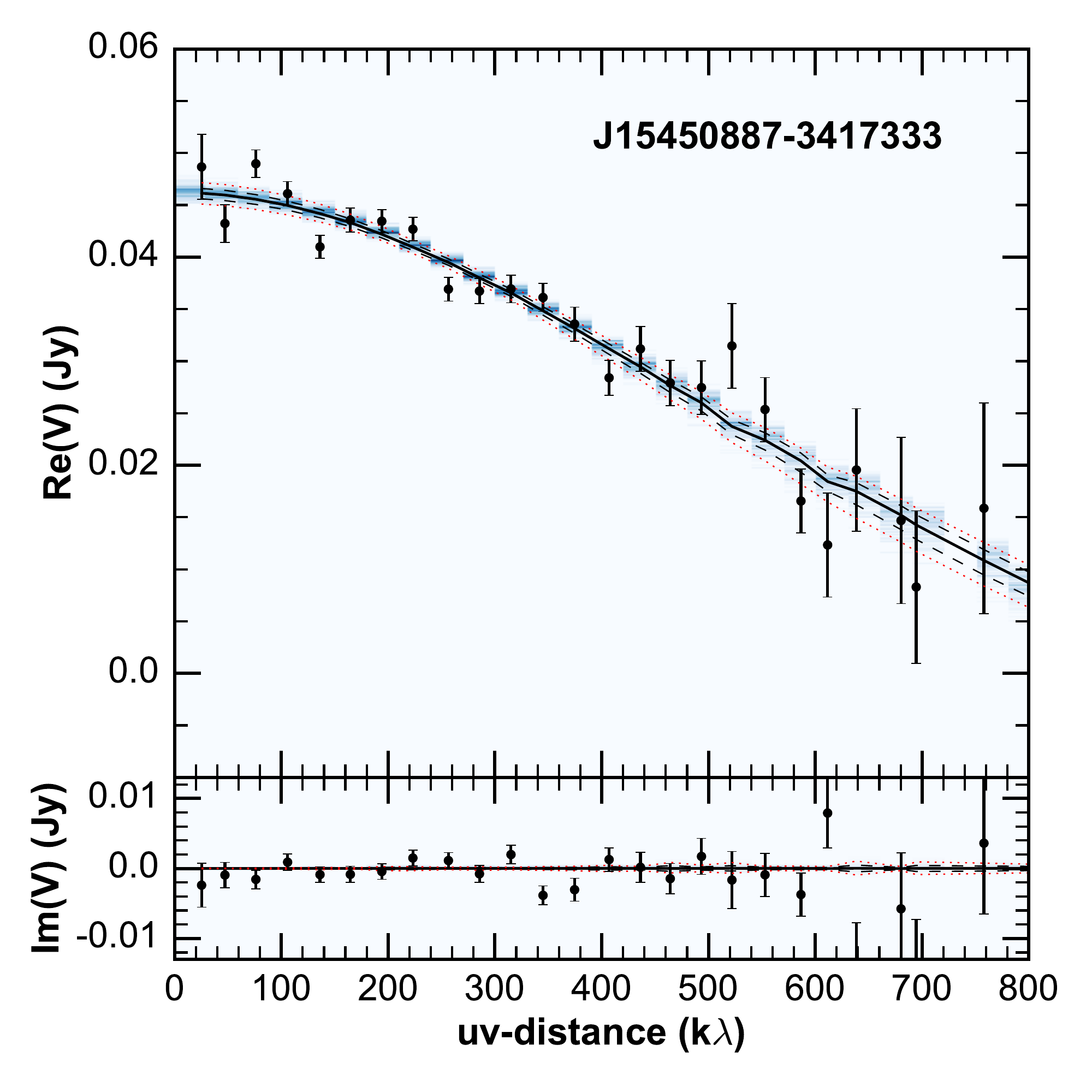}}\\[0.5cm]
\resizebox{\hsize}{!}{\includegraphics{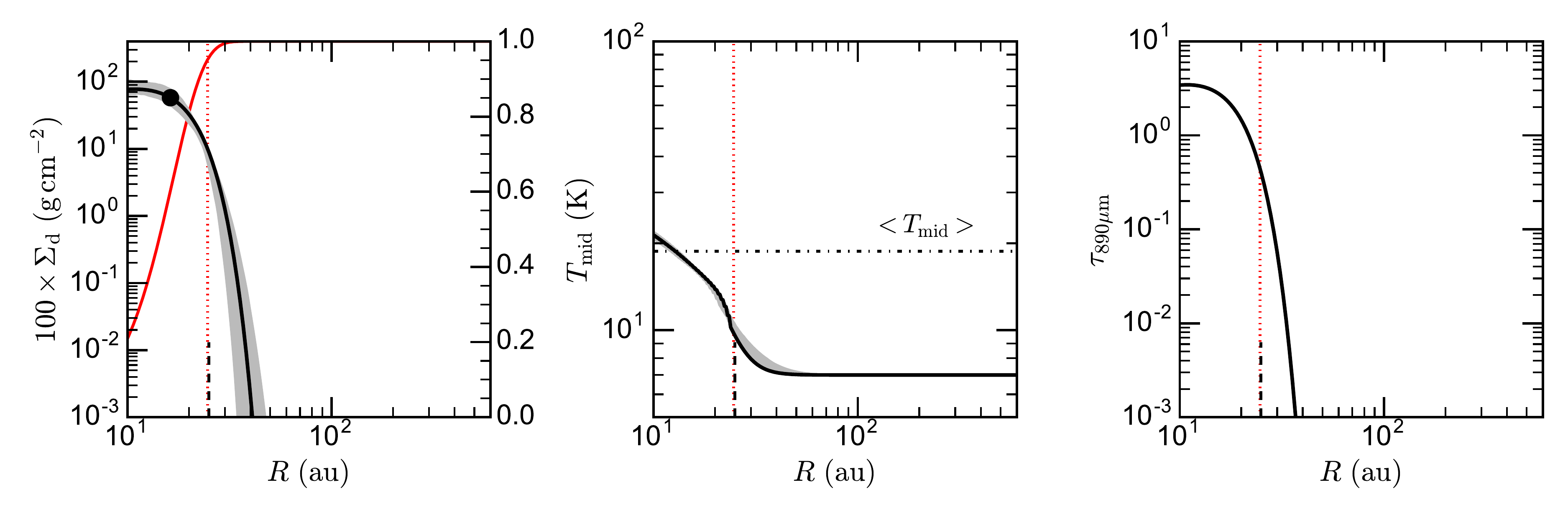}}\\[0.5cm]
\resizebox{0.8\hsize}{!}{\includegraphics{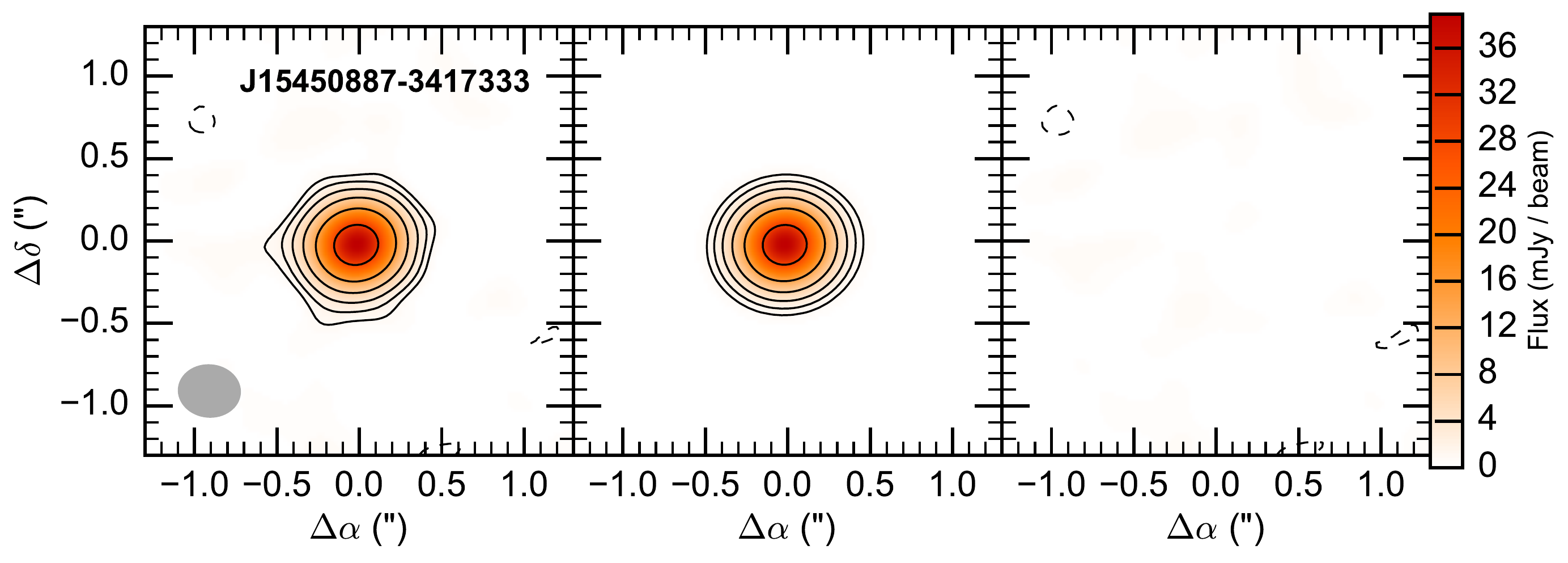}}\\[0.5cm]
\caption{Fit results for J15450887-3417333, presented as in \figref{fig:reference.fit.results}. In the images $\sigma=0.3\u{mJy/beam}$.}
\end{figure*}

\pagebreak
\begin{figure*}
\centering
\Large\textbf{Sz 68\vspace{1cm}}
\resizebox{\hsize}{!}{\includegraphics[scale=0.5]{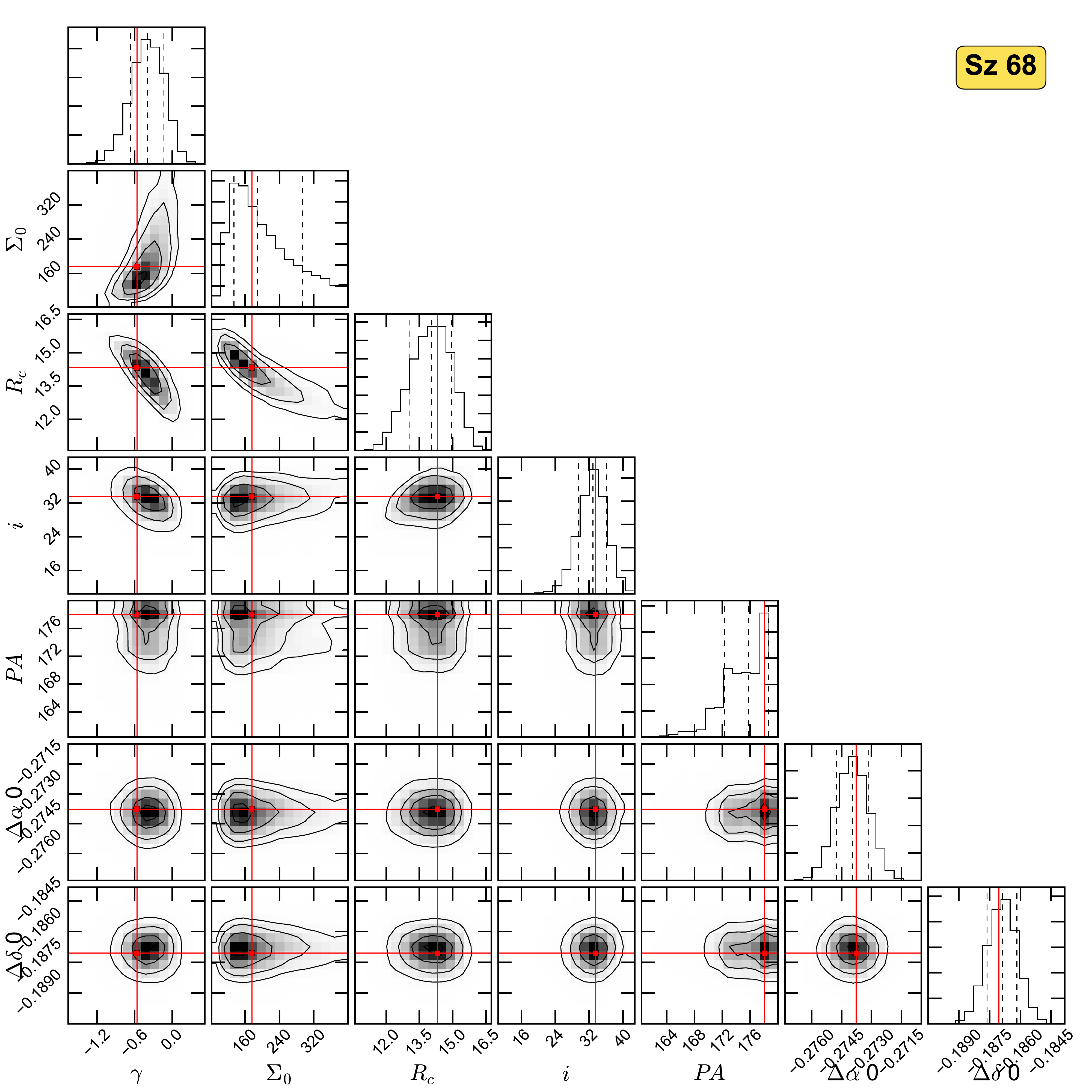}\includegraphics{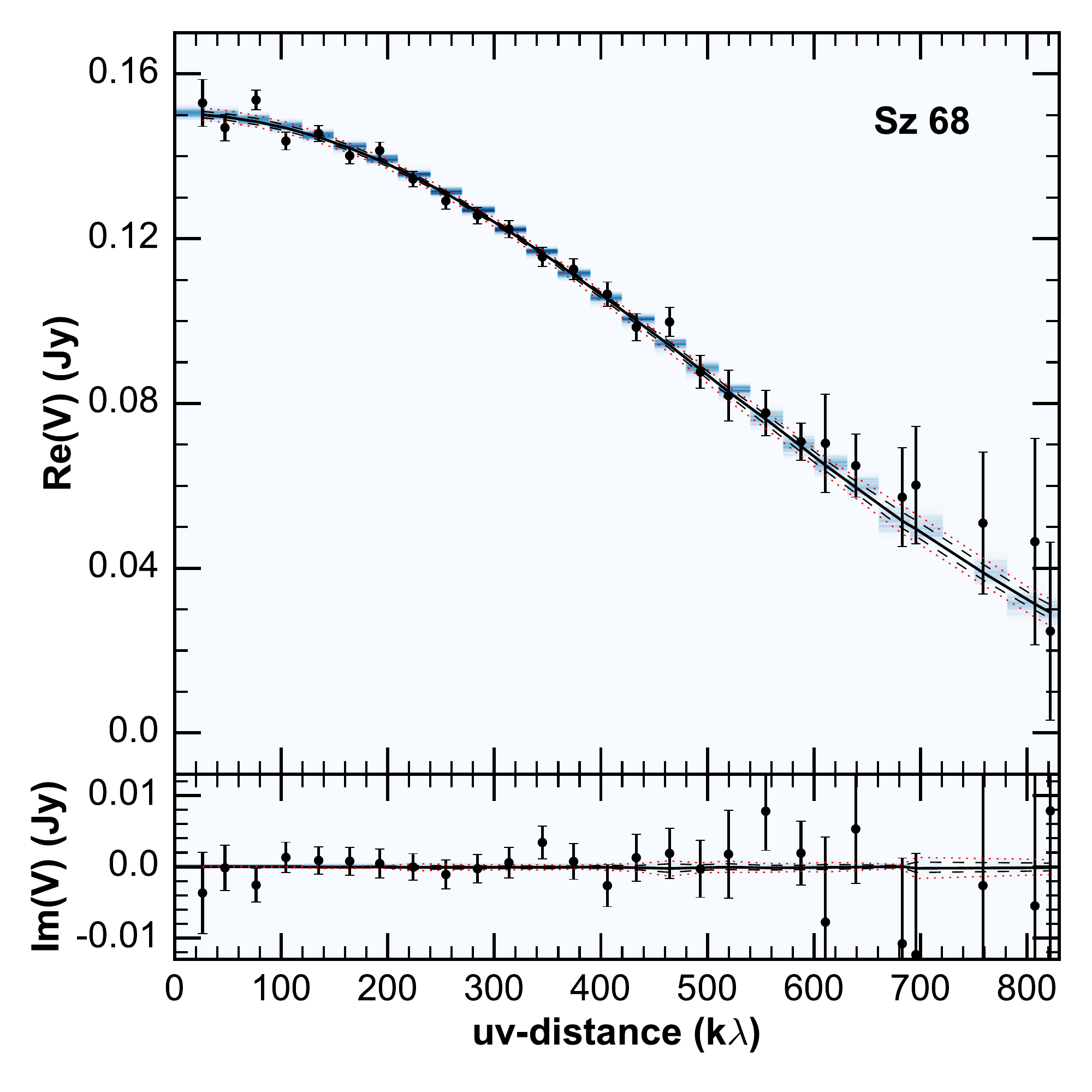}}\\[0.5cm]
\resizebox{\hsize}{!}{\includegraphics{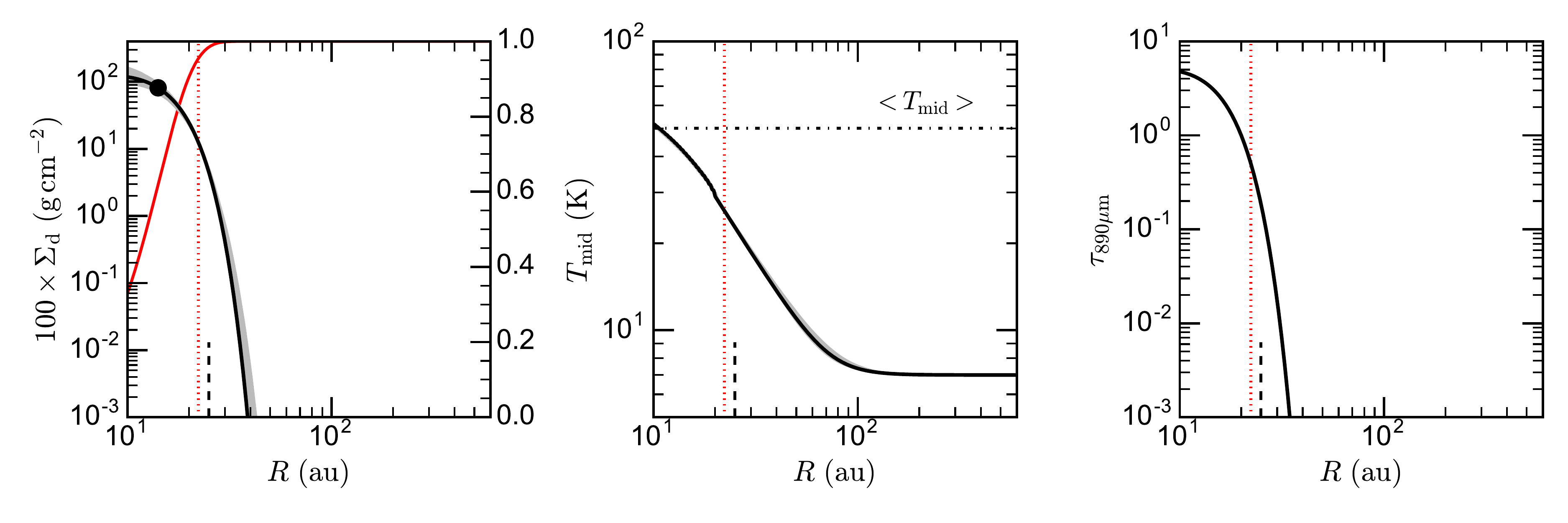}}\\[0.5cm]
\resizebox{0.8\hsize}{!}{\includegraphics{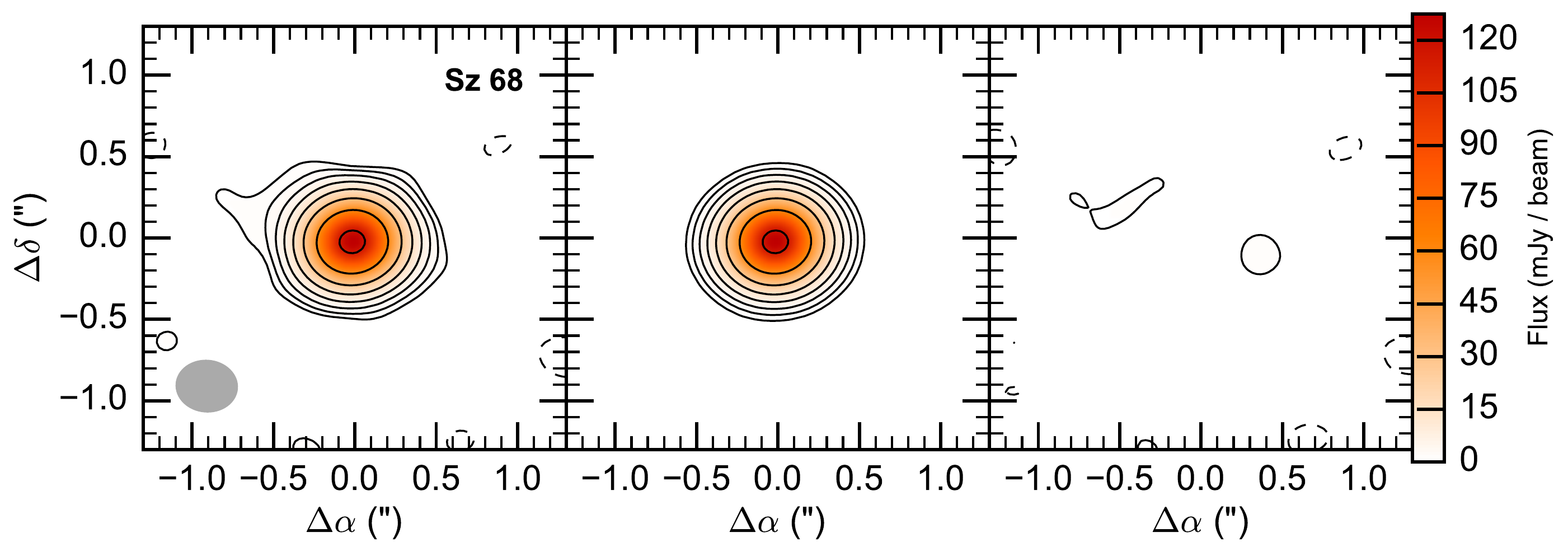}}\\[0.5cm]
\caption{Fit results for Sz 68, presented as in \figref{fig:reference.fit.results}. In the images $\sigma=0.3\u{mJy/beam}$.}
\end{figure*}

\pagebreak
\begin{figure*}
\centering
\Large\textbf{Sz 69\vspace{1cm}}
\resizebox{\hsize}{!}{\includegraphics[scale=0.5]{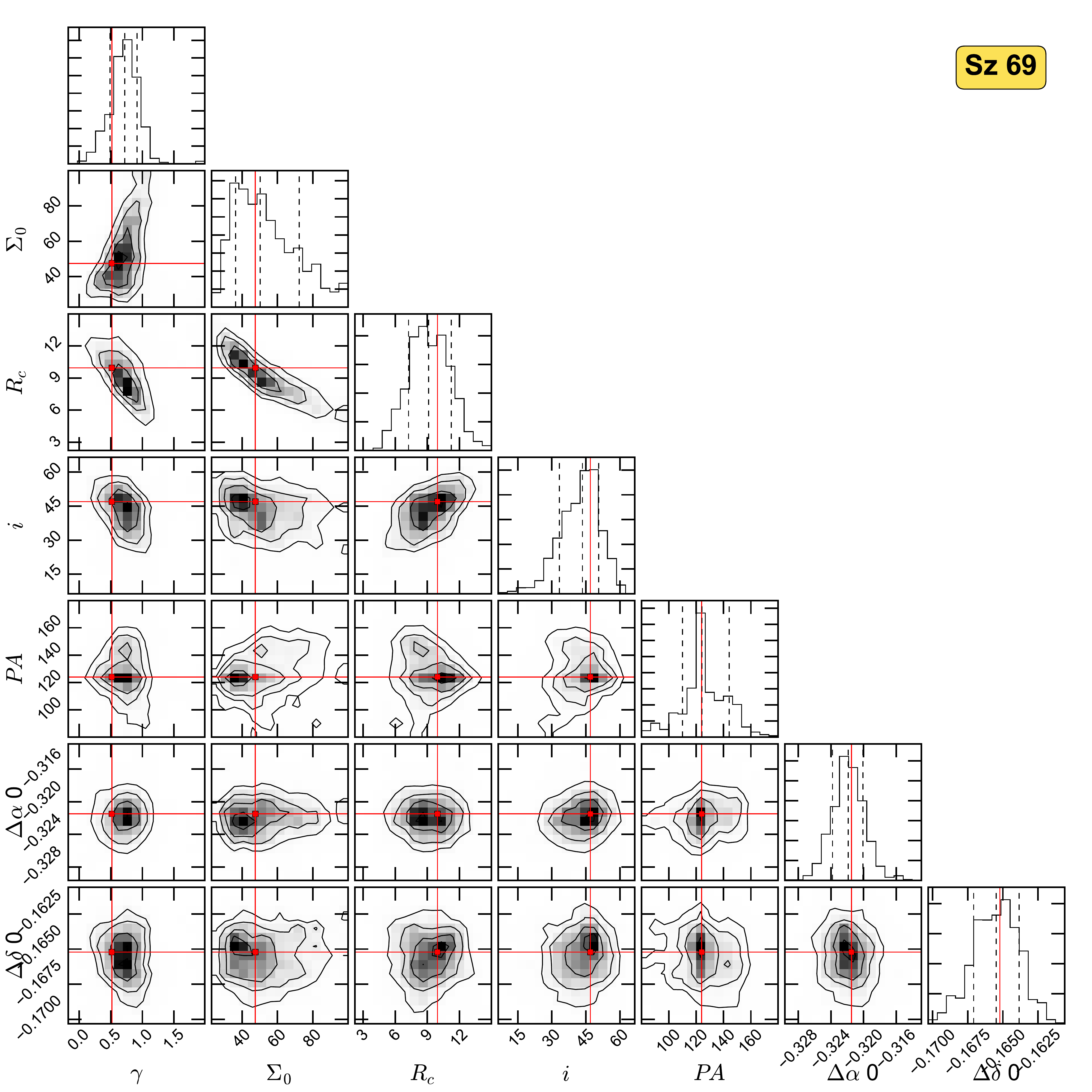}\includegraphics{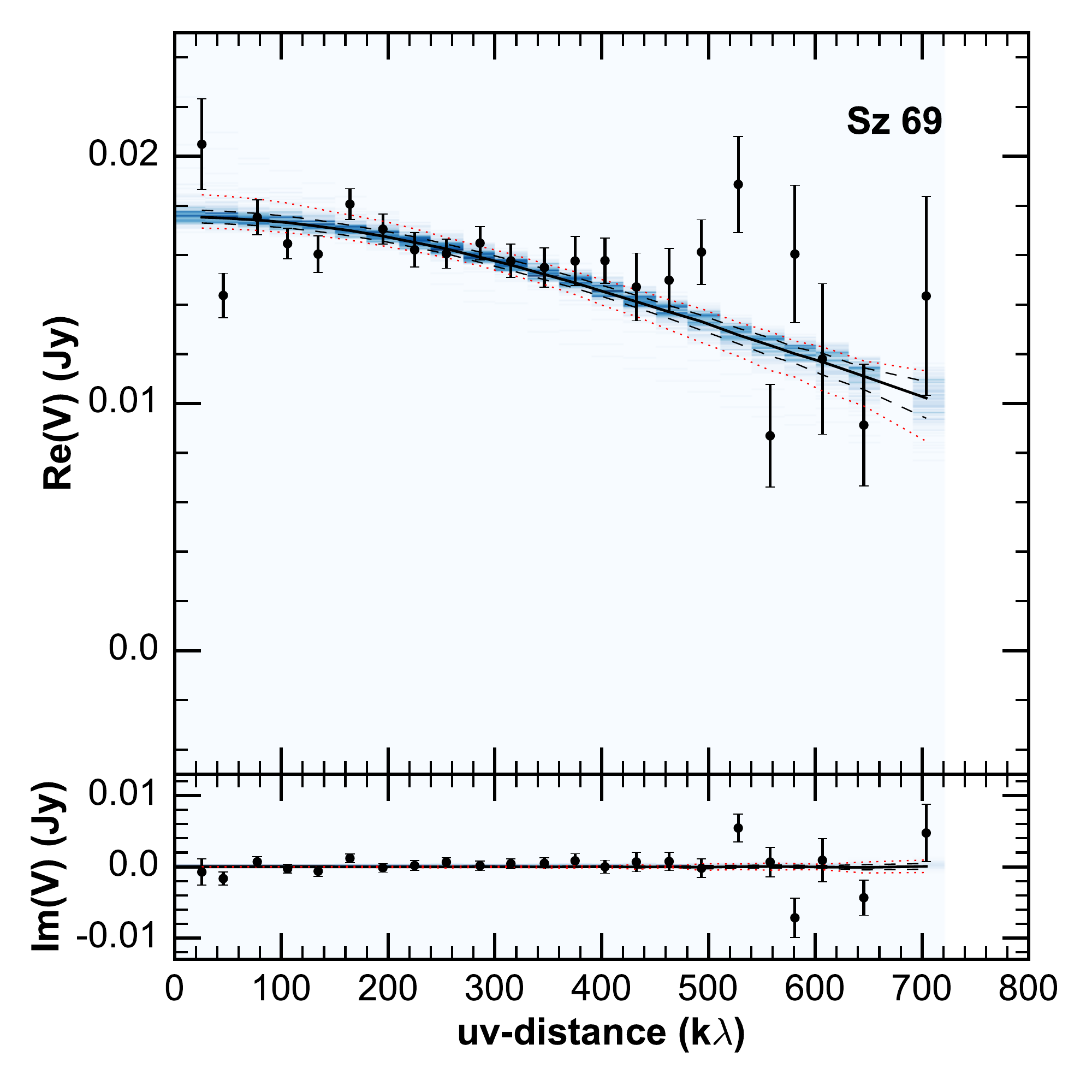}}\\[0.5cm]
\resizebox{\hsize}{!}{\includegraphics{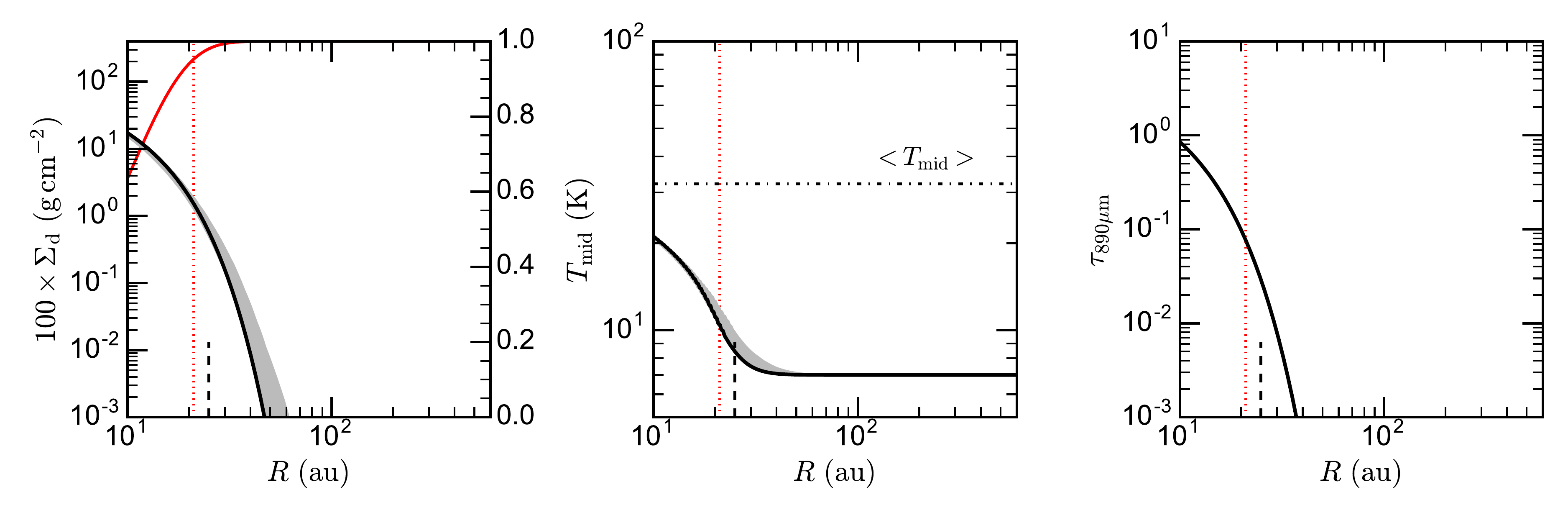}}\\[0.5cm]
\resizebox{0.8\hsize}{!}{\includegraphics{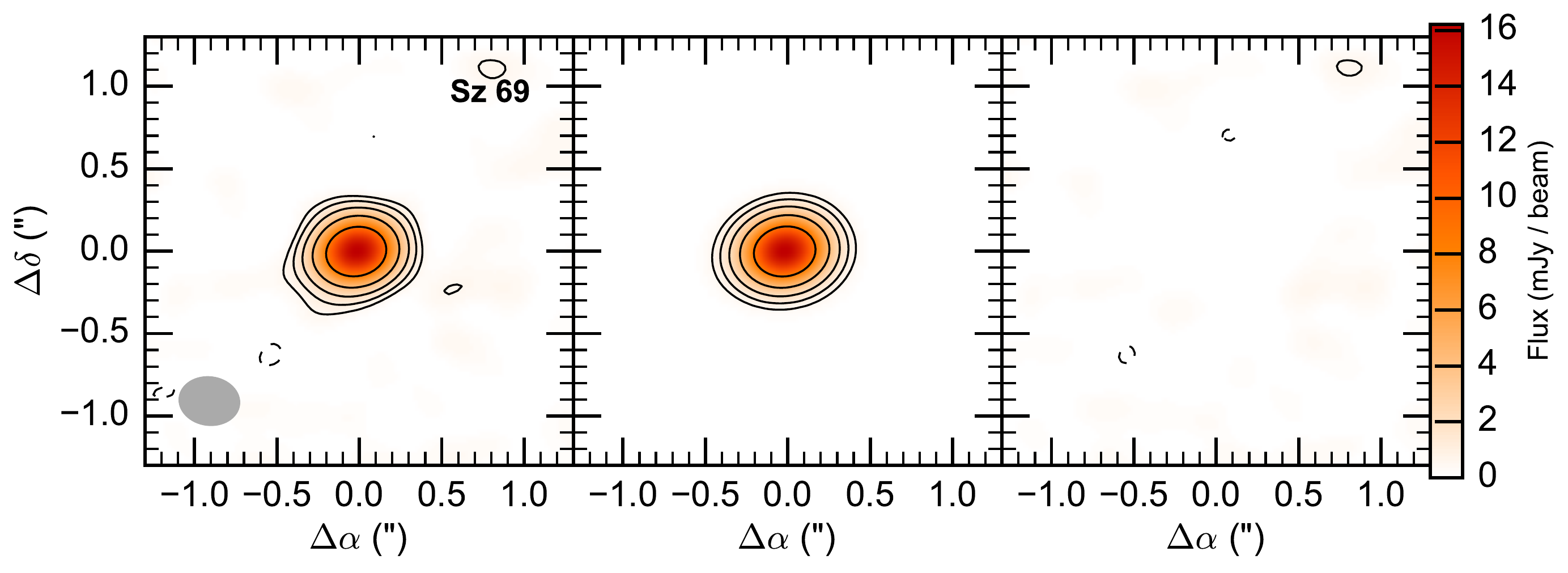}}\\[0.5cm]
\caption{Fit results for Sz 69, presented as in \figref{fig:reference.fit.results}. In the images $\sigma=0.25\u{mJy/beam}$.}
\end{figure*}

\pagebreak
\begin{figure*}
\centering
\Large\textbf{Sz 71\vspace{1cm}}
\resizebox{\hsize}{!}{\includegraphics[scale=0.5]{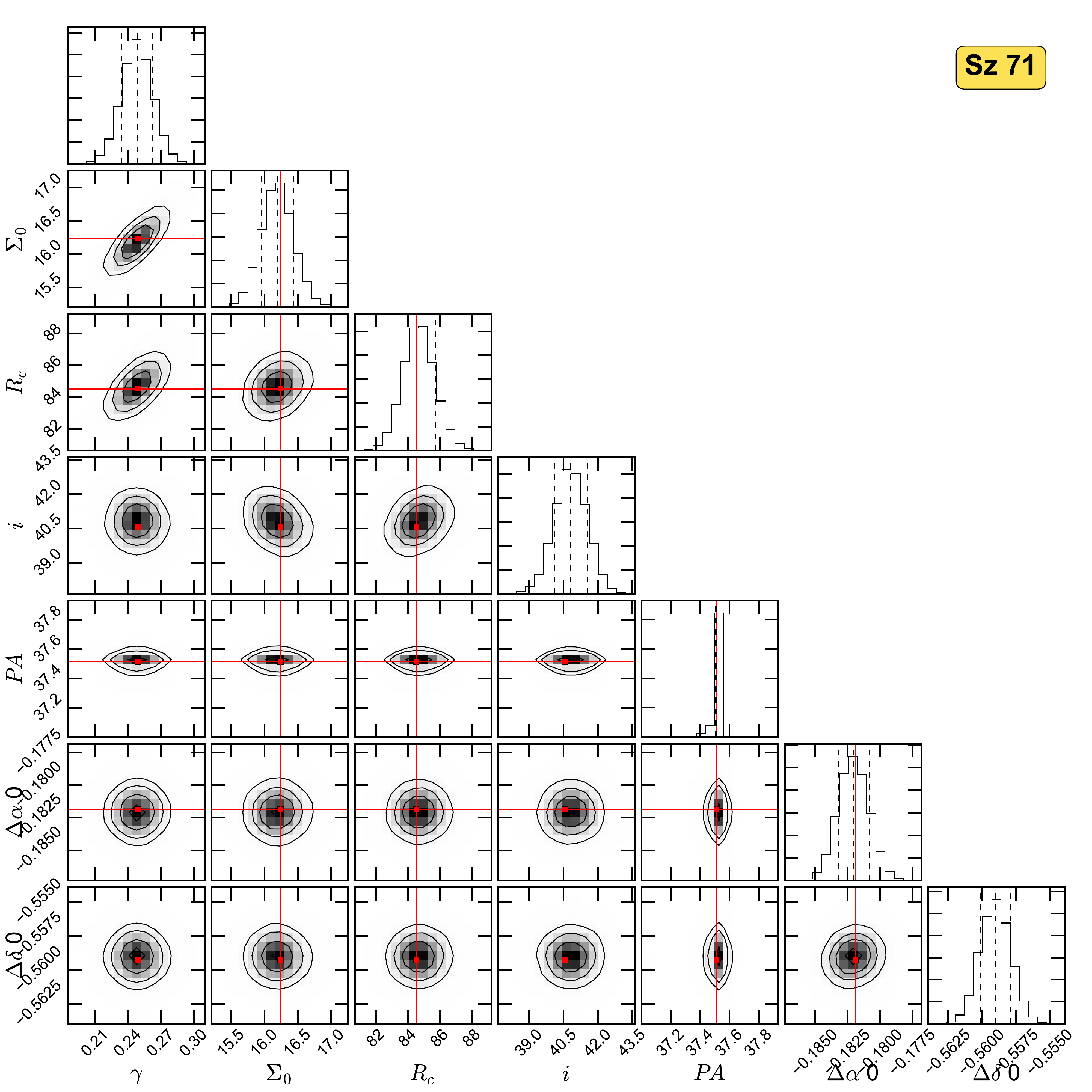}\includegraphics{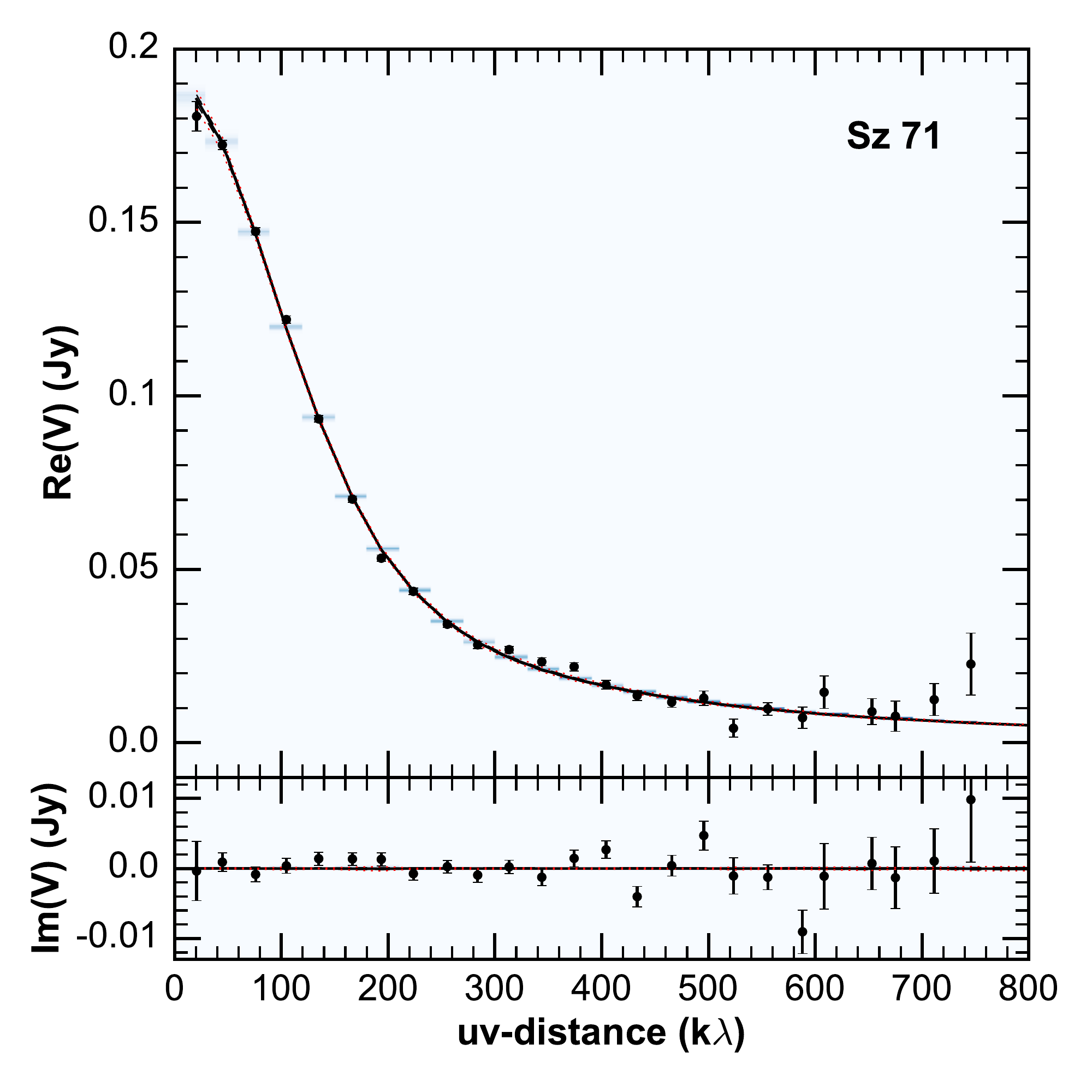}}\\[0.5cm]
\resizebox{\hsize}{!}{\includegraphics{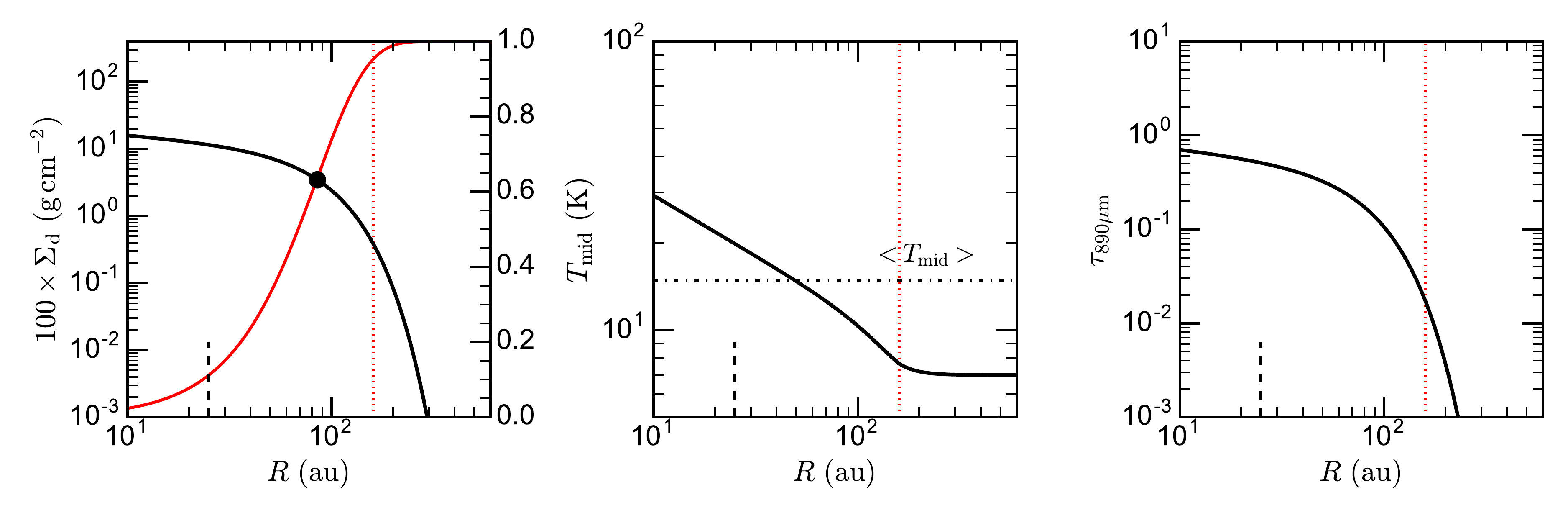}}\\[0.5cm]
\resizebox{0.8\hsize}{!}{\includegraphics{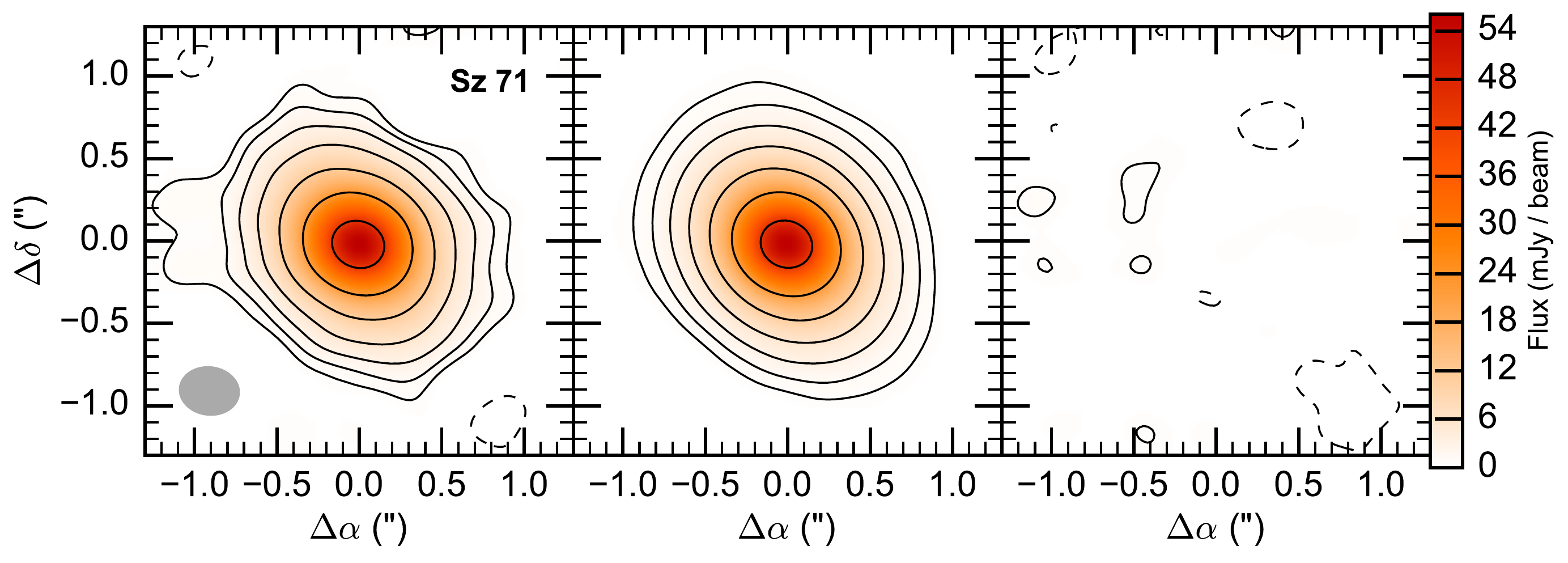}}\\[0.5cm]
\caption{Fit results for Sz 71, presented as in \figref{fig:reference.fit.results}. In the images $\sigma=0.23\u{mJy/beam}$.}
\end{figure*}

\pagebreak
\begin{figure*}
\centering
\Large\textbf{Sz 73\vspace{1cm}}
\resizebox{\hsize}{!}{\includegraphics[scale=0.5]{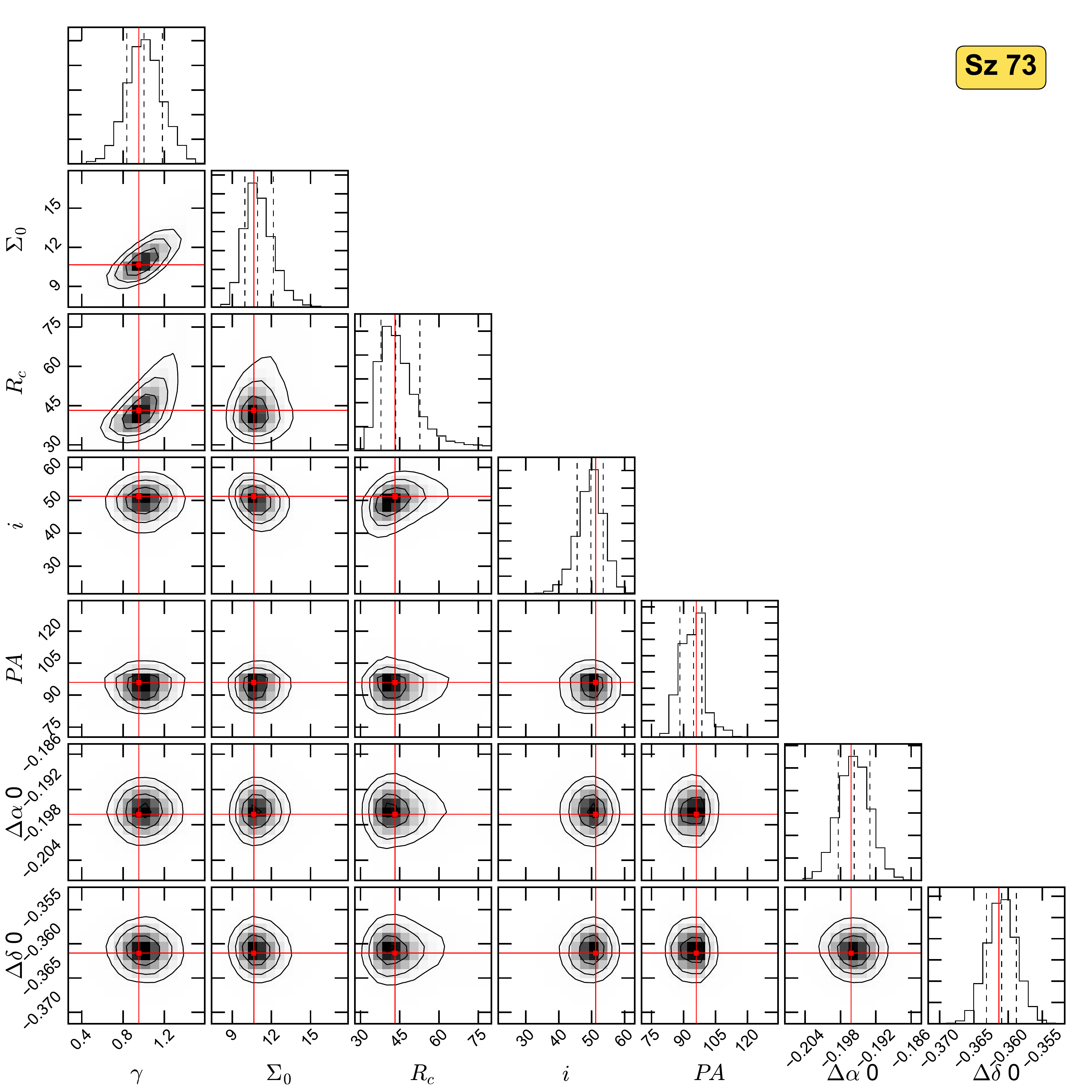}\includegraphics{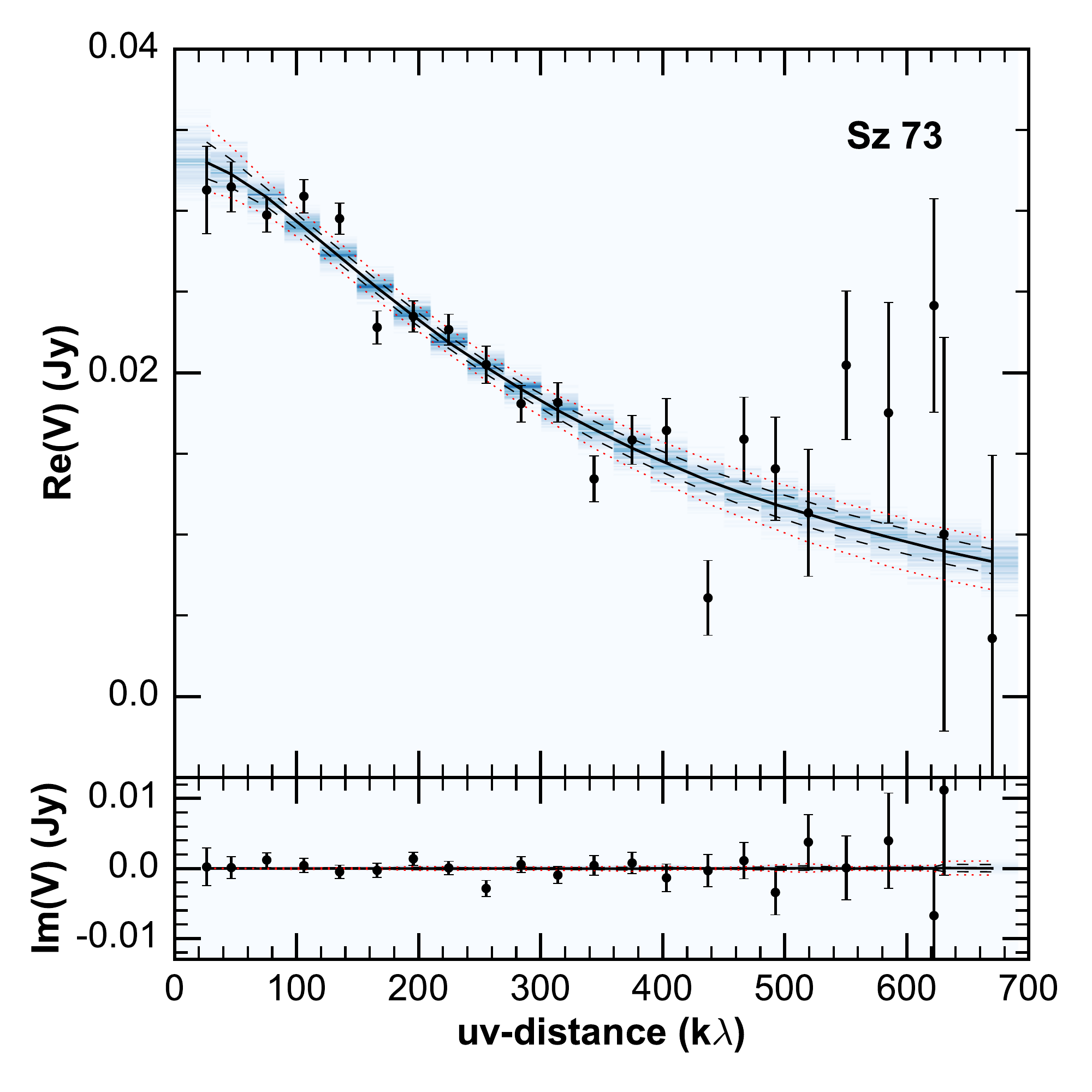}}\\[0.5cm]
\resizebox{\hsize}{!}{\includegraphics{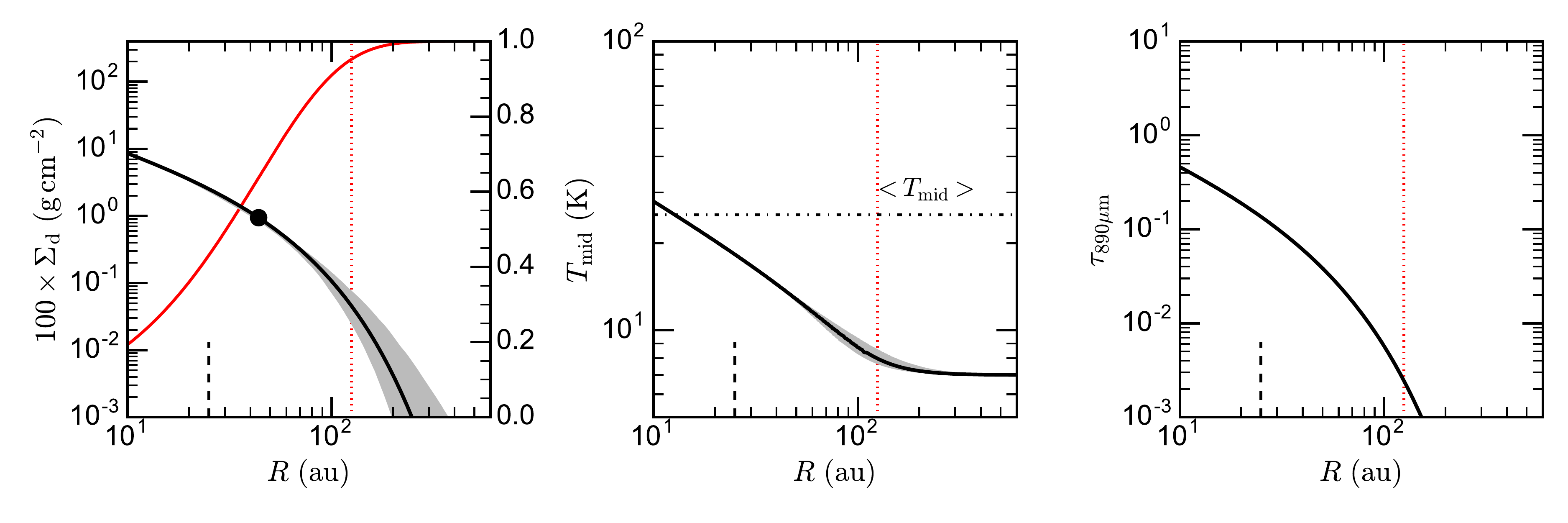}}\\[0.5cm]
\resizebox{0.8\hsize}{!}{\includegraphics{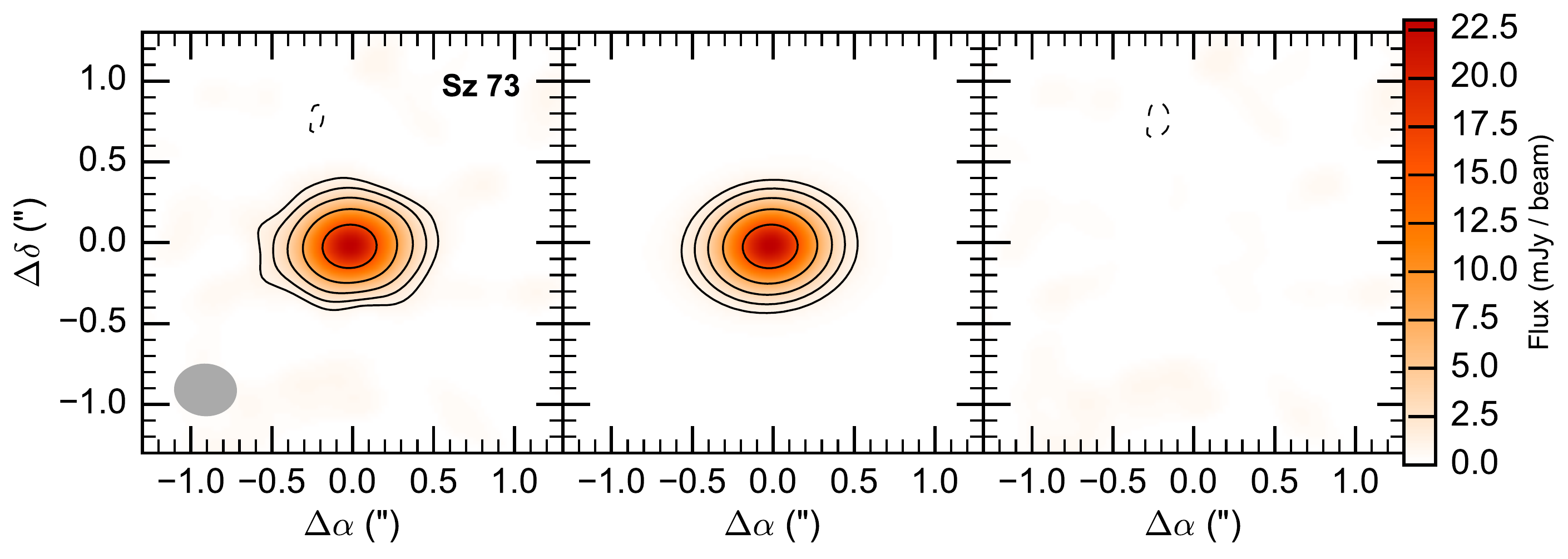}}\\[0.5cm]
\caption{Fit results for Sz 73, presented as in \figref{fig:reference.fit.results}. In the images $\sigma=0.33\u{mJy/beam}$.}
\end{figure*}

\pagebreak
\begin{figure*}
\centering
\Large\textbf{IM Lup\vspace{1cm}}
\resizebox{\hsize}{!}{\includegraphics[scale=0.5]{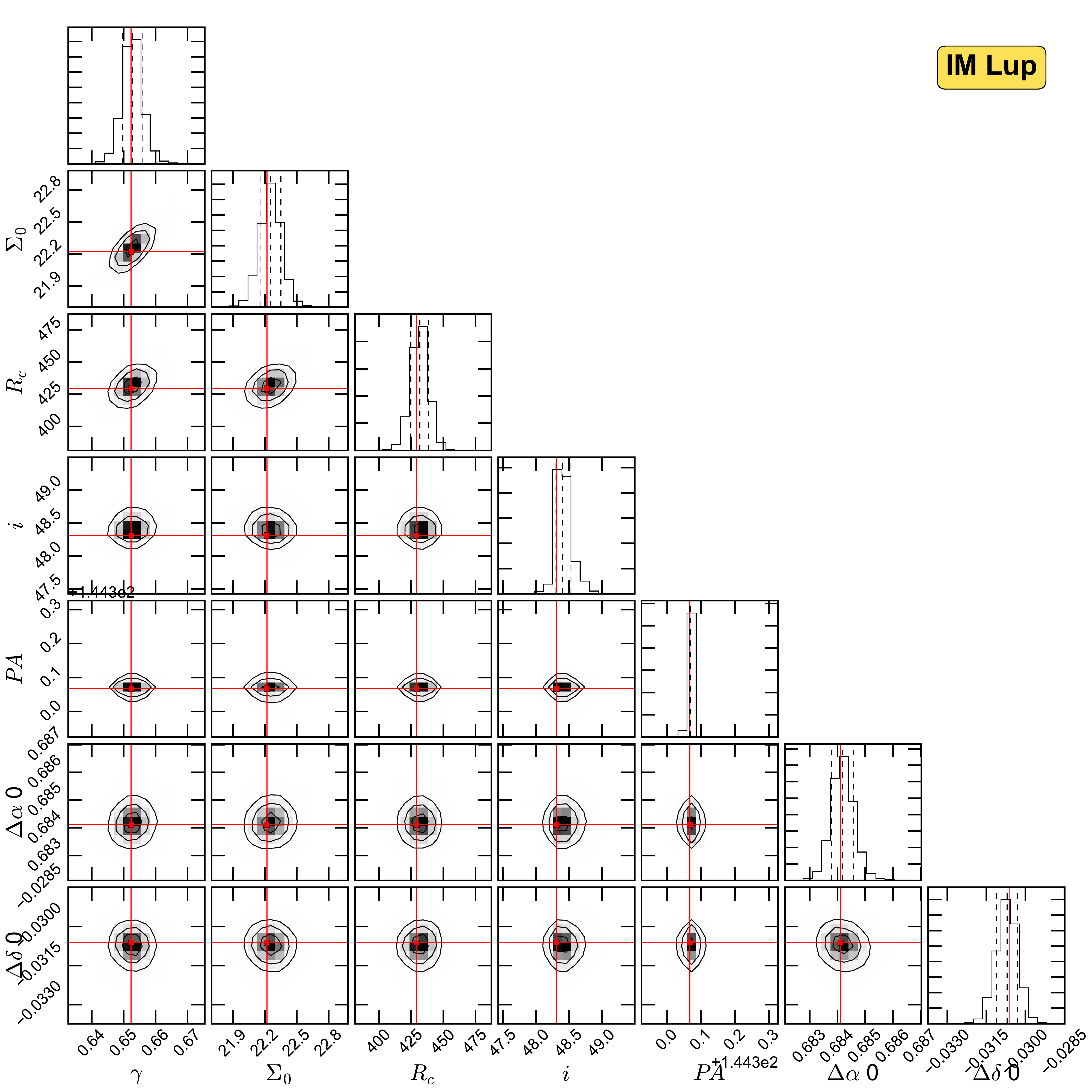}\includegraphics{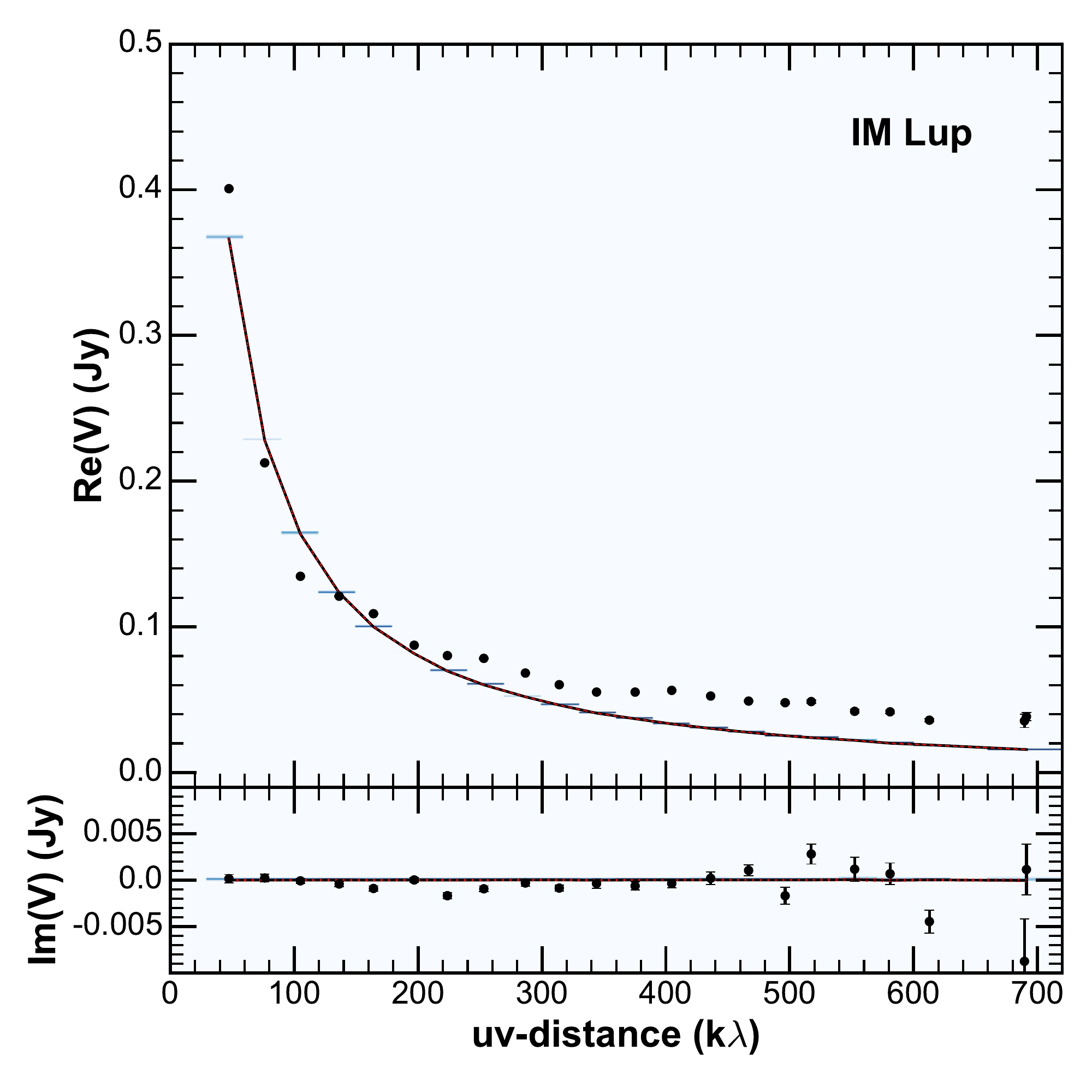}}\\[0.5cm]
\resizebox{\hsize}{!}{\includegraphics{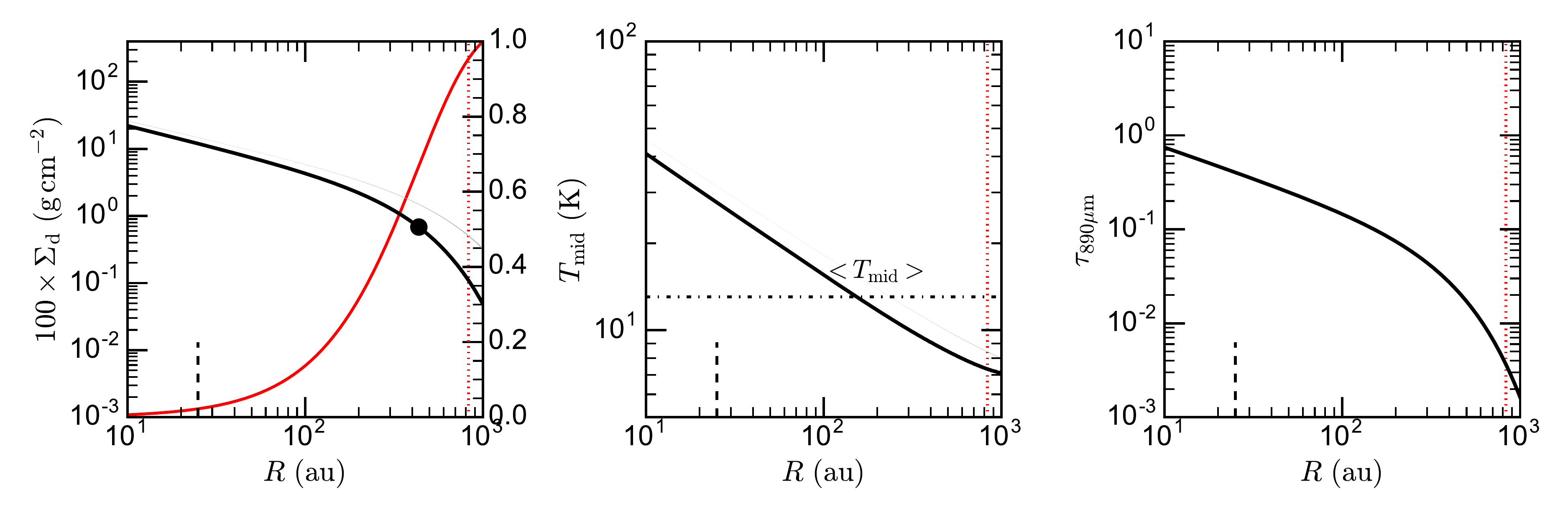}}\\[0.5cm]
\resizebox{0.8\hsize}{!}{\includegraphics{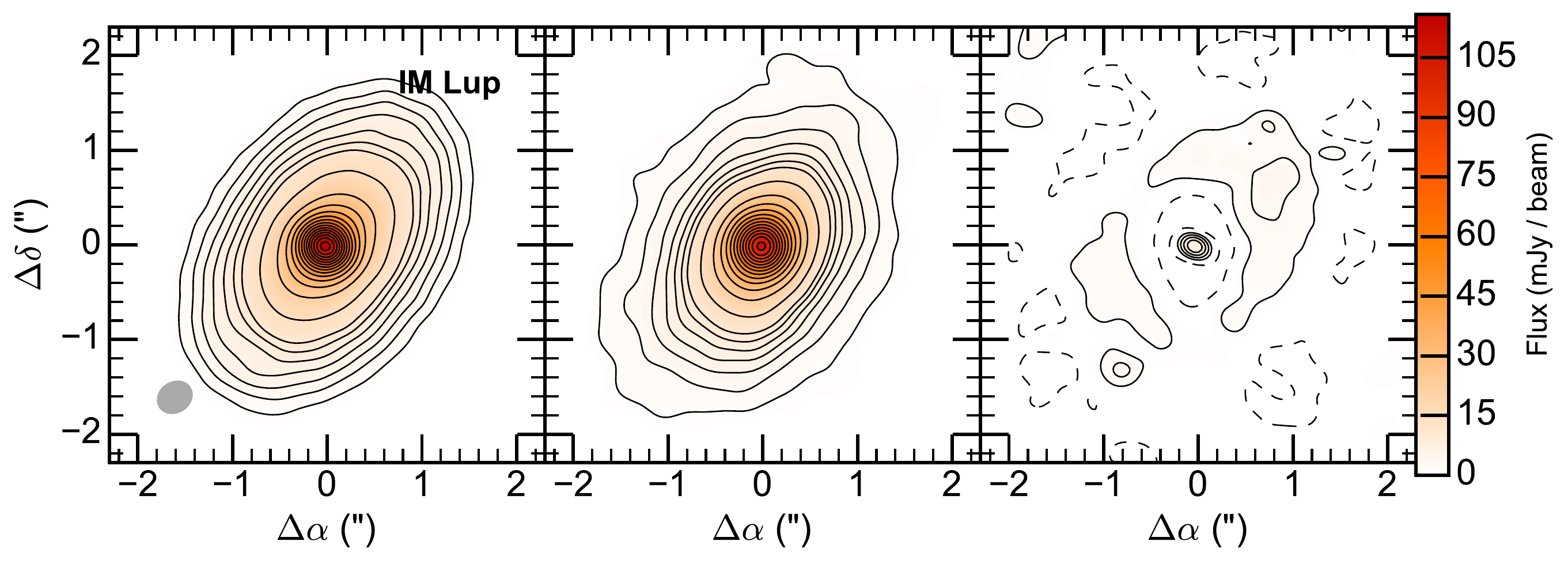}}\\[0.5cm]
\caption{Fit results for IM Lup, presented as in \figref{fig:reference.fit.results}. In the images $\sigma=0.6\u{mJy/beam}$.}
\end{figure*}

\pagebreak
\begin{figure*}
\centering
\Large\textbf{Sz 83\vspace{1cm}}
\resizebox{\hsize}{!}{\includegraphics[scale=0.5]{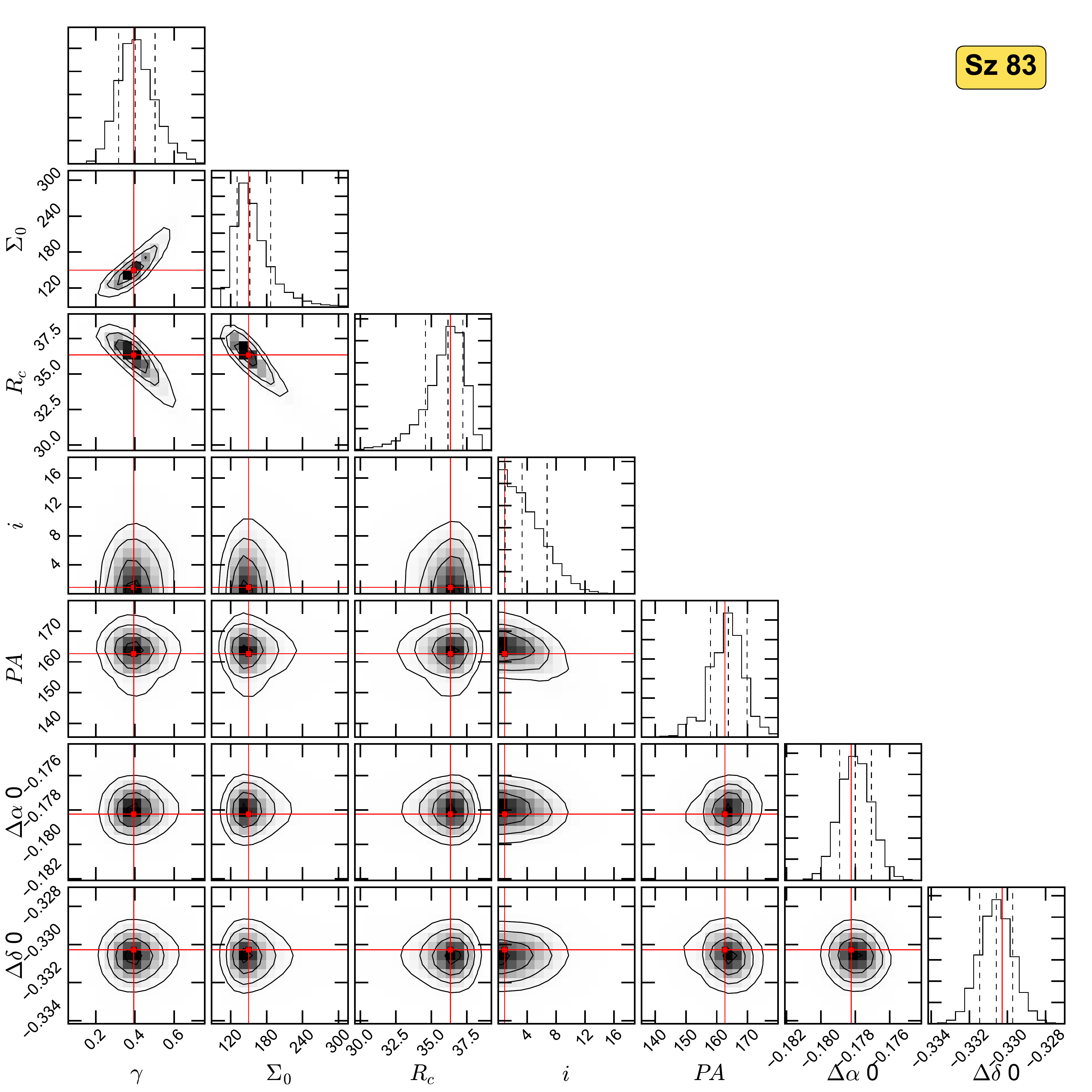}\includegraphics{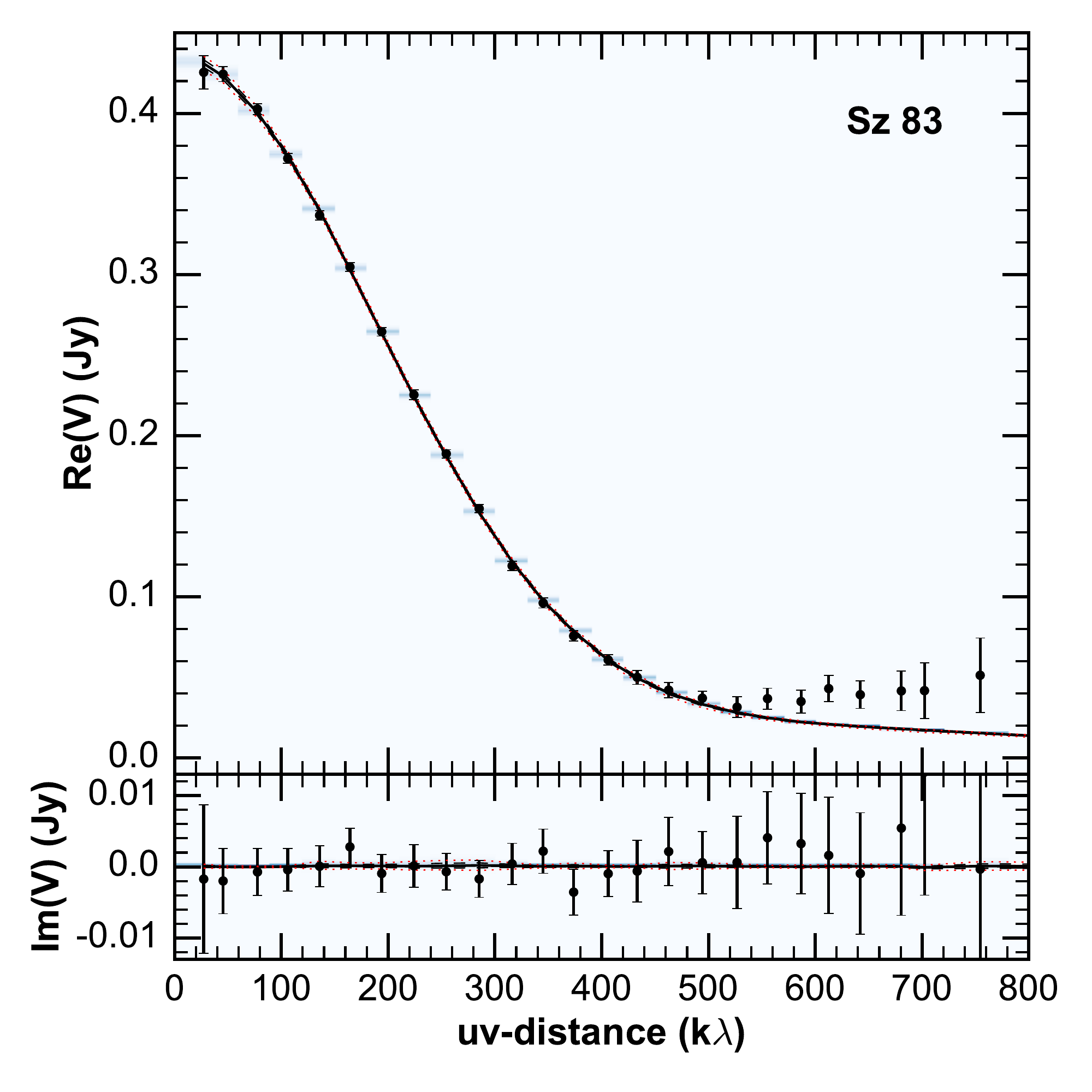}}\\[0.5cm]
\resizebox{\hsize}{!}{\includegraphics{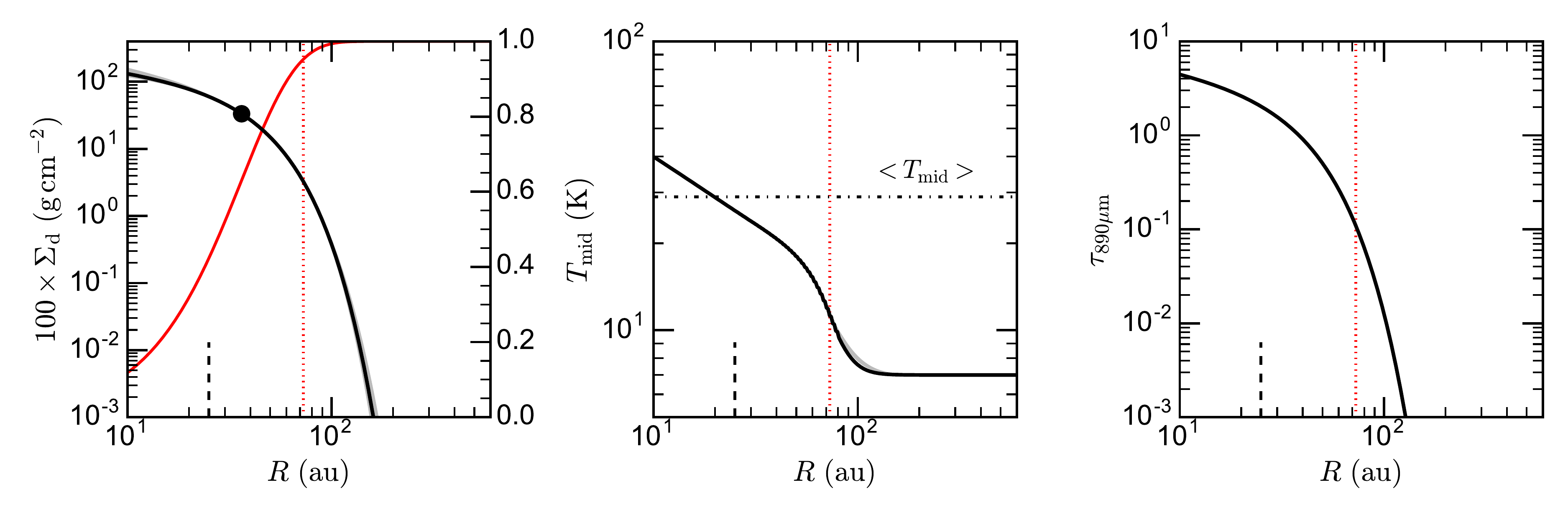}}\\[0.5cm]
\resizebox{0.8\hsize}{!}{\includegraphics{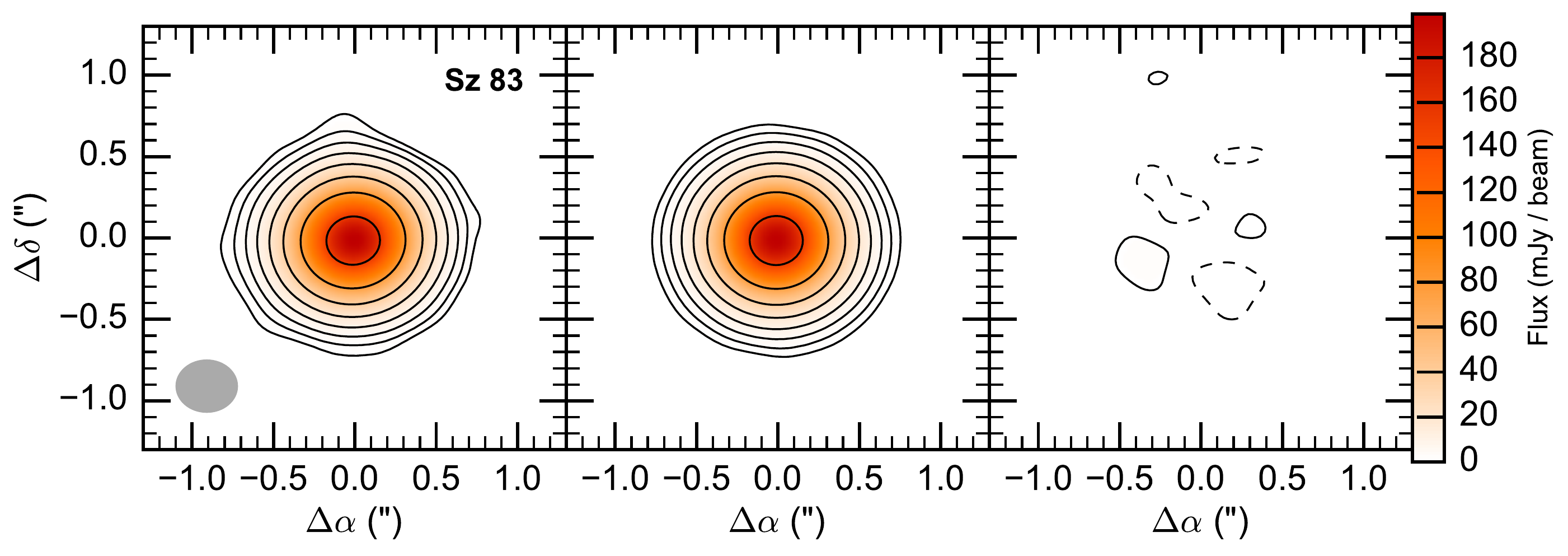}}\\[0.5cm]
\caption{Fit results for Sz 83, presented as in \figref{fig:reference.fit.results}. In the images $\sigma=0.4\u{mJy/beam}$.}
\end{figure*}

\pagebreak
\begin{figure*}
\centering
\Large\textbf{Sz 84\vspace{1cm}}
\resizebox{\hsize}{!}{\includegraphics[scale=0.5]{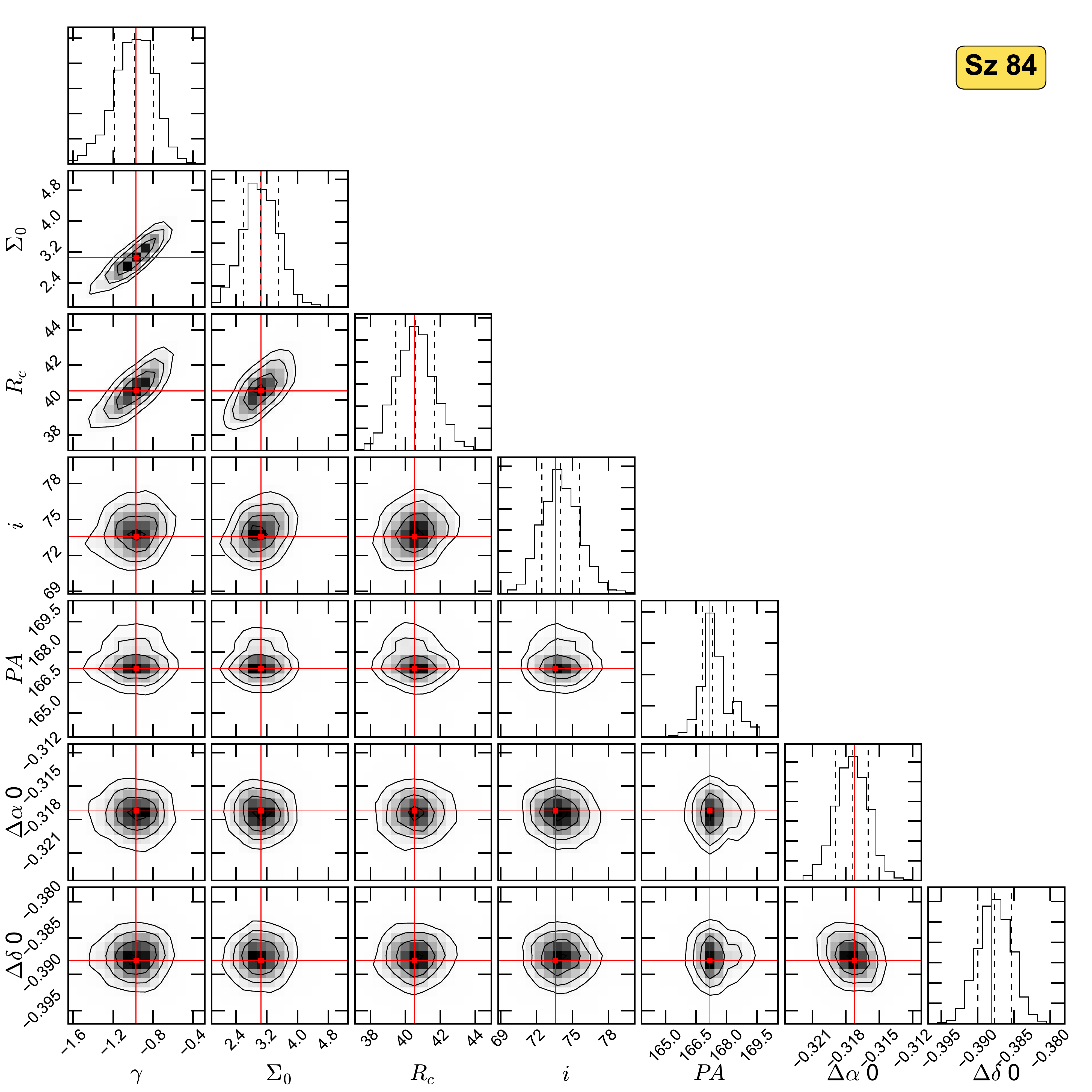}\includegraphics{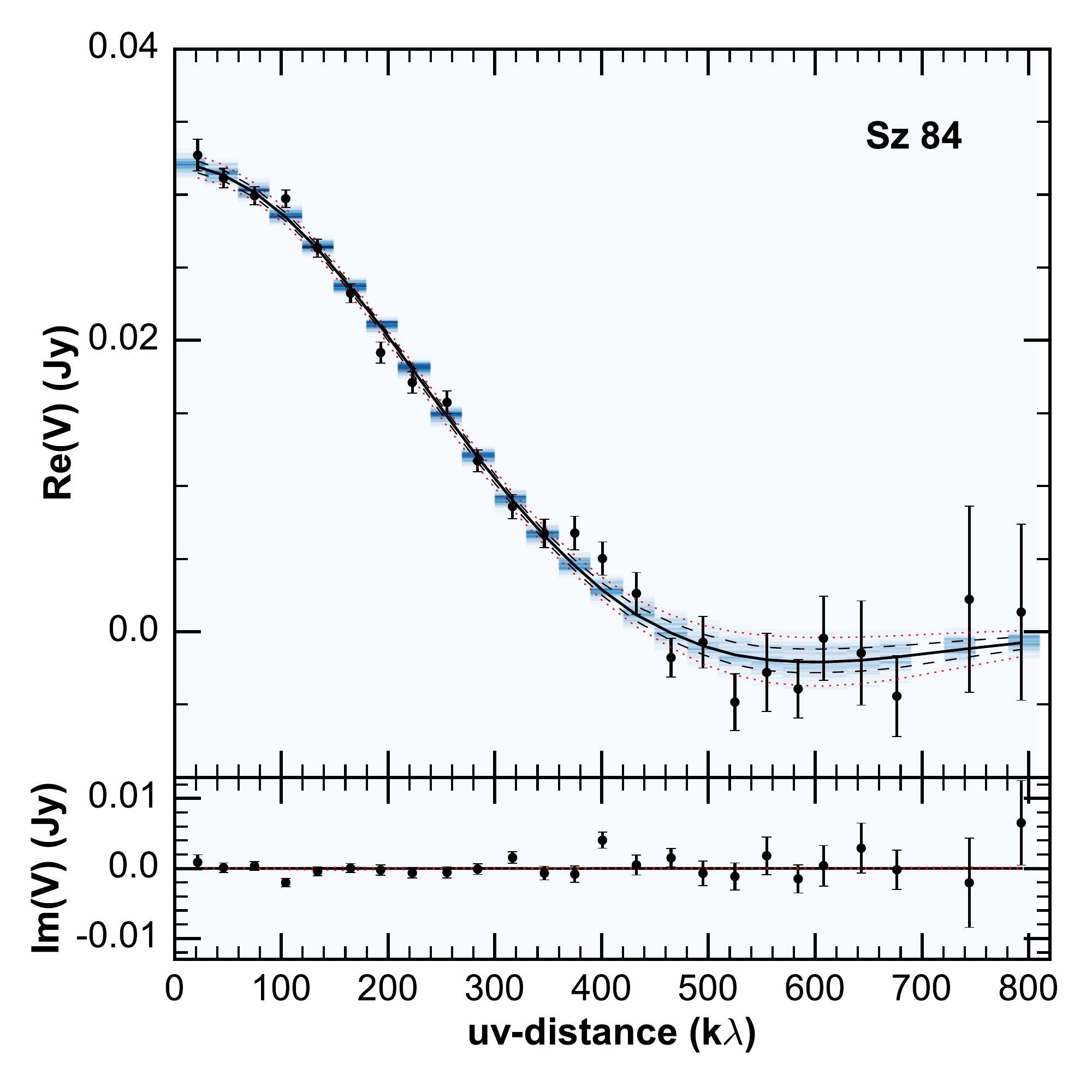}}\\[0.5cm]
\resizebox{\hsize}{!}{\includegraphics{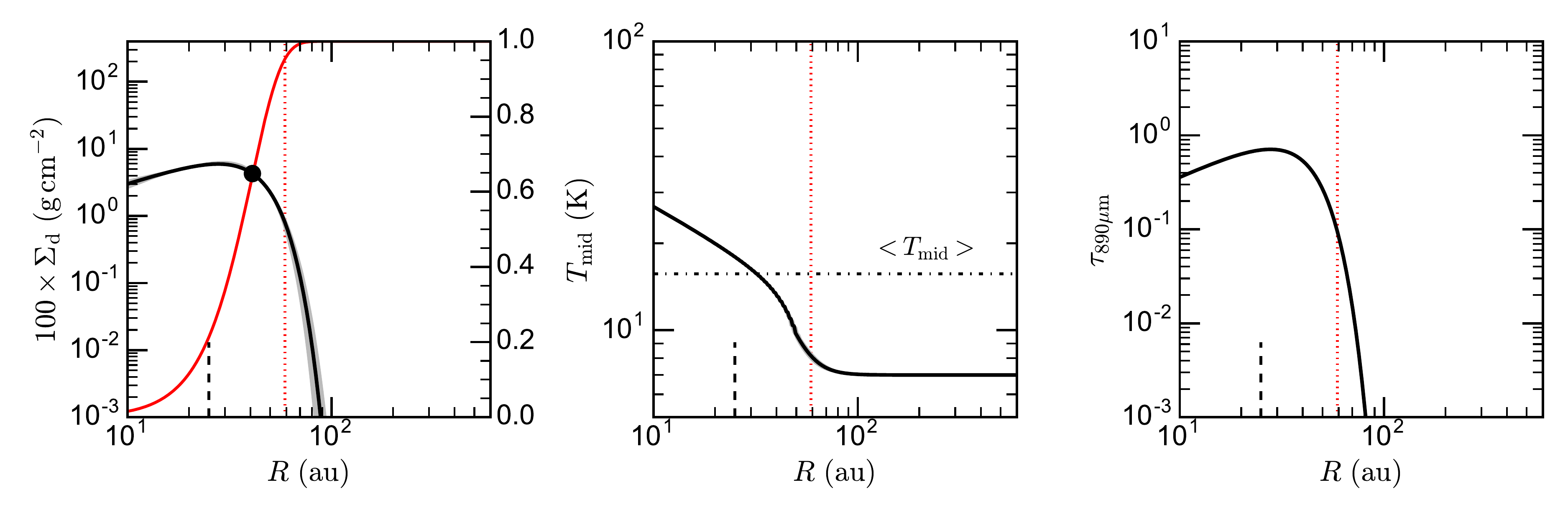}}\\[0.5cm]
\resizebox{0.8\hsize}{!}{\includegraphics{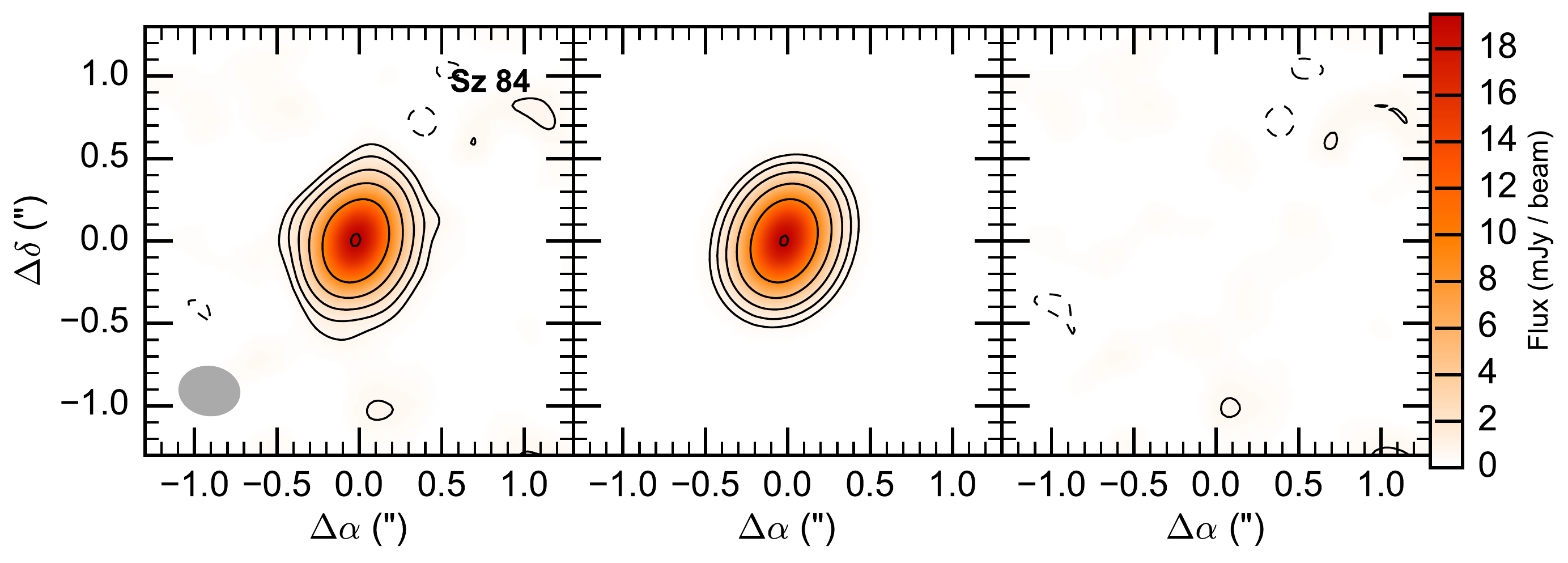}}\\[0.5cm]
\caption{Fit results for Sz 84, presented as in \figref{fig:reference.fit.results}. In the images $\sigma=0.2\u{mJy/beam}$.}
\end{figure*}

\pagebreak
\begin{figure*}
\centering
\Large\textbf{Sz 129\vspace{1cm}}
\resizebox{\hsize}{!}{\includegraphics[scale=0.5]{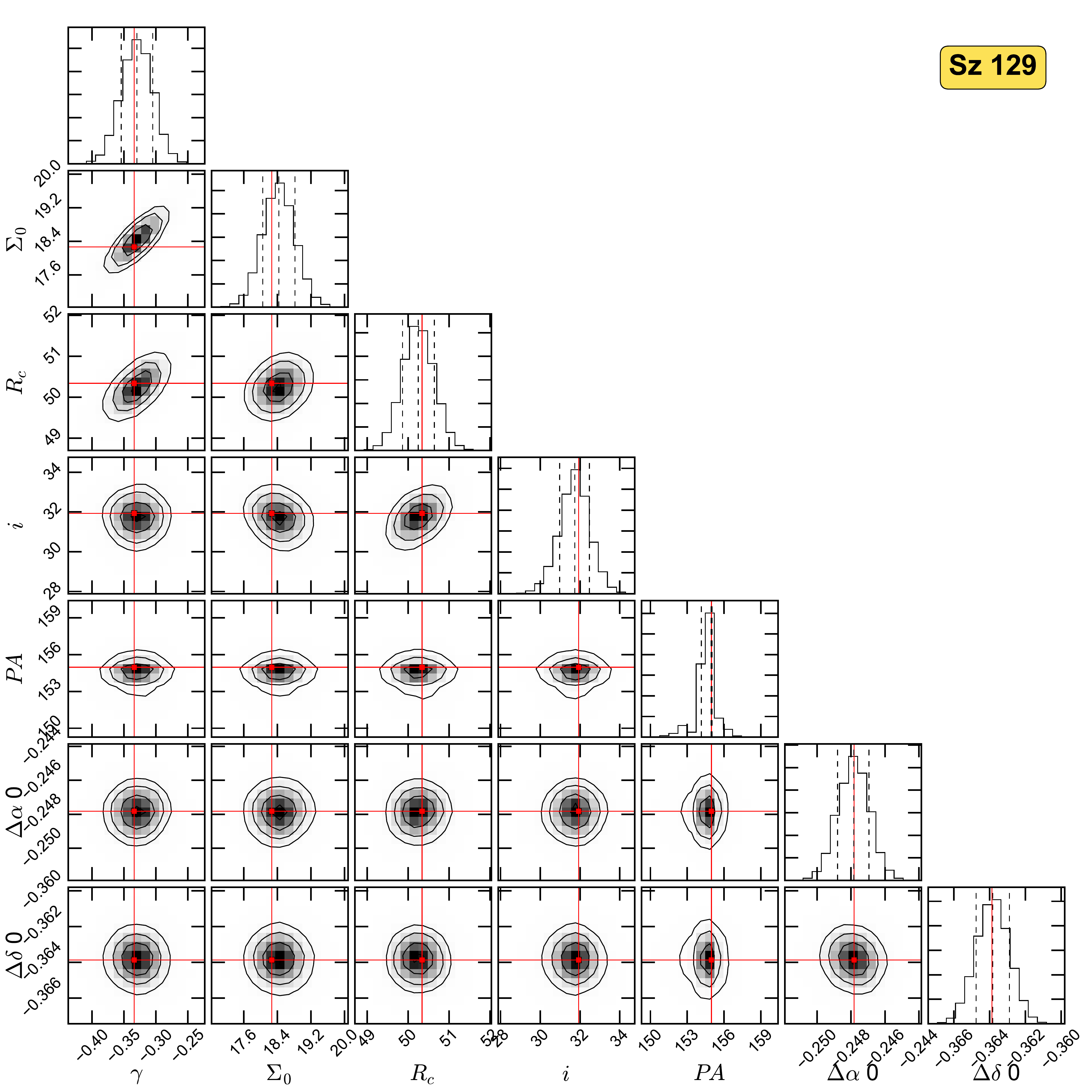}\includegraphics{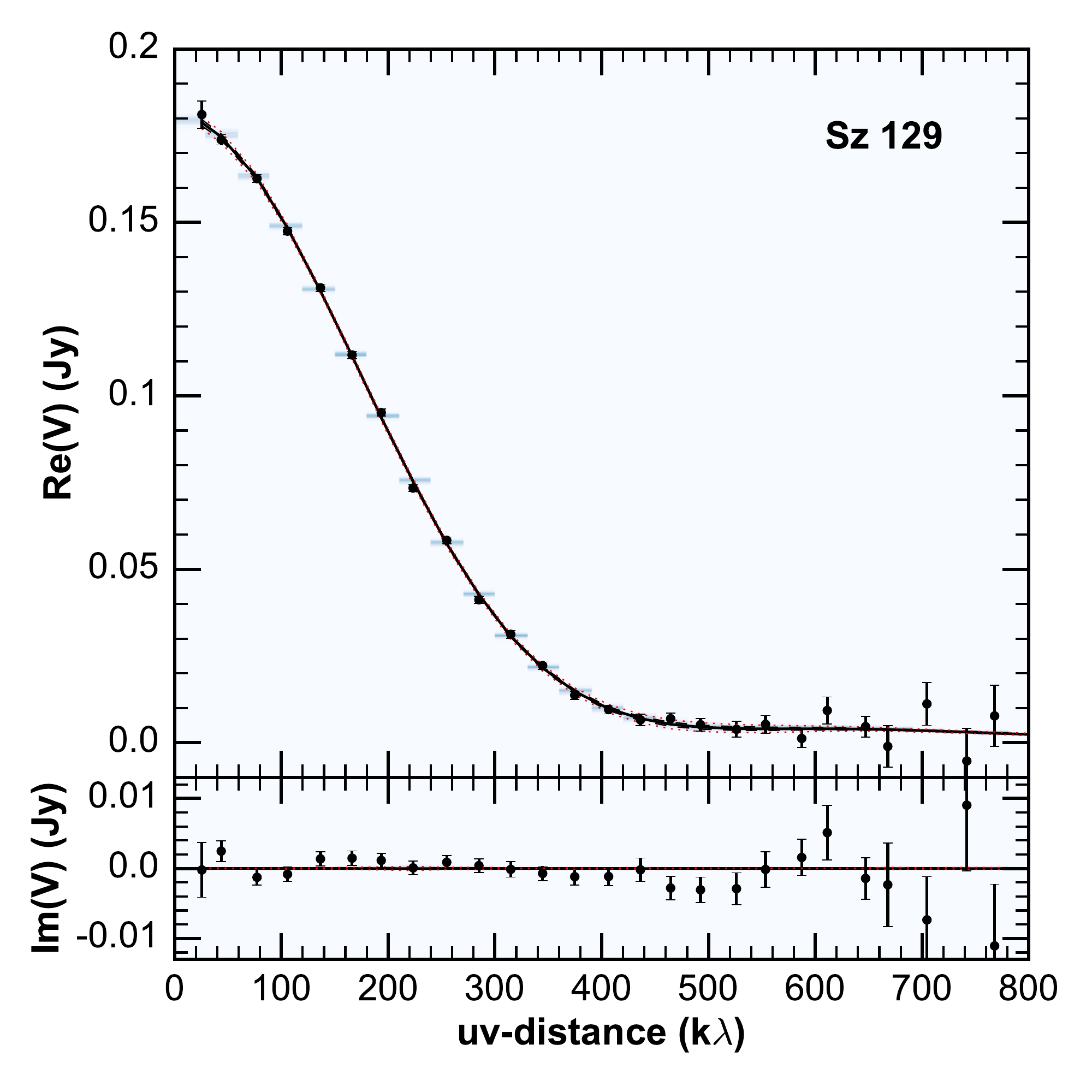}}\\[0.5cm]
\resizebox{\hsize}{!}{\includegraphics{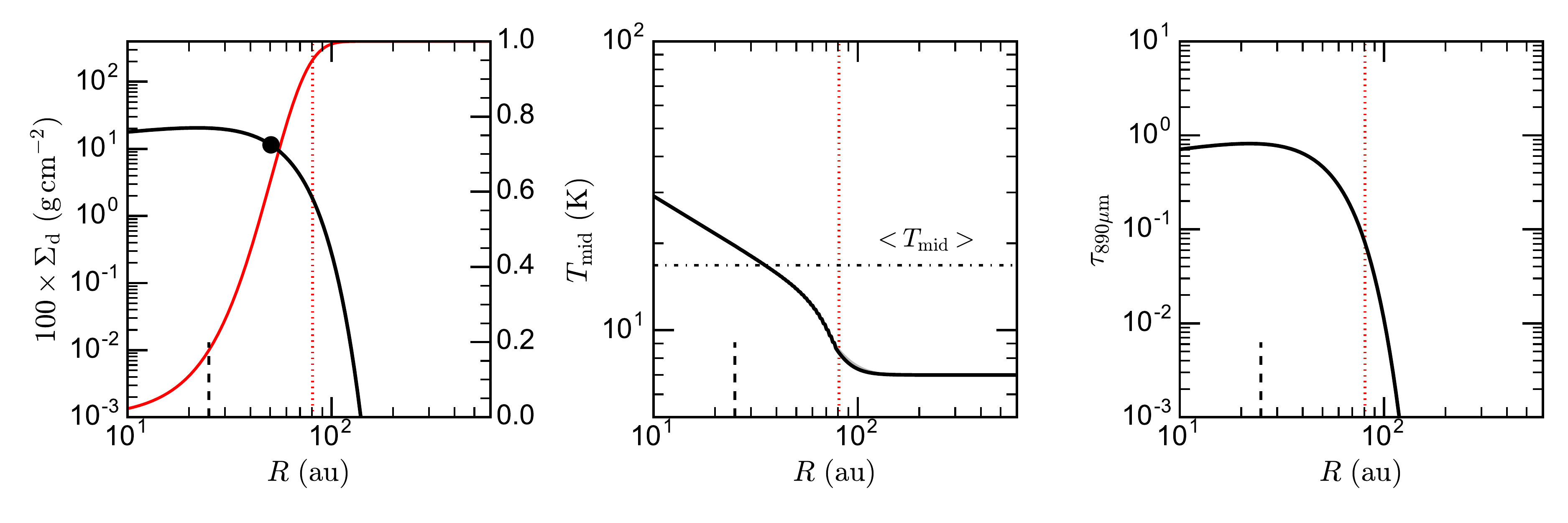}}\\[0.5cm]
\resizebox{0.8\hsize}{!}{\includegraphics{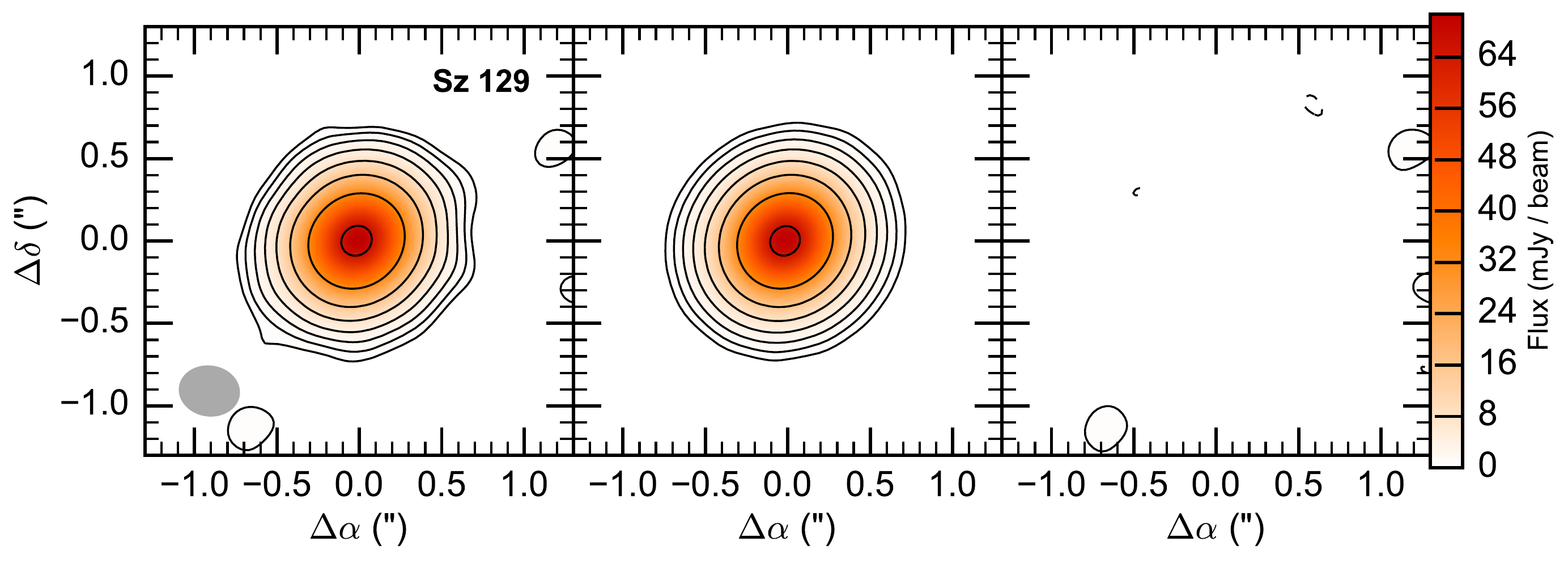}}\\[0.5cm]
\caption{Fit results for Sz 129, presented as in \figref{fig:reference.fit.results}. In the images $\sigma=0.2\u{mJy/beam}$.}
\end{figure*}

\pagebreak
\begin{figure*}
\centering
\Large\textbf{J16000236-4222145\vspace{1cm}}
\resizebox{\hsize}{!}{\includegraphics[scale=0.5]{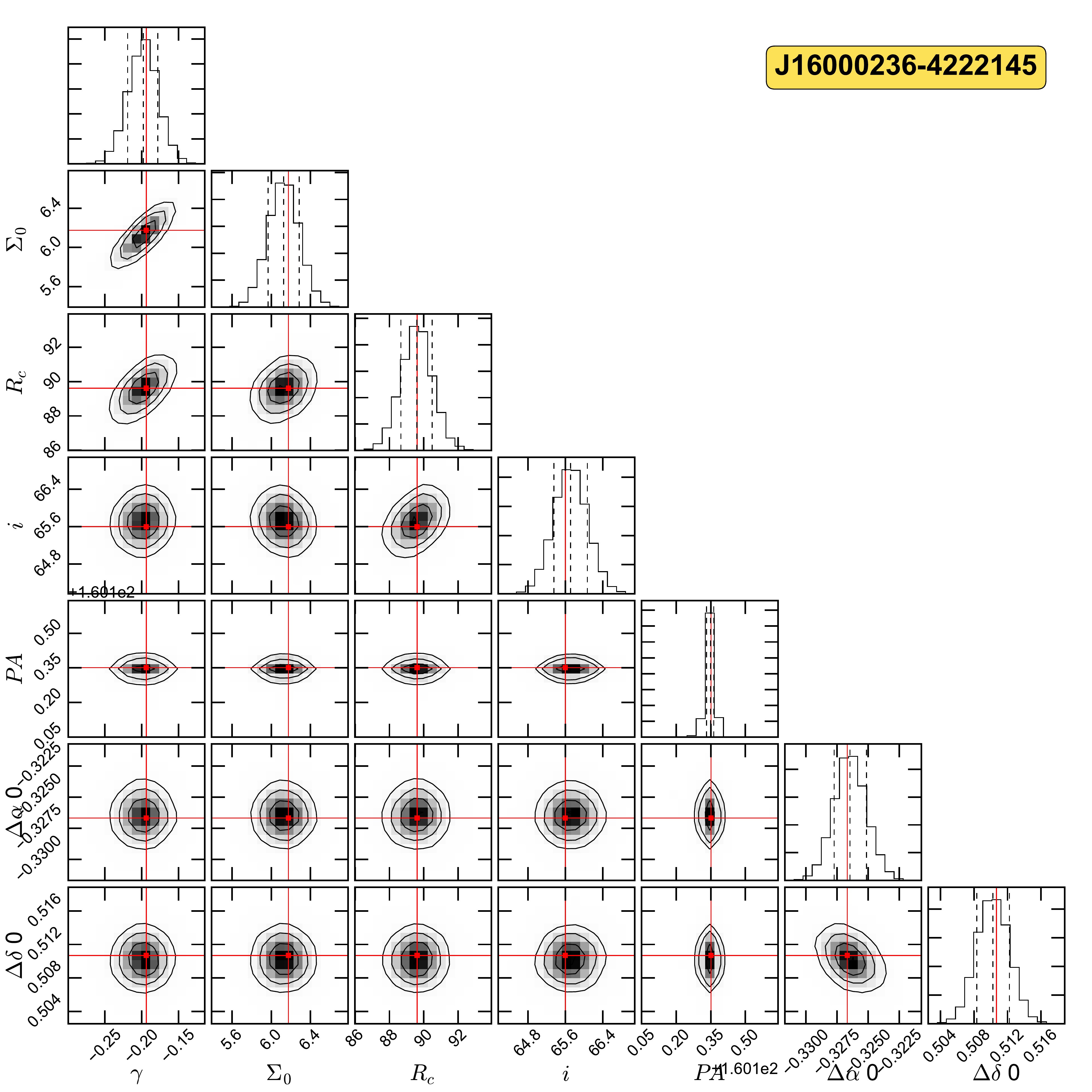}\includegraphics{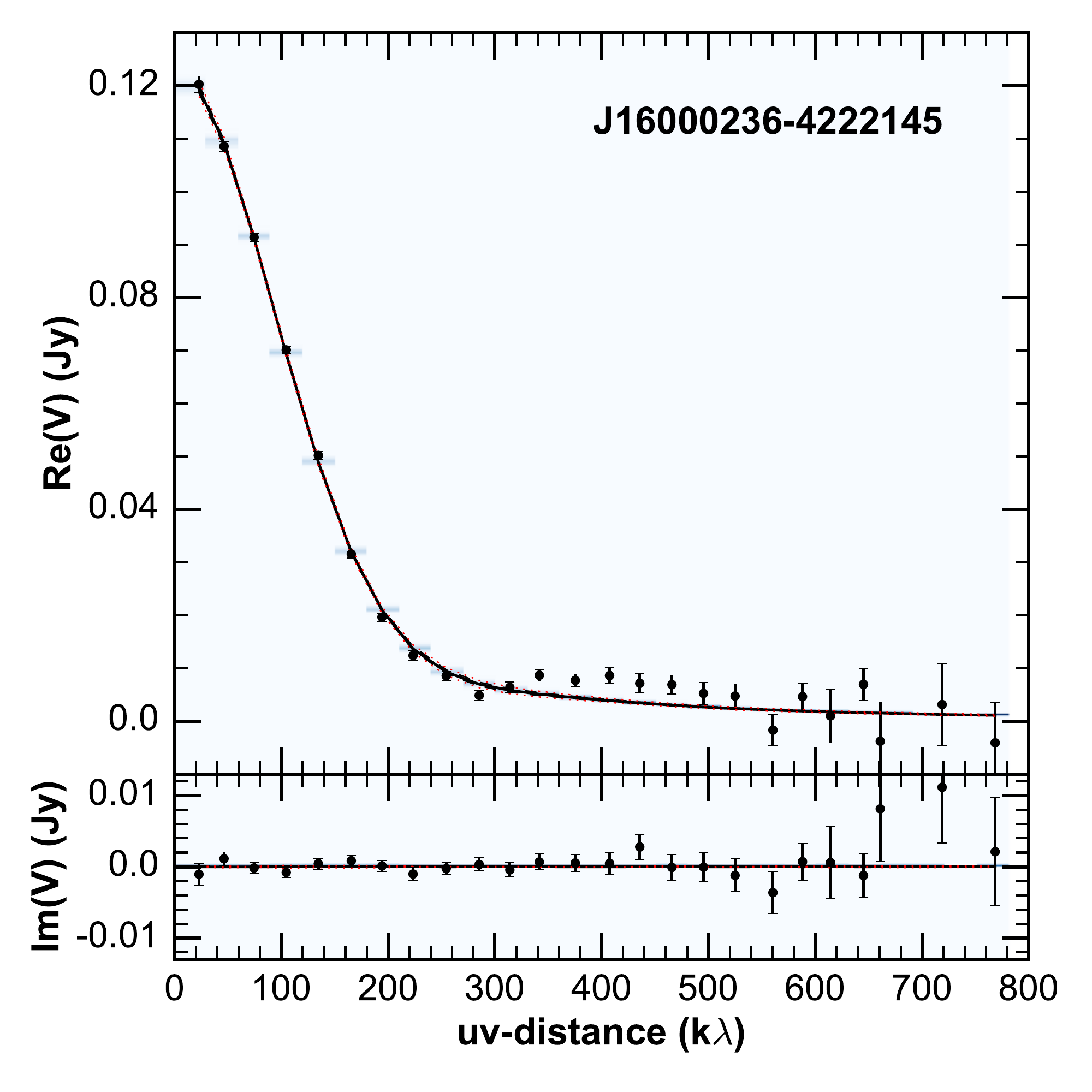}}\\[0.5cm]
\resizebox{\hsize}{!}{\includegraphics{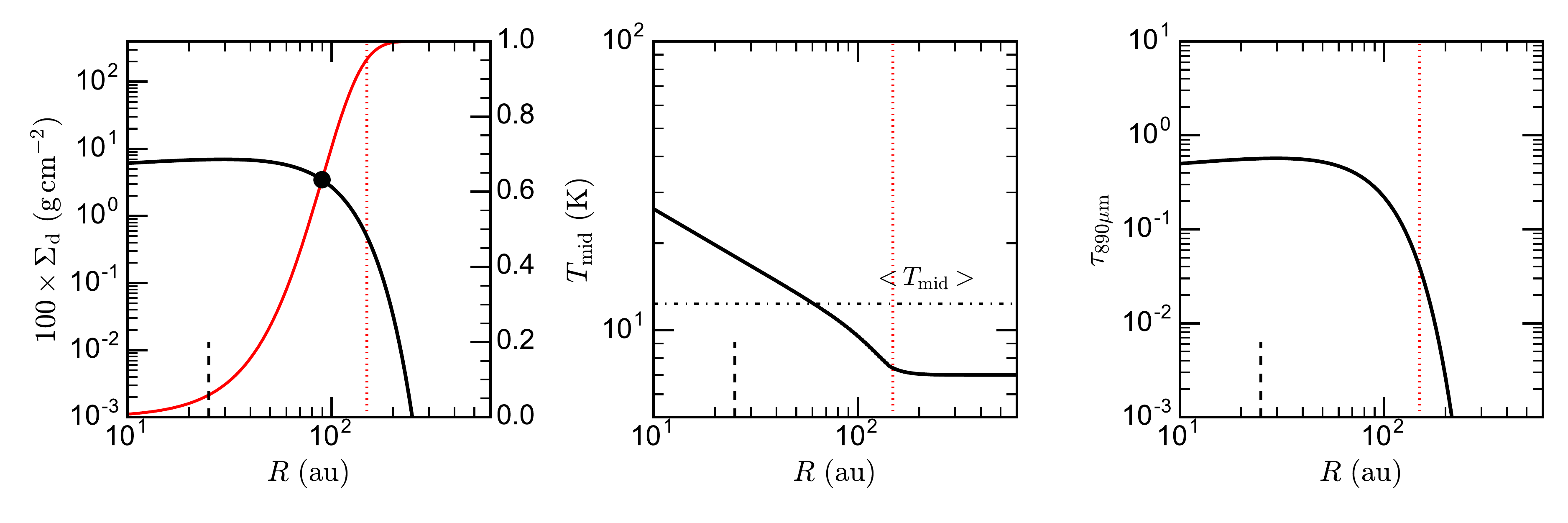}}\\[0.5cm]
\resizebox{0.8\hsize}{!}{\includegraphics{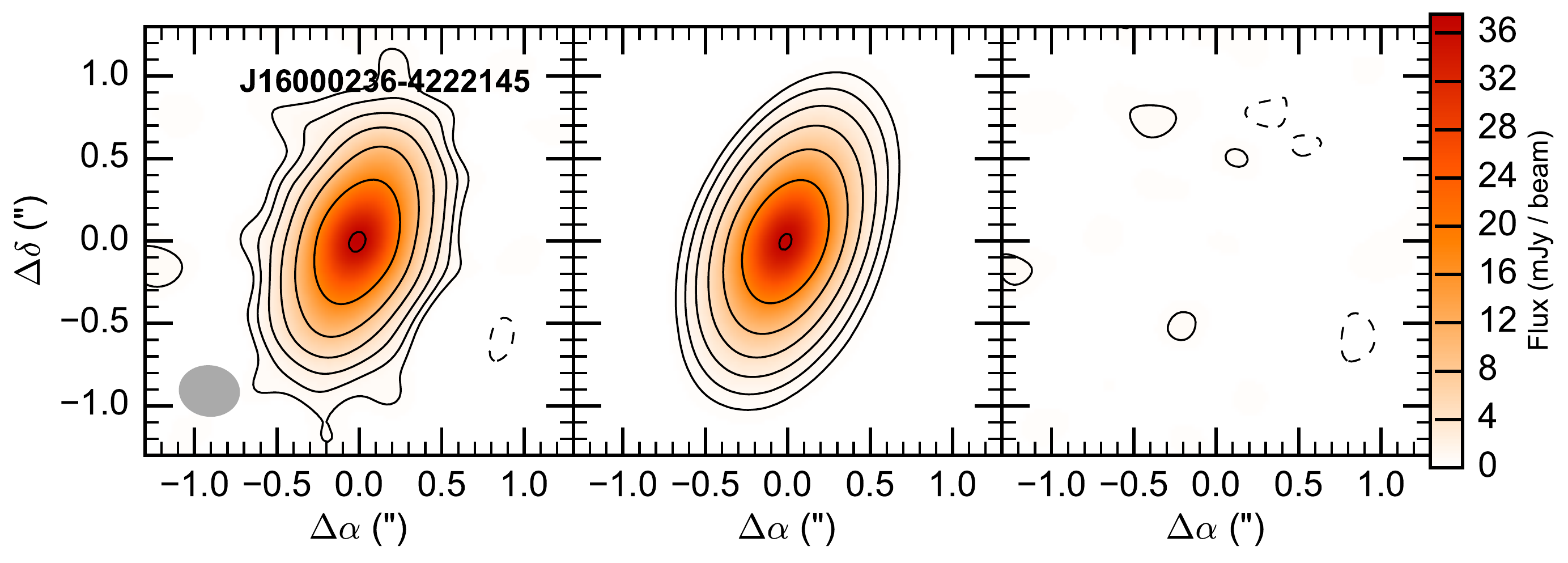}}\\[0.5cm]
\caption{Fit results for J16000236-4222145, presented as in \figref{fig:reference.fit.results}. In the images $\sigma=0.25\u{mJy/beam}$.}
\end{figure*}

\pagebreak
\begin{figure*}
\centering
\Large\textbf{MY Lup\vspace{1cm}}
\resizebox{\hsize}{!}{\includegraphics[scale=0.5]{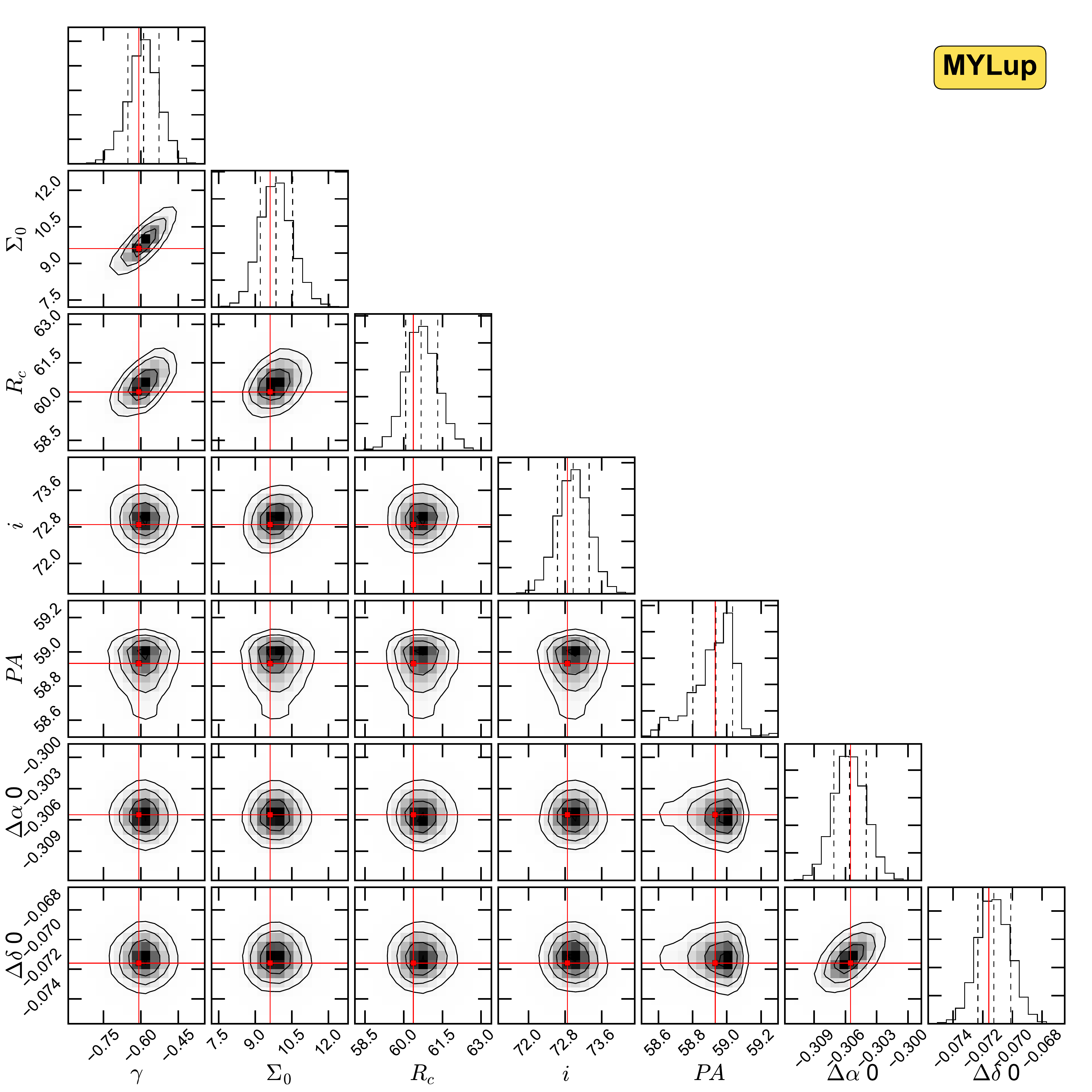}\includegraphics{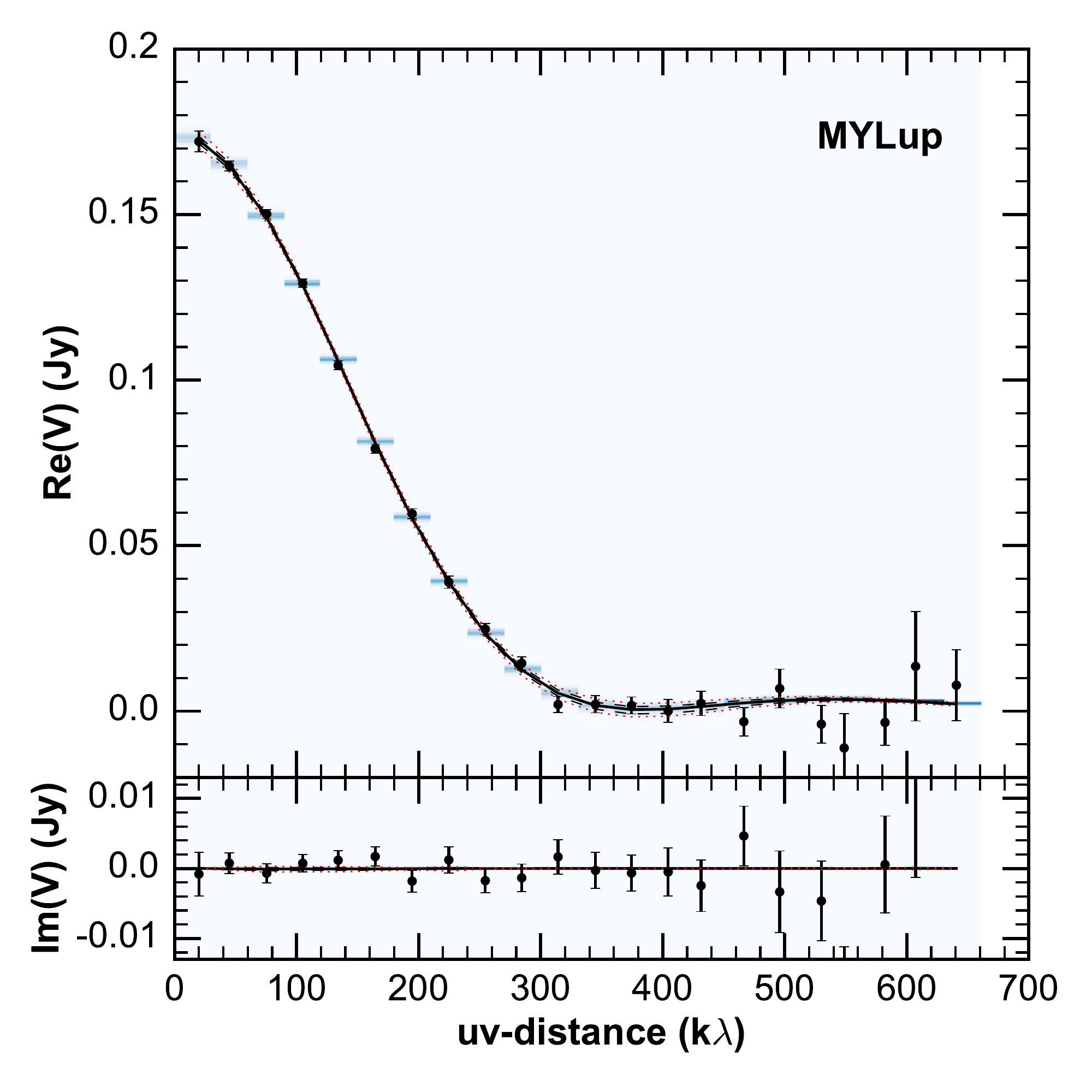}}\\[0.5cm]
\resizebox{\hsize}{!}{\includegraphics{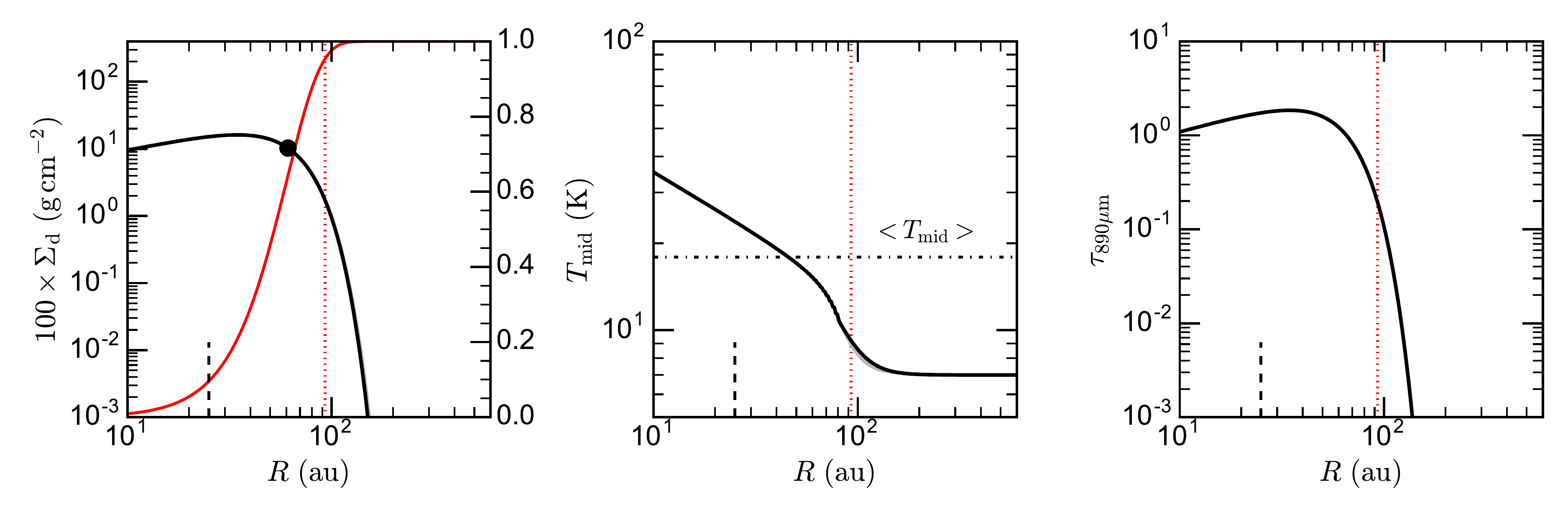}}\\[0.5cm]
\resizebox{0.8\hsize}{!}{\includegraphics{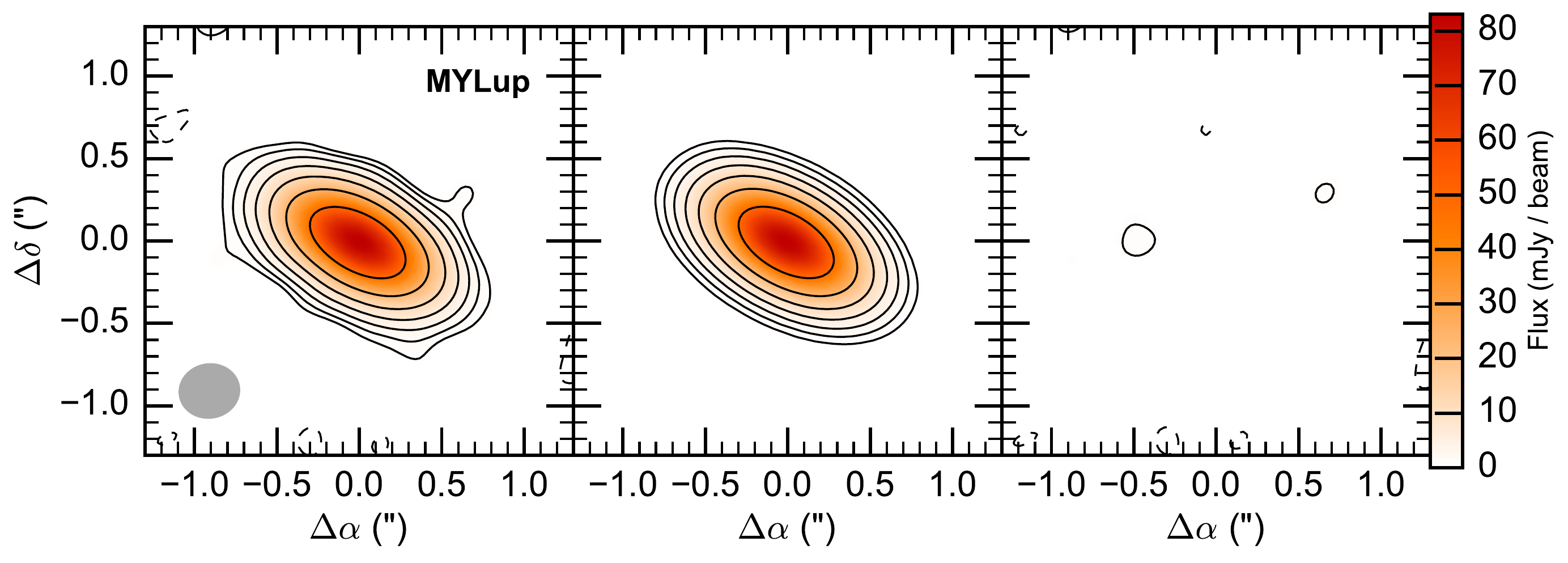}}\\[0.5cm]
\caption{Fit results for MY Lup, presented as in \figref{fig:reference.fit.results}. In the images $\sigma=0.25\u{mJy/beam}$.}
\end{figure*}

\pagebreak
\begin{figure*}
\centering
\Large\textbf{Sz 133\vspace{1cm}}
\resizebox{\hsize}{!}{\includegraphics[scale=0.5]{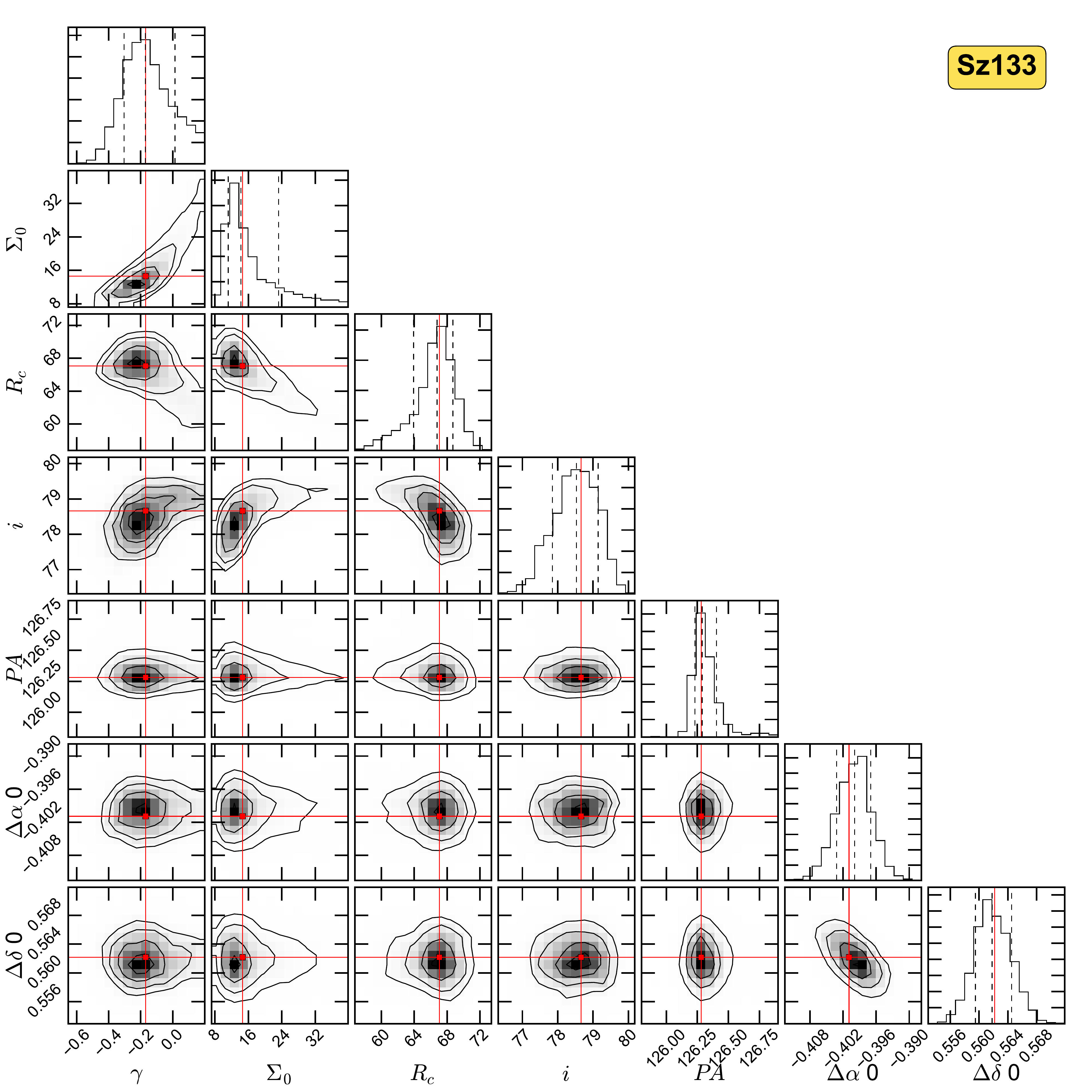}\includegraphics{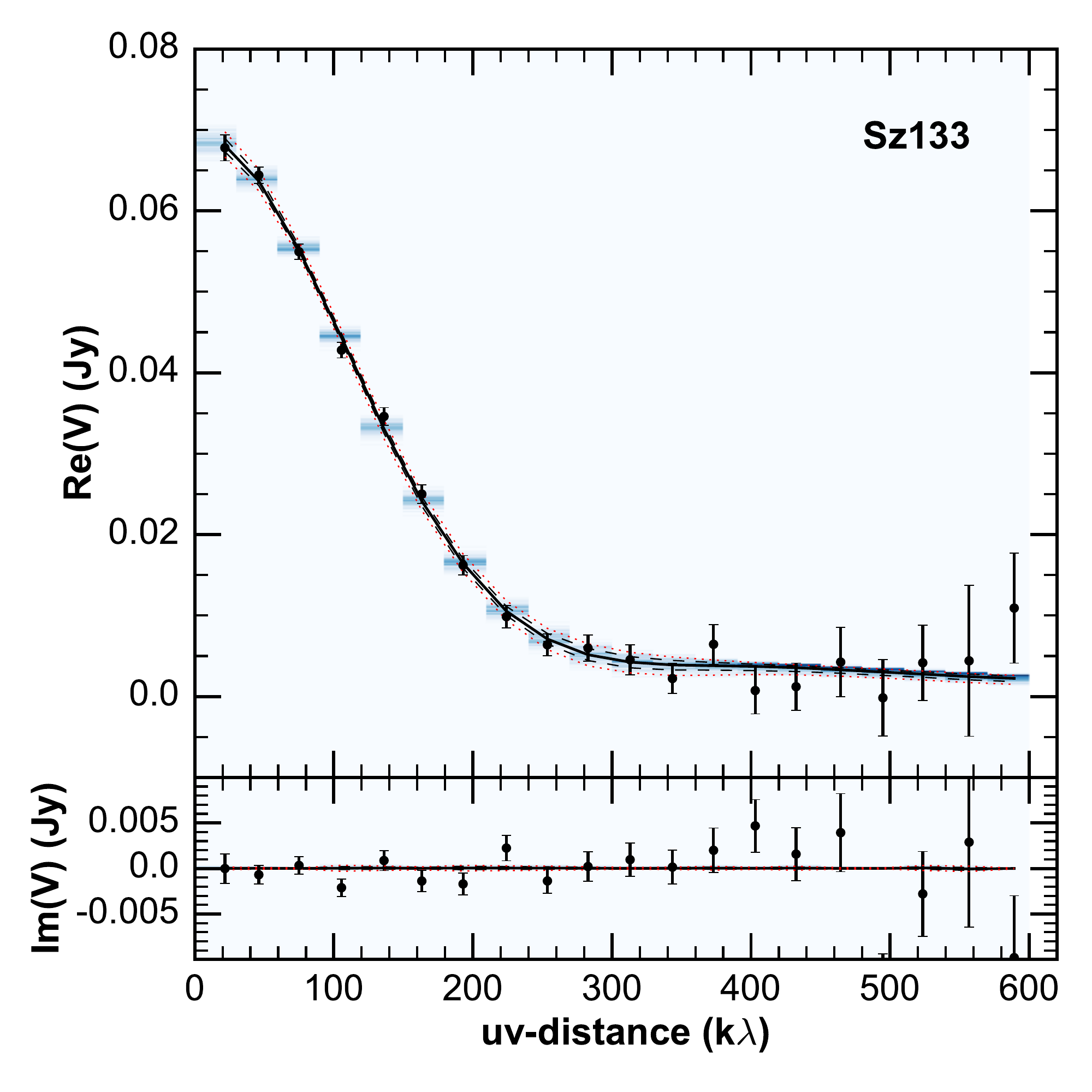}}\\[0.5cm]
\resizebox{\hsize}{!}{\includegraphics{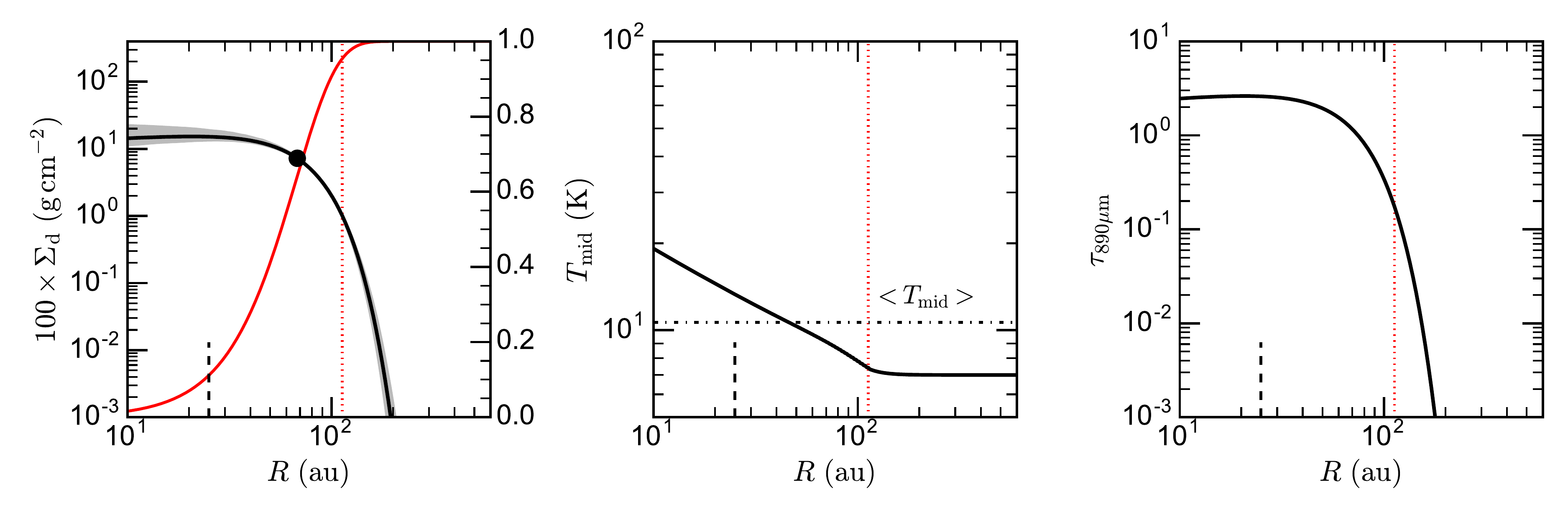}}\\[0.5cm]
\resizebox{0.8\hsize}{!}{\includegraphics{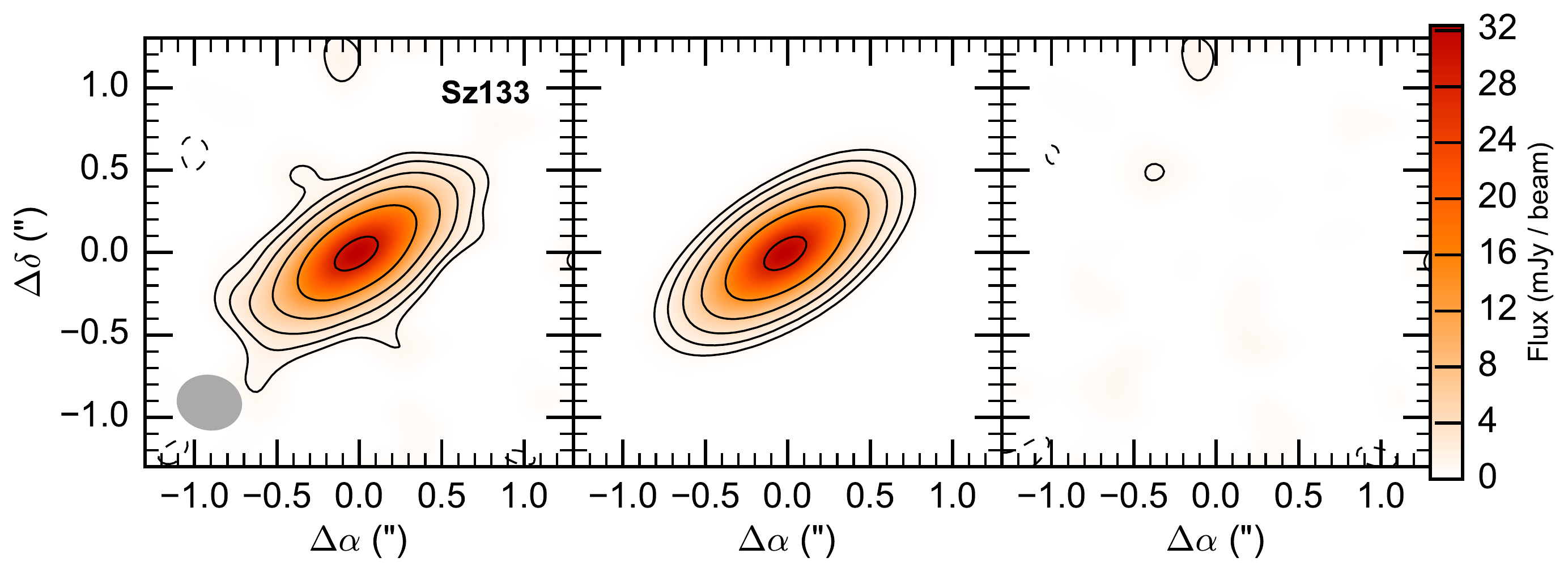}}\\[0.5cm]
\caption{Fit results for Sz 133, presented as in \figref{fig:reference.fit.results}. In the images $\sigma=0.3\u{mJy/beam}$.}
\end{figure*}

\pagebreak
\begin{figure*}
\centering
\Large\textbf{Sz 90\vspace{1cm}}
\resizebox{\hsize}{!}{\includegraphics[scale=0.5]{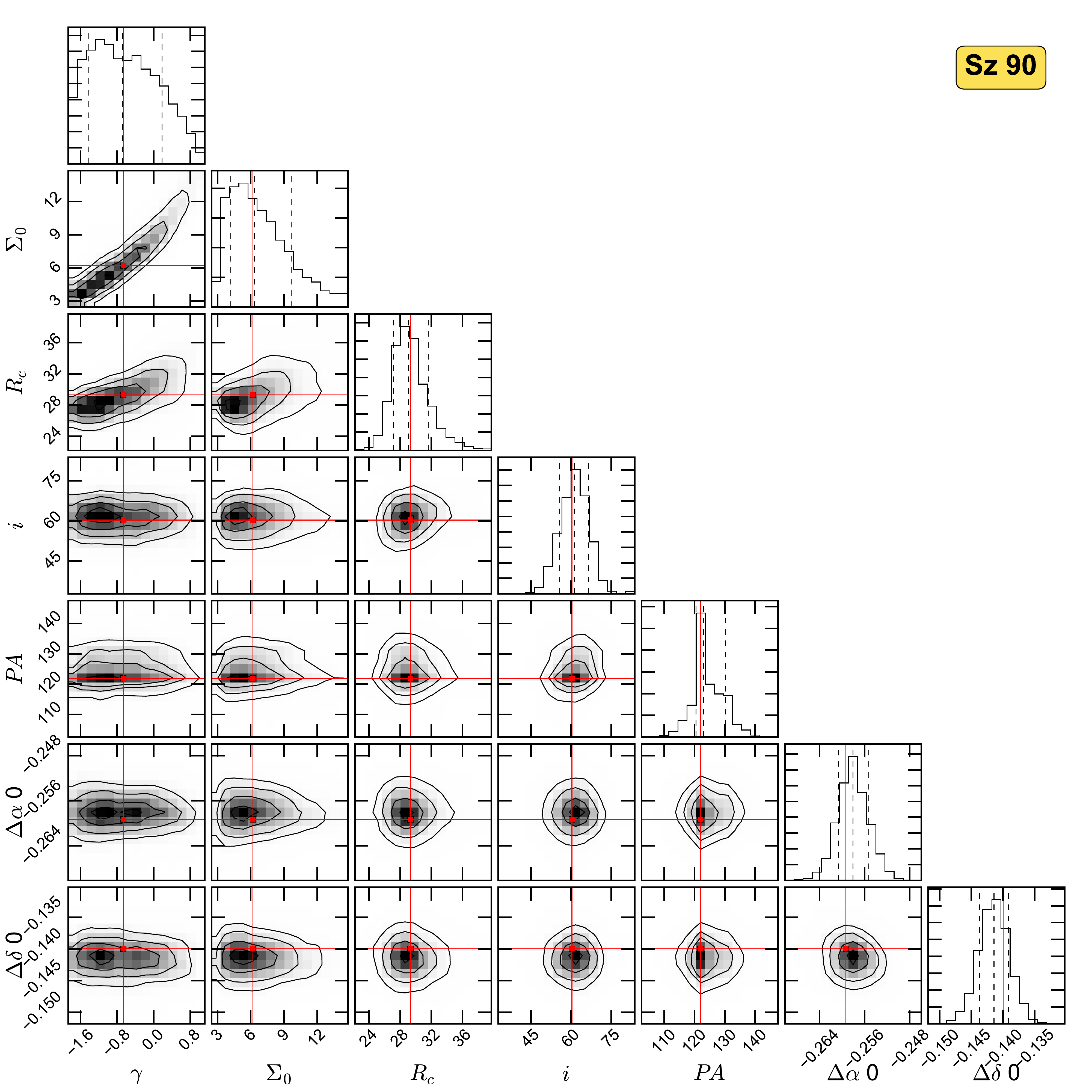}\includegraphics{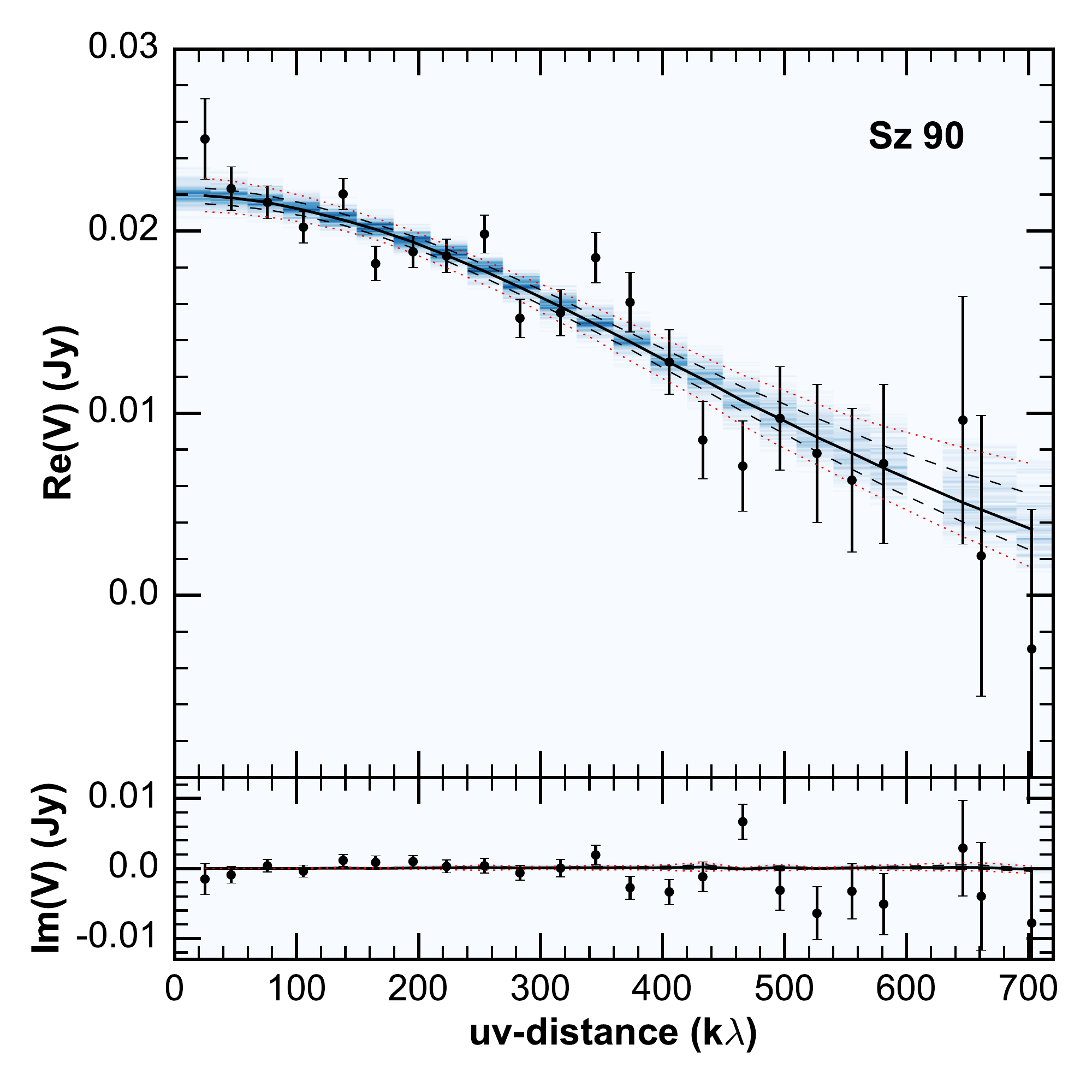}}\\[0.5cm]
\resizebox{\hsize}{!}{\includegraphics{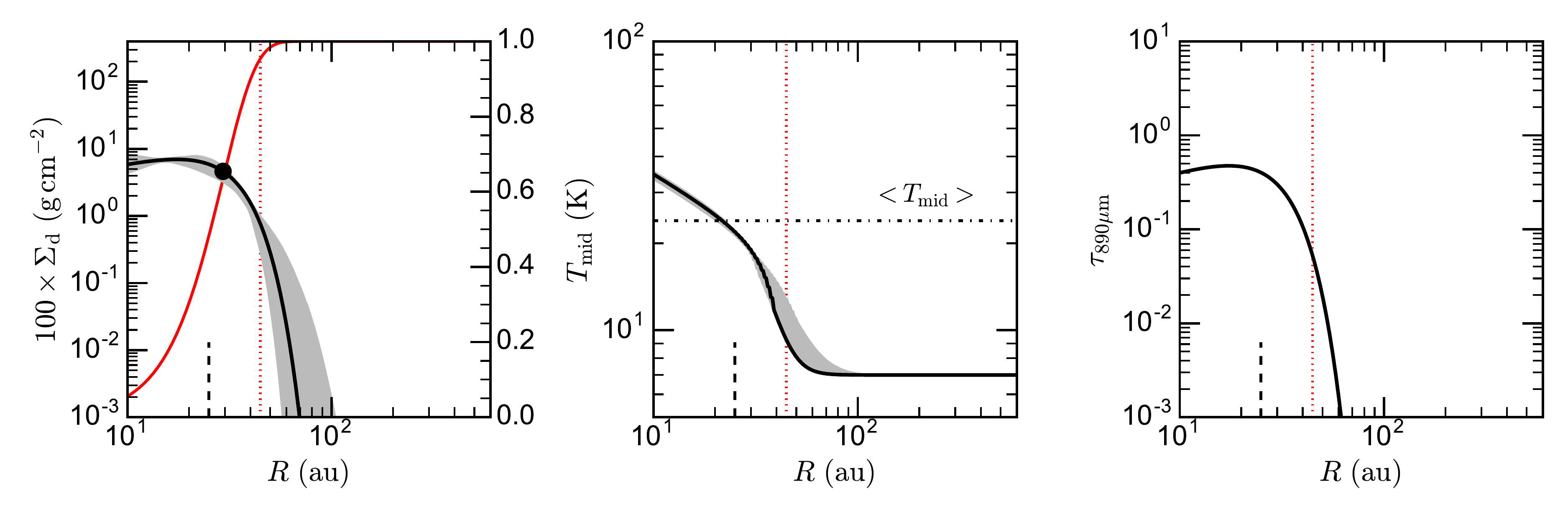}}\\[0.5cm]
\resizebox{0.8\hsize}{!}{\includegraphics{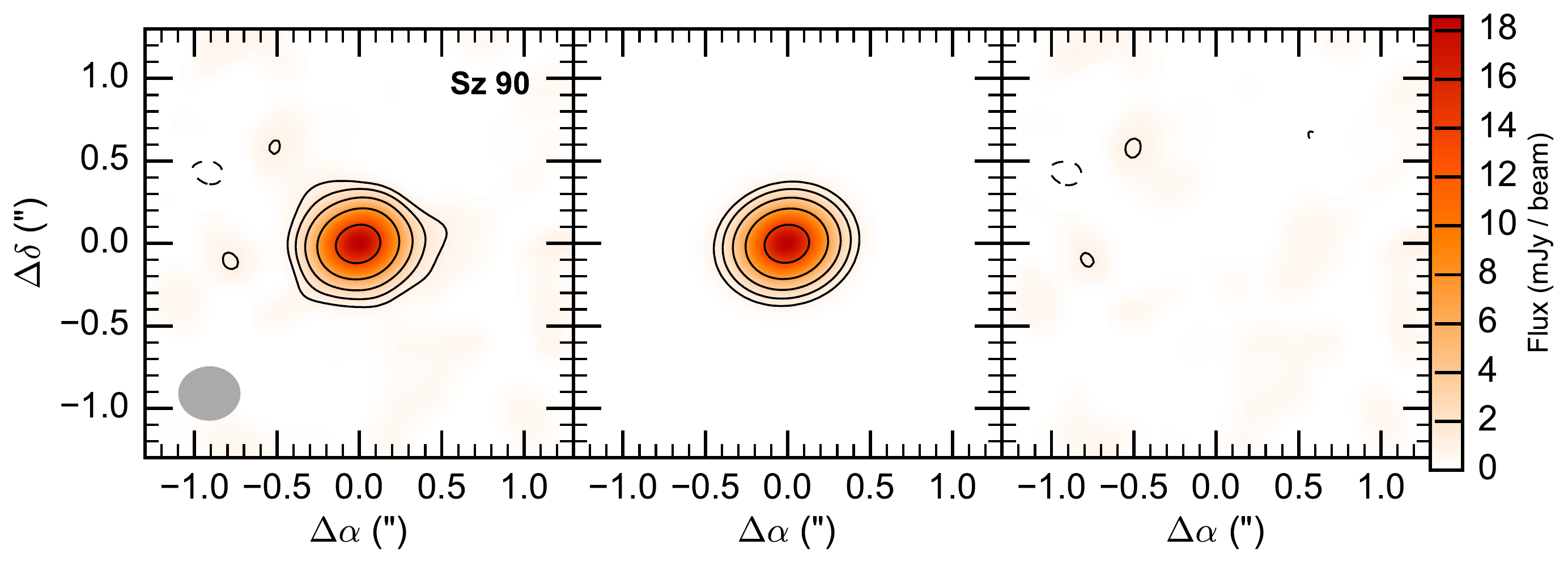}}\\[0.5cm]
\caption{Fit results for Sz 90, presented as in \figref{fig:reference.fit.results}. In the images $\sigma=0.3\u{mJy/beam}$.}
\end{figure*}

\pagebreak
\begin{figure*}
\centering
\Large\textbf{Sz 98\vspace{1cm}}
\resizebox{\hsize}{!}{\includegraphics[scale=0.5]{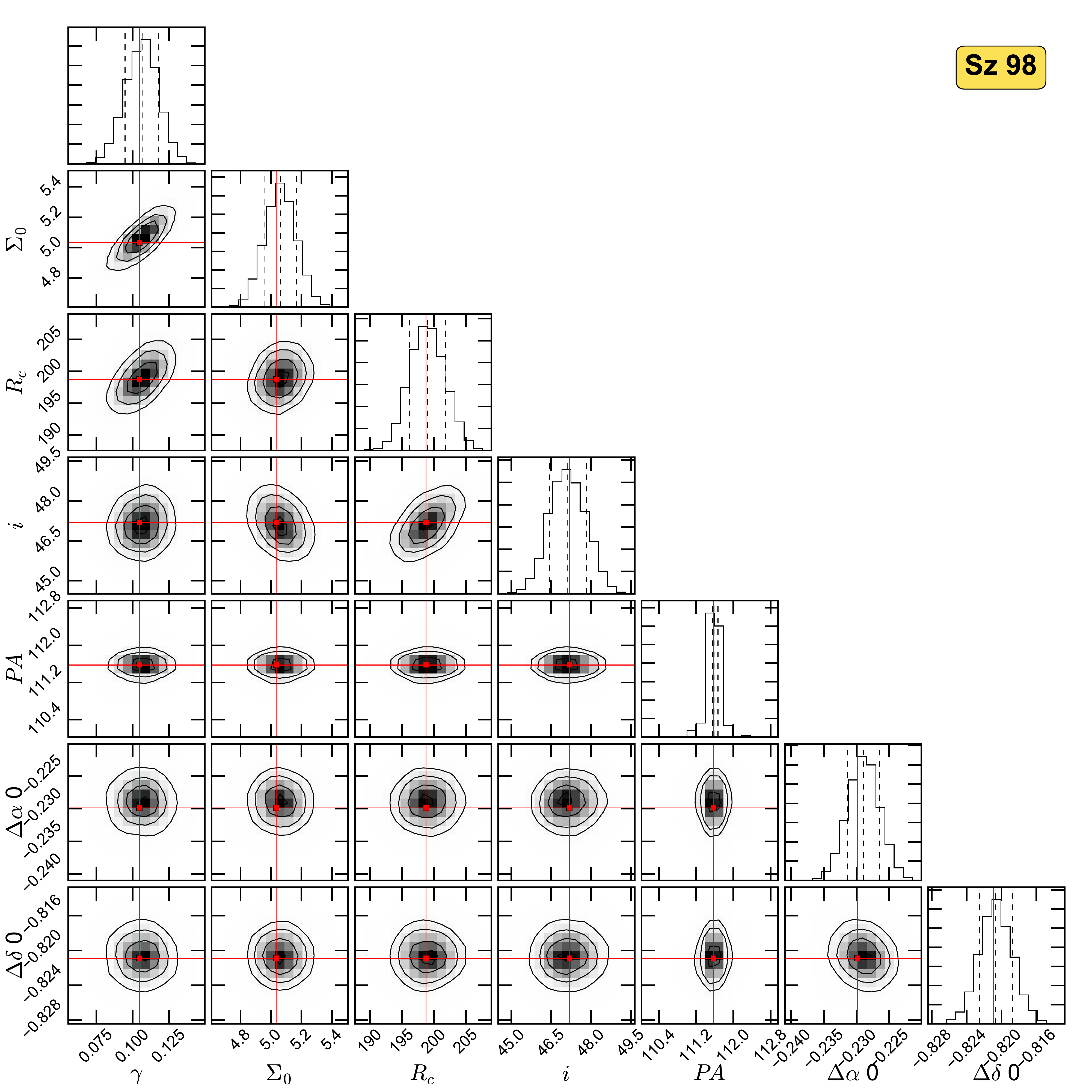}\includegraphics{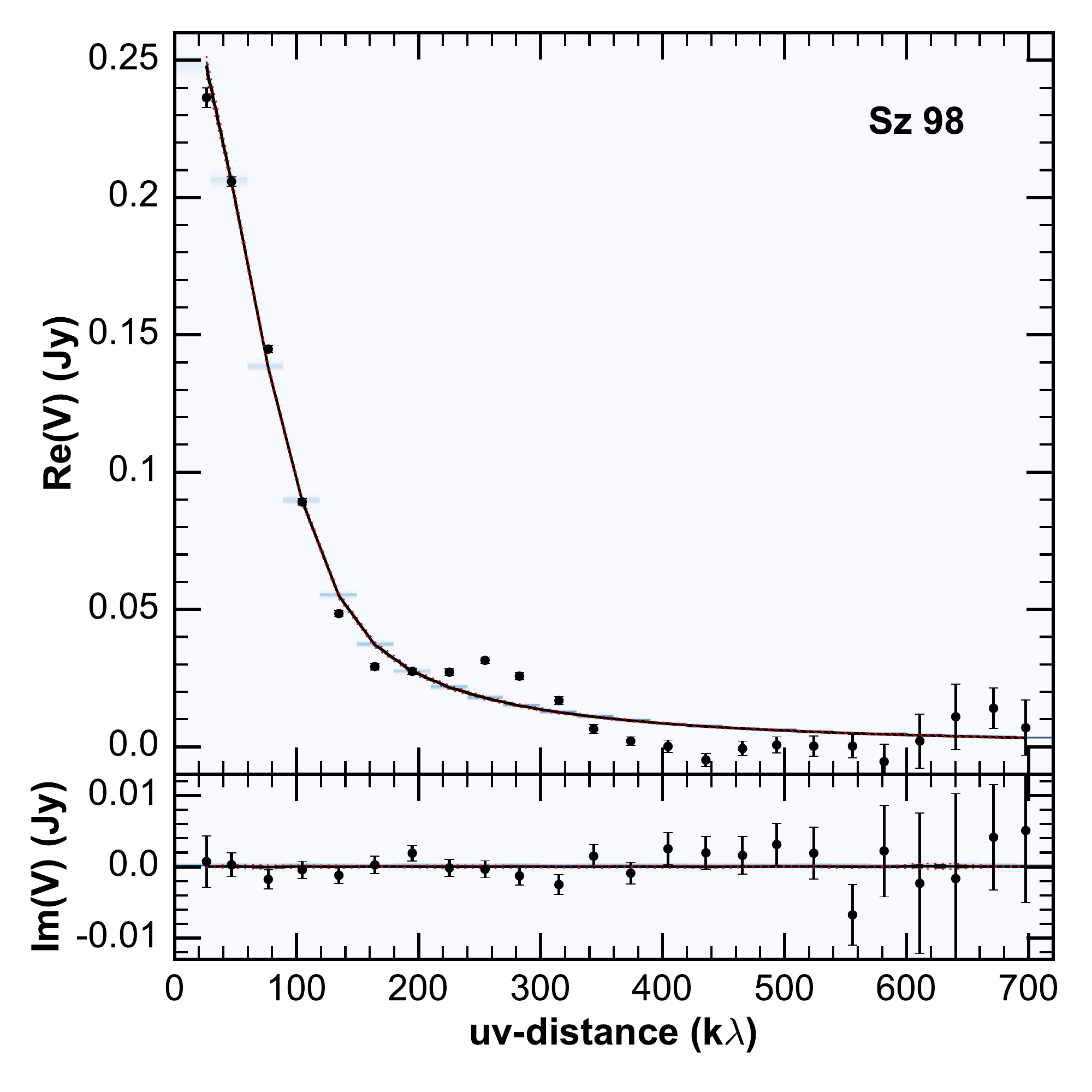}}\\[0.5cm]
\resizebox{\hsize}{!}{\includegraphics{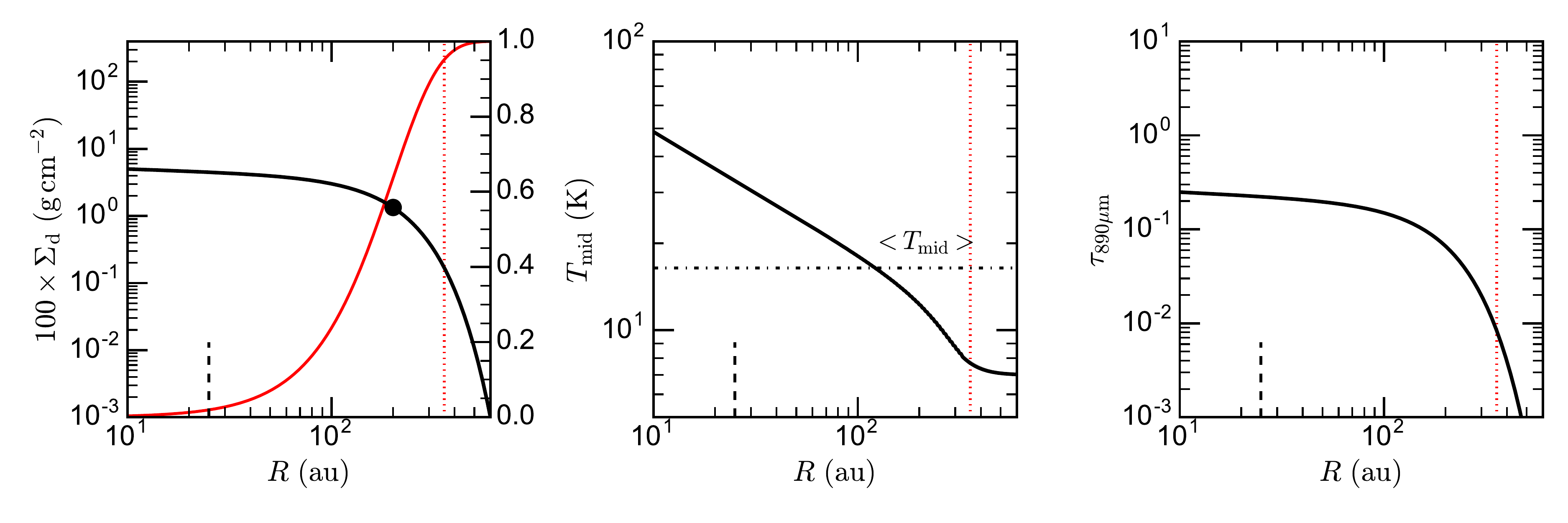}}\\[0.5cm]
\resizebox{0.8\hsize}{!}{\includegraphics{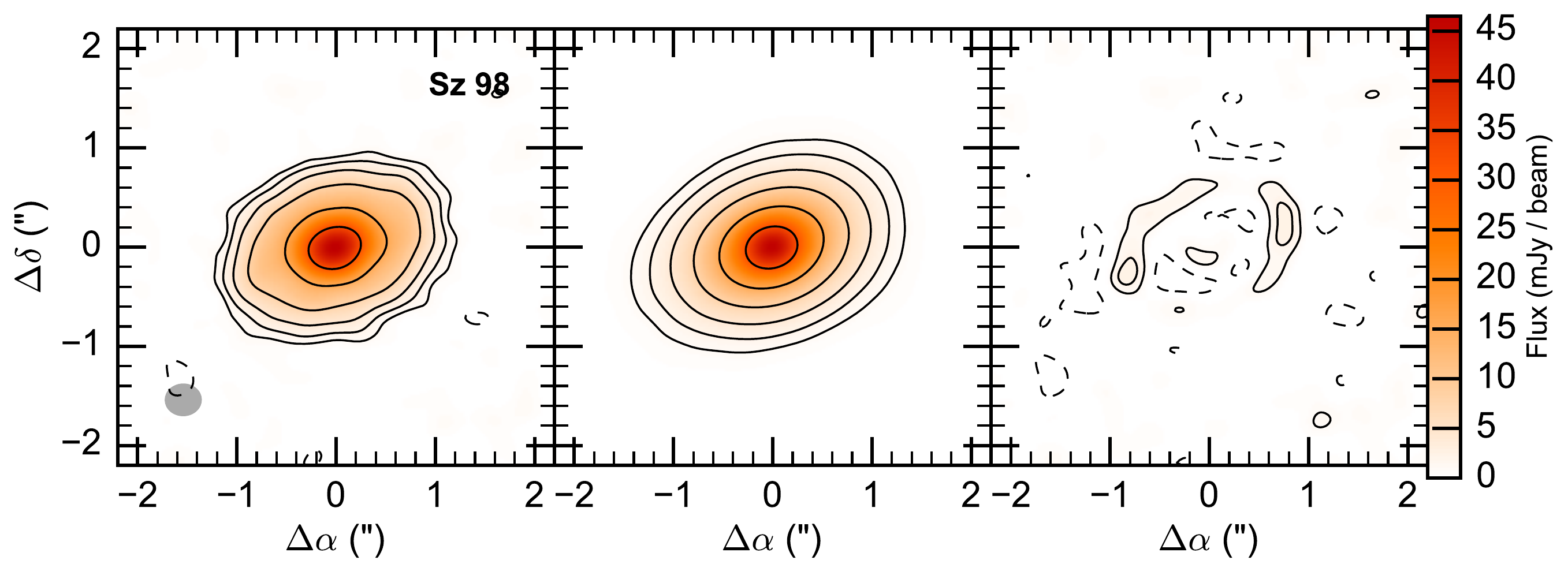}}\\[0.5cm]
\caption{Fit results for Sz 98, presented as in \figref{fig:reference.fit.results}. In the images $\sigma=0.33\u{mJy/beam}$.}
\end{figure*}

\pagebreak
\begin{figure*}
\centering
\Large\textbf{Sz 100\vspace{1cm}}
\resizebox{\hsize}{!}{\includegraphics[scale=0.5]{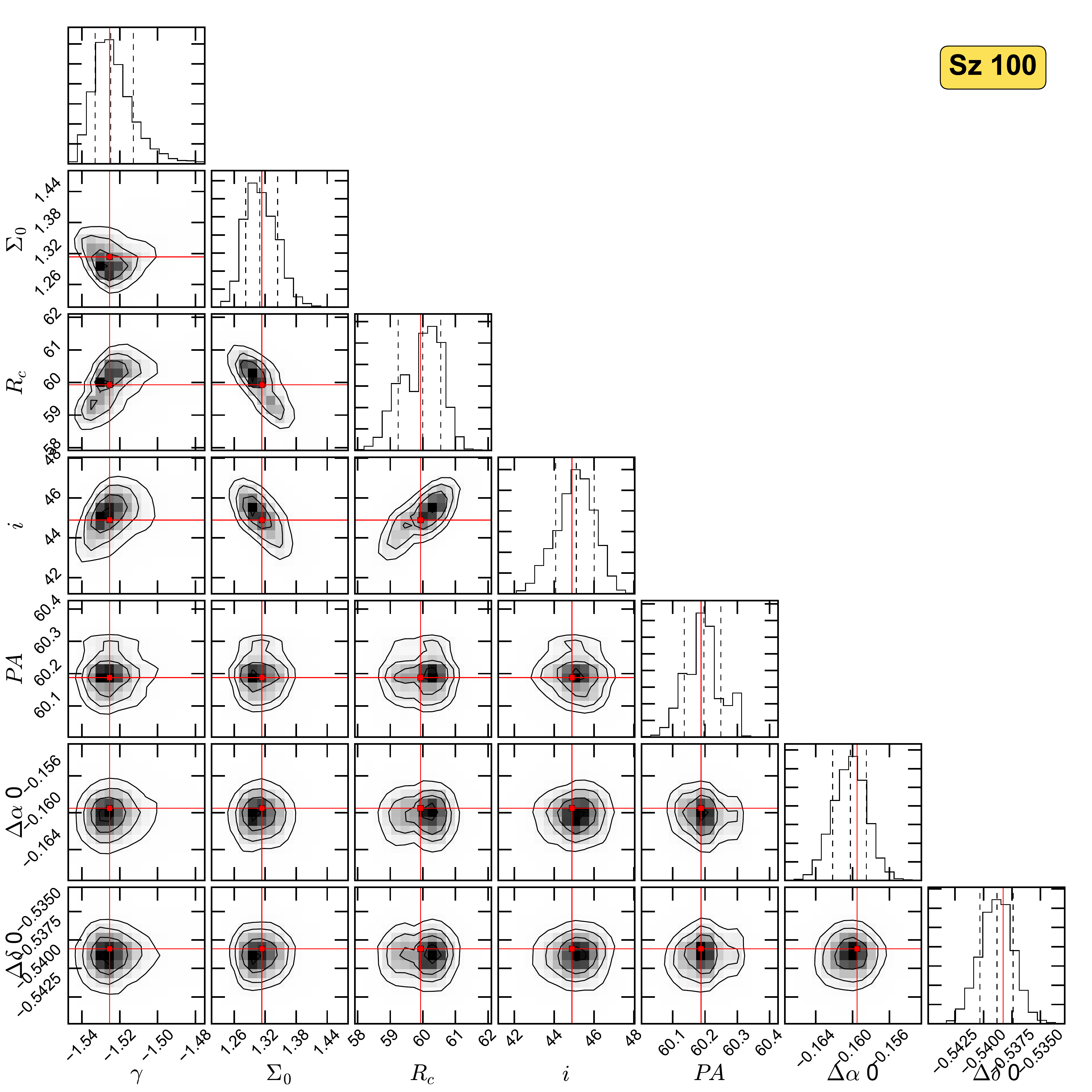}\includegraphics{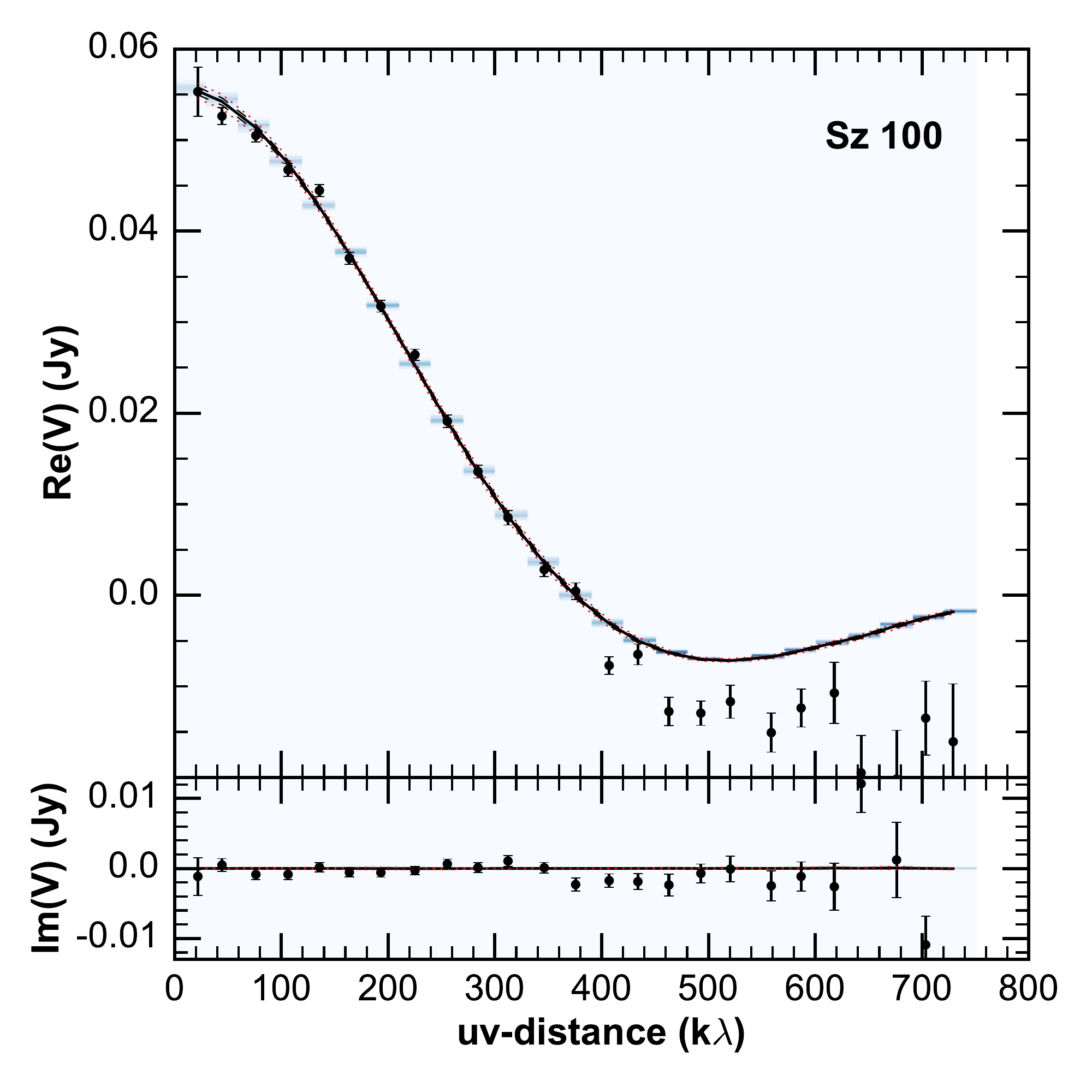}}\\[0.5cm]
\resizebox{\hsize}{!}{\includegraphics{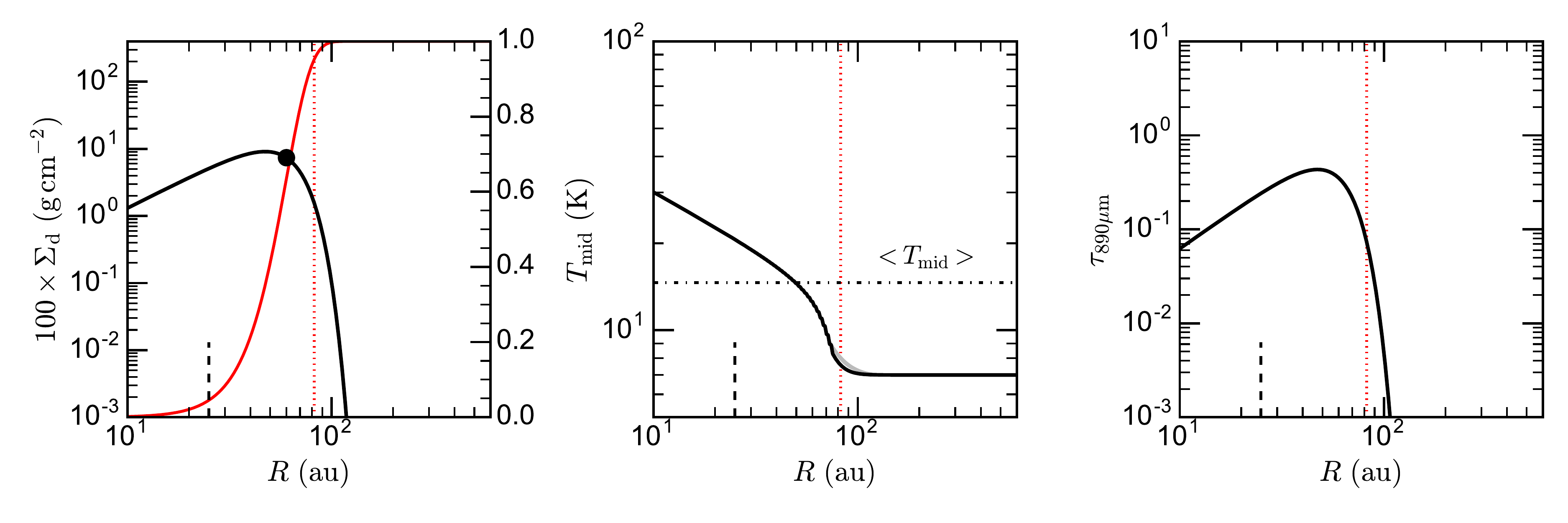}}\\[0.5cm]
\resizebox{0.8\hsize}{!}{\includegraphics{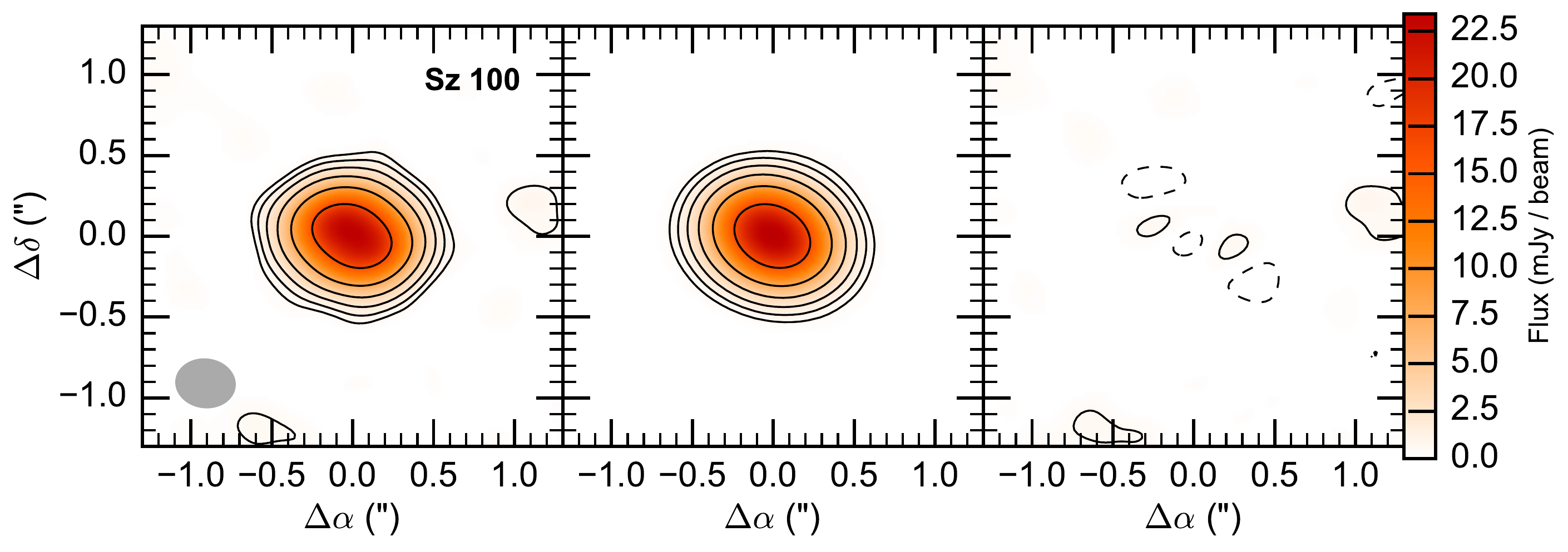}}\\[0.5cm]
\caption{Fit results for Sz 100, presented as in \figref{fig:reference.fit.results}. In the images $\sigma=0.2\u{mJy/beam}$.}
\end{figure*}

\pagebreak
\begin{figure*}
\centering
\Large\textbf{Sz 108B\vspace{1cm}}
\resizebox{\hsize}{!}{\includegraphics[scale=0.5]{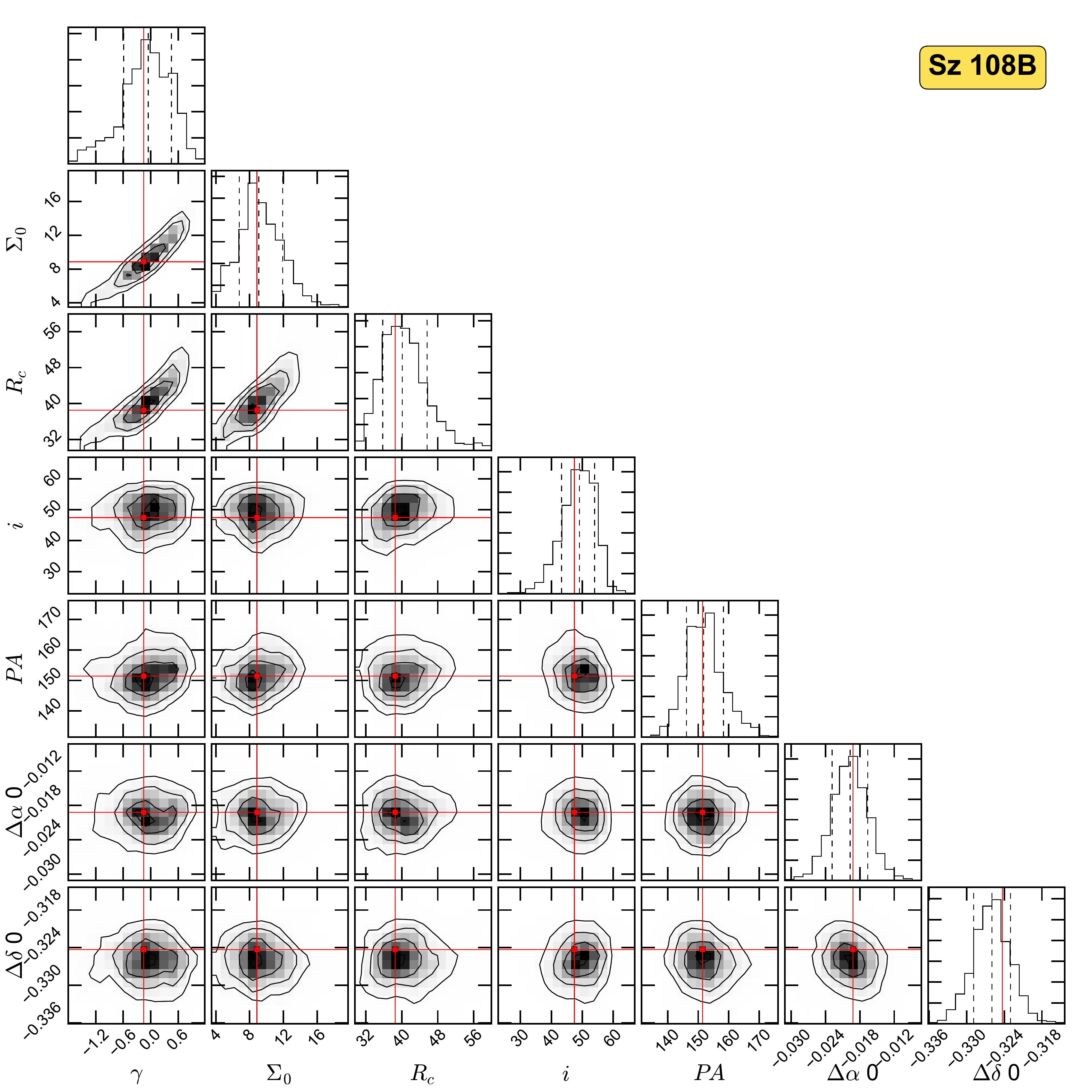}\includegraphics{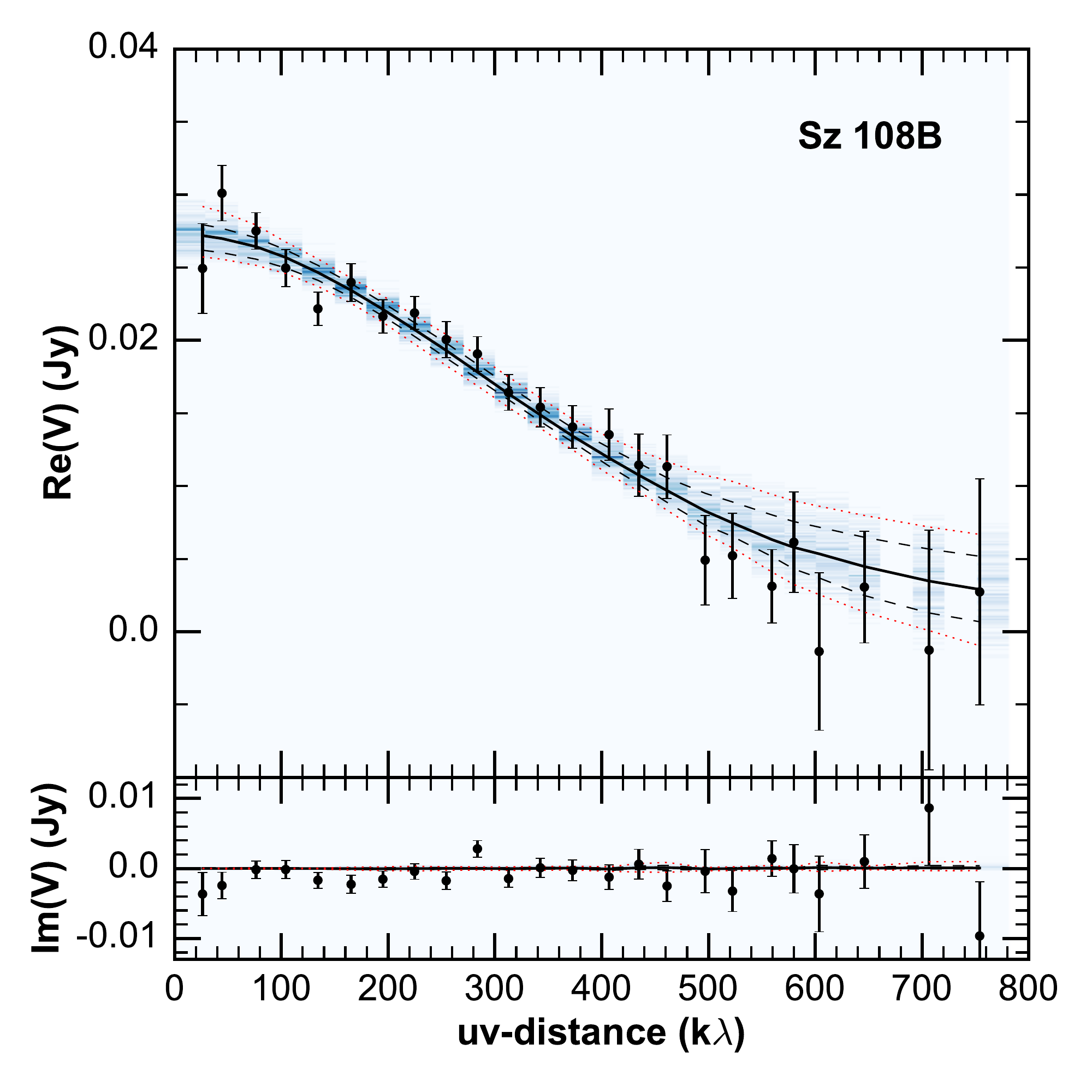}}\\[0.5cm]
\resizebox{\hsize}{!}{\includegraphics{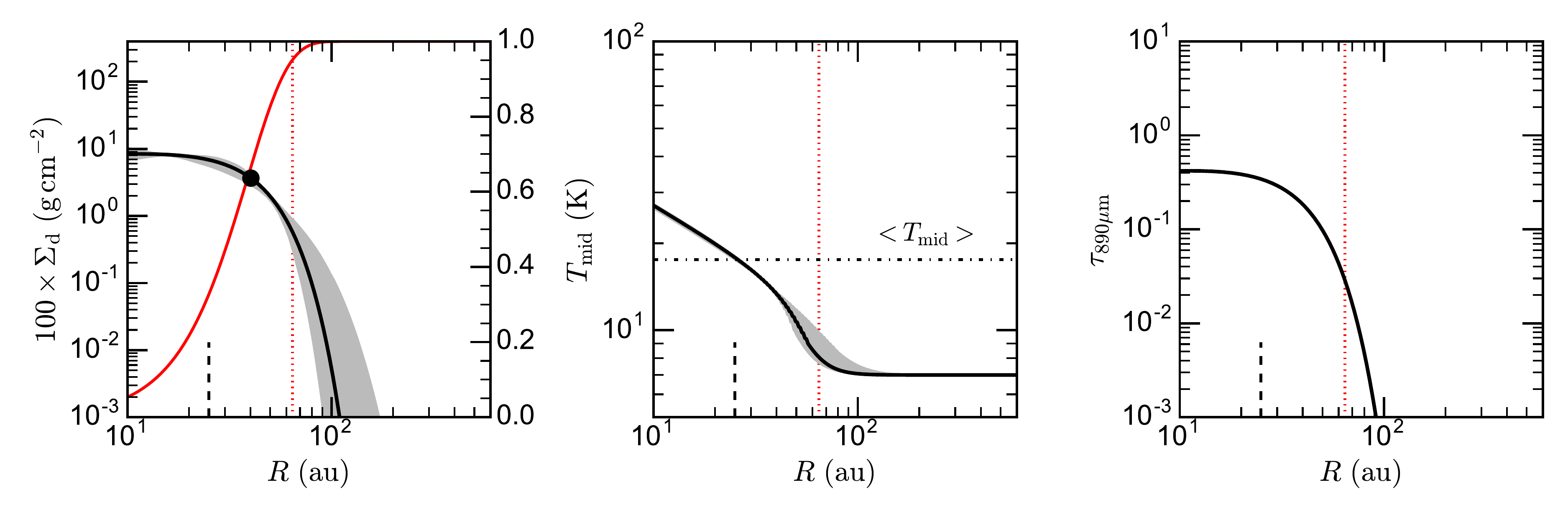}}\\[0.5cm]
\resizebox{0.8\hsize}{!}{\includegraphics{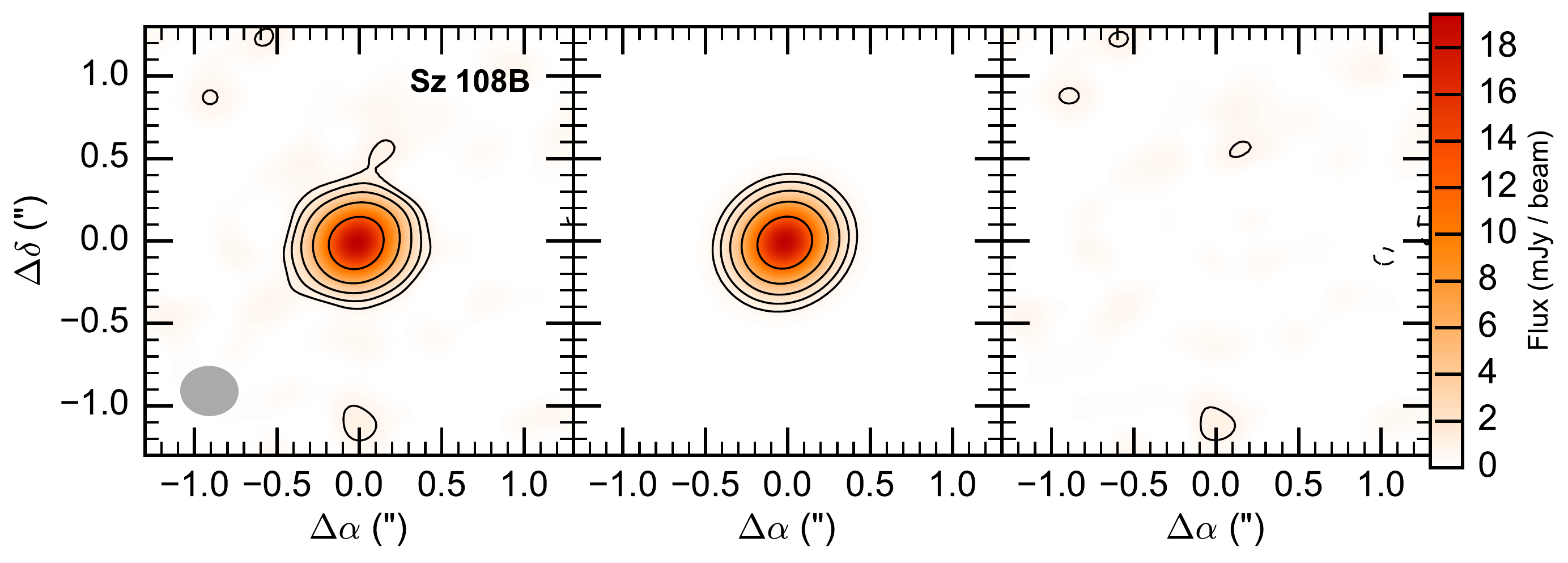}}\\[0.5cm]
\caption{Fit results for Sz 108B, presented as in \figref{fig:reference.fit.results}. In the images $\sigma=0.25\u{mJy/beam}$.}
\end{figure*}

\pagebreak
\begin{figure*}
\centering
\Large\textbf{J16085324-3914401\vspace{1cm}}
\resizebox{\hsize}{!}{\includegraphics[scale=0.5]{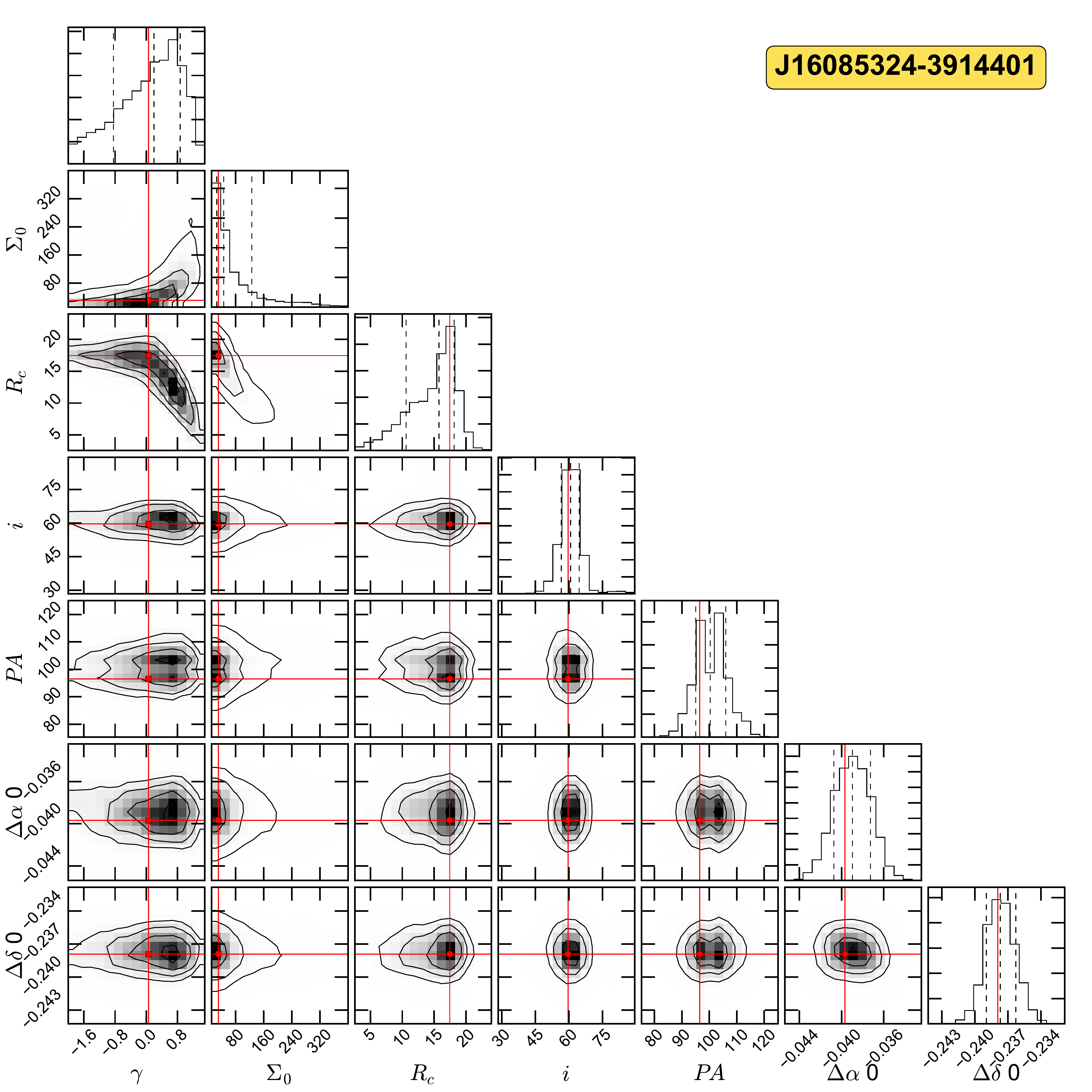}\includegraphics{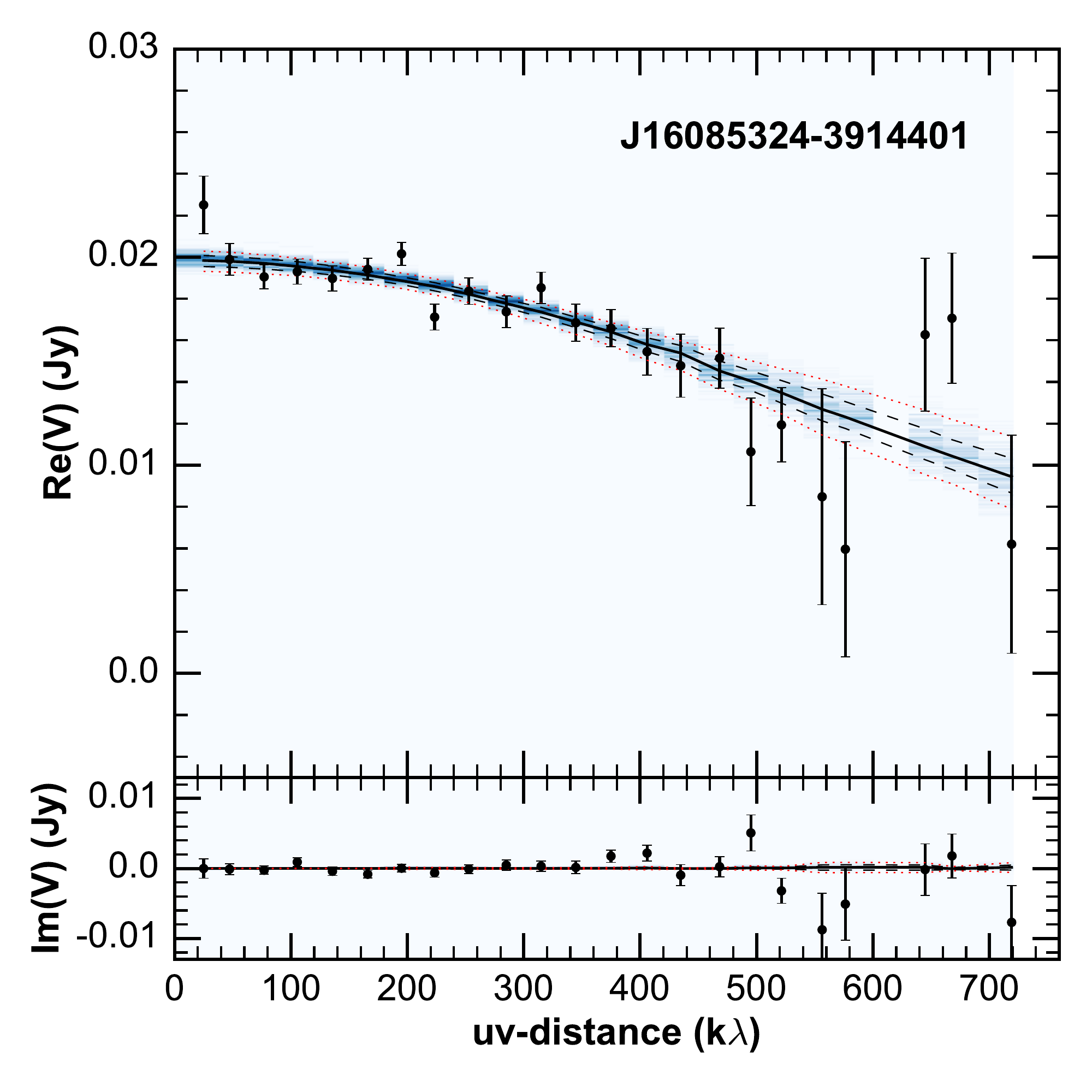}}\\[0.5cm]
\resizebox{\hsize}{!}{\includegraphics{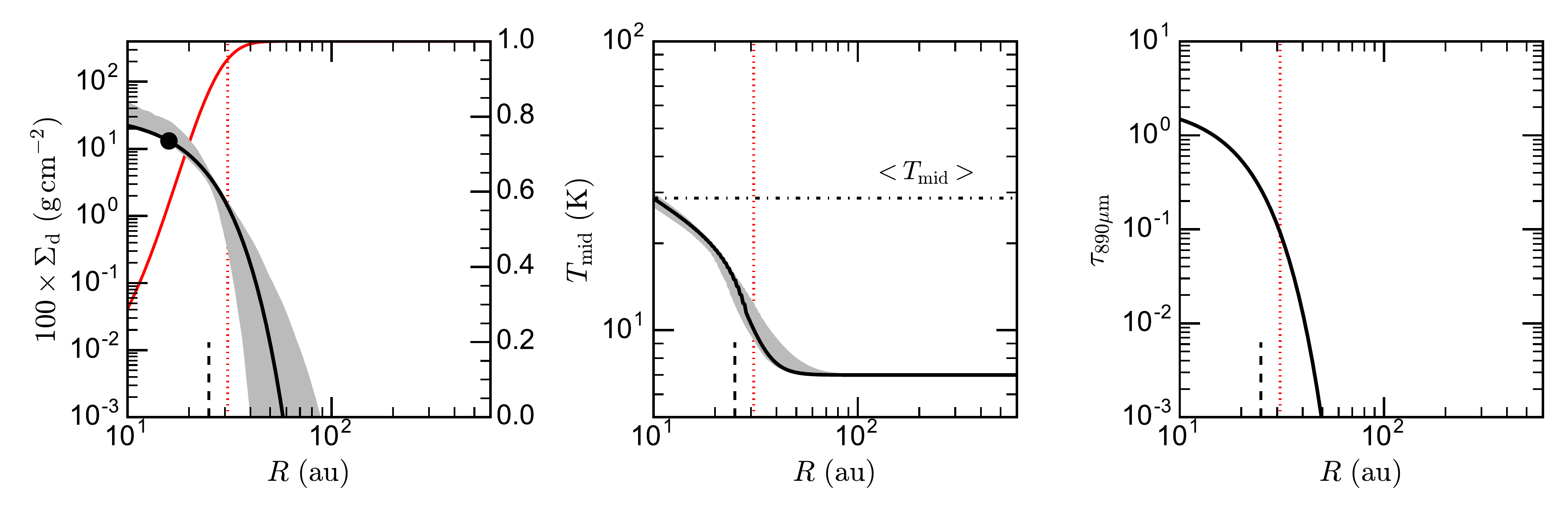}}\\[0.5cm]
\resizebox{0.8\hsize}{!}{\includegraphics{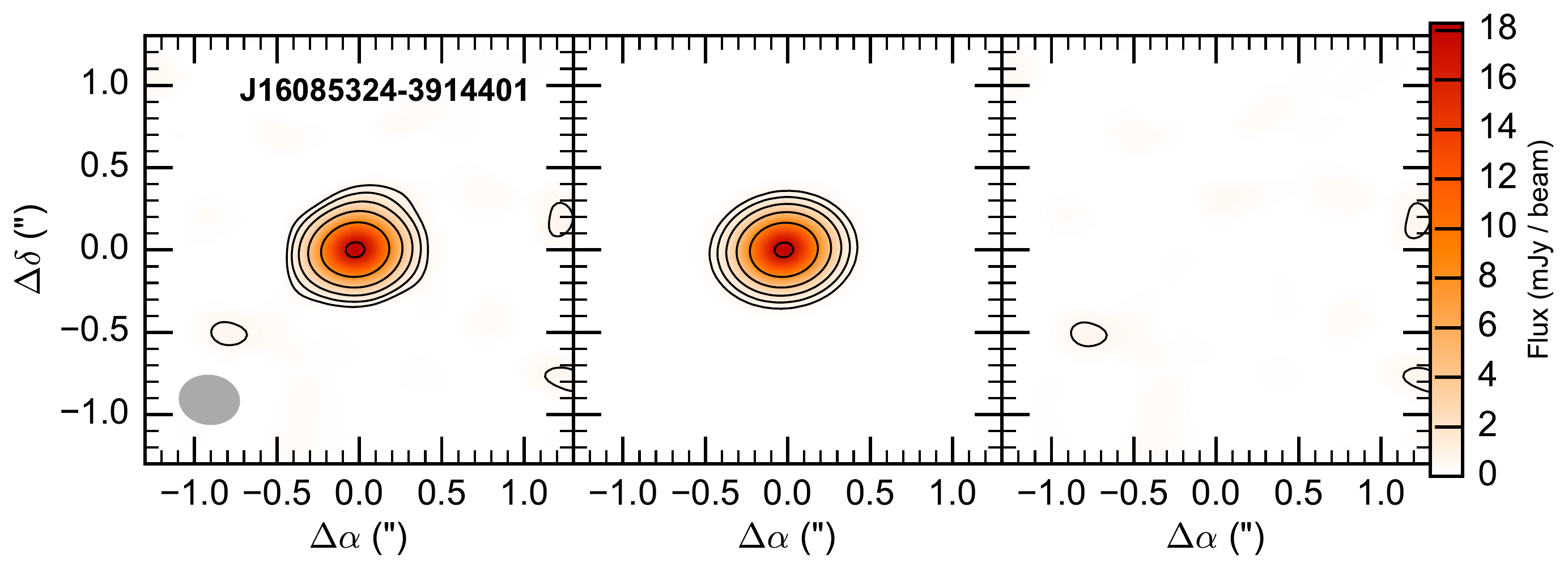}}\\[0.5cm]
\caption{Fit results for J16085324-3914401, presented as in \figref{fig:reference.fit.results}. In the images $\sigma=0.2\u{mJy/beam}$.}
\end{figure*}

\pagebreak
\begin{figure*}
\centering
\Large\textbf{Sz 113\vspace{1cm}}
\resizebox{\hsize}{!}{\includegraphics[scale=0.5]{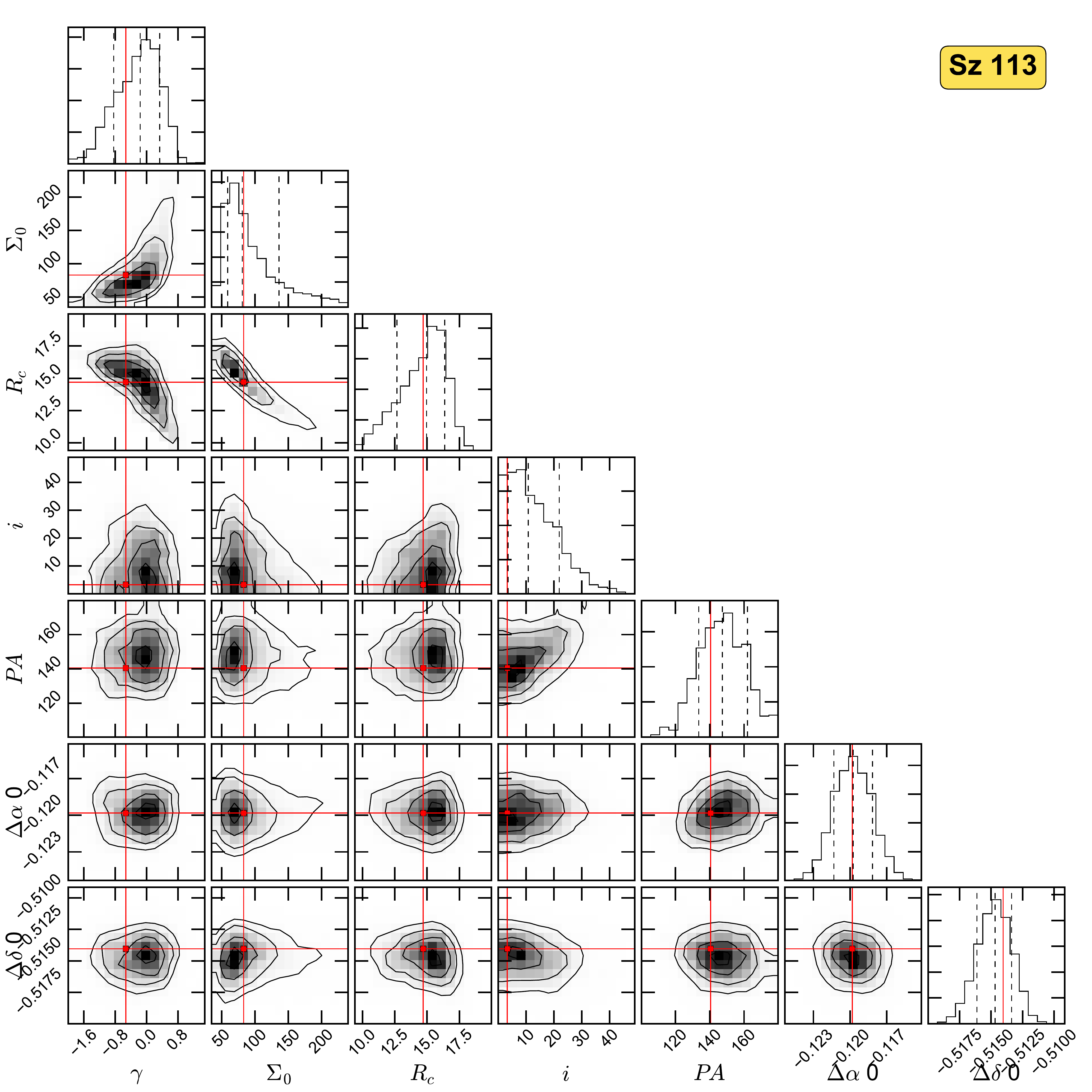}\includegraphics{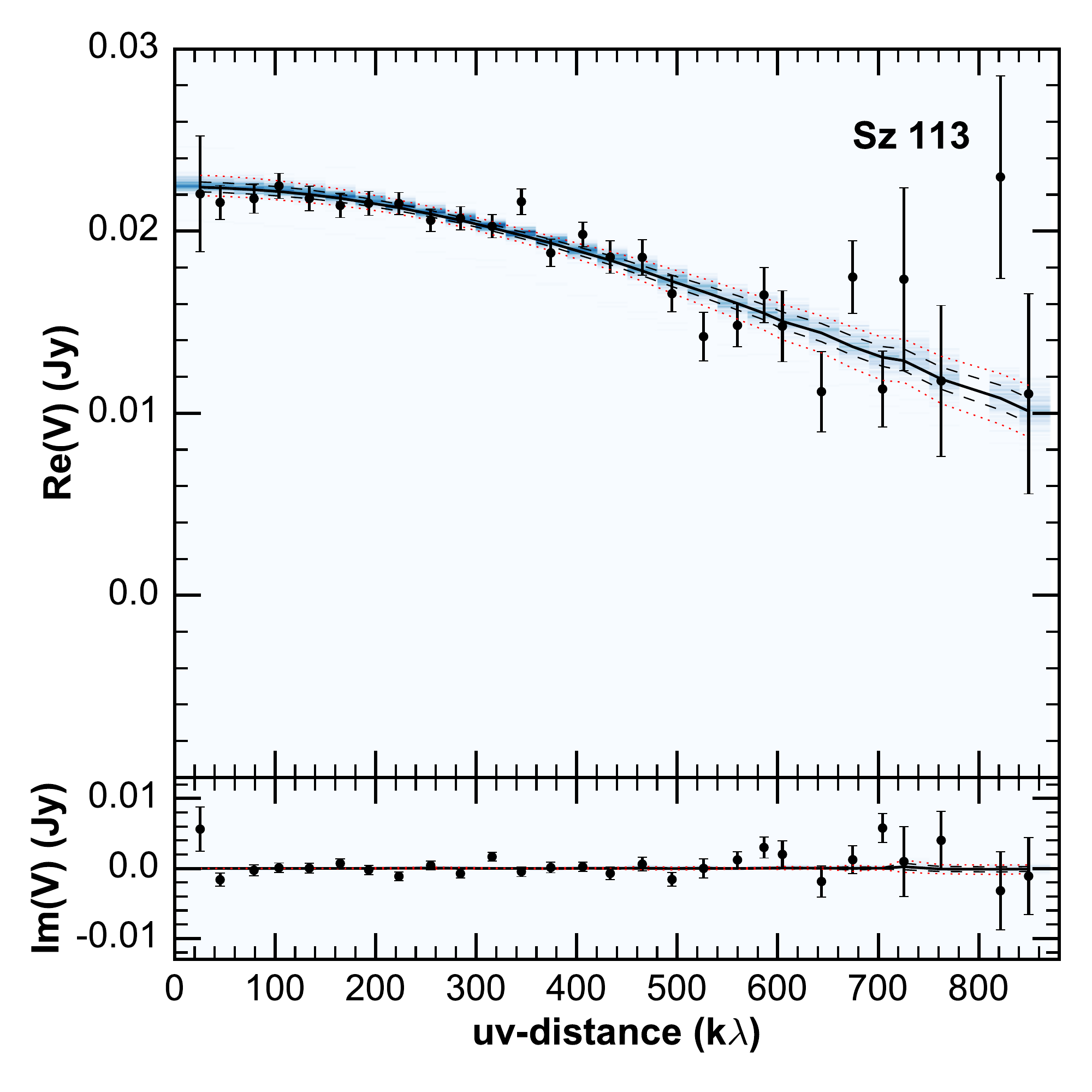}}\\[0.5cm]
\resizebox{\hsize}{!}{\includegraphics{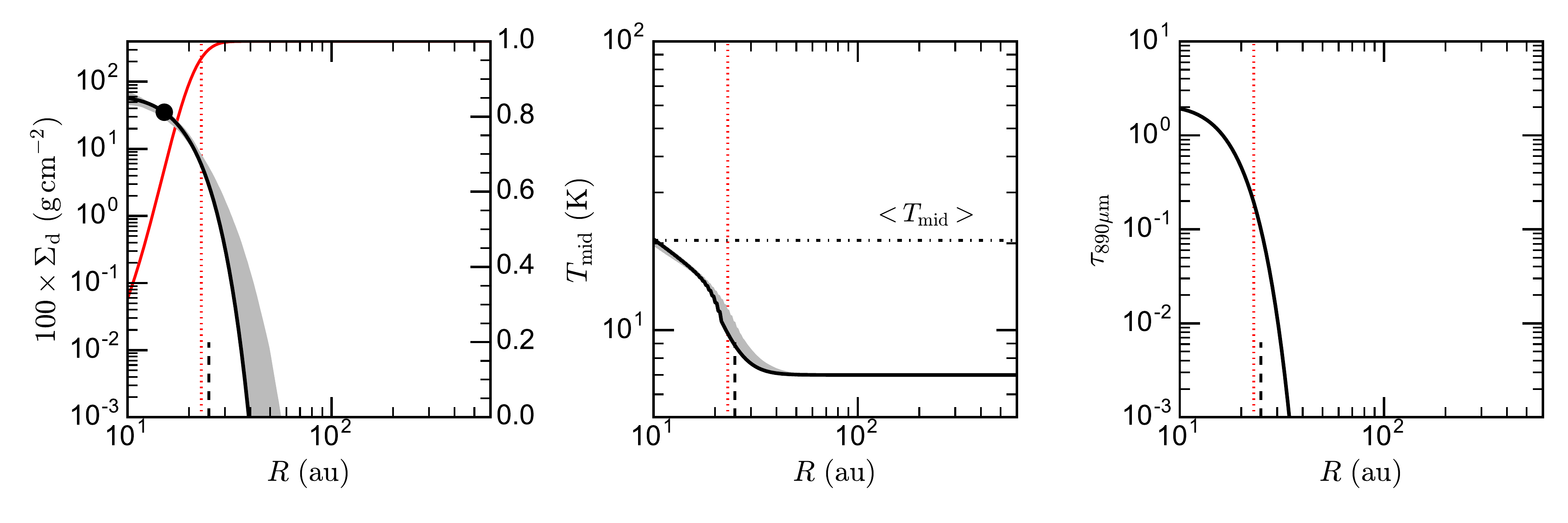}}\\[0.5cm]
\resizebox{0.8\hsize}{!}{\includegraphics{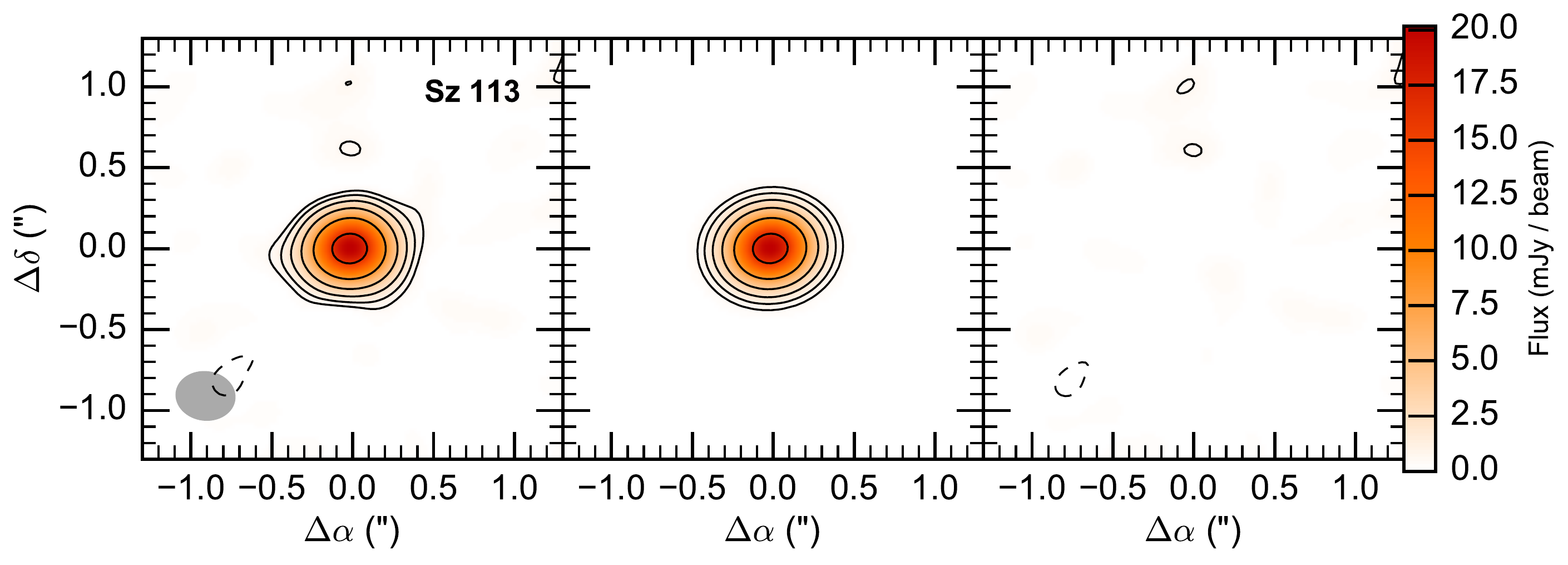}}\\[0.5cm]
\caption{Fit results for Sz 113, presented as in \figref{fig:reference.fit.results}. In the images $\sigma=0.2\u{mJy/beam}$.}
\end{figure*}

\pagebreak
\begin{figure*}
\centering
\Large\textbf{Sz 114\vspace{1cm}}
\resizebox{\hsize}{!}{\includegraphics[scale=0.5]{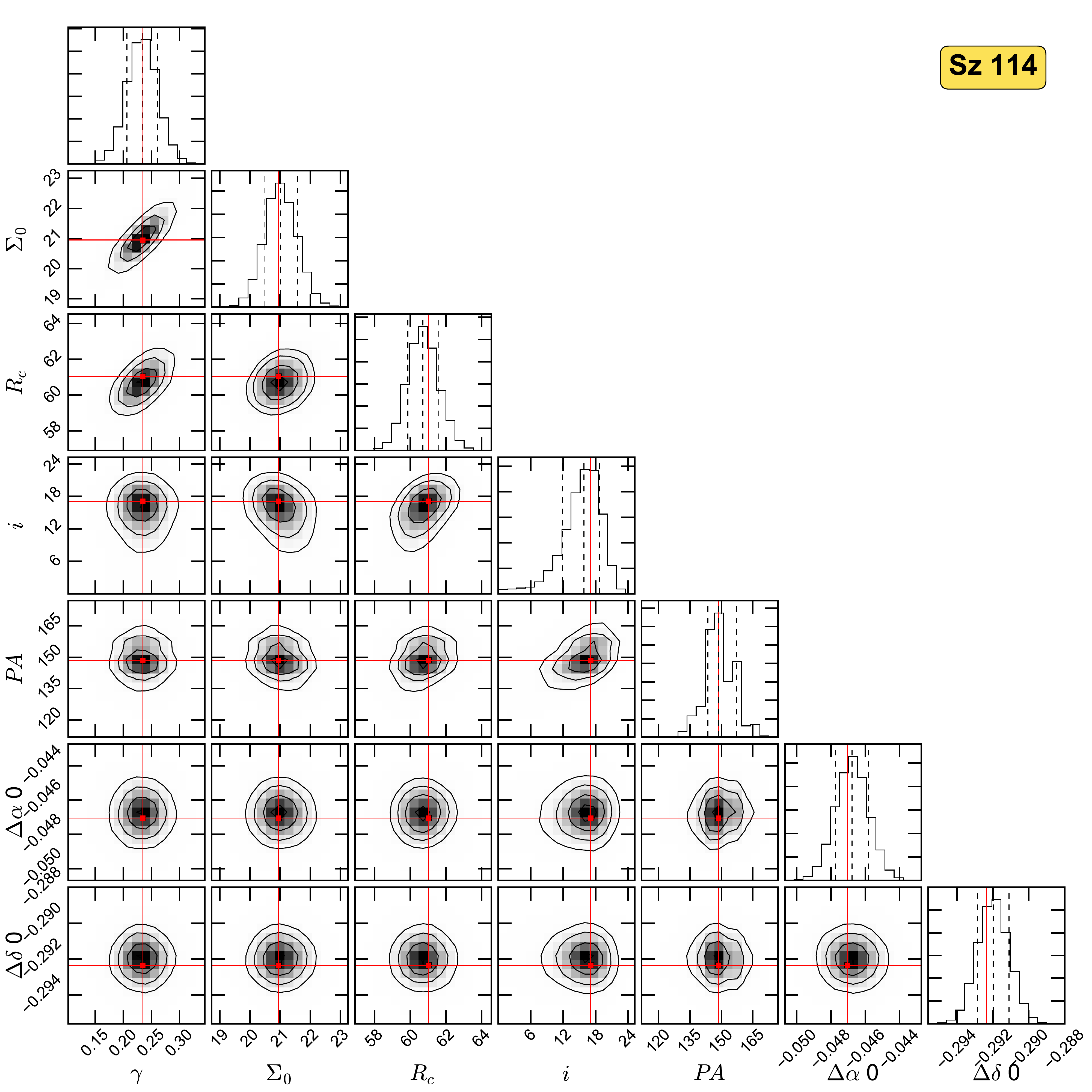}\includegraphics{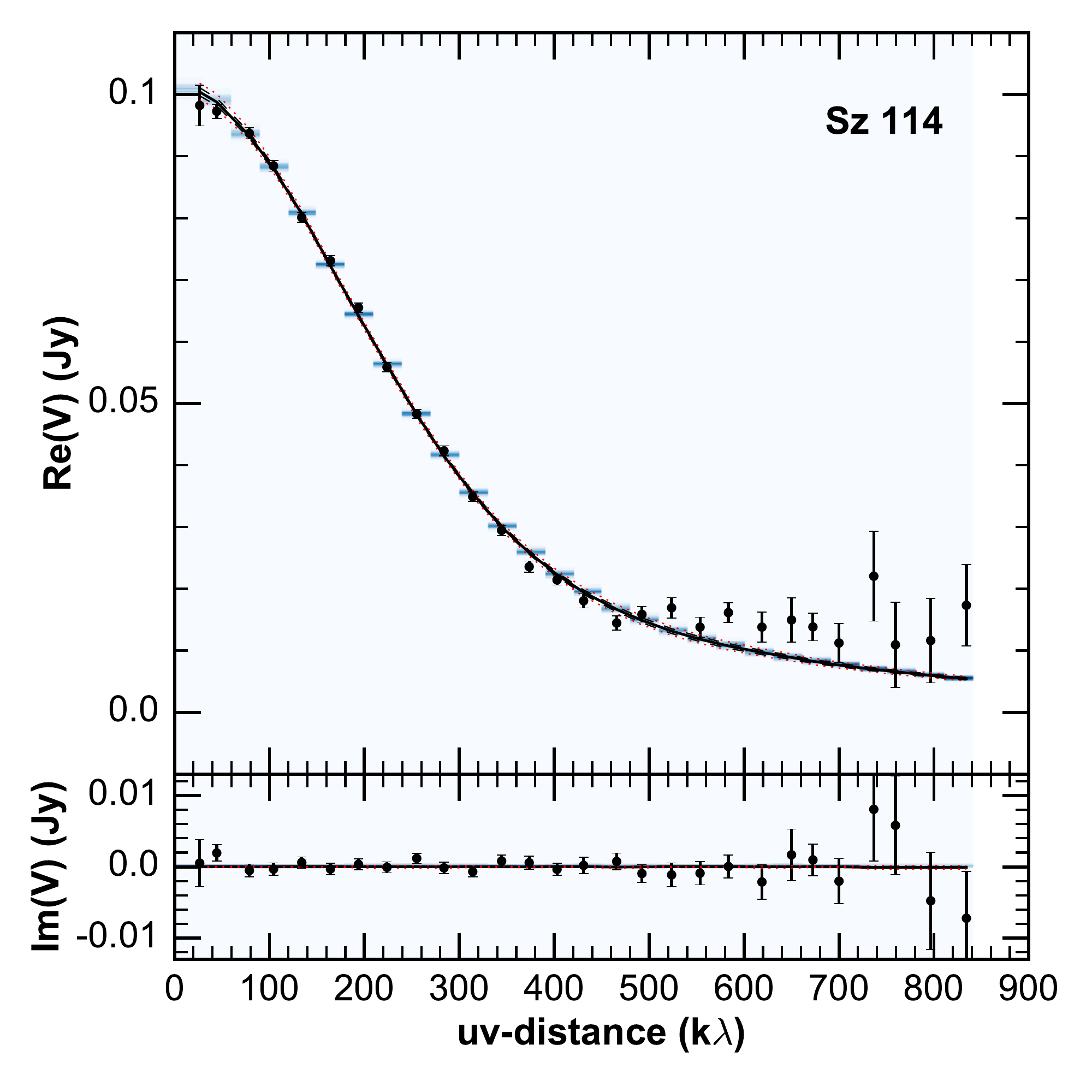}}\\[0.5cm]
\resizebox{\hsize}{!}{\includegraphics{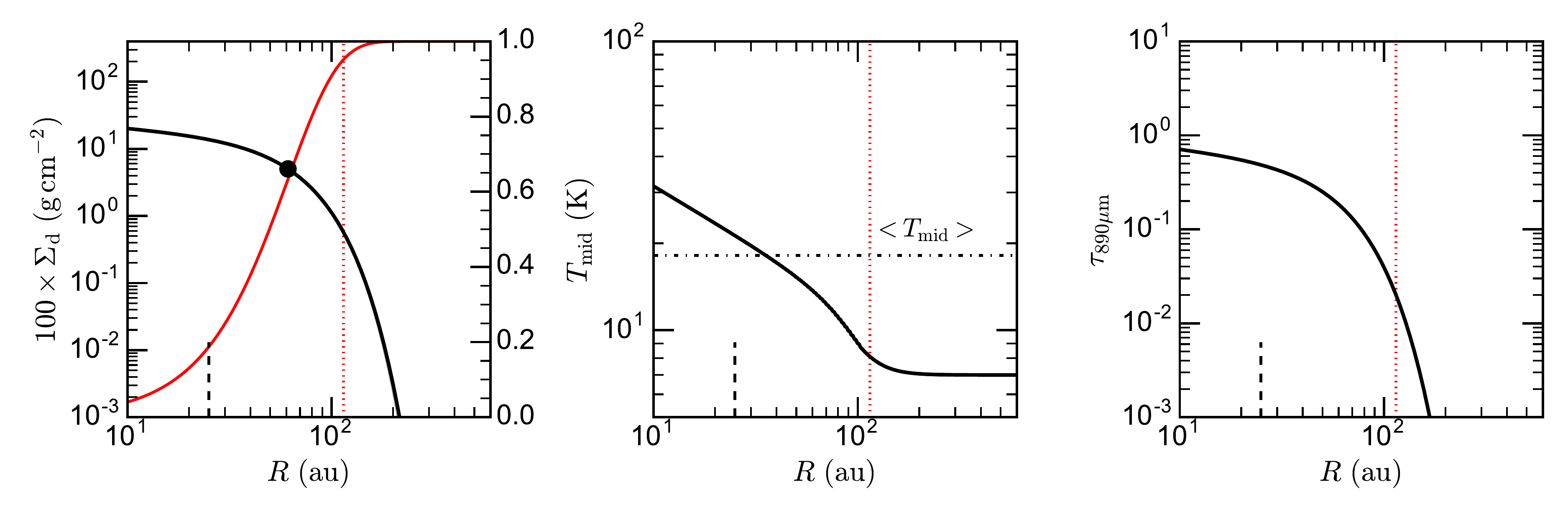}}\\[0.5cm]
\resizebox{0.8\hsize}{!}{\includegraphics{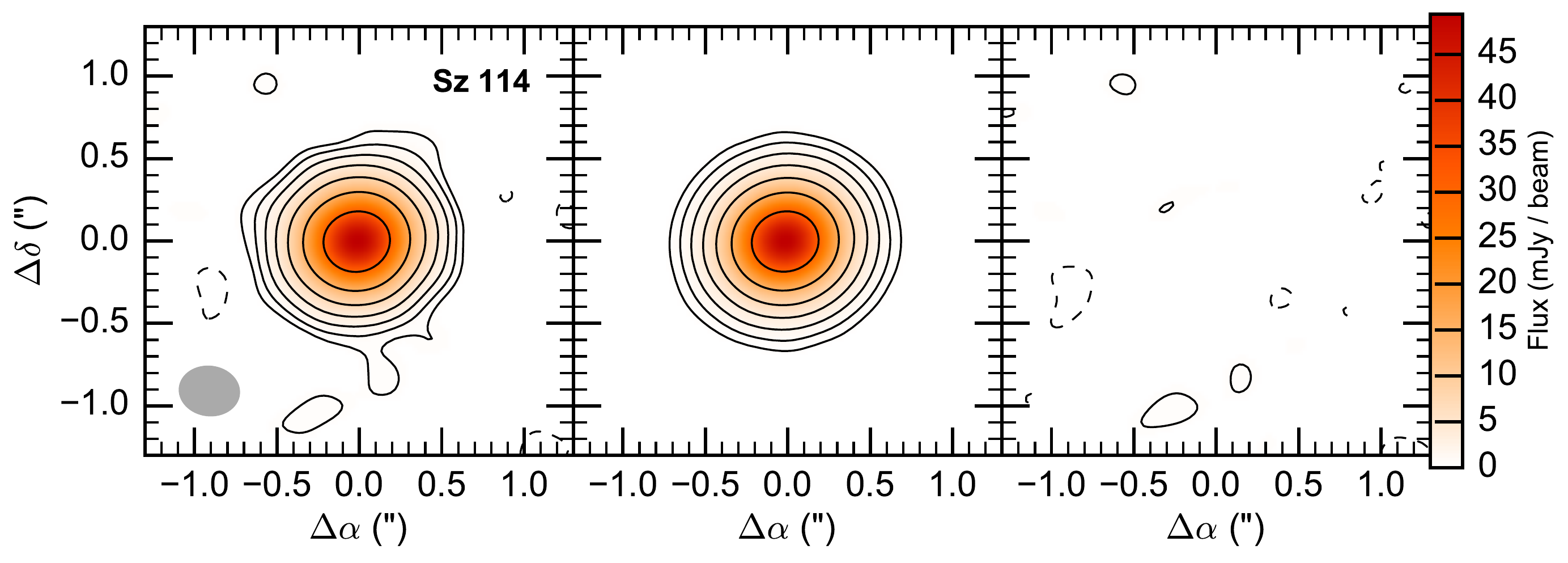}}\\[0.5cm]
\caption{Fit results for Sz 114, presented as in \figref{fig:reference.fit.results}. In the images $\sigma=0.2\u{mJy/beam}$.}
\end{figure*}

\pagebreak
\begin{figure*}
\centering
\Large\textbf{J16102955-3922144\vspace{1cm}}
\resizebox{\hsize}{!}{\includegraphics[scale=0.5]{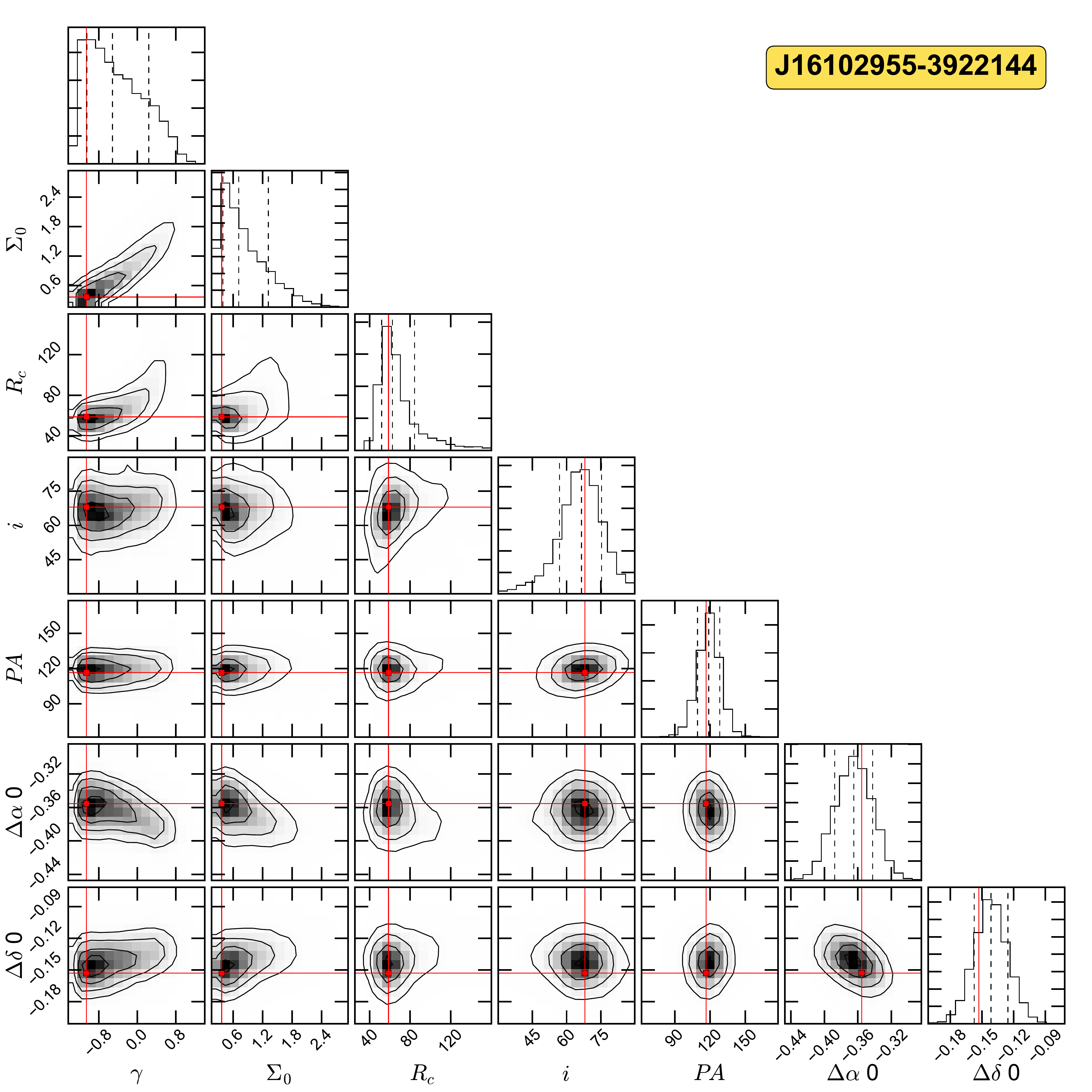}\includegraphics{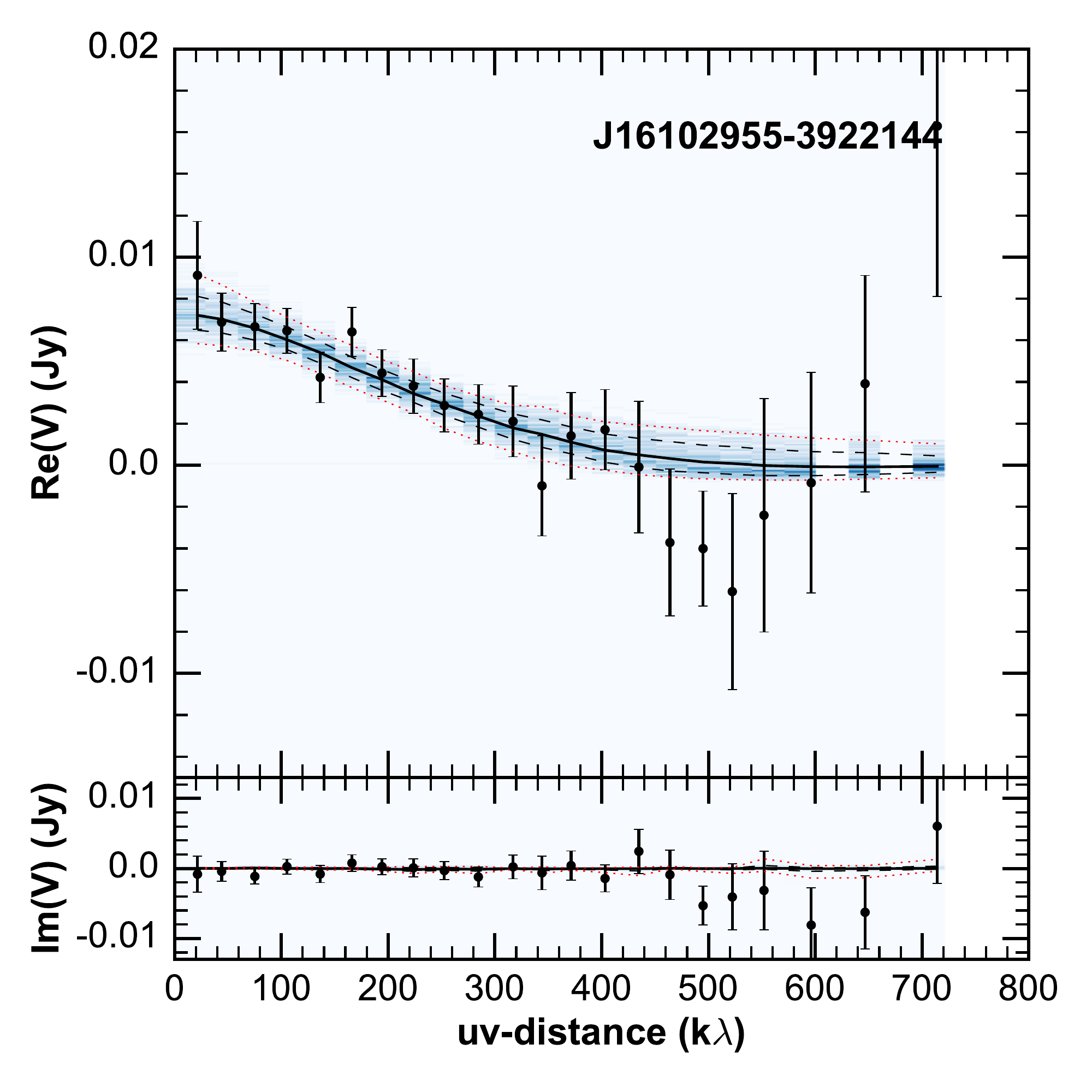}}\\[0.5cm]
\resizebox{\hsize}{!}{\includegraphics{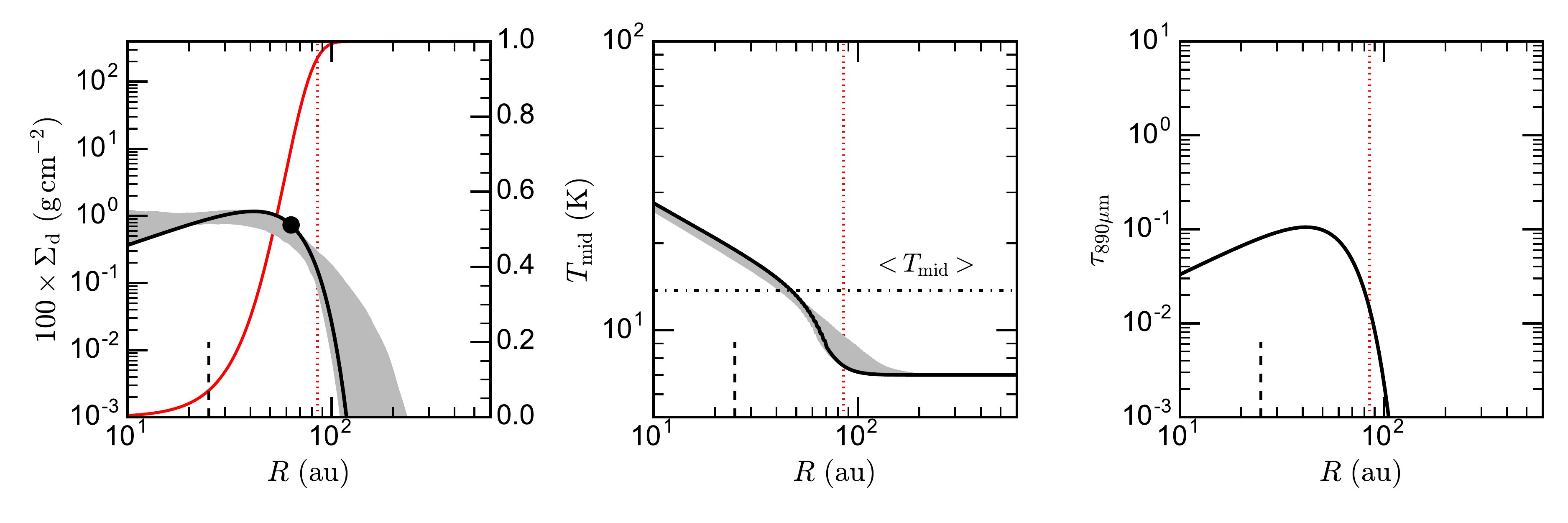}}\\[0.5cm]
\resizebox{0.8\hsize}{!}{\includegraphics{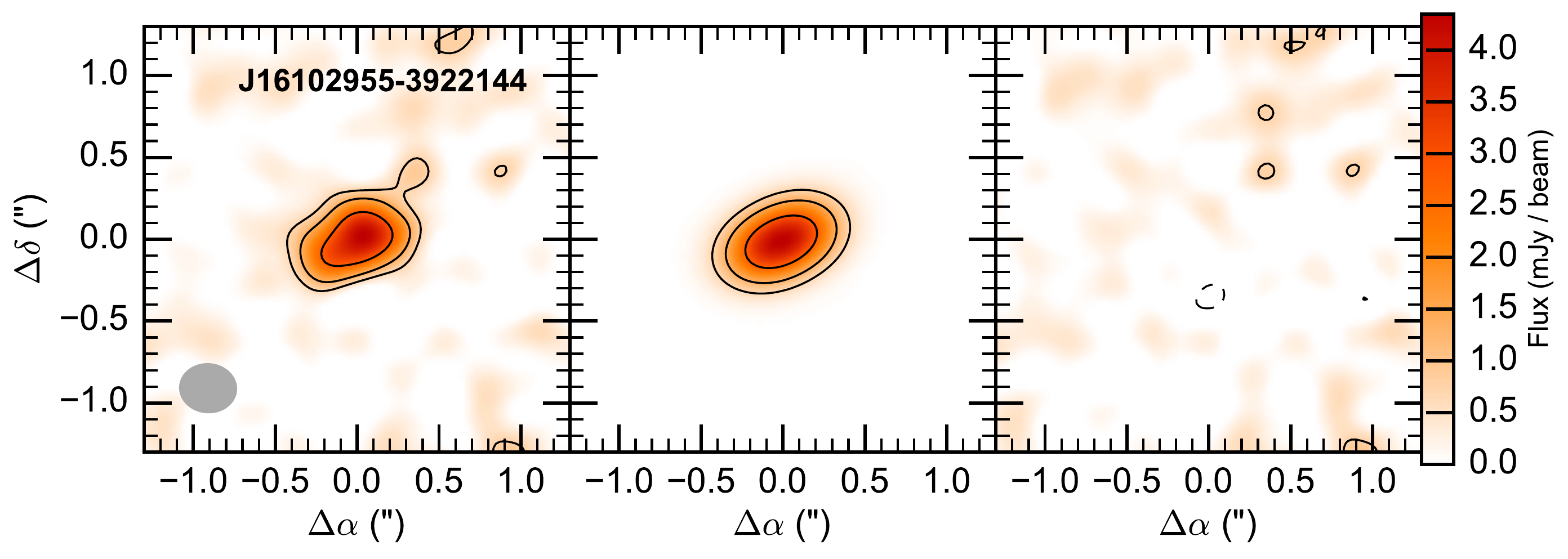}}\\[0.5cm]
\caption{Fit results for J16102955-3922144, presented as in \figref{fig:reference.fit.results}. In the maps, $\sigma=0.25\u{mJy/beam}$.}
\end{figure*}

\pagebreak
\begin{figure*}
\centering
\Large\textbf{J16124373-3815031\vspace{1cm}}
\resizebox{\hsize}{!}{\includegraphics[scale=0.5]{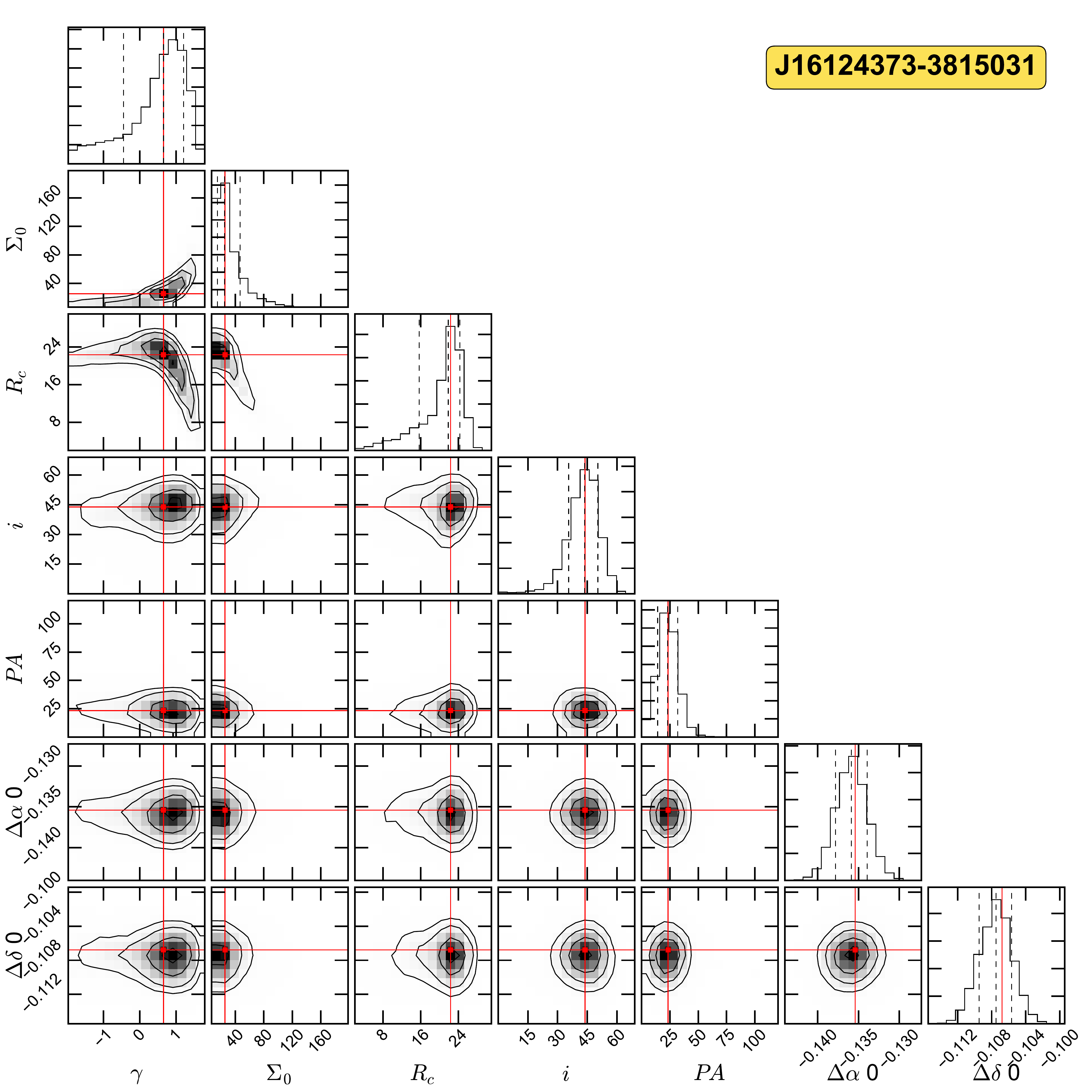}\includegraphics{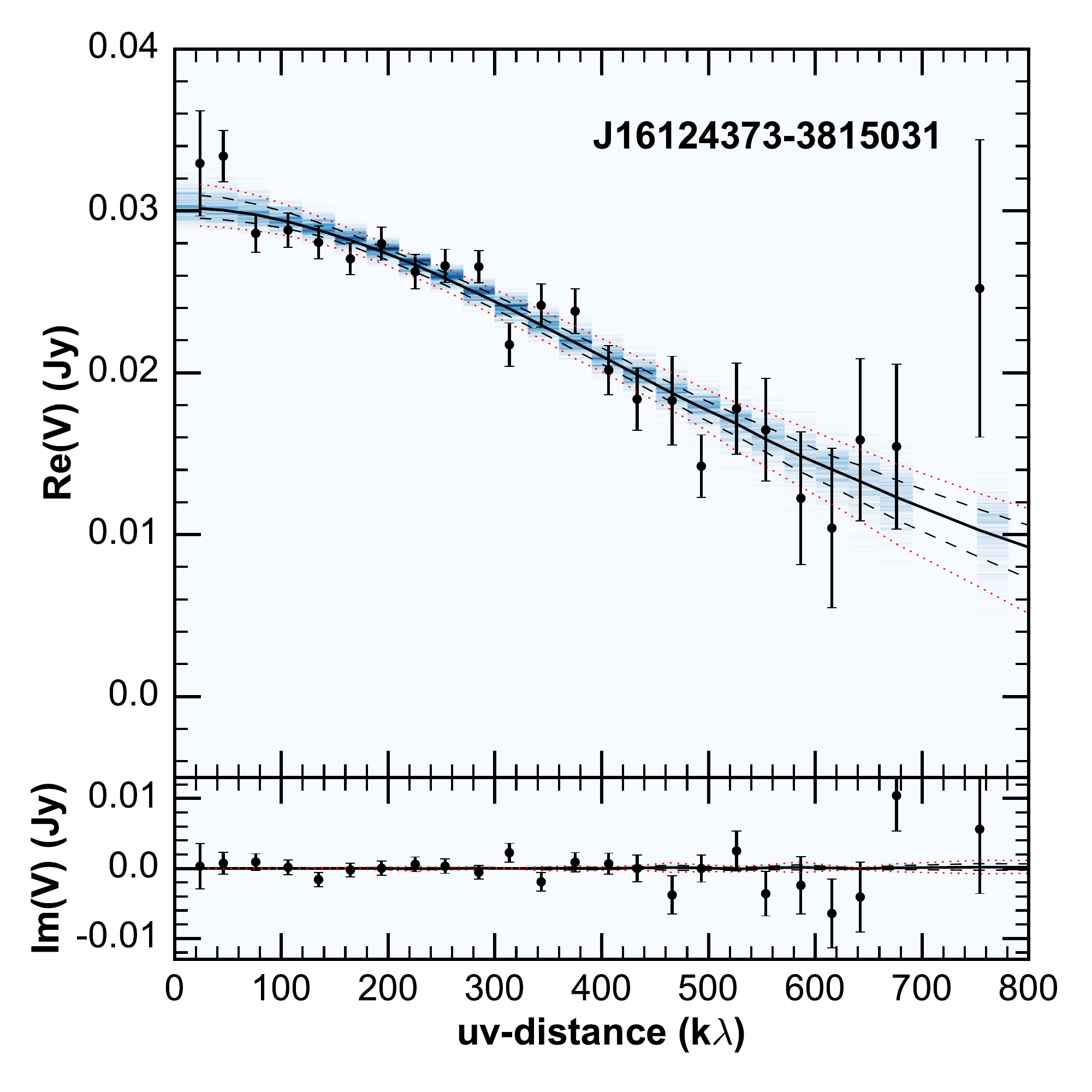}}\\[0.5cm]
\resizebox{\hsize}{!}{\includegraphics{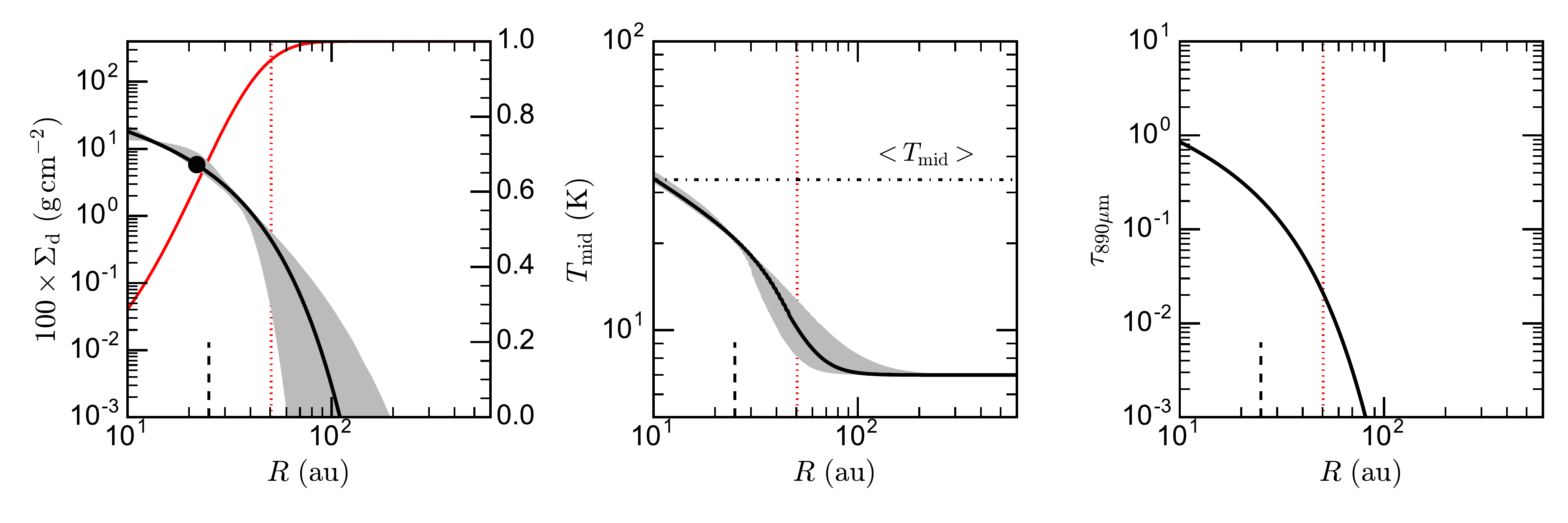}}\\[0.5cm]
\resizebox{0.8\hsize}{!}{\includegraphics{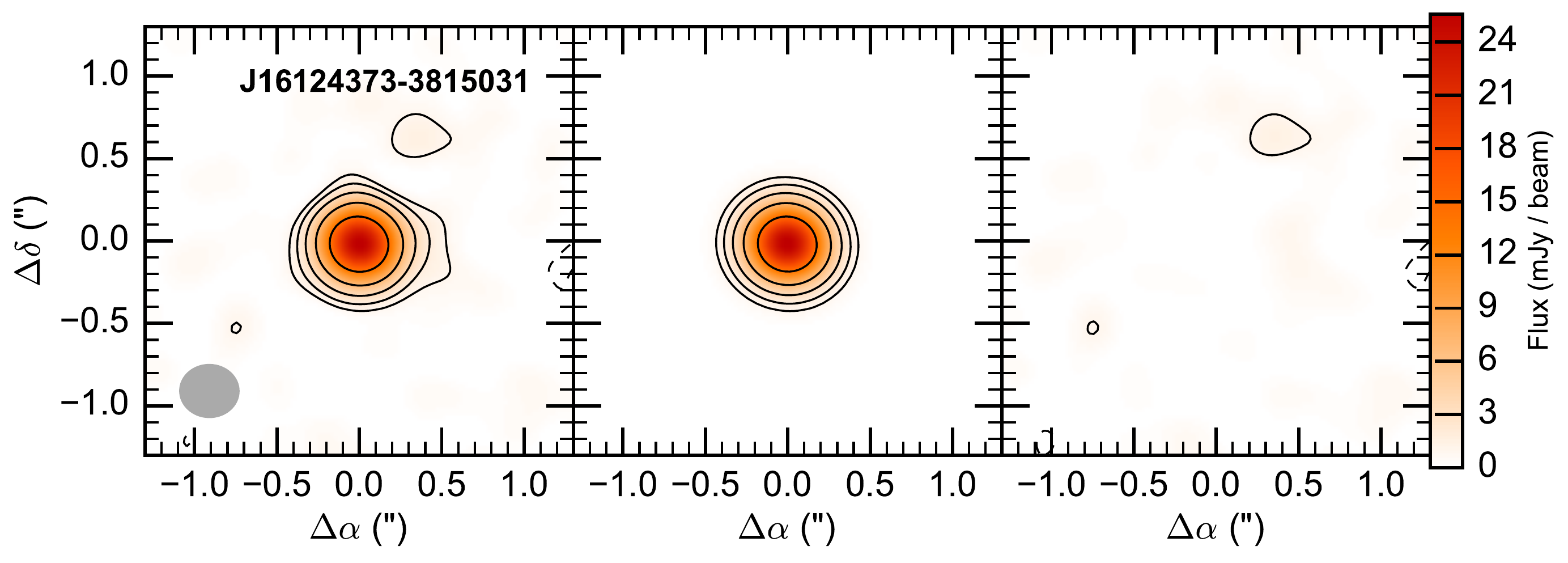}}\\[0.5cm]
\caption{Fit results for J16124373-3815031, presented as in \figref{fig:reference.fit.results}. In the images $\sigma=0.3\u{mJy/beam}$.}
\end{figure*}

\pagebreak
\section{Results of Bayesian linear regressions}
\label{app:kelly}
\begin{figure*}
\centering
\resizebox{0.49\hsize}{!}{\includegraphics{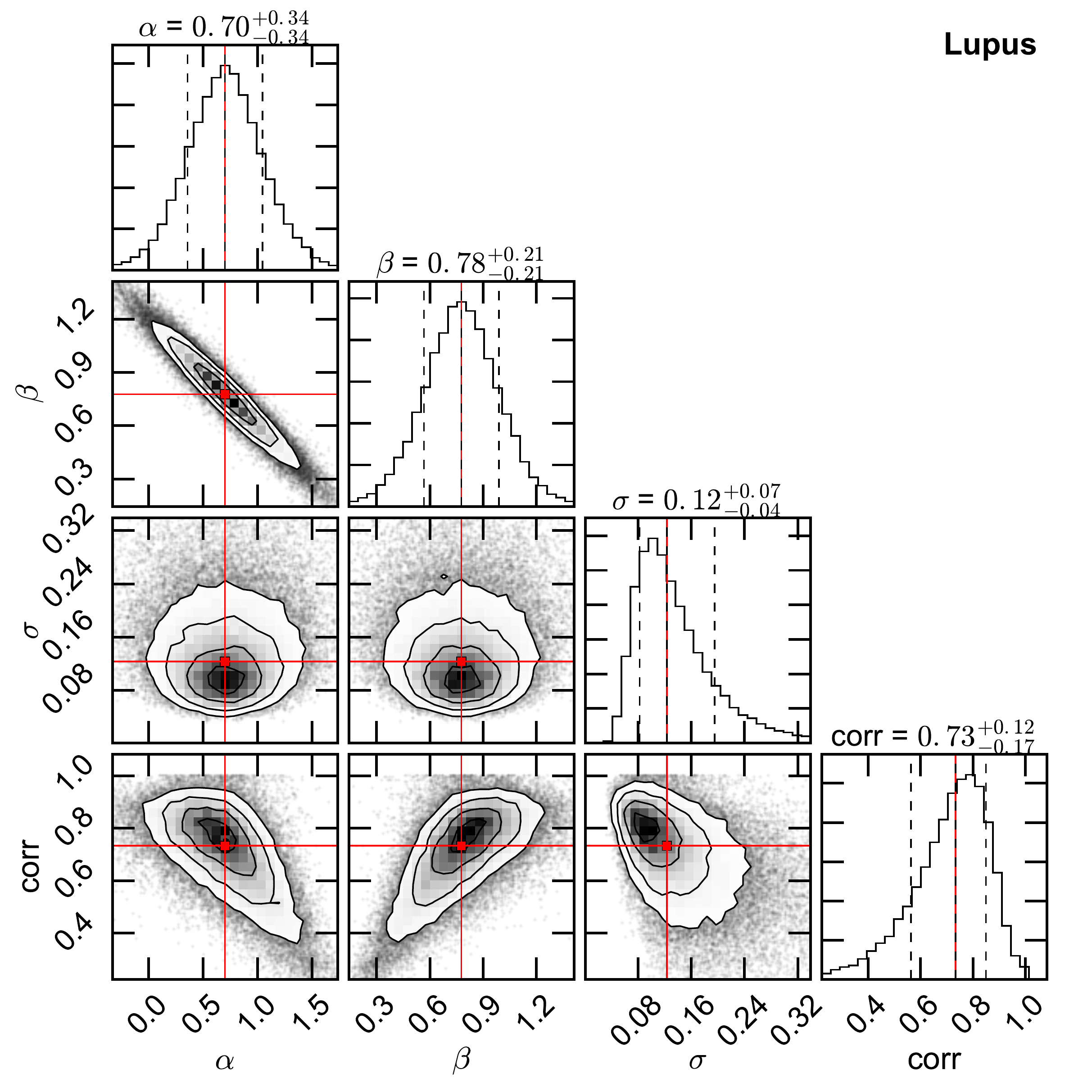}}
\resizebox{0.49\hsize}{!}{\includegraphics{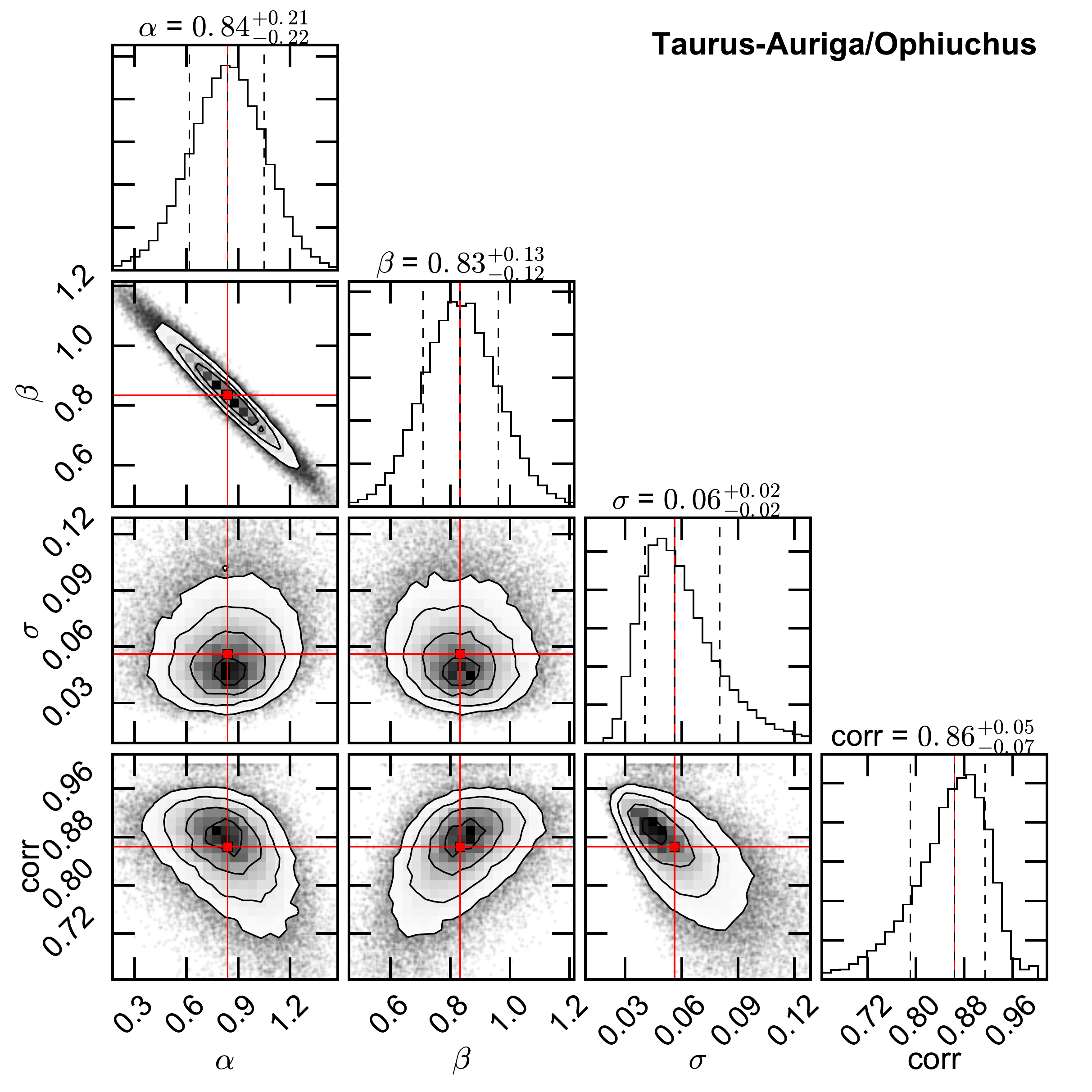}}\\
\resizebox{0.49\hsize}{!}{\includegraphics{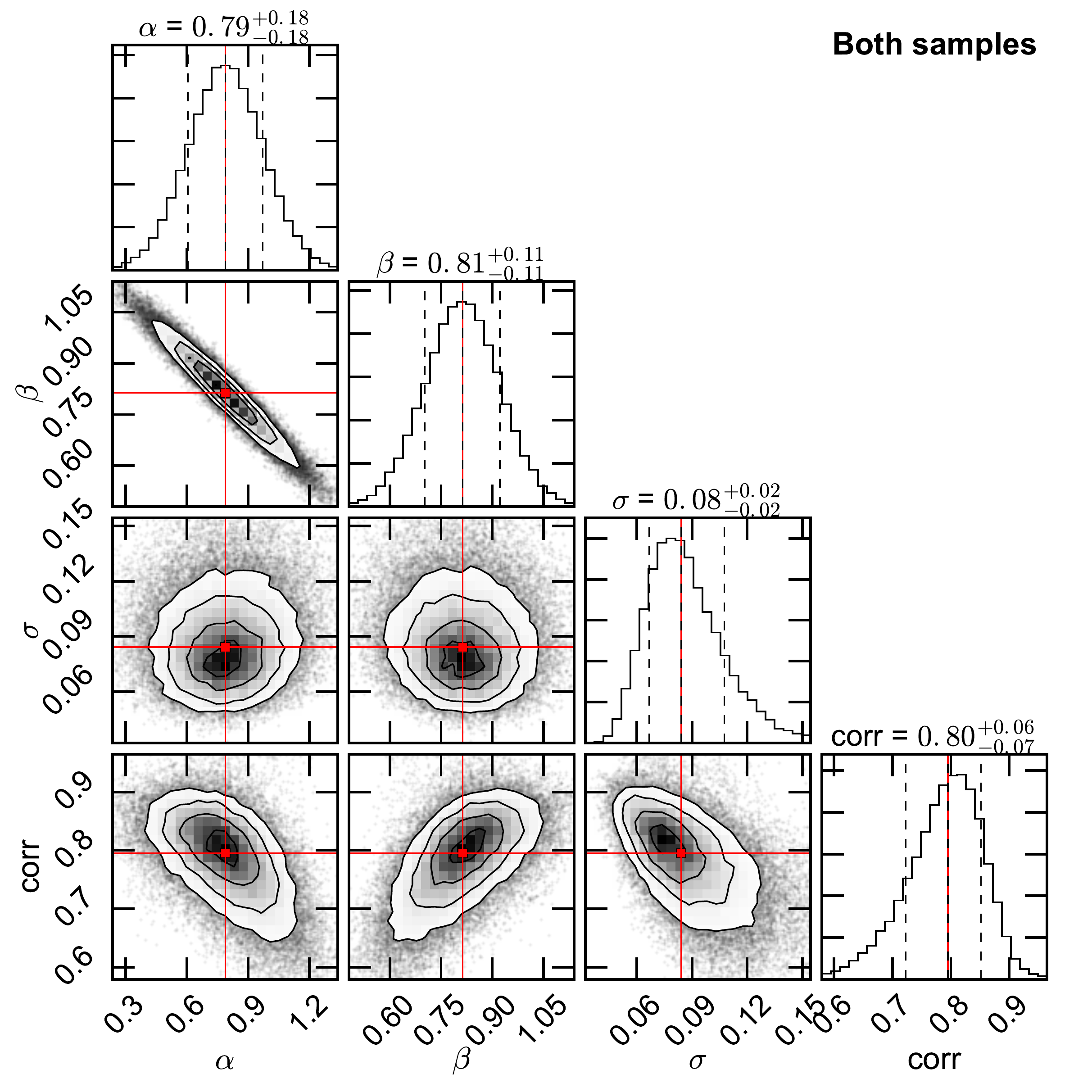}}
\caption{Staircase plots showing the posterior PDF resulting from the application of the Bayesian regression method by \citet{Kelly:2007lq} to infer a law $\log(F_{890\,\mu\mathrm{m}})=\beta \log(R_c) + \alpha$ for the Lupus sample \textit{(top left)}, for the Taurus-Auriga/Ophiuchus sample \textit{(top right)} and for both the samples together \textit{(bottom)}. Each plot shows the PDF of the intercept ($\alpha$), the slope ($\beta$), the intrinsic scatter of data points ($\sigma$) and the correlation degree (corr). The red lines highlight the posterior peak. }
\label{fig:app.kelly}
\end{figure*}

\end{appendix}

\end{document}